\documentclass[a4paper,fleqn,usenatbib]{mnras}
\usepackage{pdflscape}

\usepackage[percent]{overpic}
\usepackage{environ}
\usepackage{multirow,array,booktabs}
\usepackage{lmodern}
\usepackage[export]{adjustbox}

\newdimen\fontdim
\newdimen\upperfontdim
\newdimen\lowerfontdim
\newif\ifmoreiterations
\makeatletter
\NewEnviron{fitbox}[2]{
  \def\buildbox{%
    \setbox0\vbox{\hbox{\minipage{#1}%
      \fontsize{\fontdim}{1.2\fontdim}%
      \selectfont%
      \stuff%
    \endminipage}}%
    \dimen@\ht0
    \advance\dimen@\dp0
  }
  \def\stuff{\BODY}
  \buildbox
  \ifdim\dimen@>#2
    \loop
      \fontdim.5\fontdim 
      \buildbox
    \ifdim\dimen@>#2 \repeat
    \lowerfontdim\fontdim
    \upperfontdim2\fontdim
    \fontdim1.5\fontdim
  \else
    \loop
      \fontdim2\fontdim 
      \buildbox
    \ifdim\dimen@<#2 \repeat
    \upperfontdim\fontdim
    \lowerfontdim.5\fontdim
    \fontdim.75\fontdim
  \fi
  \loop
    \buildbox
    \ifdim\dimen@>#2
      \moreiterationstrue
      \upperfontdim\fontdim
      \advance\fontdim\lowerfontdim
      \fontdim.5\fontdim
    \else
      \advance\dimen@-#2
      \ifdim\dimen@<10pt
        \lowerfontdim\fontdim
        \advance\fontdim\upperfontdim
        \fontdim.5\fontdim
        \dimen@\upperfontdim
        \advance\dimen@-\lowerfontdim
        \ifdim\dimen@<.2pt
          \moreiterationsfalse
        \else
          \moreiterationstrue
        \fi
      \else
        \moreiterationsfalse
      \fi
    \fi
  \ifmoreiterations \repeat
  \box0
}
\makeatother

\usepackage[T1]{fontenc}
\usepackage{ae,aecompl}

\usepackage{graphicx}	
\usepackage{amsmath}	
\usepackage{amssymb}	
\usepackage{enumitem}
\usepackage [authoryear]{natbib}
\usepackage[outdir=./]{epstopdf}
\usepackage{color}

\title[VALES I: The molecular gas content in galaxies up to $z=0.35$]
      {VALES I: The molecular gas content in star-forming dusty ${\it H}$-ATLAS galaxies up to ${\mathbf{z=0.35}}$}

\author[V.~Villanueva et al.]
{\parbox{\textwidth}{\raggedright V.~Villanueva,$^{1}$
E.~Ibar,$^{1}$
T.~M.~Hughes,$^{1}$
M.~A.~Lara-L{\'o}pez,$^{2,3}$
L.~Dunne,$^{4,5}$
S.~Eales,$^{5}$
R.~J.~Ivison,$^{6,4}$
M.~Aravena,$^{7}$
M.~Baes,$^{8}$
N.~Bourne,$^{4}$
P.~Cassata,$^{1}$
A.~Cooray,$^{9,10}$
H.~Dannerbauer,$^{11,12}$
L.~J.~M.~Davies,$^{13}$
S.~P.~Driver,$^{13}$
S.~Dye,$^{14}$
C.~Furlanetto,$^{14,15}$
R.~Herrera-Camus,$^{16}$
S.~J.~Maddox,$^{4,5}$
M.~J.~Micha{\l}owski,$^{4}$
J.~Molina,$^{17}$
D.~Riechers,$^{18}$
A.~E.~Sansom,$^{19}$
M.~W.~L.~Smith,$^{5}$
G.~Rodighiero,$^{20}$
E.~Valiante$^{5}$ and
P.~van der Werf$^{21}$}\vspace{0.4cm}\\
\parbox{\textwidth}{\raggedright $^{1}$Instituto de F\'isica y Astronom\'ia, Universidad de Valpara\'iso, Avda.\ Gran Breta\~na 1111, Valpara\'iso, Chile\\
$^{2}$Instituto de Astronom\'ia, Universidad Nacional Aut\'onoma de M\'exico, A.P. 70-264, 04510 M\'exico, D.F., M\'exico\\
$^{3}$Australian Astronomical Observatory, PO Box 915, North Ryde, NSW 1670, Australia\\
$^{4}$Institute for Astronomy, University of Edinburgh, Royal Observatory, Blackford Hill, Edinburgh EH9 3HJ, UK\\
$^{5}$School of Physics and Astronomy, Cardiff University, Queens Buildings, The Parade, Cardiff CF24 3AA, UK\\
$^{6}$European Southern Observatory, Karl-Schwarzschild-Stra{\ss}e 2, 85748 Garching bei M\"unchen, Germany\\
$^{7}$N\'ucleo de Astronom\'ia, Facultad de Ingenier\'ia, Universidad Diego Portales, Av. Ej\'ercito 441, Santiago, Chile\\
$^{8}$Sterrenkundig Observatorium, Universiteit Gent, Krijgslaan 281 S9, B-9000 Gent, Belgium\\
$^{9}$Dept. of Physics \& Astronomy, University of California, Irvine, CA 92697, USA\\
$^{10}$California Institute of Technology, 1200 E. California Blvd., Pasadena, CA 91125, USA\\
$^{11}$Universidad de La Laguna, Dpto. Astrof\'isica, E-38206 La Laguna, Tenerife, Spain\\
$^{12}$Instituto de Astrof\'isica de Canarias (IAC), E-38205 La Laguna, Tenerife, Spain \\
$^{13}$International Centre for Radio Astronomy Research (ICRAR), The University of Western Australia, M468, 35 Stirling Highway, Crawley, Australia, WA 6009\\
$^{14}$School of Physics and Astronomy, University of Nottingham, NG7 2RD, UK\\
$^{15}$CAPES Foundation, Ministry of Education of Brazil, Brasilia/DF, 70040-020, Brazil\\
$^{16}$Max-Planck-Institut f\"ur Extraterrestrische Physik (MPE), Postfach 1312, 85741, Garching, Germany\\
$^{17}$Departamento de Astronom\'ia, Universidad de Chile, Casilla 36-D, Santiago, Chile\\
$^{18}$Department of Astronomy, Cornell University, 220 Space Sciences Building, Ithaca, NY 14853, USA\\
$^{19}$Jeremiah Horrocks Institute, University of Central Lancashire, Preston PR1 2HE, UK\\
$^{20}$Dipartimento di Astronomia, Universit\`{a} di Padova, vicolo Osservatorio, 3, 35122 Padova, Italy\\
$^{21}$Leiden Observatory, Leiden University, P.O.\ Box 9513, NL-2300 RA Leiden, The Netherlands}}

\date{Accepted 2017 May 26. Received 2017 May 25; in original form 2017 February 13.}

\pubyear{2017}

\begin{document}
\label{firstpage}
\pagerange{\pageref{firstpage}--\pageref{lastpage}}
\maketitle

\begin{abstract}

{
We present an extragalactic survey using observations from the Atacama
Large Millimeter/submillimeter Array (ALMA) to characterise galaxy
populations up to $z = 0.35$: the Valparaíso ALMA Line Emission Survey
(VALES). We use ALMA Band-3 CO(1--0) observations to study the
molecular gas content in a sample of 67 dusty normal star-forming
galaxies selected from the {\it Herschel} Astrophysical Terahertz Large Area
Survey ({\it H}-ATLAS). We have spectrally detected 49 galaxies at
$>5\sigma$ significance and 12 others are seen at low significance in
stacked spectra. CO luminosities are in the range of
$(0.03-1.31)\times10^{10}$\,K\,km\,s$^{-1}$\,pc$^2$, equivalent to
$\log({\rm M_{gas}/M_{\odot}}) =8.9 - 10.9$ assuming an $\alpha_{\rm
CO}$\,=\,4.6\,(K\,km\,s$^{-1}$\,pc$^{2}$)$^{-1}$, which perfectly
complements the parameter space previously explored with 
local and high-z normal galaxies. We compute the optical to CO size 
ratio for 21 galaxies resolved by ALMA at $\sim3\farcs5$ resolution (6.5 kpc), 
finding that the molecular gas is on average $\sim$\,0.6 times more compact than the 
stellar component. We obtain a global Schmidt-Kennicutt relation, given by
$\log [\Sigma_{\rm SFR}/({\rm M_{\odot} yr^{-1}kpc^{-2}})] =(1.26 \pm
0.02) \times \, \log [\Sigma_{\rm M_{H2}}/({\rm M_{\odot}\,pc^{-2}})]
- (3.6 \pm 0.2)$. We find a significant fraction of galaxies lying at 
`intermediate efficiencies' between a long-standing mode of 
star-formation activity and a starburst, specially at $\rm L_{IR}=10^{11-12} L_{\odot}$. 
Combining our observations with data taken from the literature, 
we propose that star formation efficiencies can be parameterised by $\log
\,[{\rm SFR/M_{H2}}] = 0.19 \times \,{\rm (\log \,{L_{IR}} -
11.45)}-8.26- 0.41 \times \arctan [-4.84\,(\log {\rm L_{IR}}-11.45) ]$.
Within the redshift range we explore ($z<0.35$), we identify a rapid increase of the gas content as a function of redshift.}
\end{abstract}

\clearpage



\section{Introduction}
\label{Introduction}

Understanding the way in which galaxies form and evolve throughout
cosmic time is one of the major challenges of extragalactic
astrophysics. Recently, theoretical models adopting a $\Lambda$CDM
cosmology have been successful in probing the hierarchical
gravitational growth of dark matter haloes, which is then associated
to the large-scale structure of the observed baryonic matter
\citep[e.g. ][]{Spergel2003,Spergel2007}. On smaller scales, however,
the physical processes that control galaxy growth have intricate
non-linear dependencies that make its explanation far from trivial
(e.g.\, \citealt{Vogelsberger2014, Schaye2015, Crain2015}). One of the
key observations used to constrain galaxy formation and evolution
models is the behaviour of the cosmic star-formation rate
density. Understanding the cosmic evolution of the interplay between
the observed star formation rate (SFR), molecular gas content ($M_{\rm
  gas}$), global stellar mass content ($M_\star$) and gas-phase
metallicity ($Z$) is a major goal in this field of research. We
therefore require a detailed knowledge of the origin and the
properties of the gas reservoir that ignites and sustains the star
formation activity in galaxies at different epochs.

\begin{figure*}
\begin{center}
\includegraphics[width=14.5cm]{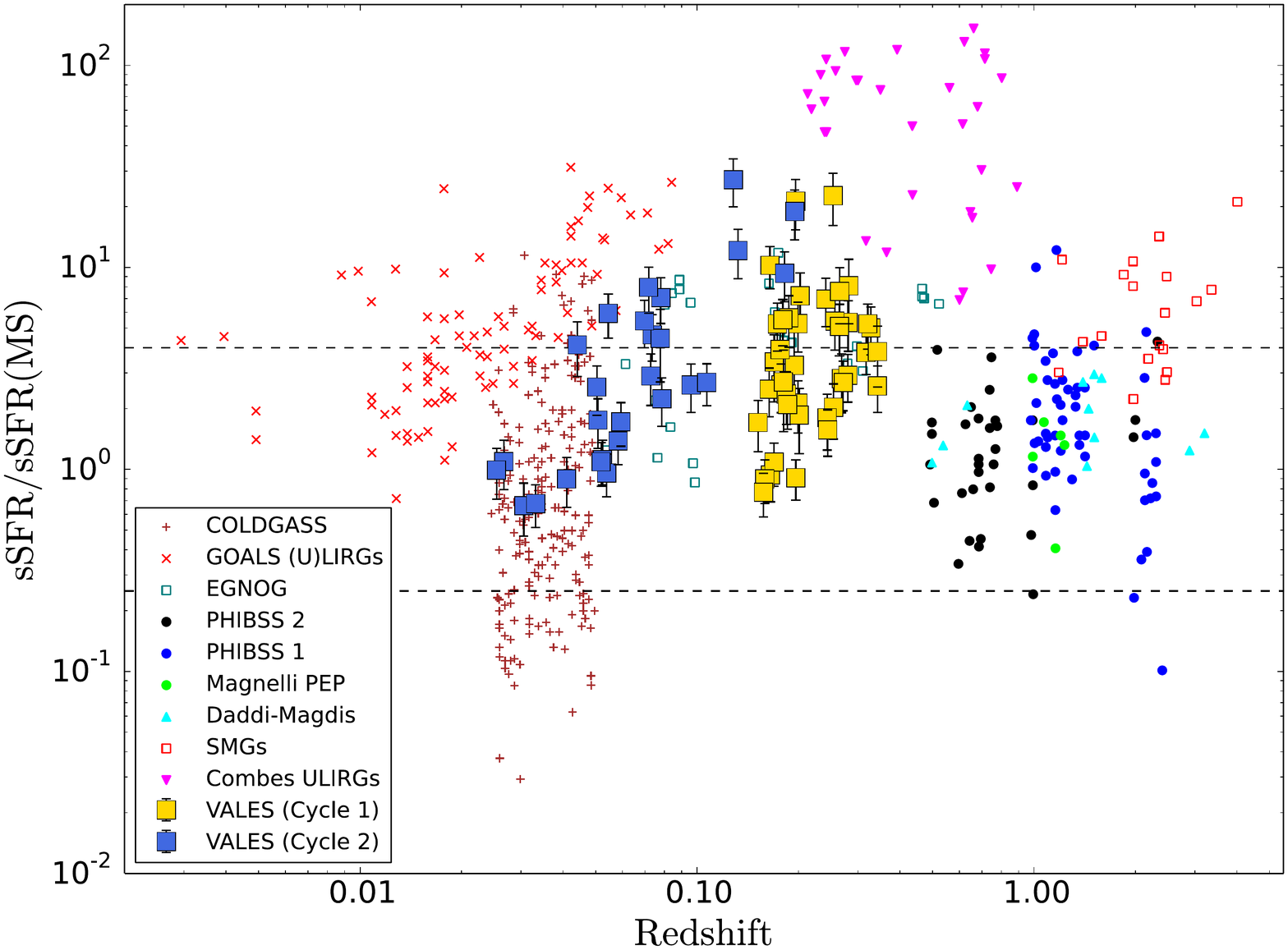}
\caption[Sample Definition]{\small{The figure shows the specific Star
    Formation Rate (sSFR, defined as $\rm sSFR = SFR / M_{\star}$),
    normalised to the one estimated for the `main sequence'
    \citep{Elbaz2011,Whitaker2012} as a function of redshift for
    different samples of galaxies detected in CO. We use
      the parametrisation of the `main sequence' made by
      \cite{Genzel2015} as $\log[{\rm sSFR}({\rm MS},z,{\rm
          M_{\star}})]=-1.12+1.14z-0.19z^{2}-(0.3+0.13z) \times({\rm
        \log M_{\star}}-10.5)$ Gyr$^-1$, where dashed black lines
      denote 0.6\,dex off this equation for star-forming galaxies. Our
    data is presented in filled squares with error bars taken from our
    ALMA Cycle-1 (yellow) and Cycle-2 (royal blue) campaigns. We
    estimate the SFR using $L_{\rm IR}(8-1000\mu{\rm m})$ extracted
    from the $H$-ATLAS data, stellar masses using MAGPHYS fits (see
    \S~\ref{SED_fitting}) -- both using the same IMF, and redshifts
    taken from the GAMA survey.  Dark red crosses are nearby galaxies
    \citep{Saintonge2011a}, red crosses are (U)LIRGs
    \citep{Howell2010}, light blue unfilled squares are $z=0.05 - 0.5$
    normal galaxies \citep{Bauermeister2013}, pink inverted triangles
    are ULIRGs at intermediate redshifts
    \citep{Combes2011,Combes2013}, blue dots are `main sequence'
    galaxies at $z=1 - 1.5$ and $2 - 2.5$
    \citep{Tacconi2010,Tacconi2013}, black dots are `main sequence'
    galaxies at $z=0.5 - 1$ and $z \sim 2$ \citep{Combes2015}, light
    blue triangles are `main sequence' galaxies at high-$z$
    \citep{Magnelli2012a}, light green dots are `main sequence'
    galaxies at $z = 0.5 - 3.2$ \citep{Daddi2010a,Magdis2012a} and red
    unfilled squares are sub-milimetre galaxies at $z=1.2 - 3.4$
    \citep{Greve2005,Tacconi2006,Tacconi2008,Bothwell2013}.  Figure
    adapted from \cite{Genzel2015}.  }}\label{SampleDefinition}
\end{center}
\end{figure*}

The accretion of gas into the potential wells of galaxies, either from
the inter-galactic medium or via galaxy-galaxy interactions, provides
the gas reservoir for ongoing and future star formation
\citep{DiMatteo2007,Bournaud2009,Dekel2009}.  Most stars form in giant
molecular clouds (GMCs), in which the majority of the mass is in the
form of molecular hydrogen (H$_2$). The lack of a permanent dipole
moment in this molecule means that direct measurements of cold H$_2$
gas are extremely difficult
\citep[e.g.][]{Papadopoulos&Seaquist1999,Bothwell2013}.  Thus, an
alternative approach to study the molecular gas content is through
observations of carbon monoxide (CO) line emission of low-J
transitions (e.g.\ $J=2-1$ or $J=1-0$) -- the best standard tracer of
the total mass in molecular gas \citep[M$_{\rm H_2}=\alpha_{\rm
    CO}\,L'_{\rm CO(1-0)}$; e.g.][]{Bolatto2013}. Even though this
tracer has been historically used as a tracer of the molecular gas
mass, the $^{12}$C$^{16}$O($J=1-0$) [hereafter CO(1--0)] emission line
is optically thick, hence the dynamics of the system becomes critical
for converting luminosities into masses
\citep{Solomon&VandenBout2005}.  For instance, in the case of a merger
where dynamical instabilities are large and the system is not virialised,
Doppler-broadening could affect the line profiles and the emitting
regions could be more dispersed throughout the inter stellar medium
(ISM), thus enhancing the CO emission compared to that from a
virialised system of the same mass \citep[][]{Downes&Solomon1998}.  In
dense, optically-thick virialised GMCs, it is found that $\alpha_{\rm
  CO}\sim5$\,M$_{\odot}$\,(K\,km\,s$^{-1}$\,pc$^2$)$^{-1}$, whereas
$\alpha_{\rm
  CO}\sim0.8\,$M$_{\odot}$\,(K\,km\,s$^{-1}$\,pc$^{2}$)$^{-1}$ in more
dynamically disrupted systems, such as in Ultra Luminous Infrared
Galaxies (ULIRGs; \citealt{Downes1998}). On the other hand,
$\alpha_{\rm CO}$ may be boosted in low-metallicity environments due
to a lack of shielding dust that enhances photo-dissociation of the CO
molecule \citep{Wolfire2010,Narayanan2012}. For instance,
\cite{Narayanan2012} find a parametrisation of $\alpha_{\rm CO}$ in
terms of gas metallicity, where $\alpha_{\rm CO}\propto Z^{-0.65}$
(mixing both low and high-$z$ galaxies), similar to that found by
\cite{Feldmann2012}. A higher redshifts, a flatter slope has been suggested (\citealt{Genzel2012}).

Recent observations taken with the \textit{Herschel Space
  Observatory}\footnote{\textit{Herschel} is an ESA space observatory
  with science instruments provided by European-led Principal
  Investigator consortia with an important participation from NASA.}
\citep{Pilbratt2010} of local star-forming galaxies suggest the
existence of at least two different mechanisms triggering the star
formation. Taking into account the $\rm L_{\rm FIR}/ M_{H2}$ ratio
(where $\rm L_{\rm FIR}$ is the far-IR luminosity) as a tracer of the star-formation efficiency, 
\cite{GraciaCarpio2011} find an unusual point at ${\rm \sim
  80\,L_{\odot}\,M^{-1}_{\odot}}$ at which average properties of
the neutral and ionised gas change significantly, this observation 
is broadly consistent with a scenario of a highly compressed and more
efficient mode of star formation that creates higher ionisation
parameters that cause the gas to manifest in low line to continuum
ratios. This value is similar to the one at which \cite{Genzel2010}
and \cite{Daddi2010} claim a transition to a more efficient
star-formation mode, above the so-called `main-sequence' for
star-forming galaxies (e.g.\ \citealt{Elbaz2011}). The different
mechanisms controlling the star-formation activity are thought to be
the product of dynamical instabilities, where higher efficiencies are
seen in more compact and dynamically disrupted systems, such as in Ultra Luminous. 
Over the last few years, significant efforts have been made to
characterise the star formation activity of normal and starburst
galaxies at low-z (e.g.\ \citealt{Saintonge2011a, Howell2010,
  Bauermeister2013,Bothwell2014}). The construction of large samples
of galaxies with direct molecular gas detections (via CO emission) has
remained a challenge. Beyond the local Universe, CO detections are
limited to the most massive/luminous yet rare galaxies.  For example,
\cite{Braun2011} report detections of the CO($J=1-0$) transition for
11 ULIRGs with an average redshift of $z=0.38$. For these ULIRGs, the
molecular gas mass as a function of look-back time demonstrates a
dramatic rise by almost an order of magnitude from the current epoch
out to 5\,Gyr ago. In addition, \cite{Combes2011} presented 18
detected ULIRGs at $z\sim0.2-0.6$ for CO(1--0), CO(2--1) and CO(3--2)
with an average CO luminosity of $\rm {L}'_{\rm CO(1-0)}=2 \times
10^{10}$\,K\,km\,s$^{-1}{\rm pc}^{2}$, finding that the amount of gas
available for a galaxy quickly increases as a function of
redshift. Moreover, \cite{Magdis2014} presented the properties of 17
{\it Herschel}-selected ULIRGs ($L_{\rm IR} > 10^{11.5} L_{\odot}$) at
$z=0.2-0.8$, showing that the previously observed evolution of ULIRGs at
those redshifts is already taking place by $z\sim0.3$. Nevertheless,
the observation of `normal' galaxies at these redshifts (and beyond)
has so far been, at least, restricted.

The advent of the Atacama Large Millimeter/submillimeter Array (ALMA)
opens up the possibility to explore the still unrevealed nature of the
`normal' star forming galaxies (SFGs) at low/high-z redshift. In this
work, we exploit the {\it Herschel} Astrophysical Terahertz Large Area
Survey ({\it H}-ATLAS\footnote{\url{http://www.h-atlas.org/}};
\citealt{Eales2010}) and the state-of-the-art capabilities of ALMA to
characterise the CO($1-0$) line emission ($\nu_{\rm rest}=115.271$
GHz) of `normal' star-forming and mildly starburst galaxies up to
$z=0.35$.  This paper is organised as
follows. Section~\ref{S2_Observations} explains the sample selection,
observing strategy and data reduction. In Section~\ref{S3_Results}, we
present the main results and the implications of these new ALMA
observations to the global context of galaxy evolution. Our conclusion
is summarised in Section~\ref{S4_Conclusion}. Throughout this work, we
assume a $\Lambda$CDM cosmology adopting the values $H_{\rm
  0}=70.0$\,km\,s$^{-1}$\, Mpc$^{-1}$, $\Omega_{\rm M} = 0.3$ and
$\Omega_\Lambda=0.7$ for the calculation of luminosity distances and
physical scales\footnote{We use Ned Wright's online calculator
  \url{http://www.astro.ucla.edu/\~wright/CosmoCalc.html.}}.

\begin{table*}
\begin{center}
\resizebox{\linewidth}{!}{ 

\begin{tabular}{  >{\bfseries}c*7{c}}
\hline
\hline
Project ID & Target Names & Observation Date & Flux & Bandpass & Phase & PWV & Number of \\
 & & & Calibrator & Calibrator & Calibrator & [mm] & Antennas \\

\hline
 \multirow{15}{*}{2012.1.01080.S} &  \multicolumn{1}{|c|}{HATLASJ090633.6+001526, HATLASJ090223.9$-$001639, HATLASJ091157.2+014453, } &  \multirow{4}{*}{January 1$^{\rm st}$, 2014} & \multirow{4}{*}{Calisto} & \multirow{4}{*}{J0522--3627 } &\multirow{4}{*}{J0811+0146} & \multirow{4}{*}{4.217} & \multirow{4}{*}{27} \\ 
 & \multicolumn{1}{|c |}{HATLASJ091420.0+000509, HATLASJ090120.7+020223 , HATLASJ085616.0+005237, }   \\
 &  \multicolumn{1}{|c |}{HATLASJ085957.9+015632, HATLASJ085750.1+012807, HATLASJ091956.9+013852, }  \\
 &  \multicolumn{1}{|c |}{HATLASJ092232.2+002708, HATLASJ085623.6+001352, HATLASJ085828.4+012211 } \\ \cline{2-8}
 
 &  \multicolumn{1}{|c|}{HATLASJ113858.4$-$001630, HATLASJ121206.2$-$013425,  HATLASJ114343.9+000203, } & \multicolumn{1}{c|}{} &  \multirow{5}{*}{Mars} & \multirow{5}{*}{J1229+0203} & \multirow{5}{*}{J1229+0203} & \multicolumn{1}{|c}{} & \multicolumn{1}{|c}{}  \\ 
 & \multicolumn{1}{|c |}{HATLASJ121141.8$-$015730, HATLASJ114540.7+002553,  HATLASJ114625.0$-$014511, } & \multicolumn{1}{c|}{December 27$^{\rm th}$, 2013} & & & &\multicolumn{1}{|c}{6.084} & \multicolumn{1}{|c}{}\\ 
 &  \multicolumn{1}{|c |}{HATLASJ121427.3+005819, HATLASJ113740.6$-$010454,  HATLASJ121908.7$-$010159, } & \multicolumn{1}{c|}{------------------------------------------------}  & & & &\multicolumn{1}{|c}{-----------} & \multicolumn{1}{|c}{26}  \\
 &  \multicolumn{1}{|c |}{HATLASJ114702.7+001207, HATLASJ115141.3$-$004240, HATLASJ121446.4$-$011155, } & \multicolumn{1}{c|}{December 29$^{\rm th}$, 2013}  & & & &\multicolumn{1}{|c}{5.757} & \multicolumn{1}{|c}{} \\
 &  \multicolumn{1}{|c |}{HATLASJ121253.5$-$002203, HATLASJ115317.4$-$010123,  HATLASJ115039.5$-$010640 } & \multicolumn{1}{c|}{}  & & & &\multicolumn{1}{|c}{} & \multicolumn{1}{|c}{}  \\ \cline{2-8}

 &  \multicolumn{1}{|c|}{HATLASJ142517.1+010546, HATLASJ141008.0+005107, HATLASJ142057.9+015233, } &  \multirow{5}{*}{March 9$^{\rm th}$, 2014} &\multirow{5}{*}{Ceres, Titan}  & \multirow{5}{*}{J1337--1257} & \multirow{5}{*}{J1410+0203} & \multirow{5}{*}{2.223} & \multirow{5}{*}{25} \\ 
 & \multicolumn{1}{|c |}{HATLASJ144218.7+003615, HATLASJ141925.3$-$011129, HATLASJ141522.0+004413, }   \\
 &  \multicolumn{1}{|c |}{HATLASJ141908.5+011313, HATLASJ144515.0+003907, HATLASJ140649.0$-$005646, }  \\
 &  \multicolumn{1}{|c |}{HATLASJ142208.8+005428, HATLASJ140912.3$-$013454, HATLASJ144129.5$-$000901, } \\
 &  \multicolumn{1}{|c |}{HATLASJ144116.2+002723 } \\

\hline

 \multirow{11}{*}{2013.1.00530.S} &  \multicolumn{1}{|c|}{HATLASJ085356.4+001255, HATLASJ085828.6+003813, HATLASJ085340.7+013348, } &  \multicolumn{1}{c|}{}  & \multirow{5}{*}{J0750+125}  & \multirow{5}{*}{J0750+1231} & \multirow{5}{*}{J0909+0121} & \multirow{5}{*}{2.223} & \multirow{5}{*}{40}  \\ 
 & \multicolumn{1}{|c |}{HATLASJ090005.0+000446, HATLASJ085405.9+011130, HATLASJ085112.9+010342, } & \multicolumn{1}{c|}{}   \\
 &  \multicolumn{1}{|c |}{HATLASJ083745.1$-$005141, HATLASJ090949.6+014847, HATLASJ090532.6+020222, } & \multicolumn{1}{c|}{} &  \\
 &  \multicolumn{1}{|c|}{HATLASJ085346.4+001252, HATLASJ083601.5+002617, HATLASJ084428.4+020350, }  & \multicolumn{1}{c|}{} \\
 &  \multicolumn{1}{|c |}{ HATLASJ091205.8+002655}  & \multicolumn{1}{c|}{}  \\ \cline{2-2} \cline{4-8}  
 
  &  \multicolumn{1}{|c|}{HATLASJ084139.6+015346, HATLASJ084350.8+005534, HATLASJ084305.1+010855, } & \multicolumn{1}{c|}{January 24$^{\rm th}$, 2015}    & \multirow{4}{*}{J0750+125} & \multirow{4}{*}{J0739+0137} & \multirow{4}{*}{J0901--0037} & \multirow{4}{*}{5.463} & \multirow{4}{*}{40} \\ 
 & \multicolumn{1}{|c |}{HATLASJ085450.2+021208, HATLASJ083831.8+000044, HATLASJ085111.4+013006, }  & \multicolumn{1}{c|}{}  \\
 &  \multicolumn{1}{|c |}{HATLASJ084428.4+020659, HATLASJ084907.1$-$005138, HATLASJ085234.3+013419, }  & \multicolumn{1}{c|}{}   \\
&  \multicolumn{1}{|c |}{ HATLASJ085748.0+004641}  & \multicolumn{1}{c|}{}  \\ \cline{2-2} \cline{4-8} 
 
  &  \multicolumn{1}{|c|}{HATLASJ084217.9+021223, HATLASJ085836.0+013149, HATLASJ084630.9+005055, } & \multicolumn{1}{c|}{}   & \multirow{2}{*}{Ganymede} & \multirow{2}{*}{J0909+0121} & \multirow{2}{*}{J0901--0037} & \multirow{2}{*}{5.553} & \multirow{2}{*}{39} \\ 
 & \multicolumn{1}{|c |}{HATLASJ090750.0+010141 }  & \multicolumn{1}{c|}{}    \\
 \hline

\end{tabular}}
\caption[Table_1]{\small{The table shows the way in which our targets were observed during the Cycle 1 and Cycle 2 campaigns.}}\label{Table_1}
\end{center}
\end{table*}

\section{Observations}
\label{S2_Observations}

\subsection{{\it H}-ATLAS sample}
\label{HATLAS_sample}

The galaxies presented in this paper have been selected from the
equatorial fields of the {\it H}-ATLAS survey ($\sim160$\, deg$^{2}$;
\citealt{Valiante2016}) and observed during ALMA Cycle-1 and Cycle-2
(programs 2012.1.01080.S \& 2013.1.00530.S; P.I.\ E.~Ibar). All
galaxies have a $>3\sigma$ detection with both the Photoconductor
Array Camera and Spectrometer (PACS) at 160\,$\mu$m and the Spectral
and Photometric Imaging Receiver (SPIRE) at 250\,$\mu$m, i.e.\ they
are detected near the peak of the spectral energy distribution (SED)
of a normal and local star-forming galaxy. All galaxies have been
unambiguously identified in the Sloan Digital Sky Survey (SDSS;
\citealt{adelman-mccarthy2008}) presenting a significant probability
for association ({\sc reliability} $R>0.8$;
\citealt{Smith11,Bourne16}).  The optical counterparts to the {\it
  Herschel}-detected galaxies all have high-quality spectra from the
Galaxy and Mass Assembly survey
(GAMA\footnote{\url{http://www.gama-survey.org/}}; {\sc
  z\_qual\,$\ge$\,3}; \citealt{Liske2015,Driver2016}).

\begin{figure*}
  \centering
  \includegraphics[height=12.5cm]{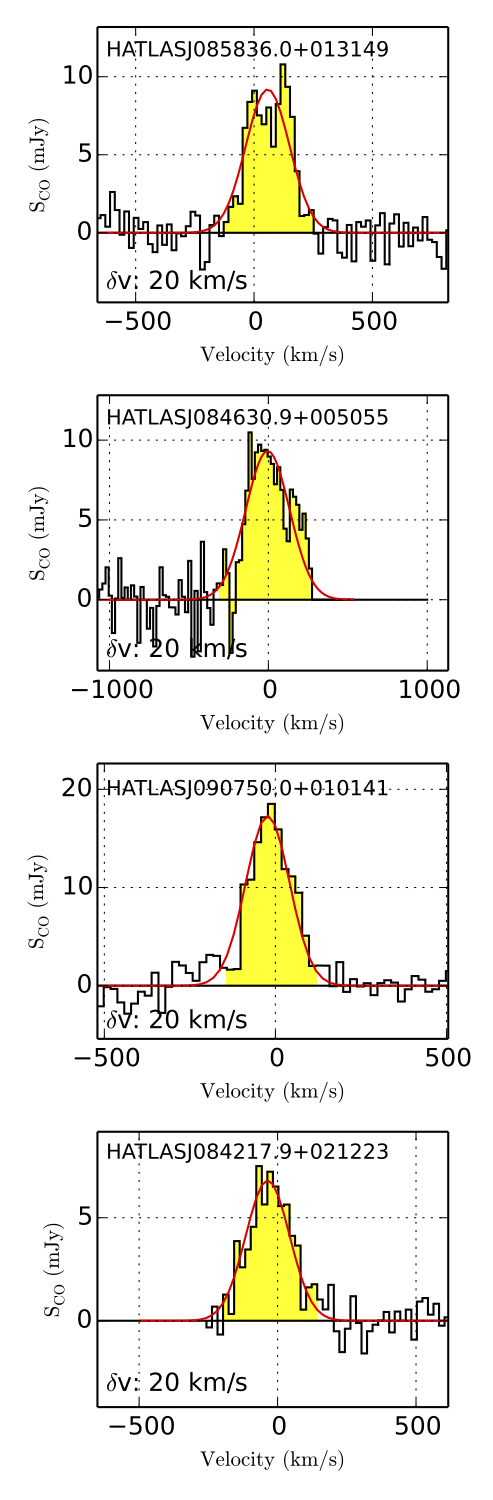} 
  \includegraphics[height=12.45cm]{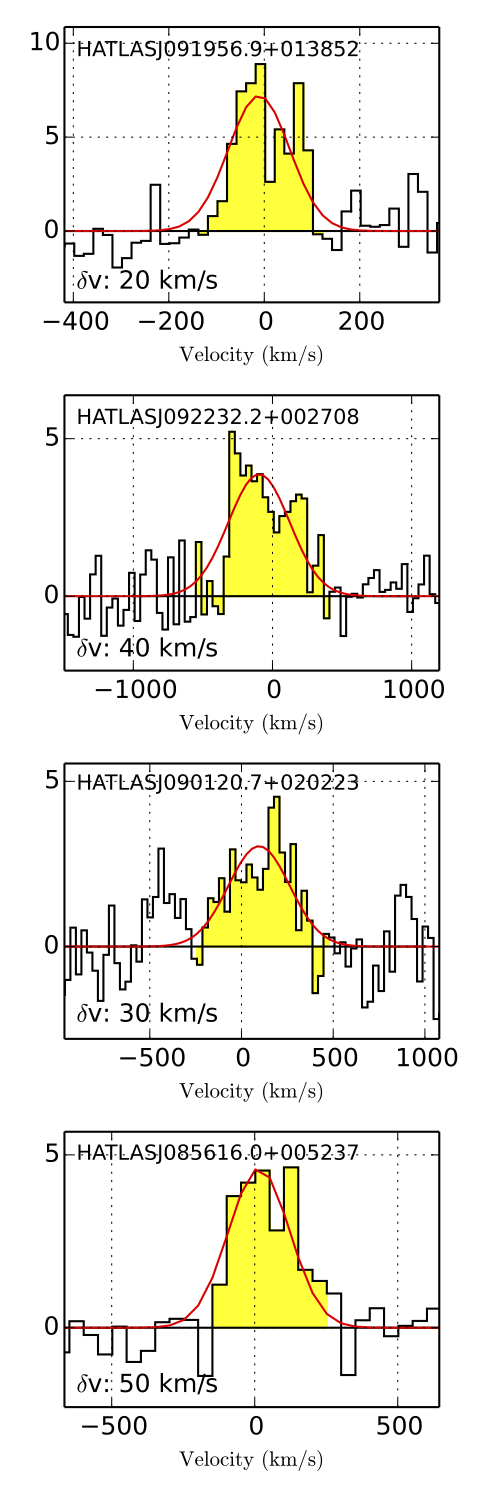} 
  \includegraphics[height=12.4cm]{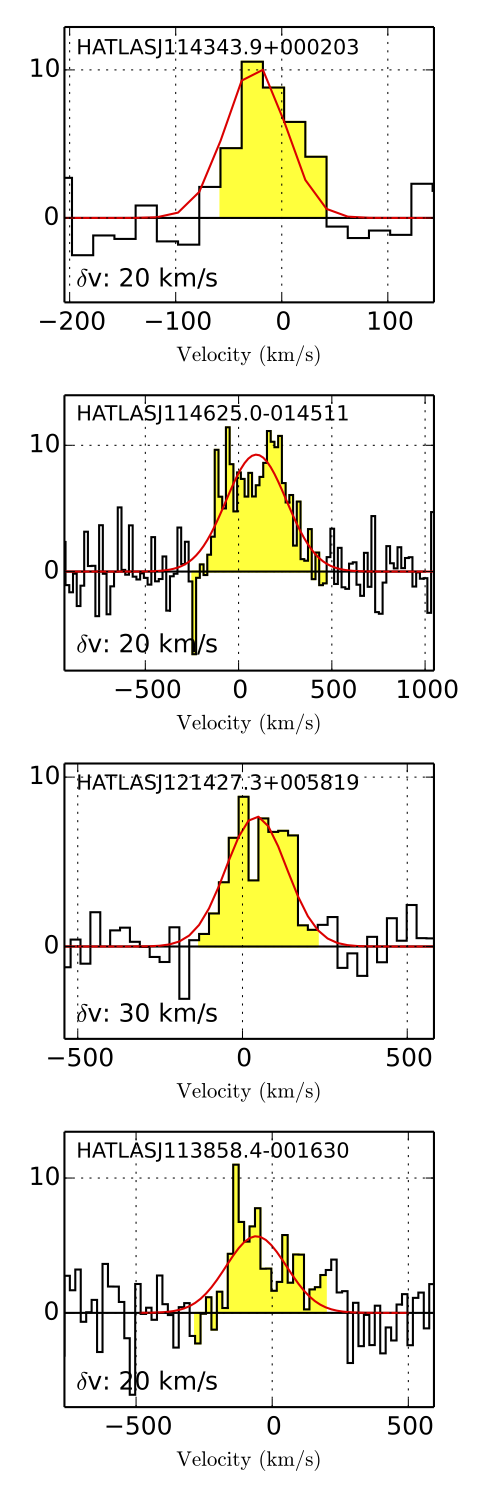} 
  \includegraphics[height=12.37cm]{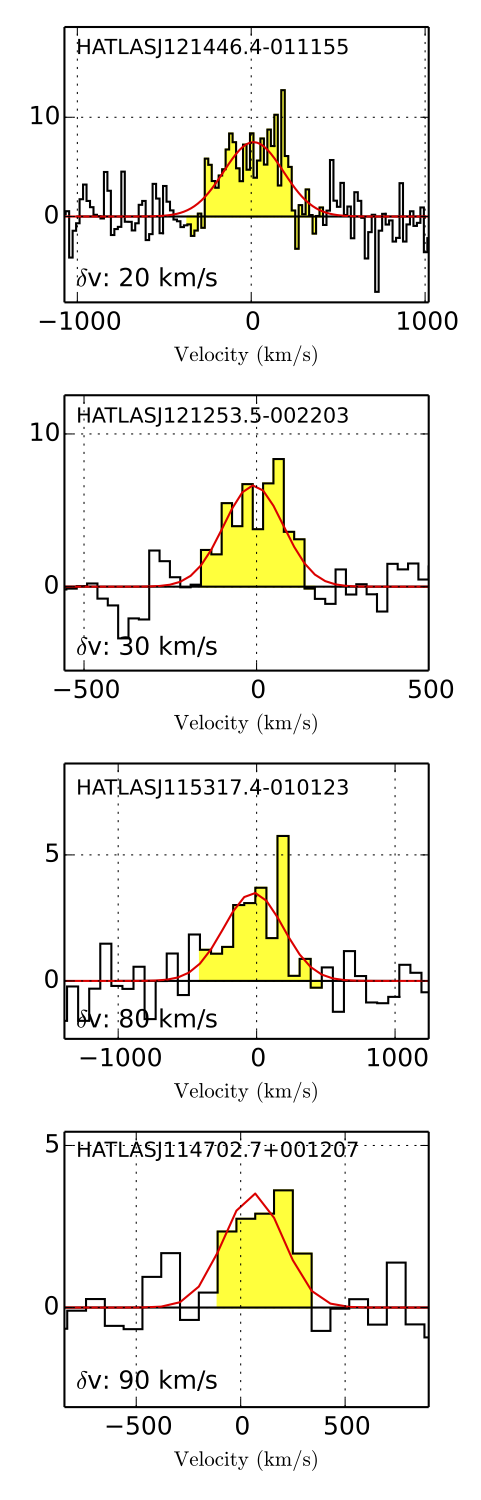} 
  \caption[Spectra 1]{The observed CO(1--0) spectra for spectrally
    detected galaxies centred on spectroscopic redshifts taken from
    GAMA. The emission is spectrally binned ($\delta_{\nu}$)
    differently in order to maximise the number of channels with
    signal above a 5.0$\sigma$ significance. The yellow color
    indicates the spectral range we have used to derive velocity
    integrated flux densities. The red lines show best-fitting single
    Gaussian profiles to the spectra (see Table~\ref{Table}).}
  \label{Spectra_1}
\end{figure*}

\begin{figure*}
  \centering
  \includegraphics[height=12.5cm]{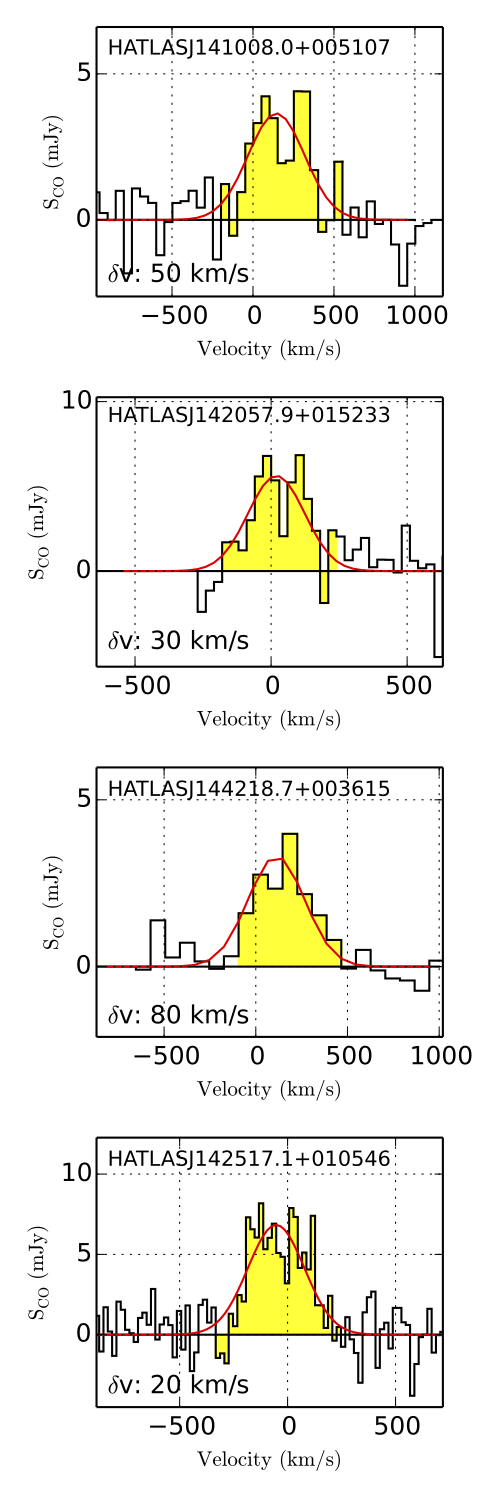}
  \includegraphics[height=12.45cm]{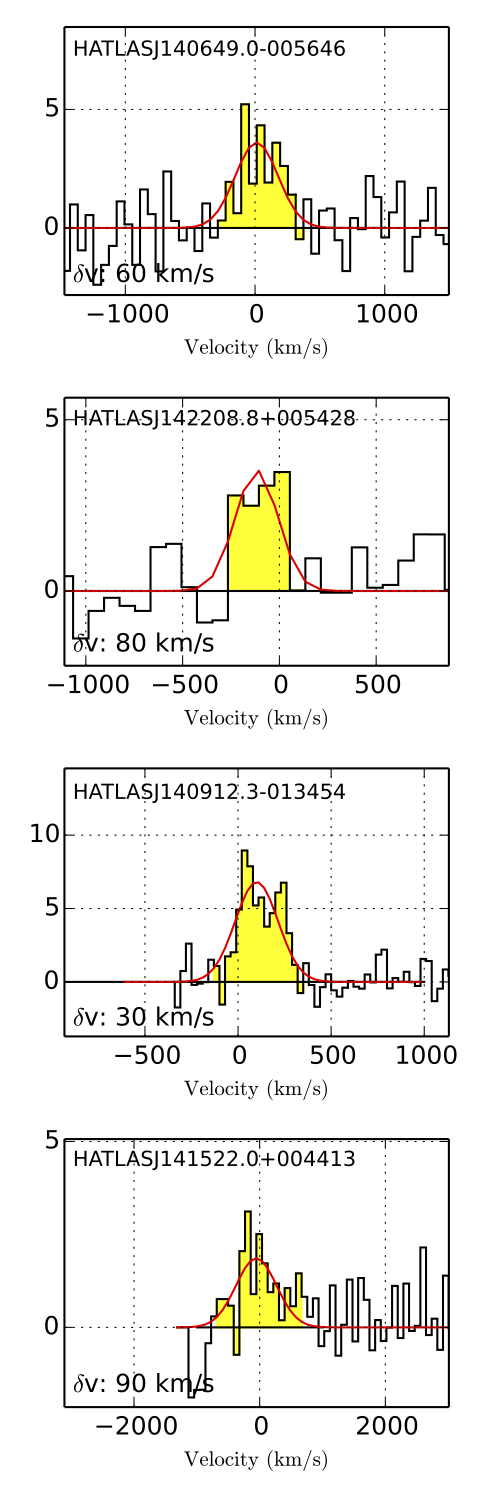}
  \includegraphics[height=12.45cm]{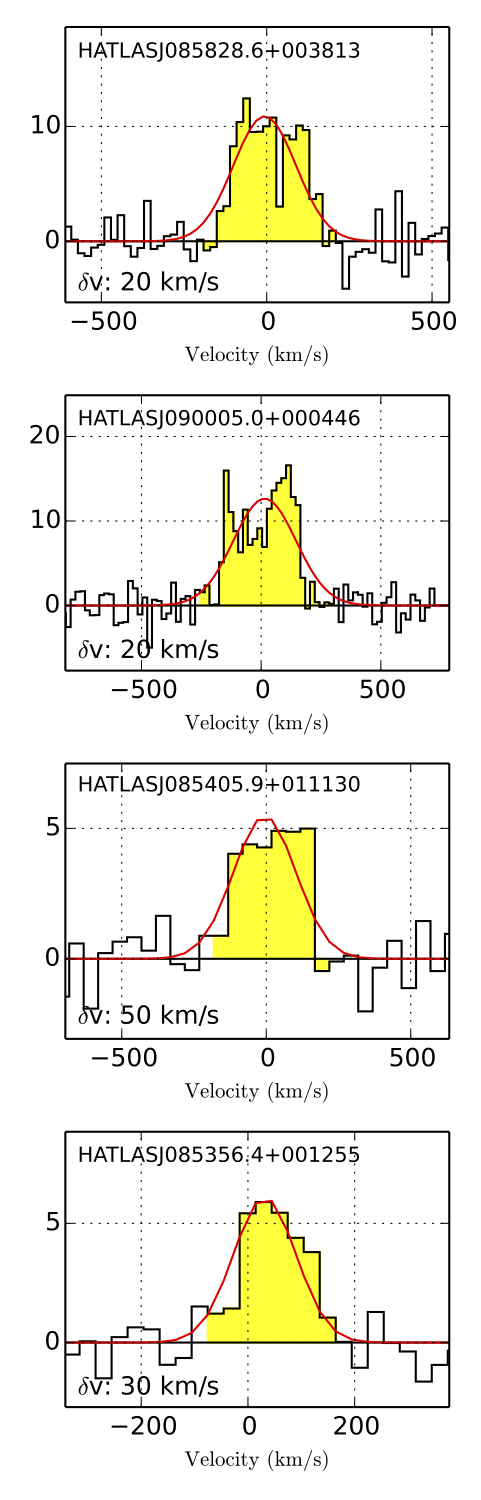}
  \includegraphics[height=12.4cm]{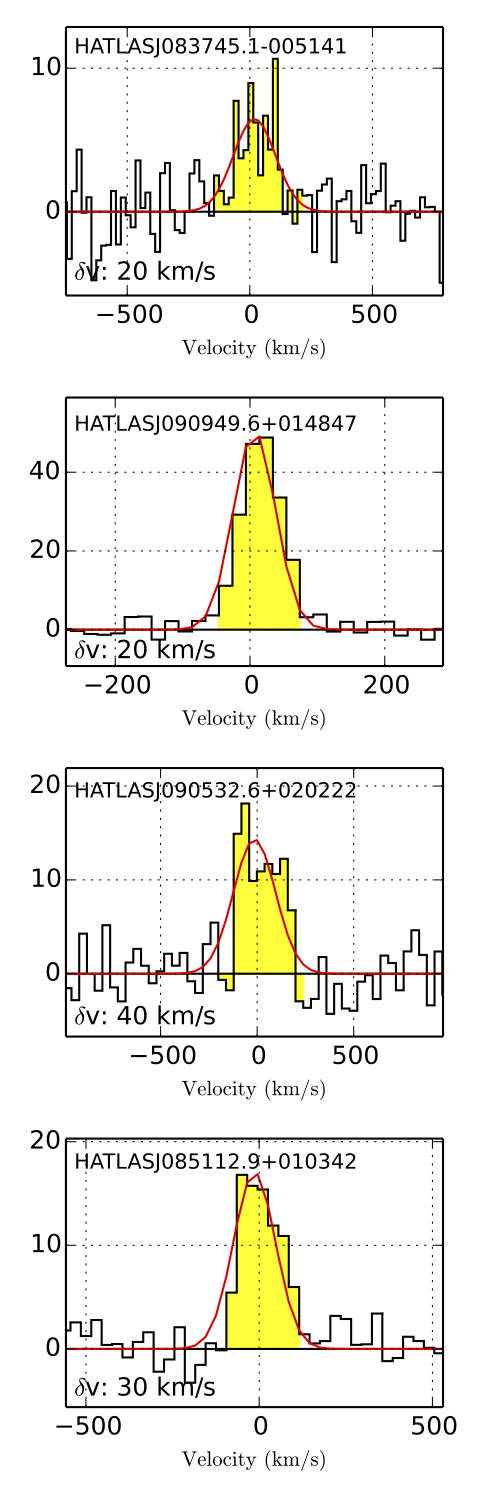}\\
                  {\textbf{Figure~\ref{Spectra_1}.} continued}
\end{figure*}

Slightly different selection criteria were used in each cycle to
construct the list of ALMA targets. In Cycle-1, we selected a
representative sample of 41 galaxies with the following criteria: 
$0.15 < z < 0.35$ (the upper threshold in redshift corresponds to the limits at
which the CO(1--0) line moves out of frequency range covered by Band-3
of ALMA); $S_{\rm 160\mu m}>100\,$mJy; SDSS sizes {\sc
  isoa}\,$<10\farcs0$; and a reduced $\chi^2<1.5$ when fitting the
far-IR/submm SED using a modified black body (following a similar
approach as in \citealt{Ibar2013}). On the other hand, in Cycle-2 we
targeted 27 galaxies that have previous {\it Herschel} PACS [C\,{\sc
    ii}] spectroscopy as shown by \citet{Ibar2015} and so added the
following criteria: $0.02<z<0.2$ (the threshold is defined by the
point where the [C\,{\sc ii}] is redshifted to the edge of the PACS
spectrometer); $S_{\rm 160\mu m}>150$\,mJy; Petrosian SDSS radii
smaller than $15\farcs0$; sources do not have $>3\sigma$ PACS
160\,$\mu$m detections within 2 arcmin (to ensure reliable on-off sky
subtraction).

Combining Cycle-1 and Cycle-2 observations, we construct one of the
largest samples of CO($1-0$) detected galaxies at $0.02<z<0.35$ (see
Fig.~\ref{SampleDefinition}). We highlight that some of the main
advantages of our sample over previous studies of far-IR-selected
galaxies are: (1) we cover fainter $L_{\rm 8-1000\mu m} \approx
10^{10-12} L_{\odot}$ and less massive $M_{\rm dust} \approx 1.5
\times 10^{7-8} {\rm M}_{\odot}$ ranges than $IRAS$--selected samples,
i.e.\ our samples are not significantly biased towards powerful ULIRGs
that potentially have complex merger morphologies as those described
by \cite{Braun2011} and \cite{Combes2011}; (2) the sample selection
dominated by the 160\,$\mu$m and 250\,$\mu$m photometry gives
relatively low dust temperature estimates ($25<T_{\rm dust}/{\rm
  K}<60$) and reduces (but not entirely) the well known bias towards
high dust temperatures evidenced in 60\,$\mu$m-selected
$IRAS$--samples \citep[see discussion
  by][]{Gao&Solomon2004,Kennicutt2009}; (3) the wealth of ancillary
data already available for all the sources
\citep{Driver2016,Bourne16}; and (4) the redshift range puts galaxies
far enough so galaxies can be imaged with a single ALMA pointing in
Band-3 -- it does not require large mosaicking (using the Atacama
Compact Array) campaigns as in more local galaxy samples.  These
reasons enable us to address our science goals using a much simpler
but wider parameter space for the diagnostics of interest (see
Fig.~\ref{SampleDefinition}).

\begin{figure*}
  \centering  
  \includegraphics[height=15.39cm]{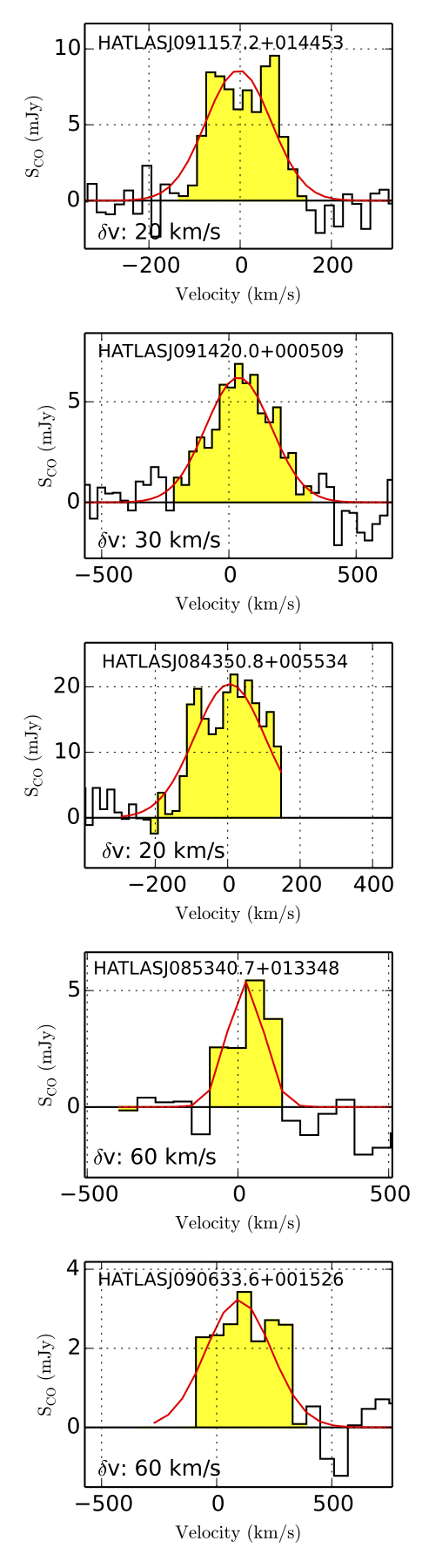}
  \raisebox{3.05cm}{\includegraphics[trim=0.0cm 0 0 0,height=12.3cm]{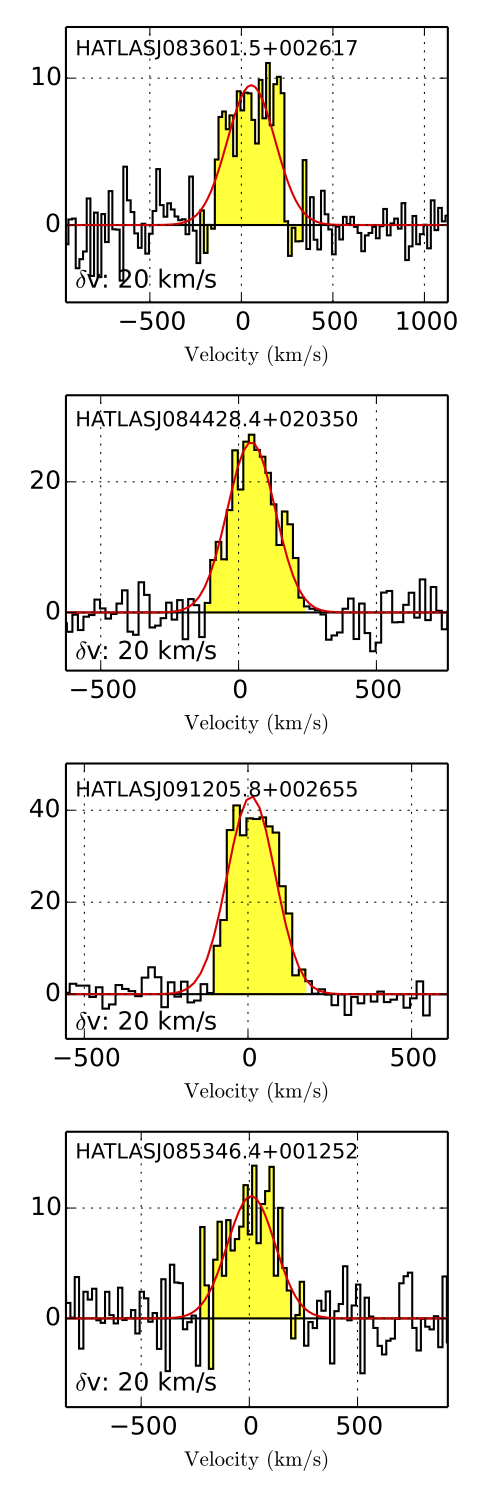}}
  \raisebox{3.05cm}{\includegraphics[trim=0.0cm 0 0 0,height=12.3cm]{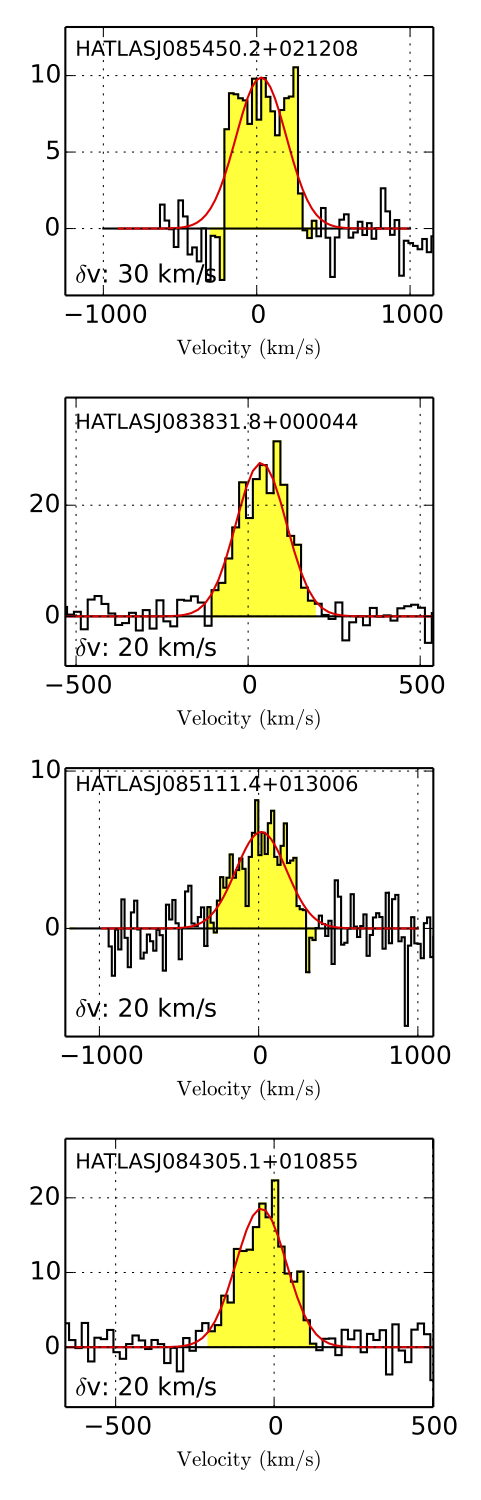}}
  \raisebox{3.05cm}{\includegraphics[trim=0.0cm 0 0 0,height=12.3cm]{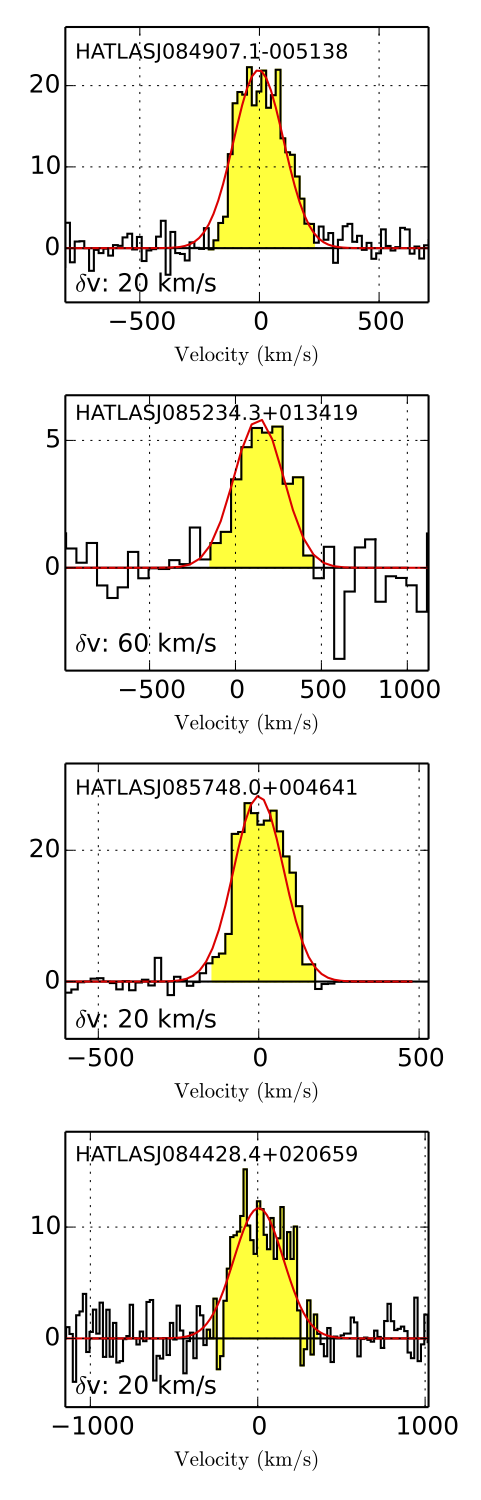}}\\ 
                  {\textbf{Figure~\ref{Spectra_1}.} continued}
\end{figure*}	
	
\subsection{Observational strategy}

ALMA Cycle-1 observations were taken in Band-3 between December 2013
and March 2014 (see Table~\ref{Table_1}), spending approximately 3 to
9 minutes on-source in each source.  Scheduling blocks (SBs) were
designed to detect the CO($1-0$) emission line down to a root mean
square ({\sc rms}) of $1.5$\,mJy\,beam$^{-1}$ at 50\,km\,s$^{-1}$
channel width and at $\sim$ 3$''$ - 4$''$ resolution (the most compact
configuration). On the other hand, Cycle-2 observations were taken in
Band-3 on January 2015 and SBs were designed to observe the CO(1--0)
emission line but down to $2$\,mJy\,beam$^{-1}$ at
$30$\,km\,s$^{-1}$. Even though ALMA is not specifically designed as a
`survey-like' telescope, we setup our experiment to minimise the
number of spectral tunings needed to observe all sources
independently. We make use of the fact that our targets come from
three equatorial {\it H}-ATLAS/GAMA fields which are $\sim 4\times14$
deg$^{2}$ size, providing large numbers of galaxies at similar
redshifts. We modified the `by-default' approach provided by the ALMA
Observing Tool (OT) by setting source redshifts to zero, but fixing
the spectral windows (SPW) manually in order to cover the widest
possible spectral range, i.e.\ redshift range. We optimised the
central frequency position of the SPWs (over $\sim7.5\,$GHz) to
maximise the number of sources with the CO(1--0) line redshifted into
the ranges covered by our SPWs. This observing strategy allowed us to
spectrally resolve the CO(1--0) emission in 49 galaxies (see
Fig.~\ref{Spectra_1}; $\sim$\,70\% of the whole sample), while in 12
others we see low signal to noise emission in collapsed spectra
(moment$-0$).

\subsection{Data reduction}
\label{Data_Reduction}

A summary of all ALMA observations are shown in
Table~\ref{Table_1}. To process all observations in a standardised
way, we developed a common pipeline within the {\sc{Common Astronomy
    Software
    Applications}}\footnote{\url{http://casa.nrao.edu/index.shtml}}
({\sc casa} version 4.4.0). Based on the standard pipeline for data
processing, we designed our own structured pipeline for calibration,
concatenation and imaging. The structure was designed in modules,
taking into account the vast amount of data and high flexibility at
the time to flag corrupted data. When a science goal has more than one
observation, we re-calibrate the phase calibrator to an average flux
density (usually variations are seen at $\lesssim$15\%) and bootstrap
this scaling to the targets before concatenating the observations. The
bandpass, flux and phase calibrators for each data set can be seen in
Table \ref{Table_1}.

In the first instance, imaging was performed using the task {\sc
  clean} at different spectral resolutions (from 20 to
100\,km\,s$^{-1}$ in steps of 10\,km\,s$^{-1}$). We sought the
resolution that provided the highest number of non-cleaned point-like
detections $>5.0\sigma$ within the data-cube (R.A.-Dec.-$\nu$) near
the expected source position. If the source was undetected, then we
created the cube at 100\,km\,s$^{-1}$ channel width. After choosing
the best resolution, we ran task {\sc clean} again but this time
applying a primary beam correction, manually cleaning the CO line
emission down to a threshold of $3.0\sigma$, and choosing an image
size of $256\times256$ pixels with roughly 5 pixels (semi-major axis)
per synthesised beam full width half maximum ({\sc fwhm}). We used the
optically-derived spectroscopic redshifts ($z_{\rm spec}$) of each
source in a barycentric velocity reference frame. The final cubes
were created using natural weighting, resulting in image cubes 
with typical synthesised beams of  3$"$ - 4$"$. The physical sizes
for each source, i.e.  the deconvolved major-axes (in kpc) are given
in Table~\ref{Table}.

\subsection{Source properties}
\label{Source_Properties}

\subsubsection{CO emission}
\label{CO_emission}
		
We get an average {\sc rms} level of $1.6$\,mJy\,beam$^{-1}$ (at
50\,km\,s$^{-1}$) for both Cycle-1 and Cycle-2.  We identify 49
galaxies (out of 67) with a $>5\sigma$ peak line detection in at least
one spectral channel (from 10 to 100\,km\,s$^{-1}$ in all binned). For the 49
spectrally detected galaxies we determine the central frequency
($\nu_{\rm obs}$) of the CO emission line by using a single Gaussian
fit to the spectra. We found that central frequencies are in agreement 
and within the scatter of the expected GAMA's optical redshifts (see column
$\rm v_{obs}$ in Table~\ref{Table}). The fitted {\sc fwhm} of the CO line in
our sample covers a range of $67 - 805$\,km\,s$^{-1}$. All the spectra
with spectrally resolved CO signal are displayed in
Fig.~\ref{Spectra_1}, whereas non-detections are summarised in
Table~\ref{Table}.

The velocity integrated CO flux densities ($S_{\rm CO}\Delta \rm v$ in
units of Jy\,km\,s$^{-1}$) were obtained by collapsing the data cubes
between $\pm$\,1$\times$\,{\sc fwhm} centred on the line (see yellow
range shown in Fig.~\ref{Spectra_1}). The 2D intensity map is then
fitted with a 2D-Gaussian for all detected sources using the task {\sc 
gaussfit} within {\sc casa}. Errors in these measurements are 
taken directory from {\sc gaussfit}'s outputs. In seven cases the CO emission is not
well fitted by a 2D-Gaussian, so we have used an irregular aperture
covering the whole extension of the emission. Errors for those
aperture measurements come from the standard deviation of fluxes
measured in random sky regions around the source. We find measurements
in the range of $2.2 - 20.8$ Jy km s$^{-1}$, with an average value of
$6.9 \pm 0.2$ Jy km s$^{-1}$. We get 21 galaxies which are spatially
resolved in CO, based on a fitted semi-major axis $\sqrt{2}$ times
larger than the major axis of the synthesised beam. 

For non-detections, we collapsed the cubes (moment 0 maps) between
$\pm$\,250\,km\,s$^{-1}$ centred at the expected observed frequency --
a range consistent with the average line width we derive for the
whole sample ($251.6 \pm 38.3$ km\,s$^{-1}$). In these stacked spectra
12 other galaxies show emission (ensuring a corrected optical and redshift
association). We provide these measurements in Table~\ref{Table}. 
These collapsed maps the {\sc rms} values range between 
$0.04 - 5.35$\,Jy\,km\,s$^{-1}$ (at 100\,km\,s$^{-1}$ channel width) with an 
average of $1.64$\,Jy\,km\,s$^{-1}$.

Some spectra show double line profiles providing 
valuable dynamical information. Our kinematic results will be 
published by {\color{blue}Molina et al. in prep.}. We stress, however, 
that our single Gaussian profiles shown in Fig.~\ref{Spectra_1} 
are to define the spectral range used to collapse the cubes, 
from which we obtain the intensity maps to extract the  
velocity integrated flux densities. We look at how much the velocity integrated flux densities could change if we use double Gaussian profiles to fit the emission lines (in 16 spectra). Collapsing the cubes between the lower and the upper {\sc fwhm} bound limits (of both Gaussians), and comparing these to those obtained from a single Gaussian fit, we obtain that fluxes decrease by a $\sim$\,5\% (on average), although with a large scatter ($\sim$\,30\%). We decide to stick to the single Gaussian fit to estimate the {\sc fwhm} to collapse the cubes.


\begin{figure}
  \includegraphics[width=8.8cm]{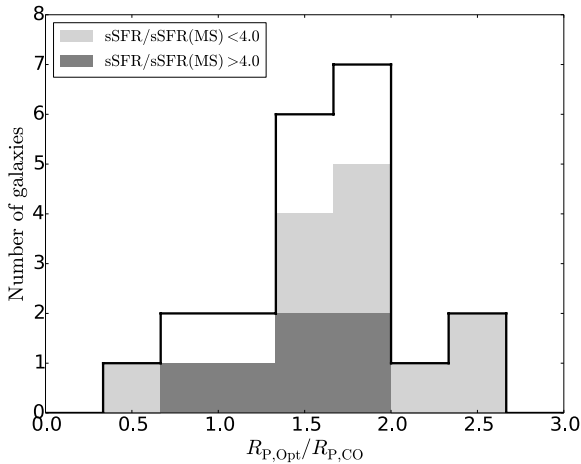}
  \caption[Histogram]{The ratio between the Petrosian radius in
    $r$-band ($R_{\rm P,Opt}$) and CO ($R_{\rm P, CO}$) for the 21
    ALMA-resolved sources. We separate our sample into two populations
    based on the relative levels of starburst activity (see
    Fig.~\ref{SampleDefinition}). We find that most of our galaxies
    are distributed around a size ratio $\sim1.3-2.0$.}
  \label{Histogram}
\end{figure}

\subsubsection{IR emission}
\label{FIR_emission}

For each galaxy, we measure the IR luminosity by fitting the
rest-frame SED constructed with photometry from $IRAS$, Wide-field
Infrared Survey Explorer (WISE), and {\it Herschel} PACS and SPIRE
instruments, using a modified black body that is forced to follow a
power law at the high-frequency end of the spectrum. The fit
constraints the dust temperature ($T_d$), the dust emissivity index
($\beta$), the mid-IR slope ($\alpha_{\rm mid-IR}$), and the
normalisation. Then we integrate the flux of the best-fitting SED
between 8 and $1000\,\mu$m to obtain the total IR luminosity
(\citealp{Ibar2013,Ibar2015}), i.e.,

\begin{equation}
  L_{\mathrm{IR}}(8-1000\,\mu{\rm m}) =
  4\pi\,D_{\rm L}^2(z)\,\int_{\nu_{\rm 1}}^{\nu_{\rm 2}}S_\nu\,d\nu .
\end{equation}

The uncertainties on the IR luminosity are obtained by randomly
varying the broadband photometry within the observational
uncertainties in a Monte-Carlo simulation (100 times). Our results
are listed in Table~\ref{Table}.

We estimate the SFR following ${\rm SFR(M_{\odot}\,yr^{-1})} =
10^{-10} \times L_{\rm IR}$ assuming a \cite{Chabrier2003} Initial
Mass Function (IMF), where $L_{\rm IR}$ is in units of $L_{\odot}$ 
\citep{Kennicutt1998}, and we assume a 1.72 factor to convert from 
a Salpeter to a Chabrier IMF.

\begin{figure*}
\begin{center}
\includegraphics[width=0.68\columnwidth]{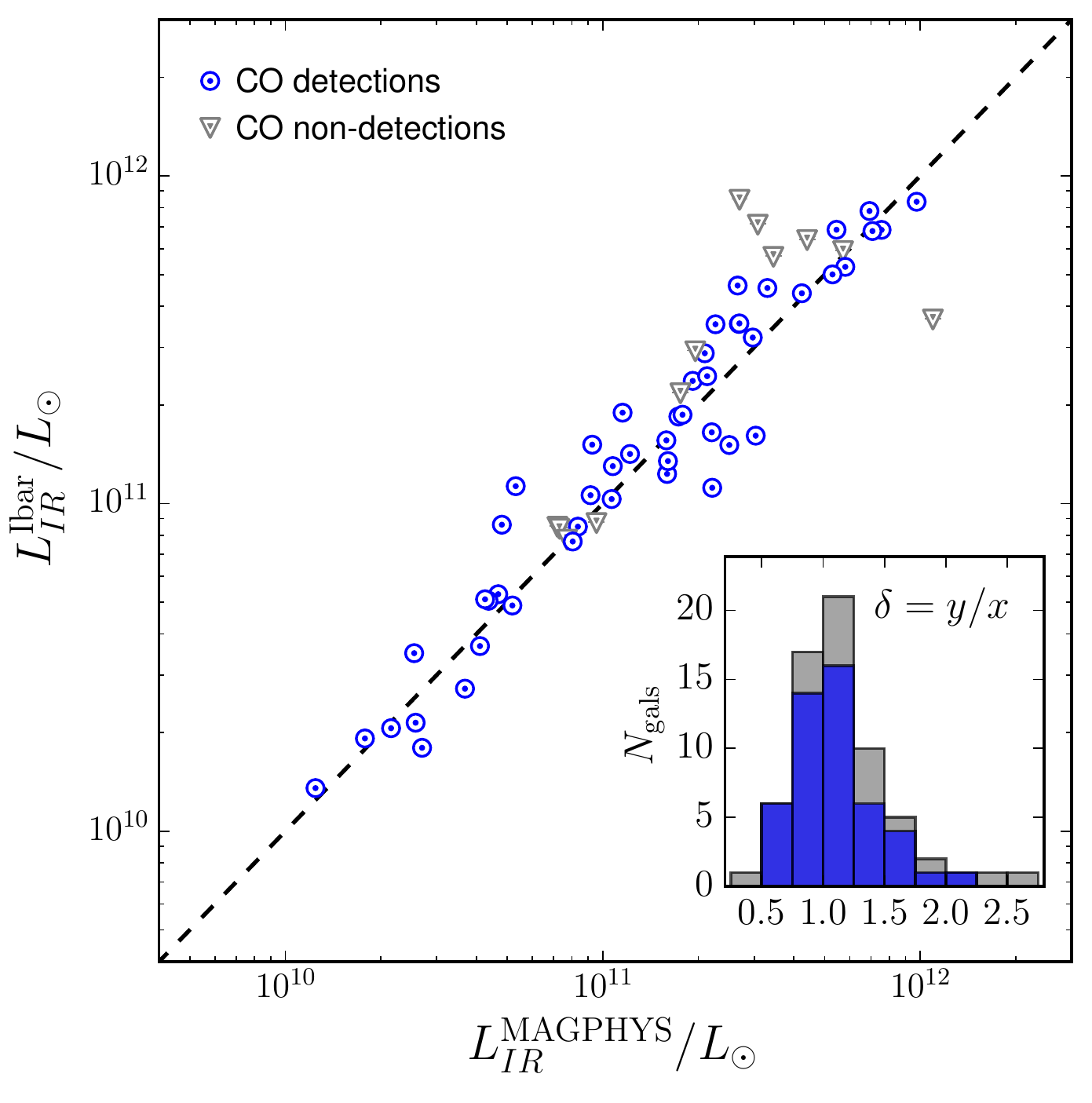}
\includegraphics[width=0.68\columnwidth]{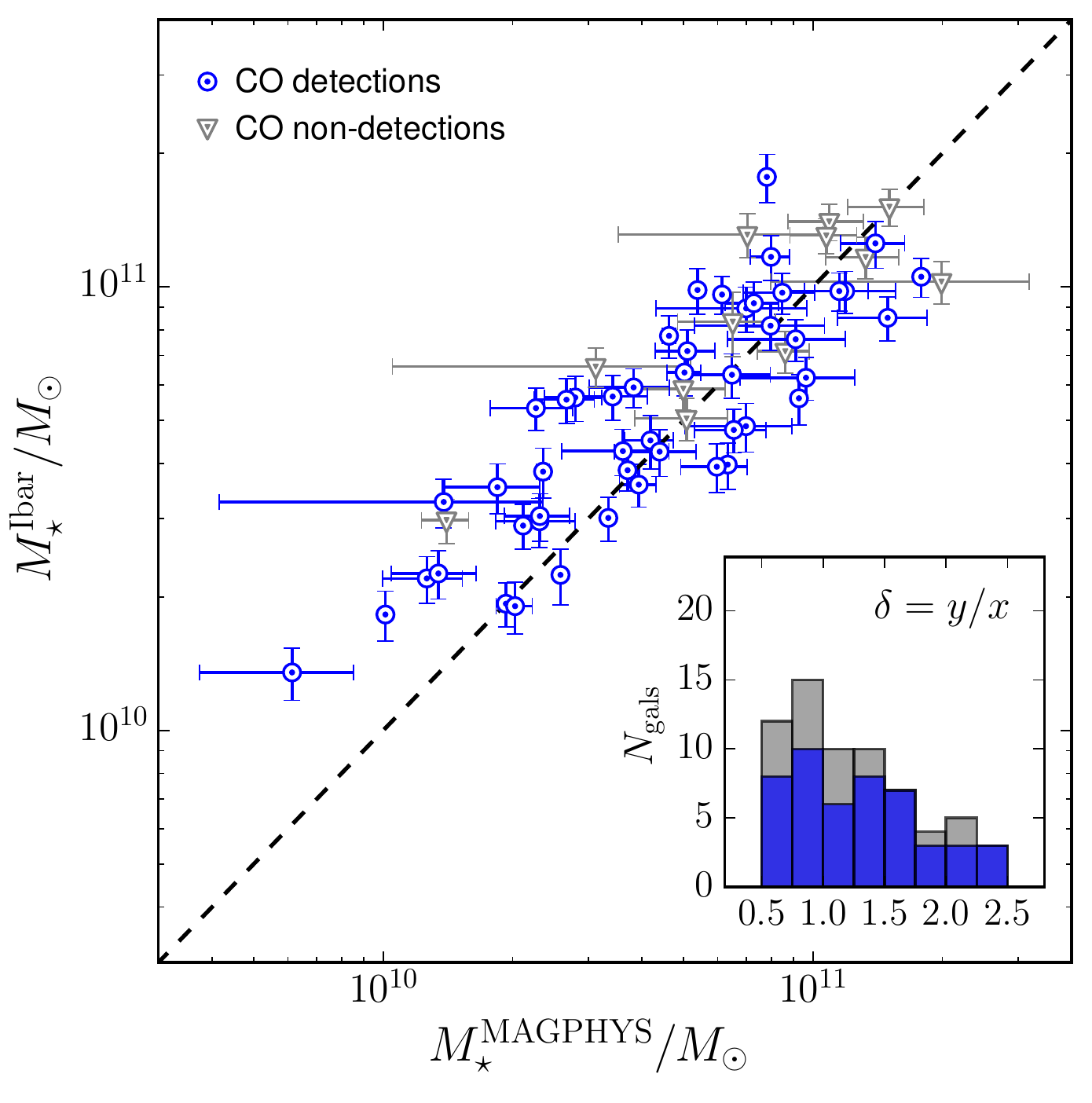}
\includegraphics[width=0.68\columnwidth]{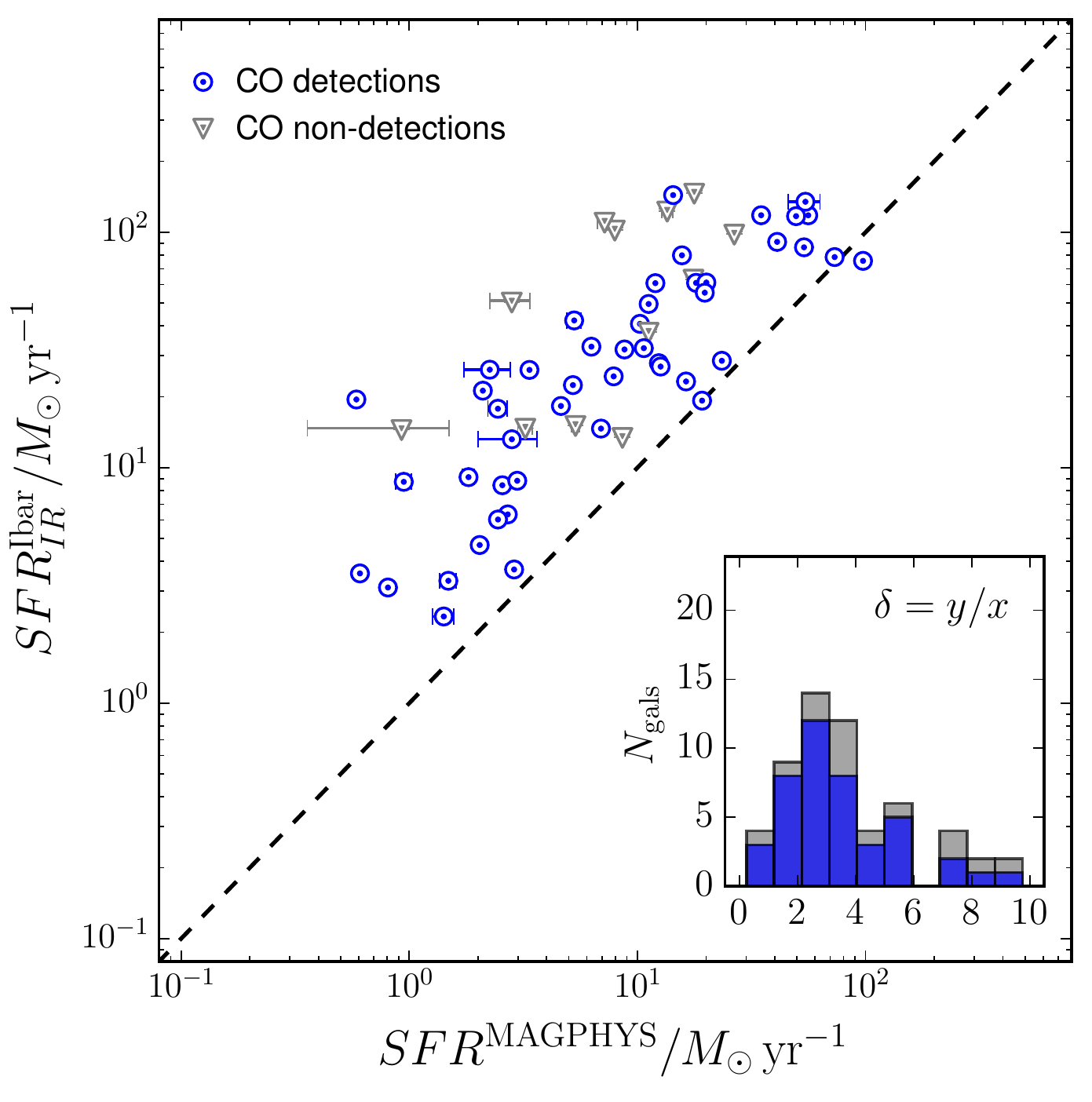}
\end{center}
\vspace{-0.5cm}
\caption{Comparing the infrared luminosity (\textit{left}),
stellar mass (\textit{middle}) and star formation rate
(\textit{right}) derived from empirical methods (see
  \citealp{Ibar2015}) with those obtained from the MAGPHYS UV--submm
  SED fitting (see \citealp{Driver2016}).}\label{fig:comparison}
\end{figure*}

\subsubsection{SED fitting}
\label{SED_fitting}

\begin{figure*}
$
\begin{array}{ccc}
\includegraphics[width=8.4cm]{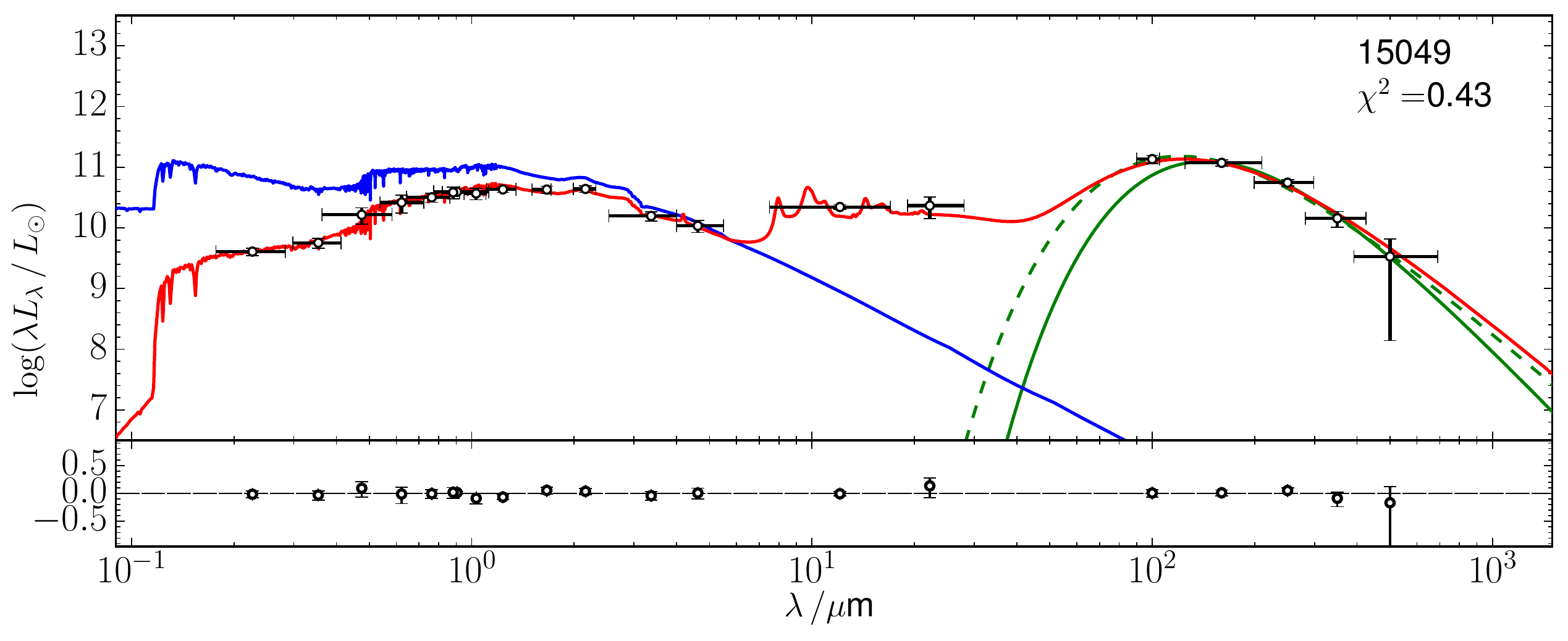} &
\includegraphics[width=5.0cm]{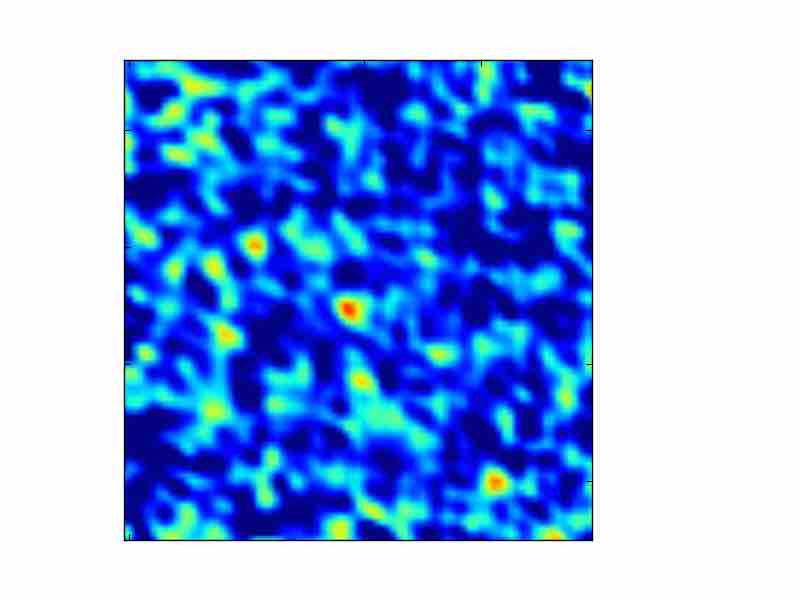} &
\hspace*{-1.2cm}\begin{overpic}[width=3.4cm]{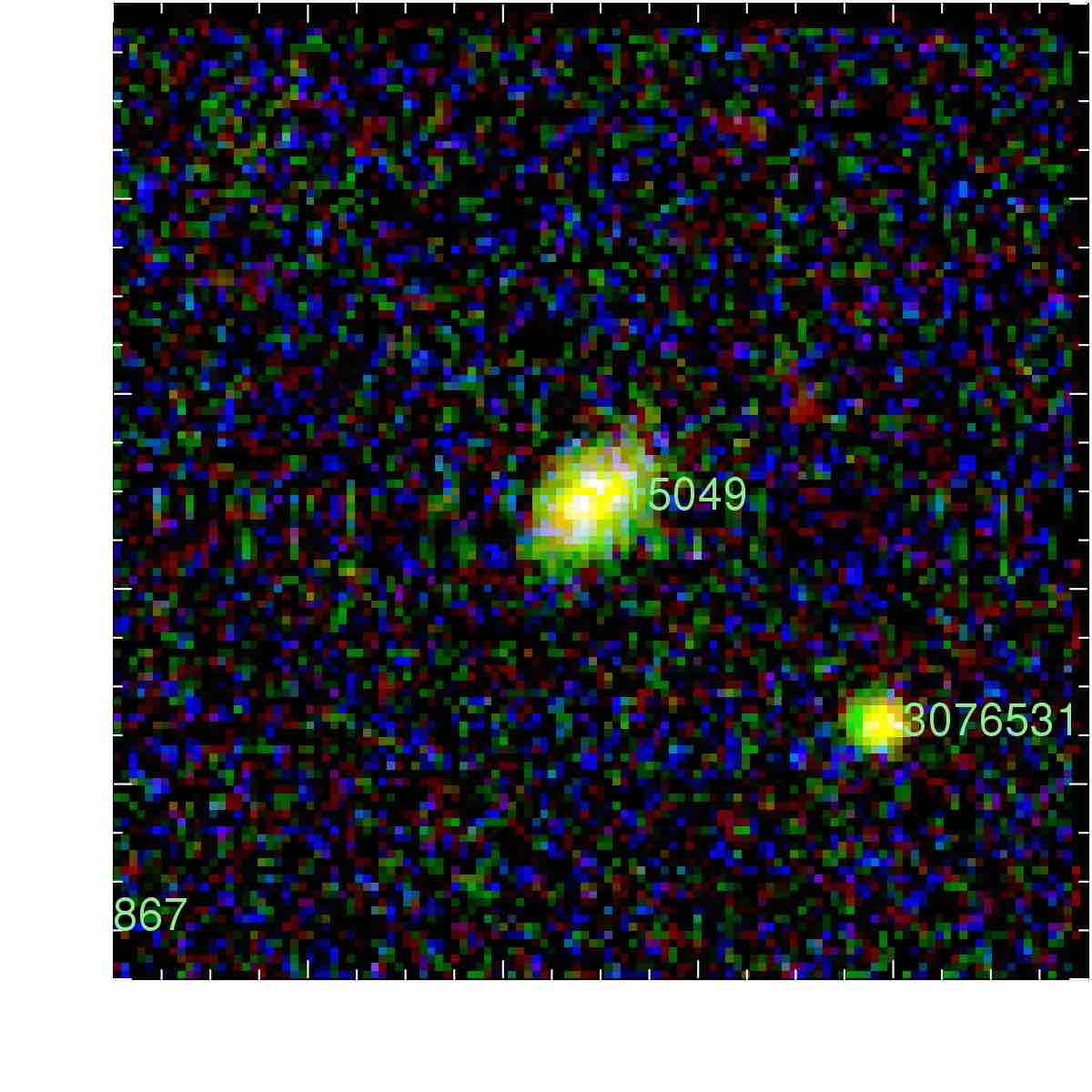} \put (9,85) { \begin{fitbox}{2.25cm}{0.2cm} \color{white}$\bf BDC$ \end{fitbox}} \end{overpic} \\ 	
	
\includegraphics[width=8.4cm]{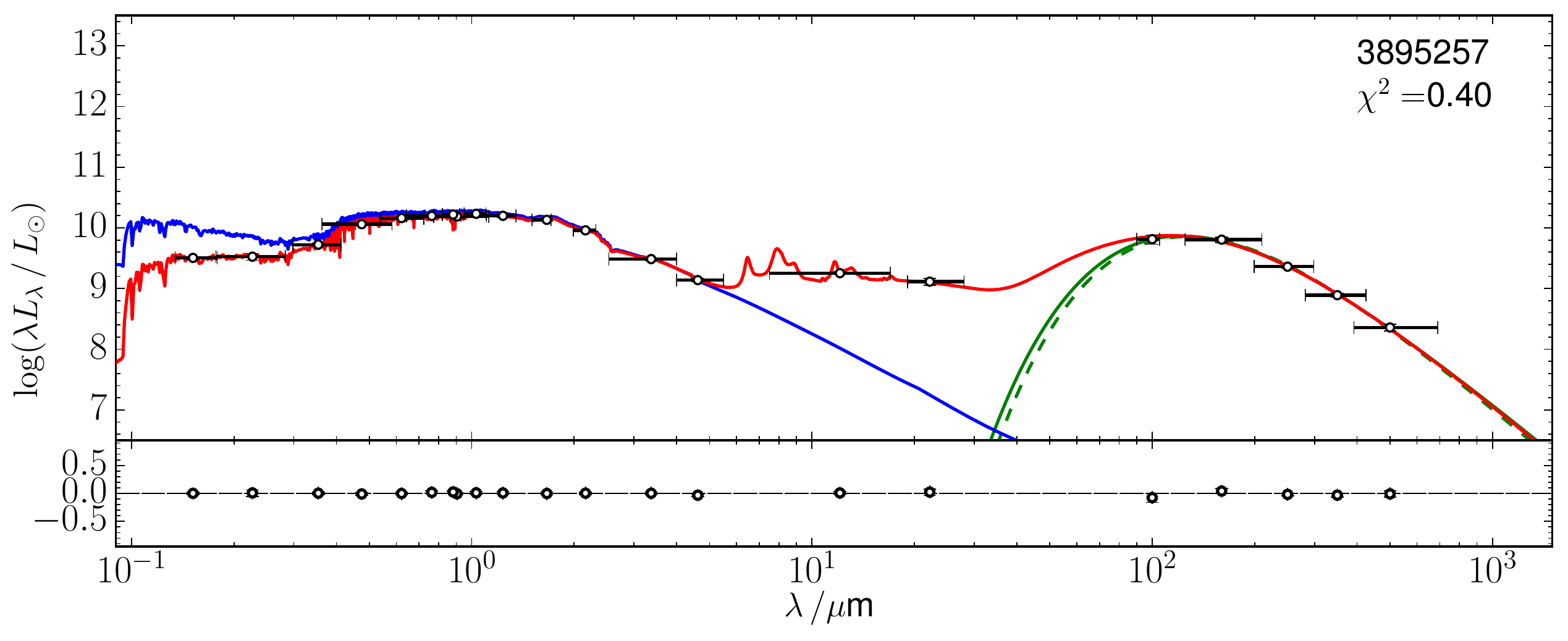} &
\includegraphics[width=5.0cm]{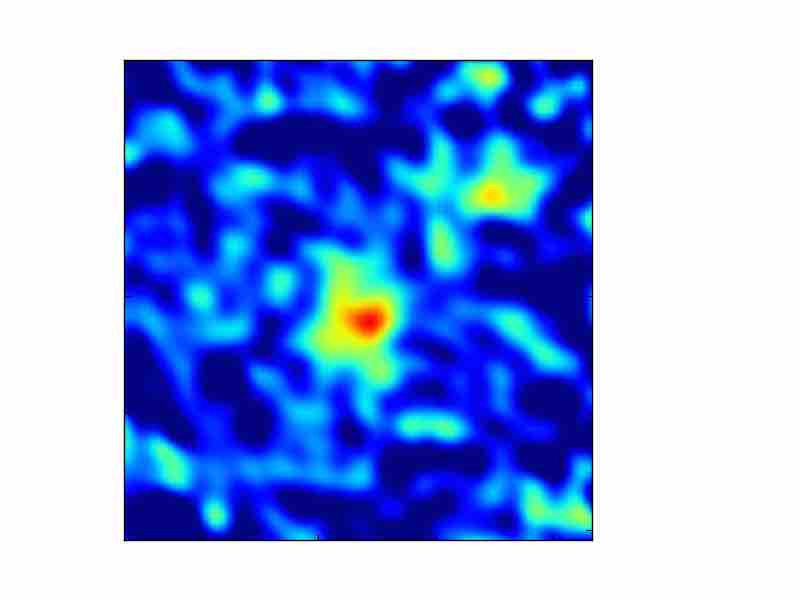} &
\hspace*{-1.2cm}\begin{overpic}[width=3.4cm]{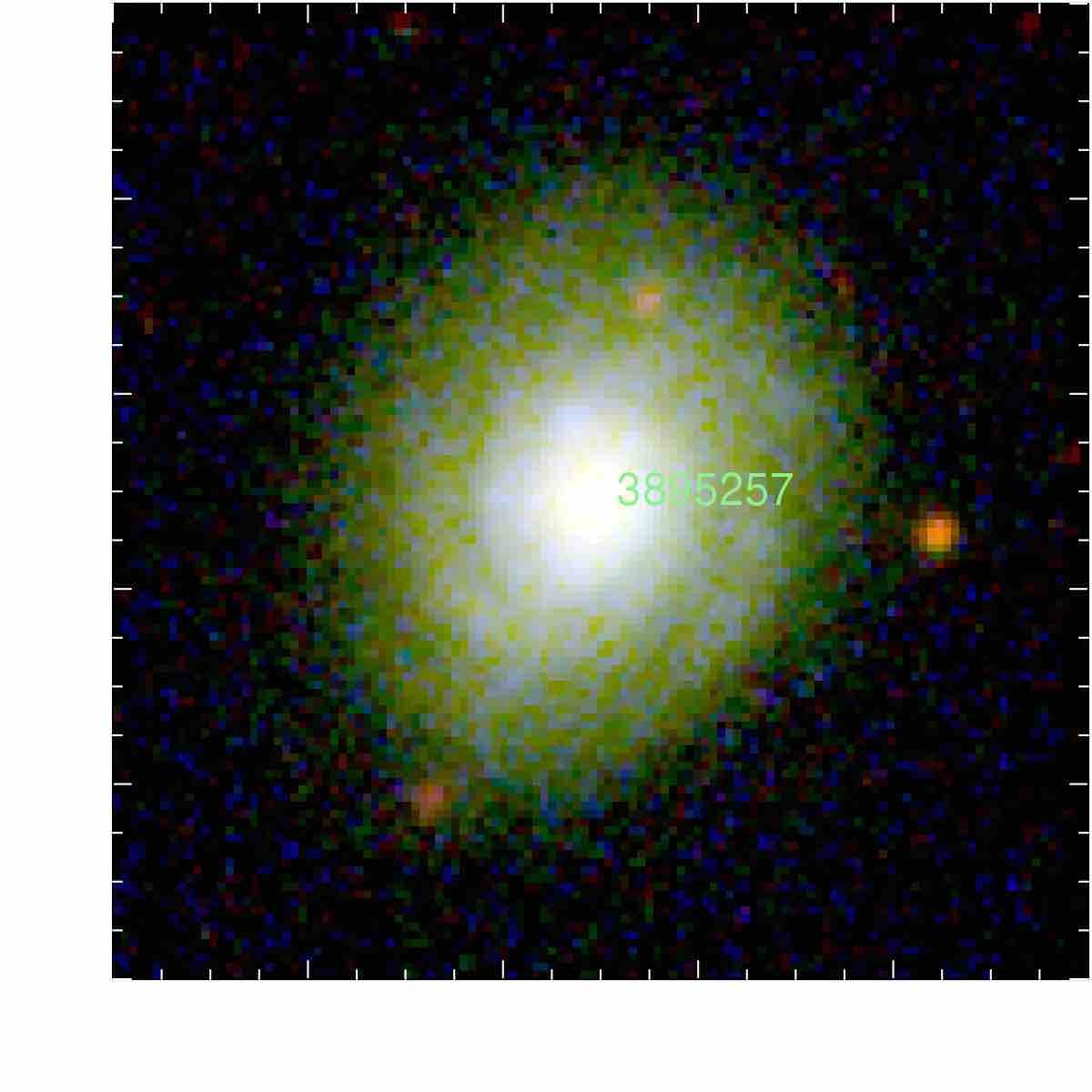} \put (9,85) { \begin{fitbox}{2.25cm}{0.2cm} \color{white}$\bf D$ \end{fitbox}} \end{overpic} \\ 

\includegraphics[width=8.4cm]{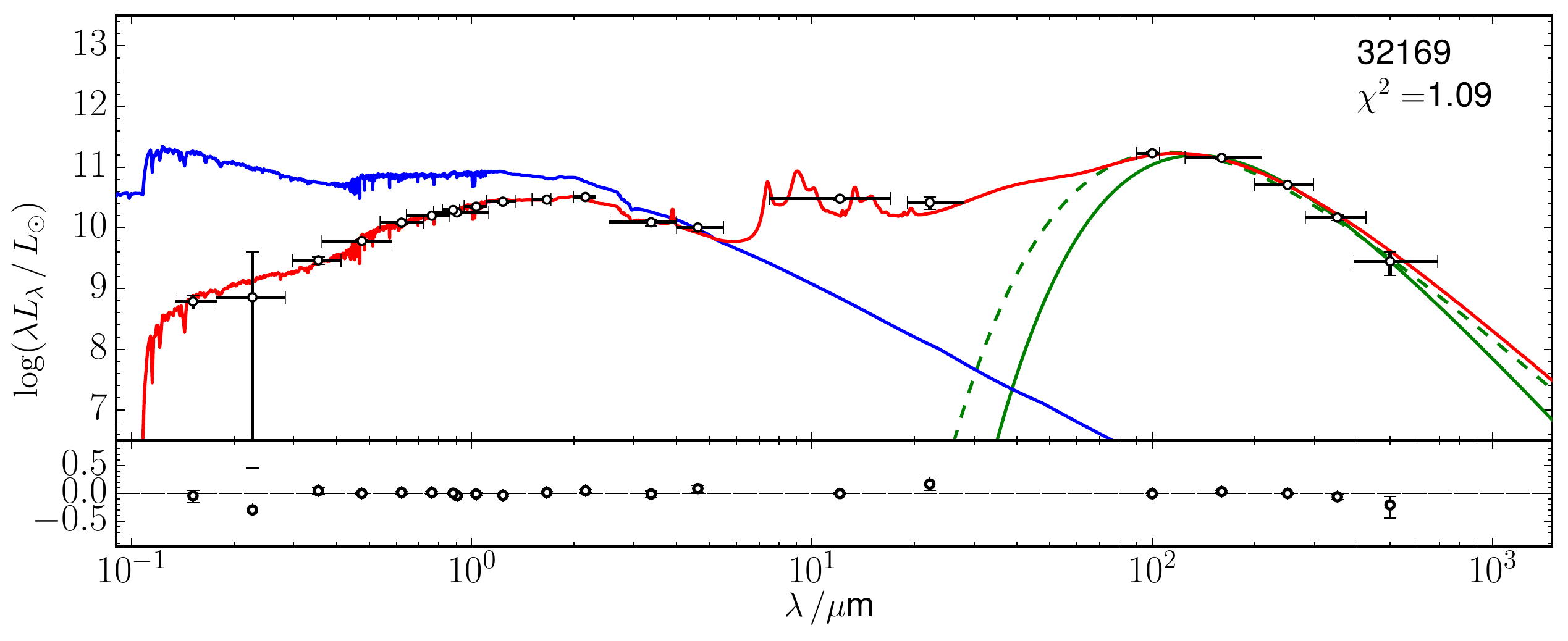} &
\includegraphics[width=5.0cm]{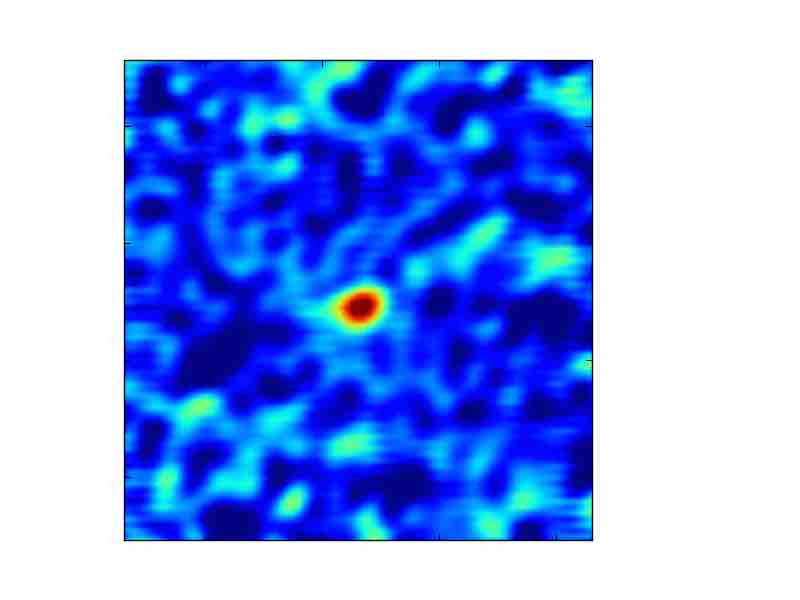} &
\hspace*{-1.2cm}\begin{overpic}[width=3.4cm]{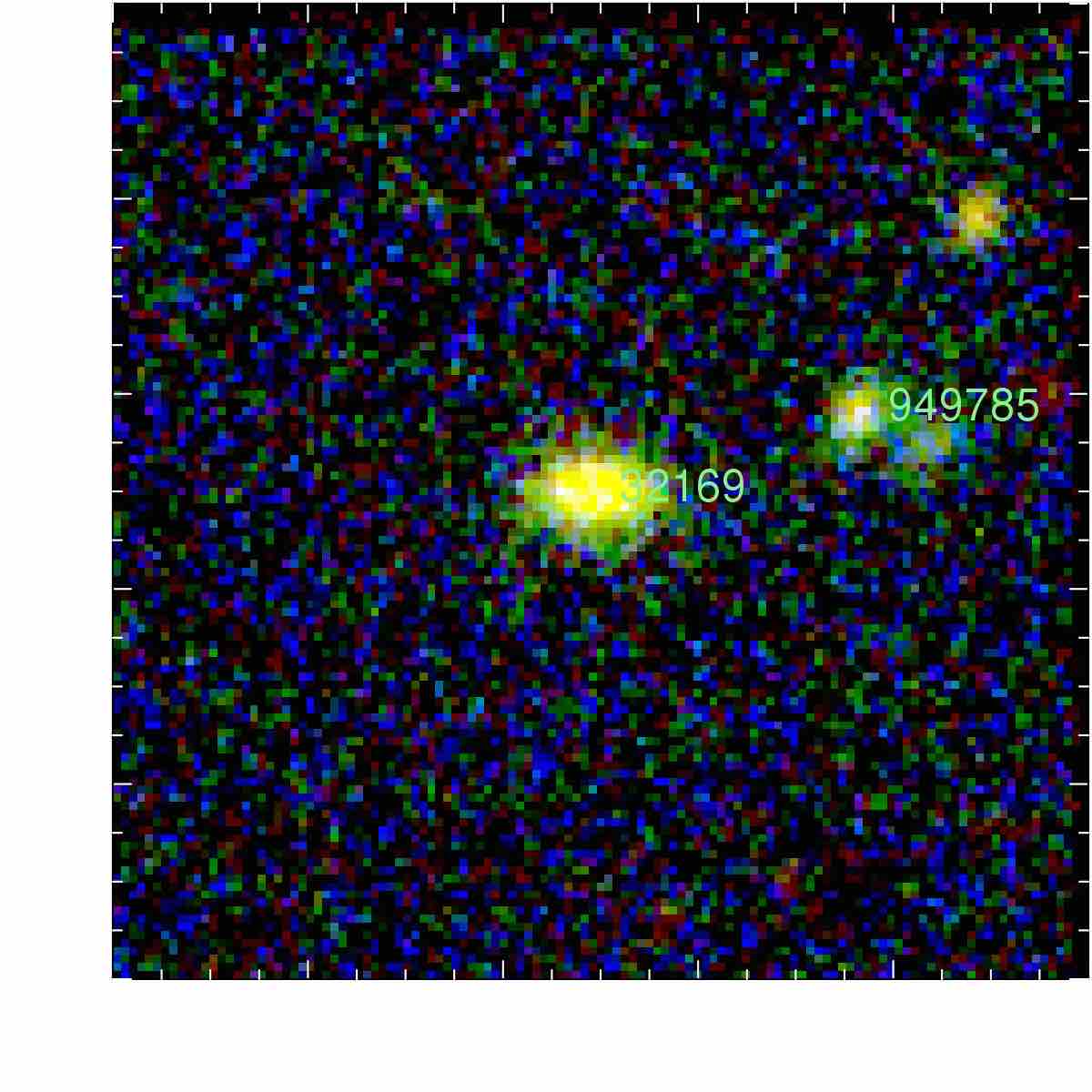} \put (9,85) { \begin{fitbox}{2.25cm}{0.2cm} \color{white}$\bf BC$ \end{fitbox}} \end{overpic} \\ 

\includegraphics[width=8.4cm]{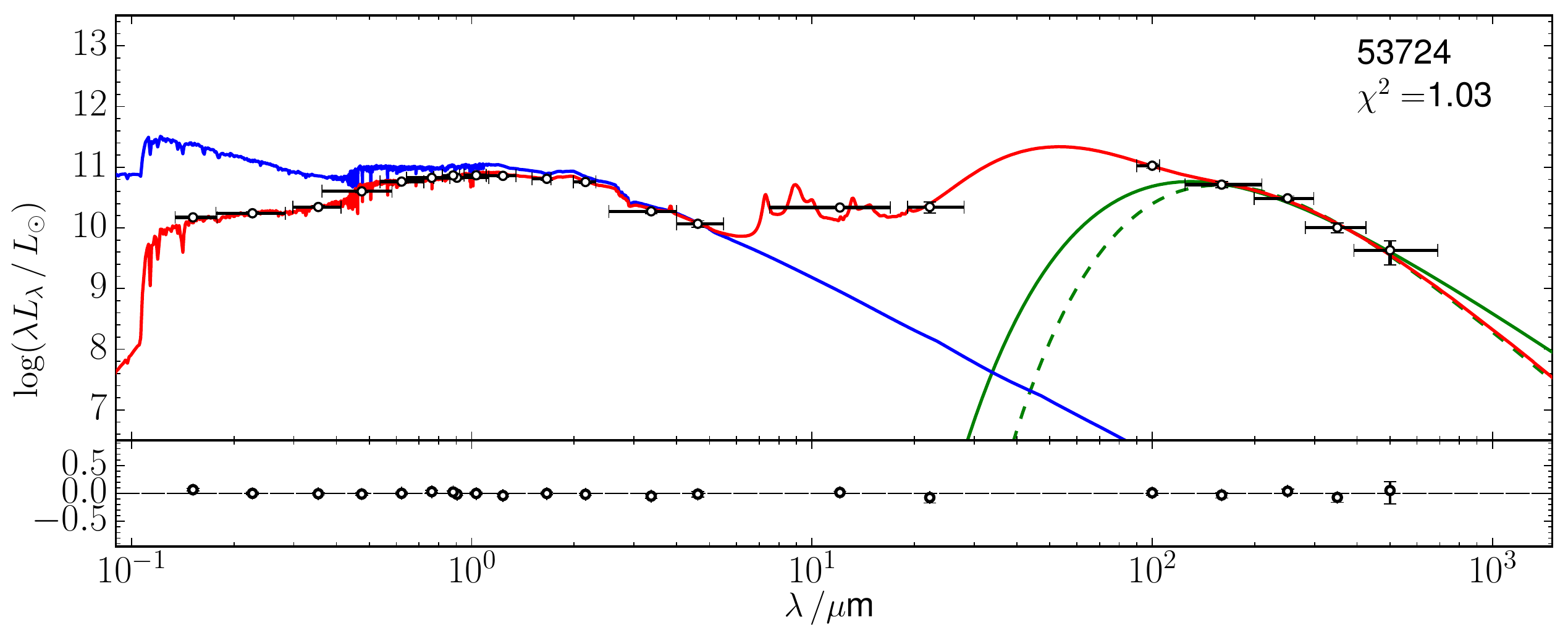} &
\includegraphics[width=5.0cm]{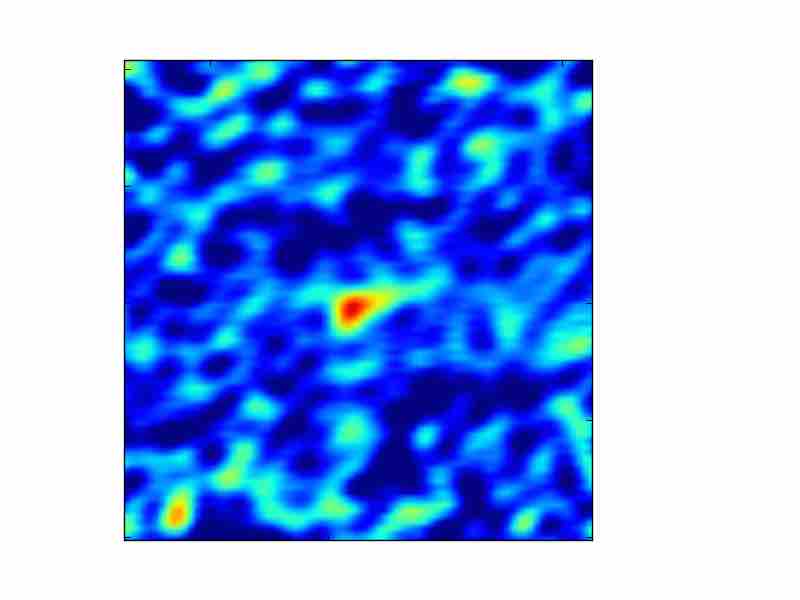} &
\hspace*{-1.2cm}\begin{overpic}[width=3.4cm]{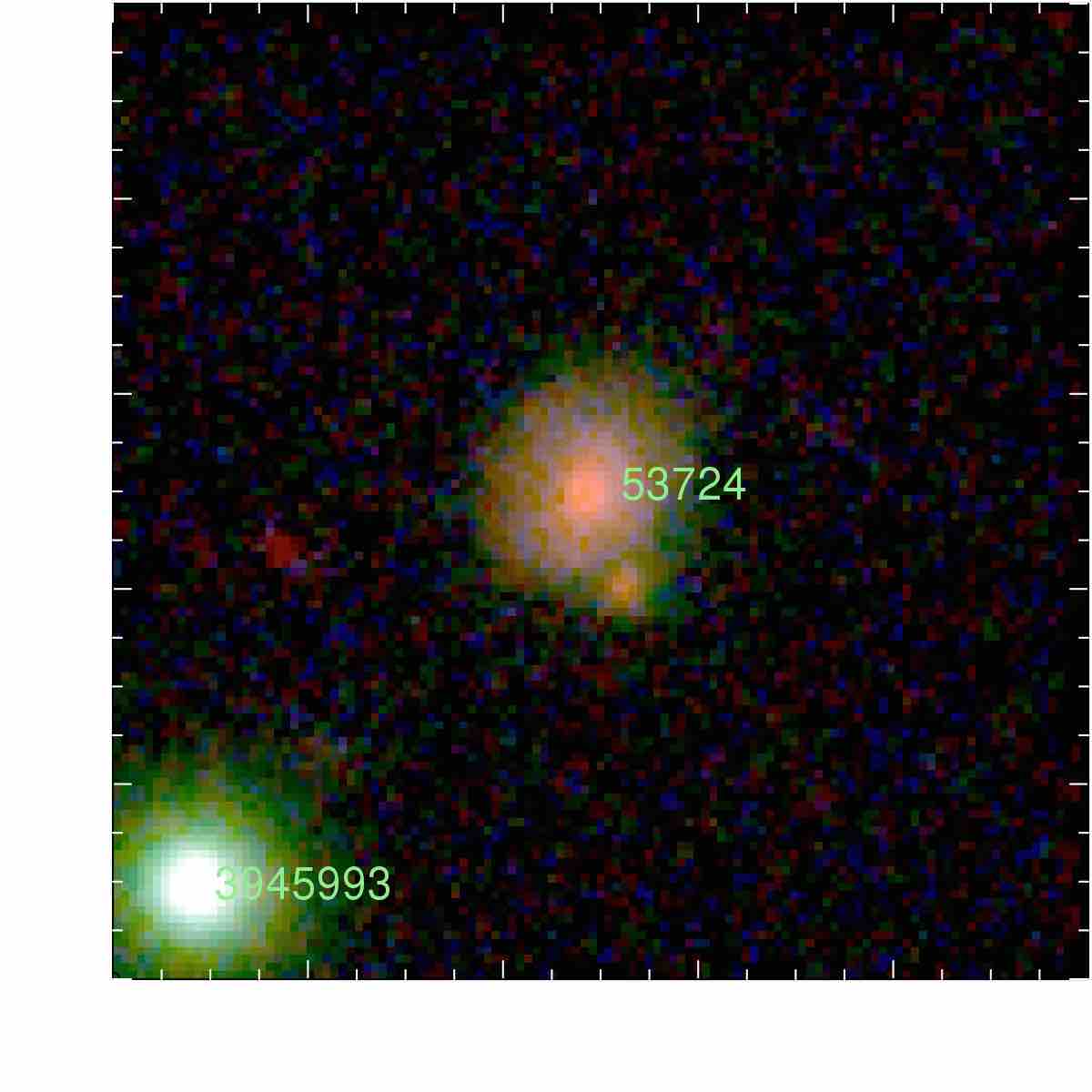} \put (9,85) { \begin{fitbox}{2.25cm}{0.2cm} \color{white}$\bf DBC$ \end{fitbox}} \end{overpic} \\

\includegraphics[width=8.4cm]{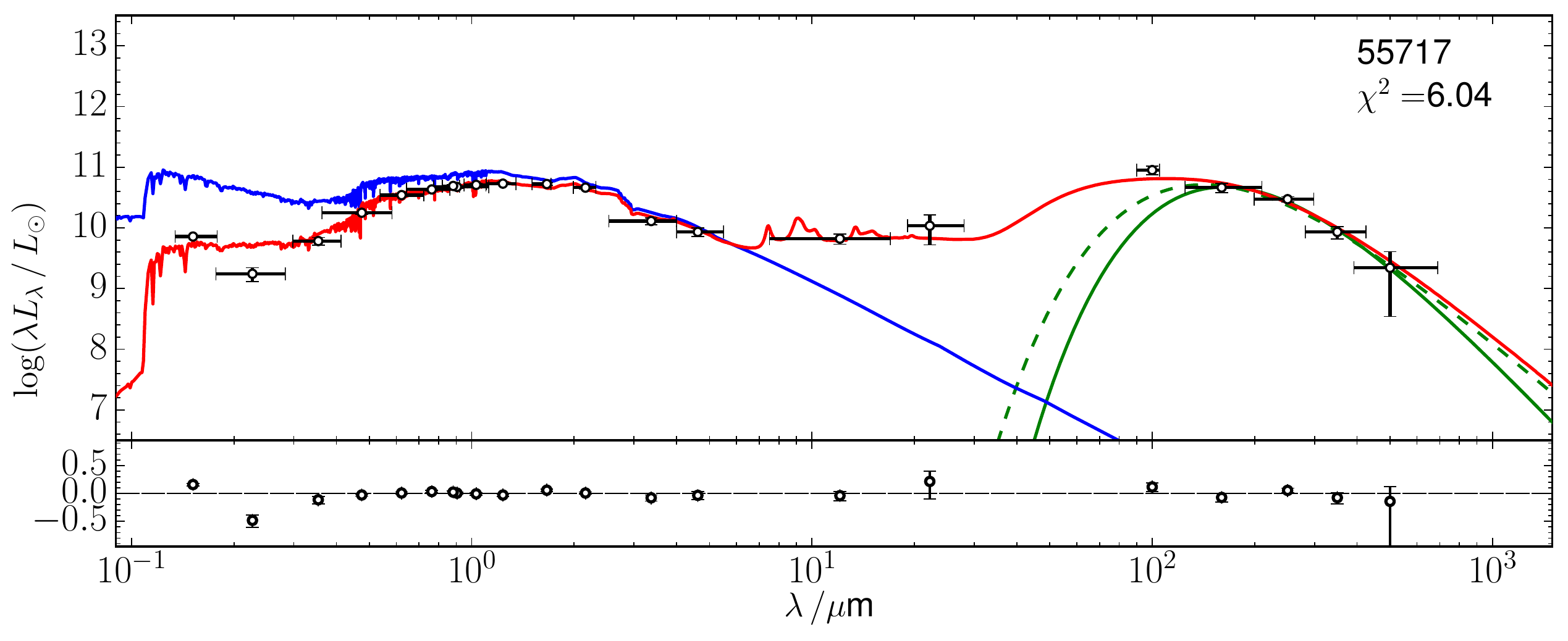} &
\includegraphics[width=5.0cm]{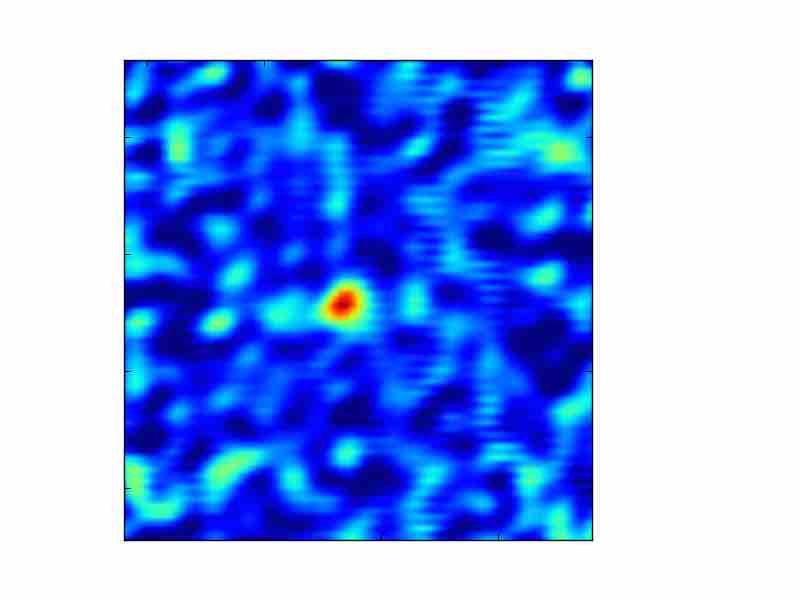} &
\hspace*{-1.2cm}\begin{overpic}[width=3.4cm]{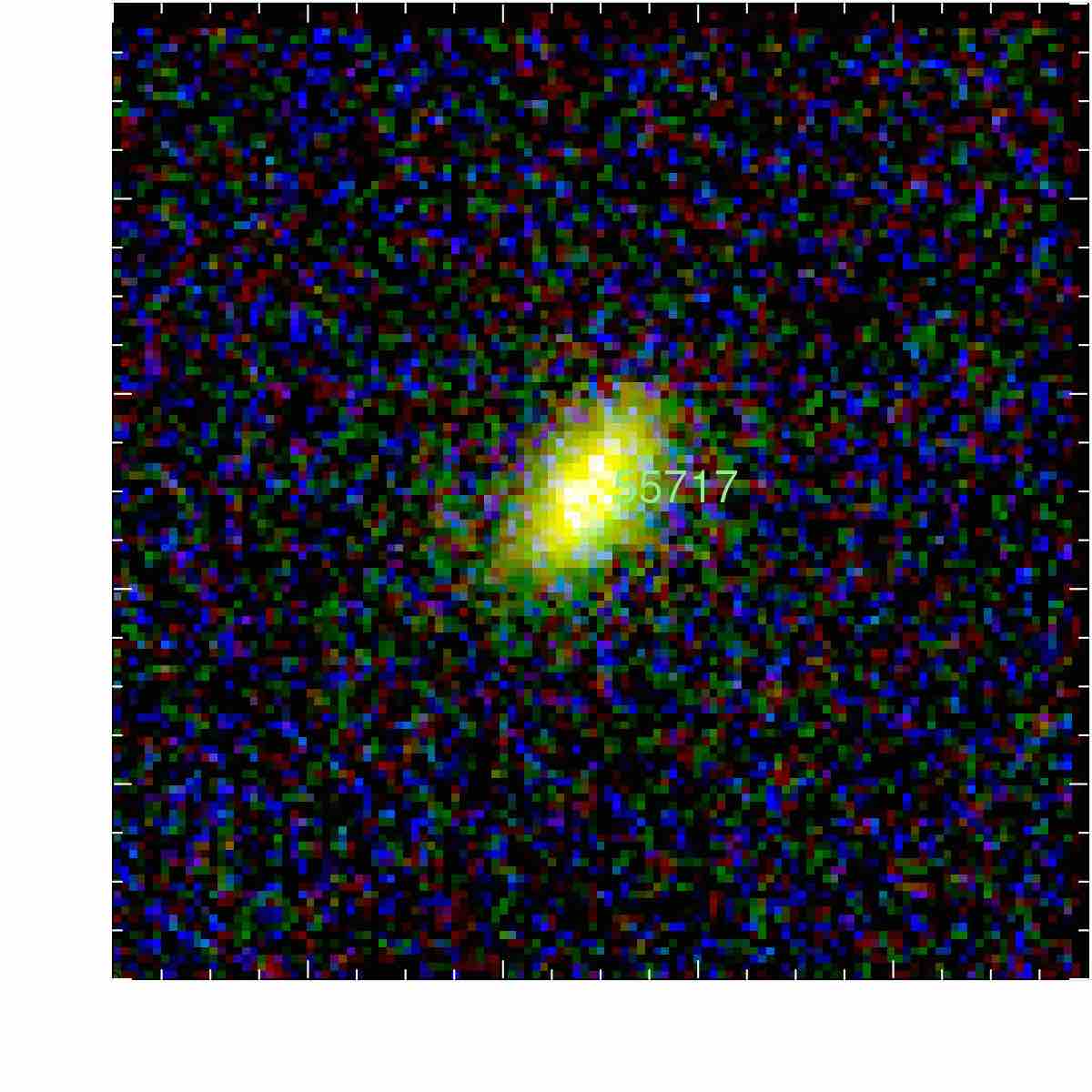} \put (9,85) { \begin{fitbox}{2.25cm}{0.2cm} \color{white}$\bf DB$ \end{fitbox}} \end{overpic} \\ 

\end{array}
$
\caption{ \textit{Left}: Panels show the observed far-UV to submm
  spectral energy distribution for detected galaxies, constructed from
  GAMA/\textit{H}-ATLAS photometry (black circles). We model the
  complete FUV--submm SED using MAGPHYS (\citealp{Dacunha2008}) to
  obtain the best-fitting SED (red line) and the unattenuated SED
  (blue line).  We also model the cold dust SED component between 100
  to 500~$\mu$m by fitting a \citet{hildebrand1983} one-component
  modified blackbody model adopting either a fixed emissivity index of
  $\beta=1.8$ (green dashed line) or $\beta$ varying as a free
  parameter (green solid line). Panels are the residuals between
  the observed and best-fitting SED model template. Each individual
  plot is labelled with the GAMA ID of the target and the $\chi^{2}$
  values corresponding to the best-fitting SED
  template. \textit{Middle}: The collapsed CO cubes.  Each panel is
  ${\rm 40}'' \times {\rm 40}''$ in size and centred on the source
  coordinates.  \textit{Right}: Multi-band images composed by VISTA
  $K$-band (red), SDSS $r$-band (green) and SDSS $u$-band (blue) with
  same size as middle panels. Green numbers are the GAMA ID for the
  objects present in the field of view.}\label{pdrdiaglit}
\end{figure*}

All of our galaxies are present in the GAMA Panchromatic Data
Release\footnote{\url{http://gama-psi.icrar.org}}
(\citealp{Driver2016}) that provides imaging for over 230 deg$^{2}$
with photometry in 21 bands extending from the far-ultraviolet to
far-infrared from a range of facilities that currently includes the
GALaxy Evolution eXplorer (GALEX), Sloan Digital Sky Survey (SDSS),
Visible and Infrared Telescope for Astronomy (VISTA), WISE, and
\textit{Herschel}, meaning that the spectral energy distribution
between 0.1--500~$\mu$m is available for each galaxy. These observed
rest-frame SEDs have all been modelled with the Bayesian SED fitting
code, MAGPHYS \citep{Dacunha2008}, which fits the panchromatic SED
from a library of optical and infrared SEDs derived from a generalised
multi-component model of a galaxy, whilst giving special consideration
to the dust--energy balance (see Fig.~\ref{pdrdiaglit}). Although
Driver et al.\ (in prep.) will present a complete catalogue and
analysis of all the GAMA SEDs modelled with MAGPHYS and the
corresponding best-fit model parameters (see also \citealt{Hughes2016}), 
in our present study we use the derived stellar masses and
their uncertainties, which we calculate from the upper (16th
percentile) and lower (84th percentile) limits of the probability
distribution function associated with the stellar mass given by the
best-fit model, as presented in Table~\ref{Table}. 

We briefly assess the quality of the fitting by comparing the stellar 
masses and IR luminosities derived from MAGPHYS to 
those estimates from our previous study presented in
\citet{Ibar2015}. Both of these parameters demonstrate satisfactory
agreement with a mean scatter of between 0.15 and 0.2 dex (see
\hyperref[fig:comparison]{Fig.~\ref*{fig:comparison}}). In contrast,
our derived star formation rates show a constant systematic offset 
across the parameter range (of a factor of 2, where 
MAGPHYS are lower than those estimates from \citealt{Ibar2015} 
using $L_{\rm IR}$), which likely arises from differences in
SFR definition/calibration (see e.g. \citealt{Kennicutt&Evans2012}). 
However, removing this systematic offset yields a mean scatter of 0.2 dex.

\begin{figure*}
  \includegraphics[width=8.8cm]{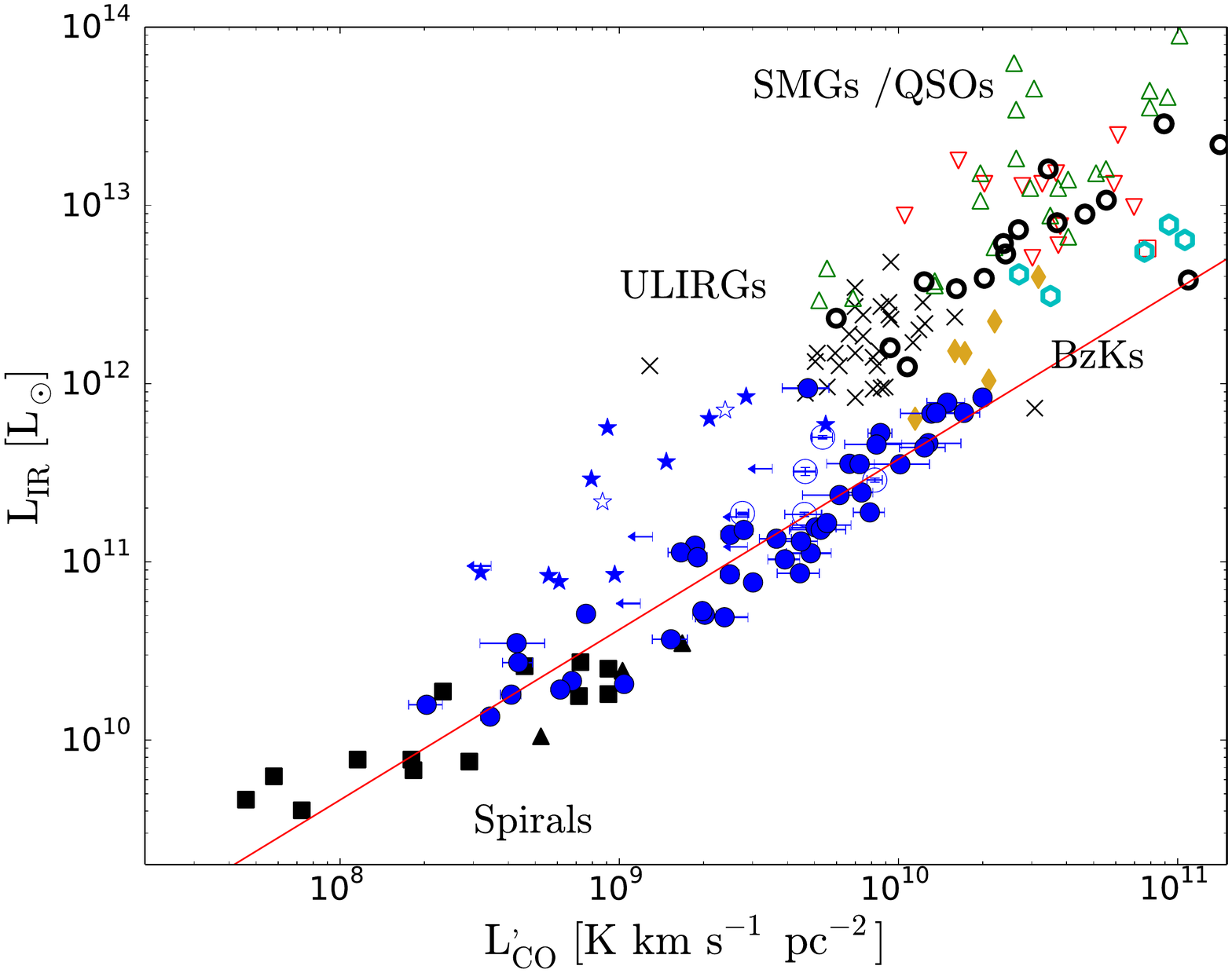}
  \includegraphics[width=8.25cm]{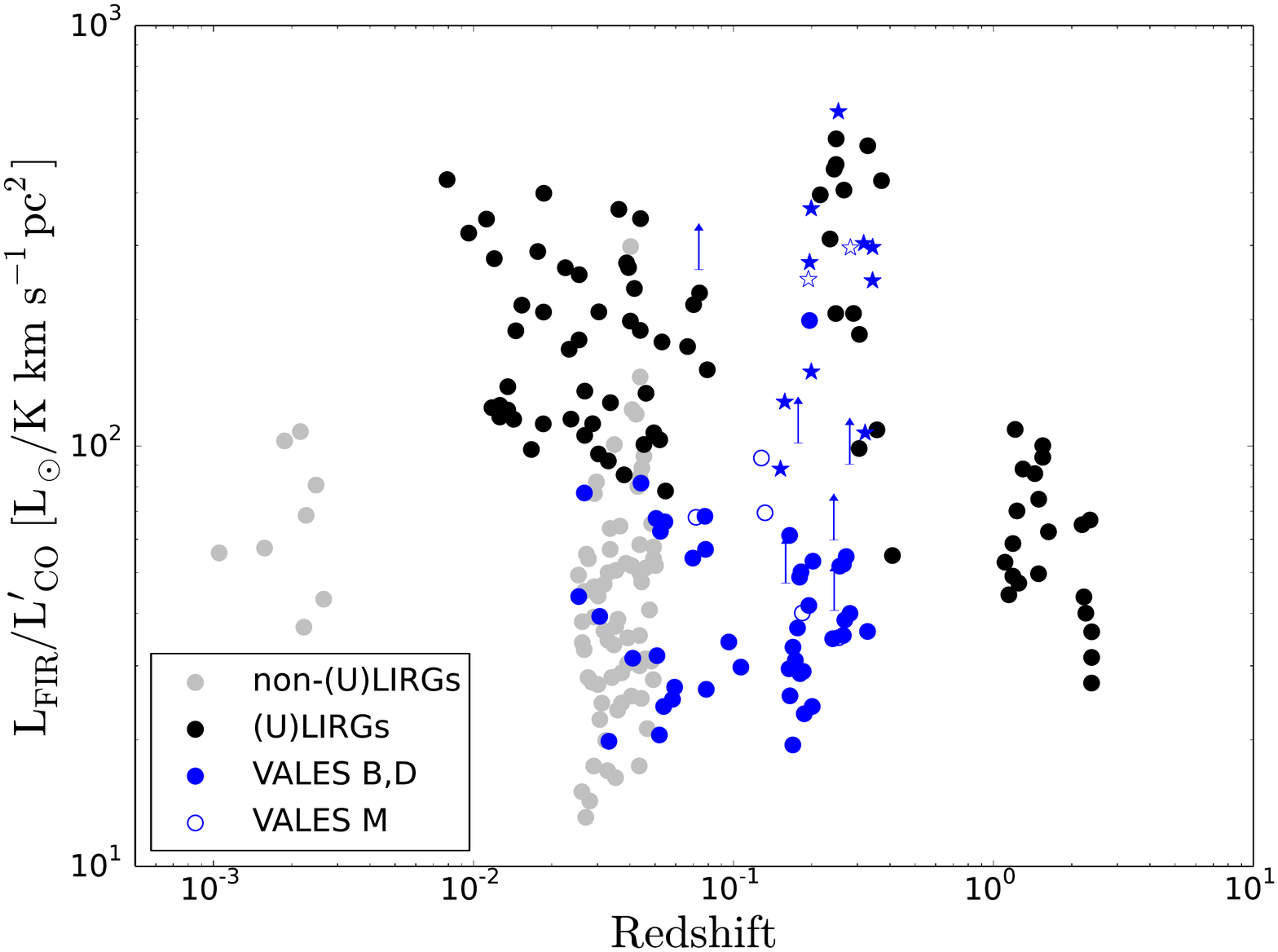}
\caption[Far-IR luminosity versus CO luminosity]{{\it Left:}
  Comparison of the CO and IR luminosities for our spectrally detected
  (blue circles), spectrally undetected but measured with low signal
  to noise in collapsed cubes (stars), and undetected (blue arrows) CO
  galaxies. Blue filled symbols are B-- and D--dominated, while blue
  unfilled symbols are M--dominated galaxies. Our sample is compared
  to other studies taken from the literature, including:
  colour-selected galaxies (BzK; \citealt{Daddi2010a}; gold
  diamons), submillimeter galaxies (SMGs; \citealt{Greve2005},
  \citealt{Daddi2009a,Daddi2009b}; red down triangles,
  \citealt{Frayer2008}; blue empty square, \citealt{Ivison2011}; 
  unfilled cyan hexagons), quasi stellar objects
  (QSOs; \citealt{Riechers2006}; \citealt{Solomon&VandenBout2005};
  green empty triangles), local ULIRGs (\citealt{Solomon1997}; black
  crosses), local spirals (\citealt{Leroy2008,Leroy2009}; black filled
  squares, \citealt{Wilson2009}; black filled triangles) and strongly
  lensed high-redshift dusty galaxies (\citealt{Aravena2016}; black
  unfilled circles). The solid line shows the best-fitting ${L}'_{\rm
    CO}$ versus $L_{\rm IR}$ relation for the detected B-- and
  D--dominated galaxies presented in this work (see
  Eqn.~\ref{LR_Lfir_Lco}). {\it Right:} $ L_{\rm IR}/{L}'_{\rm CO}$
  as function of redshift. Our galaxies (blue symbols) and other
  relevant populations classified as (U)LIRGs (black circles) and
  non-(U)LIRGs (grey circles) taken from \citet{Magdis2012a},
  including; local spirals \citep[COLDGASS][]{Saintonge2011a}; local
  ULIRGs \citep[][]{Solomon1997}; $z\sim0.3$ disks
  \citep[][]{Geach2011}, {\it IRAS} selected $z\sim 0.3$ ULIRGs
  \citep{Combes2011}; and high-$z$ star forming galaxies
  \citep[][]{Daddi2010a,Genzel2010}.}
\label{FIR_luminosity_vs_CO_luminosity}
\end{figure*}

\subsubsection{Morphological properties}
\label{Morphological_Properties}

In order to explore the morphological properties of our galaxies, we
use the GAMA Panchromatic Swarp
Imager\footnote{\url{http://gama-psi.icrar.org/psi.php}} to extract
multi-wavelength imaging from GALEX, SDSS, VISTA and WISE.  We
classify each source (based on visual inspection agreed by four 
members of our team) into three different categories according 
to the prominence of key morphological features: a Bulge (`B'), Disk (`D') and
Merger-Irregular (`M'). If the source presents more than one
morphology, we mark the first letter as the dominant morphology.  If
the source has multiple neighbouring systems, then we add `C' to
denote these `companions'. In the following, we refer to our galaxies
as `B', `D' or `M' dominated galaxies. We also note that this
morphological classification is used to define the most suitable
$\alpha_{\rm CO}$ to then compute ${\rm M}_{\rm H_2}$ (this is
discussed in \S~\ref{Correlations_between_L_FIR_&_L_CO}).

\section{Results and Discussion}
\label{S3_Results}

\subsection{Morphological description}
\label{Morphological_description}

We have made a census for the different optical/near-IR morphologies
present in our sample, according to the morphological classification
scheme explained in \S~\ref{Morphological_Properties}. From a total of
49 spectrally detected sources, we have identified 18 as B--dominated,
26 as D--dominated and 5 as M--dominated galaxies (see the morphology
column in Table~\ref{Table}). By definition, the 5 M--dominated
galaxies present signs of possible morphological disruption: 3
galaxies are clear interacting system (with two or more companions),
and 2 show traces of the late stages of merger events. In the case of
the spectrally undetected galaxies, we have identified 11 as
B--dominated, 4 as D--dominated and 3 as M--dominated galaxies. We do
not identify any clear morphological difference between CO--detected
and --undetected galaxies.

ALMA observations spatially resolve the CO emission in 21 galaxies
(see \S~\ref{CO_emission}).  We calculate the deconvolved {\sc fwhm}
of the semi-major axis ($R_{\rm FWHM}$) using the task {\sc gaussfit}
(within {\sc casa}), finding CO sizes in the range of $3\farcs4 -
15\farcs2$ ($3.7-35.0$\,kpc in physical units), usually resolved at a 
significance of $\sim7\sigma$ (median value). We compare the
optical and CO sizes by using the Petrosian radius in {\it r}-band
($R_{{\rm P,Opt}}$) and the Petrosian radius in CO ($R_{{\rm P,CO}}$),
using Eqn.~(1) from \cite{Shimasaku2001}. We find values for $R_{{\rm 
P,CO}}$ in the range of $1\farcs9 - 5\farcs3$ ($2.8 -14.0$\,kpc),
with an average of $3\farcs6$ ($6.7$\,kpc). For our sample, 
we find that the mean and scatter of the $R_{\rm P,Opt}/R_{\rm P, CO}$ 
distribution are $1.6\pm0.5$ (i.e.\ the CO emission is
typically smaller than the stellar; see
Fig.~\ref{Histogram}). Previous studies have shown a CO-to-optical
ratio of unity for `main-sequence' galaxies, locally
\citep[e.g.][]{Young1995,Regan2001,Leroy2008} and at high-{\it z}
\citep[e.g.][]{Bolatto2015,Tacconi2013}. In a different luminosity
range, \cite{Simpson2015} found that the sizes of Sub-millimetre
Galaxies (SMGs) at $z = 2.6-4$ in optical {\it HST} imaging are around
four times larger than in CO. Taking into account the typical values of $\rm
sSFR/sSFR(MS)$ for our 21 resolved galaxies, we explore
if the optical--to--CO ratio changes as a function of sSFR.  We
perform a Kolmogorov-Smirnov (KS) for both `main sequence' and
`starburst' $R_{\rm P,Opt}/R_{\rm P, CO}$ populations
(using sSFR/sSFR$_{\rm MS}=\, 4.0$ as a threshold; see the
  definition of `main-sequence' in Fig.~\ref{SampleDefinition}),
finding a 90\% probability that both populations come from the same
parent distribution (see Fig.~\ref{Histogram}). This little difference
might be a product of the small deviation seen from the main sequence
(`starburstiness') or the six spatially resolved starburst galaxies presented here.

\subsection{Correlations between $\rm {\mathbf{L_{\rm IR}}}$ and $\rm {\mathbf{{L}'_{\rm CO}}}$}
\label{Correlations_between_L_FIR_&_L_CO}

We compute the CO luminosity following \citet{Solomon&VandenBout2005},

\begin{equation} 
\rm {L}'_{\rm CO} = 3.25 \times 10^{7} S_{\rm CO} \, \Delta \rm v \, \nu^{-2}_{\rm obs} \, D^{2}_{L} \, (1+z)^{-3} [{\rm K\,km\,s^{-1}\,pc^{2}}],
\label{Solomon}
\end{equation}

\noindent
where $S_{\rm CO} \, \Delta \rm v$ is the velocity-integrated flux
density in units of Jy\,km\,s$^{-1}$, $\nu_{\rm obs}$ is the observed
frequency of the emission line in GHz, $\rm D_{\rm L}$ is the
luminosity distance in Mpc, and $z$ is the redshift. We find that the
values for $\rm {L}'_{\rm CO}$ are in the range of ($0.03 -
1.31$)\,$\times\,\,10^{10}$\, ${\rm K\,km\,s^{-1}}$\,pc$^{2}$, with a
median of ($0.3 \pm 0.1$)$\,\times\,10^{10}$\,${\rm
  K\,km\,s^{-1}}$\,pc$^{2}$ (see
Fig.~\ref{FIR_luminosity_vs_CO_luminosity}). We note that our survey
expands the parameter space explored before by previous similar
studies, such as: \citep{Combes2011} at $(0.3-7)\times
10^{10}$\,K\,km\,s$^{-1}$\,pc$^{2}$; \citep{Braun2011} at
$(4-9)\times 10^{10}$\,K\,km\,s$^{-1}$\,pc$^{2}$; \citep{Magdis2014}
at $(0.5-2)\times10^{10}$\,K\,km\,s$^{-1}$ pc$^{2}$. Based on the IR
luminosities derived from the {\it H}-ATLAS photometry
(\S~\ref{FIR_emission}), we find that the ratios between $\rm L_{\rm
  IR}$ and $\rm {L}'_{\rm CO}$ are similar to those found in normal
local star-forming galaxies \citep[e.g. COLDGASS;][]{Saintonge2011a}.
However, our galaxies have smaller $\rm L_{\rm IR}/\rm {L}'_{\rm CO}$
ratio than typical (U)LIRGs in the same redshift range by a factor of
$\sim10$ (see right panel of
Fig~\ref{FIR_luminosity_vs_CO_luminosity}). We compute a linear
regression (in log scale) to the $\rm L_{\rm IR}$ versus $\rm
{L}'_{\rm CO}$ relation for our spectrally detected B-- and
D--dominated galaxies, finding:

\begin{multline}
\log [\rm L_{\rm IR} / L_{\odot}] = (0.95 \pm 0.04) \times \log {L}'_{\rm CO}/[\rm K\,km\,s^{-1}\,pc^{2}]\, \\ + \, (2.0 \pm 0.4).
\label{LR_Lfir_Lco}
\end{multline}

\noindent
This parametrisation is within 1$\sigma$ of the value previously
presented by \cite{Daddi2010}, and supports the clear linearity
between these two quantities. However, this slope is steeper 
compared with that found by \cite{Ivison2011} in high-$z$ SMG, 
local ULIRGs and LIRGs ($\sim 0.5-0.7$).
We confirm that most of our detected galaxies (blue circles in left panel
Fig.~\ref{FIR_luminosity_vs_CO_luminosity}) follow the so-called
`sequence of disks', associated to a long-standing mode of star-formation. 
We remark, however, that if we include in the statistics those
galaxies which are not spectrally detected in CO (blue stars in left
panel Fig.~\ref{FIR_luminosity_vs_CO_luminosity}), although have low
signal to noise emission in collapsed (moment 0 maps), the
scatter in the correlation significantly increases. This indicates
that deeper observations are required to provide details for the 
co-existence of different modes of star-formation.
We suggest that within the $L_{\rm IR}=10^{11-12}\,L_\odot$ range, 
there might be a break (or a significant increment of the scatter) of the linear
relation between the CO and far-IR luminosities \citep{Sargent2014}.

If we add into the statistics all samples presented in Fig.~\ref{FIR_luminosity_vs_CO_luminosity}) in addition to our B-- and
D--dominated galaxies, we find in Eqn.~\ref{LR_Lfir_Lco} a slope and 
normalisation of ($0.99 \pm 0.02$) and ($1.7 \pm 0.2$), respectively. 
Although these parameters are in agreement with previous studies (slope 
\citep[slope $\sim1.0-1.3$; e.g.][]{Genzel2010,Daddi2010}, 
we should highlight the growing evidence that the star formation 
efficiencies increase with redshift (e.g.\ 
\citealt{Santini2014,Rowlands2014,Genzel2015,Scoville2016}), 
therefore combining galaxy samples at different epochs might be an
oversimplification of the analysis (see Fig.~\ref{FIR_luminosity_vs_CO_luminosity}).


\begin{figure*}
  \includegraphics[width=8.8cm]{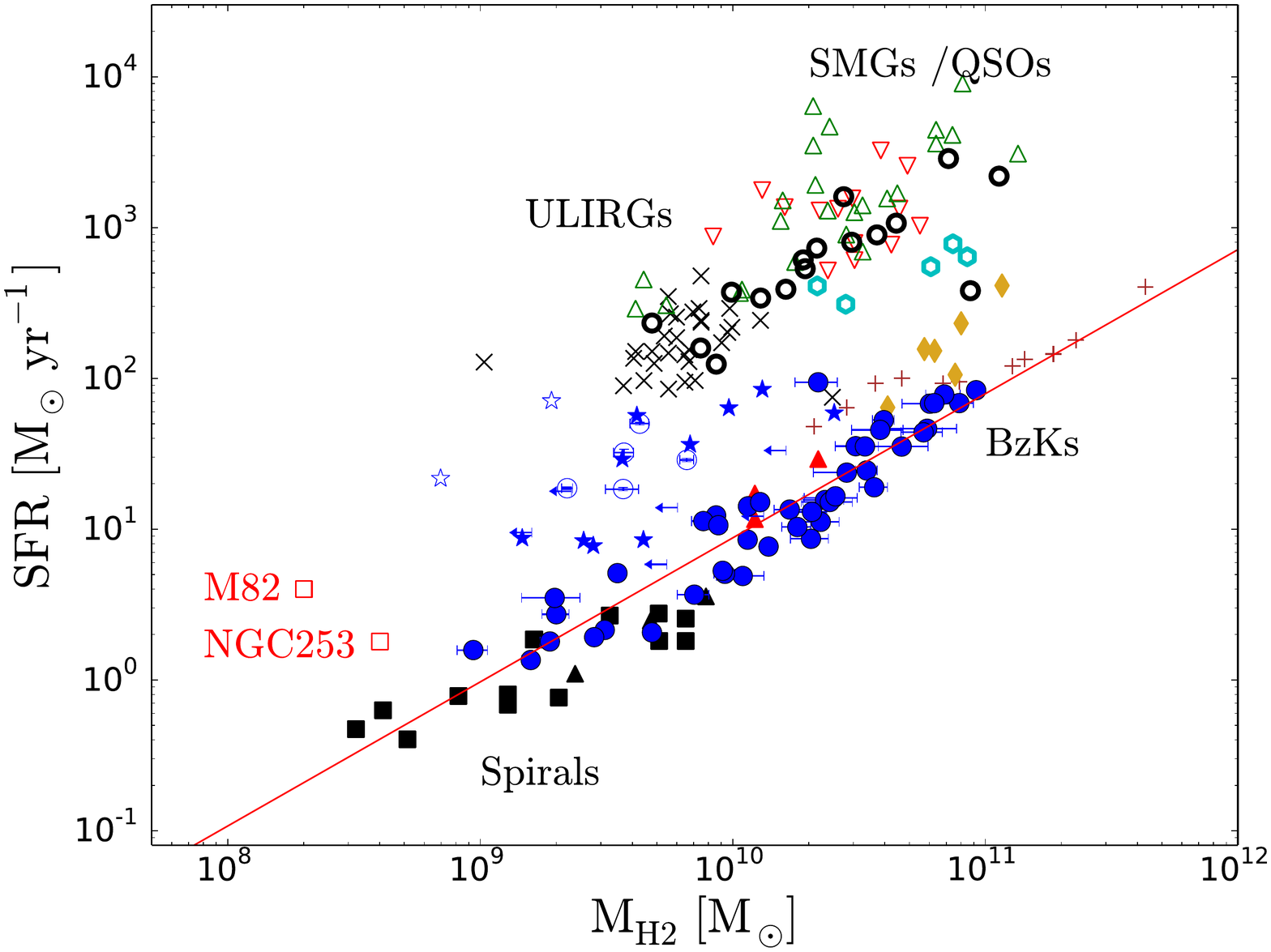}
  \includegraphics[width=8.8cm]{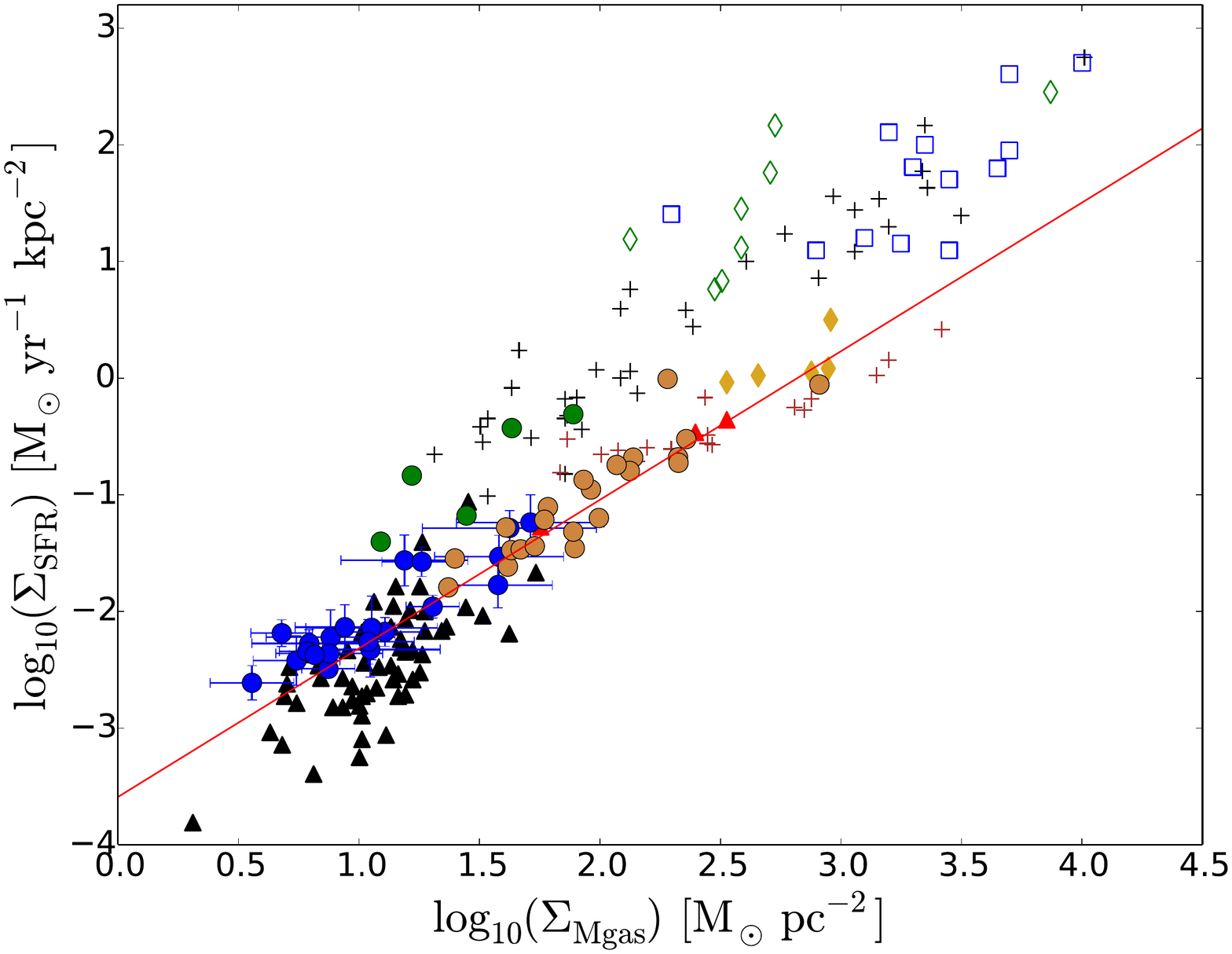}\llap{\put (-91.,15) {\includegraphics[height=2.8cm]{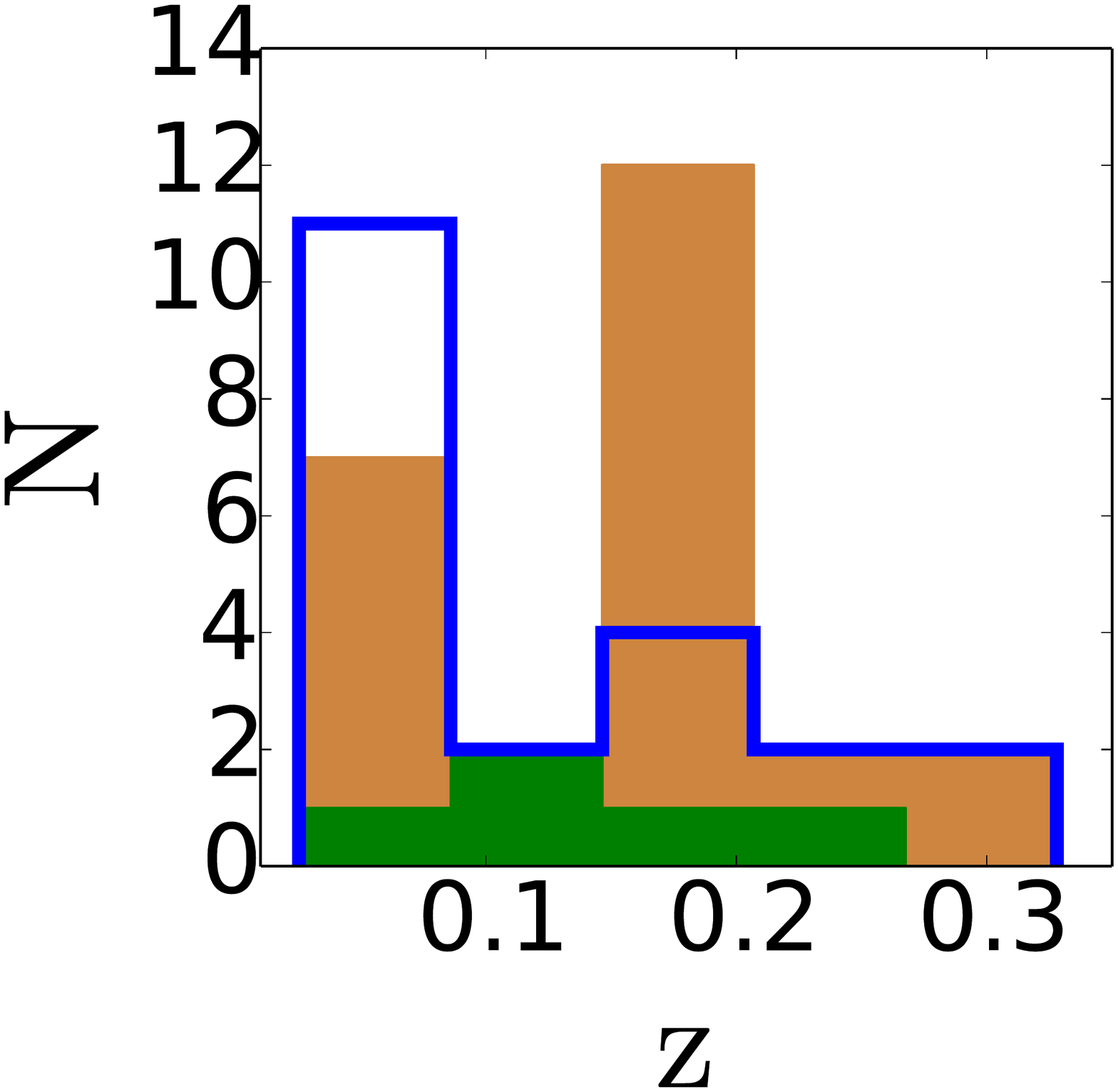}}}\llap{\put (-218.,112.) {\includegraphics[height=2.04cm]{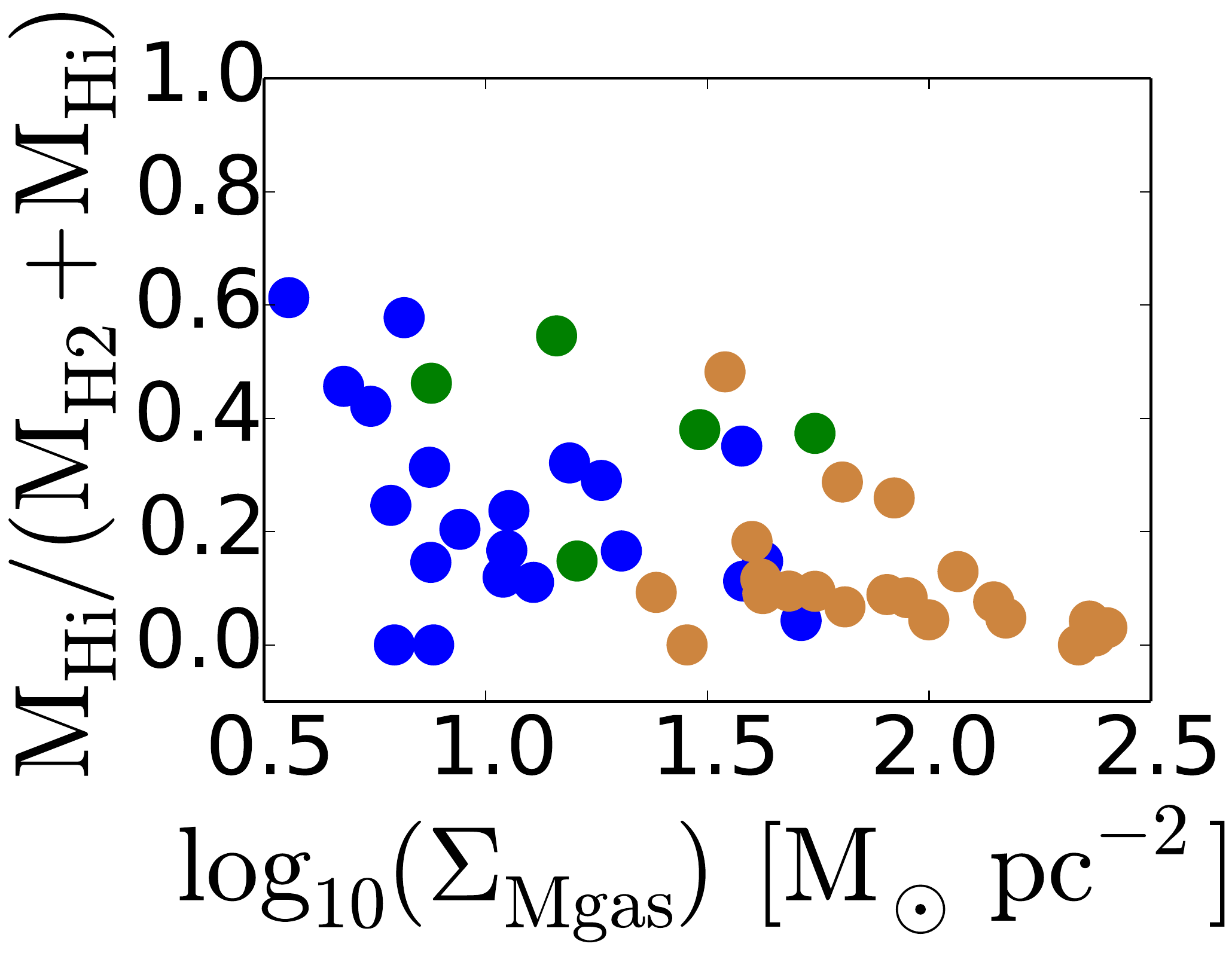}}  }

  \caption[Sigma_SFRvsSigma_Mgas]{{\it Left}: The SFR versus $\rm
    M_{\rm H2}$ for our galaxies compared to previous works taken 
    from the literature (assuming a Chabrier's IMF but considering 
    the same $\alpha_{\rm CO}$ used in the original works). We
    show our spectrally detected (blue circles), spectrally undetected
    but measured with low signal to noise in collapsed cubes (stars),
    and undetected (blue arrows) CO galaxies (see
    \S~\ref{Source_Properties} for details). Blue filled symbols are
    B-- and D--dominated galaxies, while blue unfilled symbols are
    M--dominated galaxies. Figure contains BzK galaxies
    (\citealt{Daddi2010a}; gold diamonds), $z\sim0.5$ disk
    galaxies (\citealt{Salmi2012}; red filled triangles), $z = 1-2.3$
    normal galaxies (\citealt{Tacconi2010}; brown crosses), SMGs
    (\citealt{Greve2005,Frayer2008,Daddi2009a,Daddi2009b}; red 
    down triangles, \citealt{Ivison2011}; unfilled cyan hexagons), QSOs
    (\citealt{Riechers2006}; green triangles), local ULIRGs
    (\citealt{Solomon1997}; black crosses), strongly lensed
    high-redshift dusty star forming galaxies (\citealt{Aravena2016};
    black unfilled circles), and local spirals (\citealt{Leroy2009};
    black filled squares; \citealt{Wilson2009}; black filled
    triangles). M82 and NGC~253 are also shown
    \citep{Weib2001,Houghton1997,Kaneda2009}. The solid red line is the
    best fit for the B-- and D--dominated galaxies presented in this
    work (see Eqn.~\ref{LR_Lfir_MH2}). {\it Right}: The SFR surface
    density as a function of the gas mass (atomic and molecular)
    surface density for the 21 detected sources that can be spatially
    resolved in CO (blue circles; see \S~\ref{The_Schmidt-Kennicutt_relation}), 
    23 detected B-- and D--dominated which are unresolved in CO but 
    resolved in $R$-band (brown circles), and M--dominated unresolved in CO galaxies
    (green circles). Filled triangles are (U)LIRGS and spiralx galaxies 
    from the sample of \cite{Kennicutt1998}, brown crosses are $z = 1-2.3$
    normal galaxies \citep{Tacconi2010}. The empty blue squares and 
    light green diamonds are SMGs from \cite{Bouche2007} and \cite{Bothwell2009}, 
    respectively. The red line is the best linear fit to all of our
    B-- and D--dominated galaxies (see Eqn.~\ref{LR_SigmaSFR_SigmaGASS}). 
    The inset figures at the upper left and bottom right corners are: 
    (1) the relative contribution estimated by $\rm M_{H {\textsc i}}$ 
    (following Eqn.~\ref{Zhang}) to the total gas mass ($\rm M_{H_2}+M_{H {\textsc i}}$) 
    versus $\Sigma_{\rm M_{gas}}$. This figure demonstrates the decreasing effect 
    that $\rm M_{H {\textsc i}}$ has at $\Sigma_{\rm M_{gas}} > 10  M_{\odot}$ pc$^{-2}$; 
    (2) the redshift distribution of our VALES galaxies (blue, yellow and green circles), clearly showing 
    that most of the spatially unresolved galaxies are the ones which are more distant. }
    \label{Sigma_SFRvsSigma_Mgas}
\end{figure*}
\noindent

For those spectrally identified CO galaxies, we do not identify any
clear variation of the $\rm L_{\rm IR}/{L}'_{\rm CO}$ ratio as a
function of redshift (up to $z=0.35$; see right panel of
Fig.~\ref{FIR_luminosity_vs_CO_luminosity}). Our results are
consistent with previous works that have shown a constant average $\rm
L_{\rm IR}/{L}'_{\rm CO}$ in `main-sequence' galaxies up to $z\sim
0.5$ \citep[e.g.][]{Santini2014}. The scatter on
the $\rm L_{\rm IR}/{L}'_{\rm CO}$ ratio, however, increases if non-spectrally
detected galaxies are included, an effect which is mainly dominated by
the $0.15<z<0.35$ galaxy population.

Using $\rm {L}'_{\rm CO}$, we compute the molecular gas mass ($\rm
M_{\rm H_2}$) assuming an $\alpha_{\rm CO}$ conversion factor
dependent on the morphological classification (see \S
\ref{Morphological_Properties}). We adopt $\alpha_{\rm CO}$
\,=\,4.6\,(K\,km\,s$^{-1}$\,pc$^{2}$)$^{-1}$ for the B-- and
D--dominated galaxies (which includes contribution of He;
\citealt{Solomon&VandenBout2005}), while $\alpha_{\rm CO}$
\,=\,0.8\,(K\,km\,s$^{-1}$\,pc$^{2}$)$^{-1}$ for mergers/interacting
galaxies. For B-- and D--dominated galaxies, we find $\rm M_{\rm H_2}$
values in the range of $\log({\rm M_{\rm H_2}/M_{\odot}}) = 8.9 - 10.9$ with a
median of $\log({\rm M_{\rm H_2}/M_{\odot}}) = 10.31 \pm 0.1$, while for
M--dominated galaxies values are in the range of $\log({\rm M_{\rm H_2}/M_{\odot}}) 
= 9.3 - 9.8$ with a median of $\log({\rm M_{\rm H_2}/M_{\odot}}) = 9.6 \pm 0.2$.

Performing a linear regression to the $\rm SFR$ versus $\rm M_{\rm H_2}$ 
using our B-- and D--dominated galaxies, we obtain:

\begin{multline}
\log \rm [SFR/ M_{\odot}\, yr^{-1}] = (0.95\pm0.04) \times \log (M_{\rm H_2}/[{\rm M_{\odot}}]) \\- (8.6\pm0.5).
\label{LR_Lfir_MH2}
\end{multline}

In this work, we are significantly
increase the number of previously detected galaxies at 
$\log [\rm M_{\rm H_2}/M_{\odot}] \sim 9-11$. Our sample complements the `gap'
between local spirals and `normal' high-z colour-selected galaxies
(Fig.~\ref{Sigma_SFRvsSigma_Mgas}). We note that our M--dominated
galaxies are shifted towards higher $\rm SFR$s, and closer
to the local ULIRGs described by \citet{Solomon1997}. At lower
redshifts almost all galaxies follow a tight relationship between $\rm
SFR$ and $\rm M_{H2}$, nevertheless we identify that galaxies at the
upper side of the redshift distribution ($0.15<z<0.35$) tend to show a
higher scatter in this correlation. The scatter is larger when low
signal-to-noise detections are included (see stars in left panel of
Fig.~\ref{Sigma_SFRvsSigma_Mgas}). 
If we combine our observations with all samples shown in the left panel
of Fig.~\ref{Sigma_SFRvsSigma_Mgas}), 
then in Eqn.~\ref{LR_Lfir_MH2} we obtain a slope of
$1.08\pm0.02$ with a normalisation of $-9.8\pm0.2$.

In both cases, one of the main factors controlling the scatter 
of the correlation is the different $\alpha_{\rm CO}$ conversion 
factor chosen for M--dominated galaxies. It is worth
pointing out that deeper ALMA observations to spectrally detect the CO
emission for all of our galaxies would probably confirm a population
of optically unresolved ULIRGs-like galaxies with high star formation
efficiencies (filling the `gap' between spirals and ULIRGs local
galaxies). As shown in left panel of 
Fig.~\ref{FIR_luminosity_vs_CO_luminosity} (see blue stars), this
population could significantly affect the slope and the scatter of the
correlation.

The major uncertainty in our molecular gas mass estimates
originates from the assumption of the $\alpha_{\rm CO}$ conversion
factor. Indeed, assuming a different $\alpha_{\rm CO}$ can change $\rm
M_{\rm H2}$ by over a factor of six (around 500 times higher than
observational errors). On the other hand, considering the metallicity
range of our sample \citep[$12 + \log_{10}(\rm O/H)=
  8.7-9.2$;][]{Ibar2015} and using the $\alpha_{\rm CO}$
parametrisation for star-forming galaxies made by \citep[][see their
  Eqn.~8]{Genzel2015}, we find that the $\alpha_{\rm CO}$ could vary
by a factor of four with a tendency to lower values than
$4.6$\,(K\,km\,s$^{-1}$\,pc$^{2}$)$^{-1}$. In the left panel of
Fig.~\ref{Sigma_SFRvsSigma_Mgas}, galaxies can be shifted in position
using a different $\alpha_{\rm CO}$, producing an artificial bimodal
behaviour for the star-formation activity in these galaxies. This can
clearly affect the reliability for the existence of `disk' and
`starburst' sequences. In {\color{blue} Molina et al.}\ (in prep.), 
we use kinematic arguments to confront the bimodality of the
$\alpha_{\rm CO}$ conversion factor (e.g.\ \citealt{Downes1998}, 
\citealt{Sandstrom2013}).

\subsection{The Schmidt-Kennicutt relation}
\label{The_Schmidt-Kennicutt_relation}

We introduce ${M{\rm gas} = M_{\rm H_2}+M_{\rm H {\textsc i}}}$ as the
total mass content in molecular and atomic gas. As we do not have direct 
$\rm M_{H\,{\textsc i}}$ observations, we estimate these using Eqn.~4
from \citealt{Zhang2009}:

\begin{multline}
\log [{\rm M_{H {\textsc i}}/M_{\star}}]=-1.73238\,(g-r)+ 0.215182\,\mu_{i}\\ -4.08451,
\label{Zhang}
\end{multline}

\noindent
where is {\it g} and {\it r} are the photometric magnitudes in those
filters, and $\mu_{i}$ is the $i$--band surface brightness (SDSS filters). 
This approximation provides a $0.31$\,dex scatter for the estimate. For our
sample, using Eqn.~\ref{Zhang} we find that the contribution is in 
general small (although non negligible) with a mean ratio of 
$\rm M_{H {\textsc i}}/M_{H2}\sim0.2$ (see Fig.~\ref{Sigma_SFRvsSigma_Mgas}).

Our 49 spectrally--detected CO sources have SFRs in the range of $\rm
1-94\,M_{\odot}$\,yr$^{-1}$, with a median value of $15\pm1\,{\rm
  M_{\odot}}$\,yr$^{-1}$. For those which are spatially resolved in CO
(21 in total), we estimate the SFR and ${\rm M_{gas}}$ surface
densities by dividing the measured values by the area of a two-sided
disk ($2\,\pi\,R^2_{\rm FWHM}$), where $R_{\rm FWHM}$ is the
deconvolved {\sc fwhm} of the semi-major axis measured in ALMA CO
images (see Table~\ref{Table}). In Fig.~\ref{Sigma_SFRvsSigma_Mgas} (right),
we show the Schmidt-Kennicutt relation
\citep[][]{Schmidt1959,Kennicutt1998} comparing our samples with
previous ones taken from the literature.

For those spectrally detected and spatially resolved CO galaxies, we obtain values for
$\log[\Sigma_{\rm SFR}/(\rm M_{\odot}\,yr^{-1}\,kpc^{-2})]$ in the range 
of $-2.61$ and $-1.23$\,, with a median of $-{2.18\pm0.1}$\,. Most of
the spatially resolved ones (16) are D-- dominated while the rest (5)
are B-- dominated galaxies. The B-- and D--dominated galaxies have on
average 3 times higher $\Sigma_{\rm SFR}$ than local spiral galaxies,
but around 30-70 times lower values than normal BzK galaxies at high-$z$. On
the other hand, for the same spectrally detected CO galaxies,
$\log[\Sigma_{\rm Mgas}/(\rm M_{\odot}$\,pc$^{-2})]$ values range 
between $0.55$ and $1.71$\,, with a median value of $1.04 \pm 0.29$\,. 
In terms of CO emission, we do not find any
remarkable difference between our B-- and D--dominated galaxies
(although they do have different morphological optical features); the sample has
on average 2 times greater $\Sigma_{\rm Mgas}$ than local spiral
galaxies. 

According to estimations of the molecular and atomic
gas content in nearby galaxies \citep{Bigiel&Blitz2012}, there is
a strong evidence that the atomic gas saturates in column gas densities
higher than $\sim 10 \, \rm M_{\odot}$ pc$^{-2}$. This is attributed to
a natural threshold for the atomic to molecular gas transitions
\citep{Krumholz2009,Sternberg2014}. As shown in 
Fig.~\ref{Sigma_SFRvsSigma_Mgas}, our atomic gas 
estimates decrease as a function of $\Sigma_{\rm M{gas}}$ 
(as expected by the $\rm M_{H {\textsc i}}$ saturation), although 
the large scatter dominating Eqn.~\ref{Zhang} still predicts a 
non-negligible fraction of atomic gas above 
$ \Sigma_{\rm M{gas}}>10^{10}\, \rm M_\odot$\,pc$^{-2}$-.

Something to highlight is that left and right panels of Fig.~\ref{Sigma_SFRvsSigma_Mgas} should behave similarly, nevertheless we find that most of our galaxies (blue circles) which are lying near the BzK population disappear after considering our spatially resolved CO selection criterion. To explore this further, we compare the median values of $\rm M_{H_2}$ and $R_{\rm CO}$ using resolved galaxies (in CO) at different redshift bins (centred at 0.07 and 0.2) in order to identify a possible evolution in physical size and mass of gas content. Previous studies have shown that at fixed $\rm M_{\star}$ the averaged effective radius vary as $R_{\rm eff} \propto (1 + z)^{-0.8}$ \citep{Magdis2012a}. If we assume that the stars and the molecular gas follow the same spatial distribution (which is actually not the case; see Fig.~\ref{Histogram}), the measured CO sizes are expected to decrease by a factor of $1.1$ between $z = 0.07$ and $z = 0.2$. This variation is not sufficient to explain what we observe in Fig.~\ref{Sigma_SFRvsSigma_Mgas}. 

Considering the results coming from \S~\ref{Morphological_description} to estimate $\Sigma_{\rm SFR}$ and $\Sigma_{\rm Mgas}$ for spatially unresolved CO galaxies, we used the Petrosian optical radius divided by the mean $R_{\rm P,Opt}/R_{\rm P, CO}=1.6$ ratio found for our galaxies (note that $R_{\rm FWHM}$ and $R_{\rm P, CO}$ differ by only $\sim\,$2\%). Brown and green circles in Fig.~\ref{Sigma_SFRvsSigma_Mgas} correspond to CO-unresolved B--/D--dominated and M--dominated galaxies, respectively. The inset panel in Fig.~\ref{Sigma_SFRvsSigma_Mgas} shows clearly that most of the spatially unresolved CO galaxies are those which are more distant (at $0.15 < z < 0.35$). This analysis demonstrates that our sample perfectly complements the parameter space in the Schmidt-Kennicutt relation that joins the local spiral galaxies with those `normal' at high-$z$.

Using our data, we perform a linear regression in
Fig.~\ref{Sigma_SFRvsSigma_Mgas} using all B--/D--dominated galaxies
which are spatially resolved in CO (21 sources) and unresolved in CO
but resolved in the optical (23 sources). Being aware of the possible
biases introduced by our H\,{\sc i} estimates, we provide a
parametrisation for two cases; $M_{\rm H2}$ and $
M_{\rm gas}$,

\begin{equation}
\frac{\Sigma_{\rm SFR}}{{\rm M_{\odot}\, yr^{-1}kpc^{-2}}} =
\begin{cases} 
  $($1.16  \pm 0.05$)$ \times \, \log $[$\Sigma_{\rm M_{gas}}/{\rm M_{\odot}\,pc^{-2}}$]$  \\ \quad\quad- (3.3 \pm 0.1)\\
  $($1.27 \pm 0.05$)$ \times \, \log $[$\Sigma_{\rm M_{H2}}/{\rm M_{\odot}\,pc^{-2}}$]$  \\ \quad\quad- (3.6 \pm 0.1)
\end{cases}
\label{LR_SigmaSFR_SigmaGASS}
\end{equation}

These results are consistent with previously analyses using 
ensembles of clumps composing galaxies at 
$z=1-2$ (e.g.\ \citealt{Freundlich2013, Genzel2013}) and 
star-forming disks with near-solar metallicities \citep[slope $\sim 1.0 - 1.3$; e.g.][]{Bigiel2008,Kennicutt2008,Schruba2011,Leroy2008}. Presenting
Eqn.~\ref{LR_SigmaSFR_SigmaGASS} for the molecular and total gas helps
to see the way the $\rm M_{\rm H {\textsc i}}$ could affect our
results. In particular we find that the slope is flatter when $\rm
M_{gas}$ are used. We highlight that given that these galaxies have
been selected in the far-IR, our results are not significantly
affected by the assumptions in geometrical modelling of the dust as
previous in \cite{Genzel2013} and \cite{Freundlich2013} studies.

If we take into account the samples of galaxies belonging to the
sequence of disks shown in right panel of
Fig.~\ref{Sigma_SFRvsSigma_Mgas} (including $z=1.0 - 2.3$ normal
galaxies, BzK and $z \sim 0.5$ disks, spiral galaxies and both our
spatially resolved and unresolved B--/D--dominated galaxies) in our
linear regression of Eqn.~\ref{LR_SigmaSFR_SigmaGASS}, we obtain 
a slope of $(1.26 \pm 0.02)$ and a
normalisation of $(3.6 \pm 0.2)$. Mentioned before, this should be
taken with caution as there is growing evidence for a cosmic
evolution of the star formation efficiency, effective radius and 
gas content \citep[e.g.][]{Magdis2012a}, implying that the
combination of samples at different epochs might be mixing
intrinsically different populations.

\begin{figure*}
  \includegraphics[width=8.65cm]{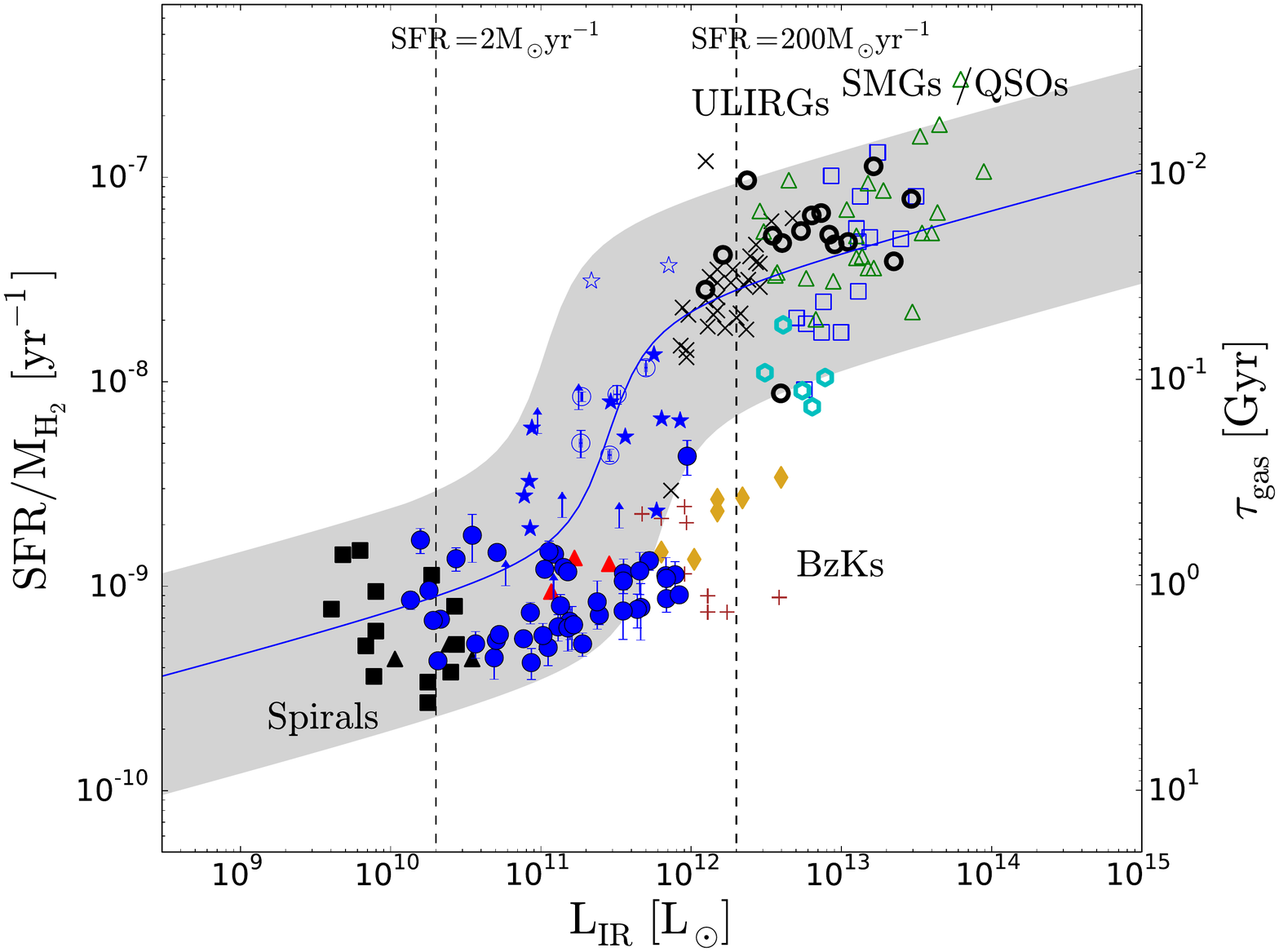} 
  \includegraphics[width=8.8cm]{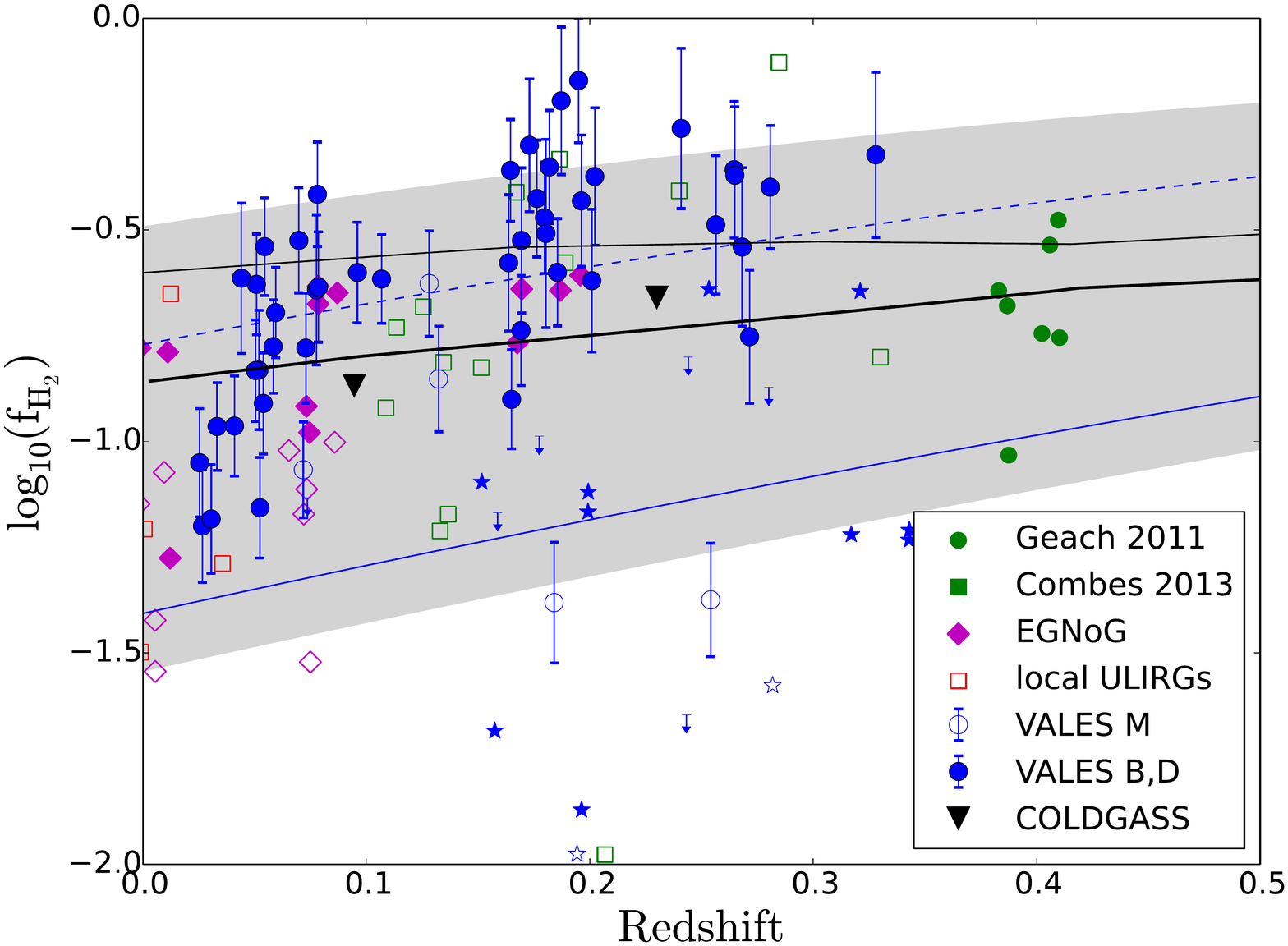} \\
  \caption[EGNoG]{{\it Left}: $\rm SFE$ versus IR luminosities
    following the same symbols and colours shown in the left panel of
    Fig.~\ref{FIR_luminosity_vs_CO_luminosity}. Blue solid line is the
    best-fitted parametrisation considering all the galaxy samples at $z<0.4$
    presented in the figure (see Eqn.~\ref{SFR_Mh2}) while the light
    grey area is the region within a 1$\sigma$ scatter of
    $0.5$\,dex. This parametrisation suggests a break on the star
    formation efficiency at $L_{\rm IR}=10^{11-12}\,L_\odot$. 
    {\it Right}: Molecular gas fraction
    ($f_{\rm M_{H2}}$) versus redshift for different samples of
    star-forming galaxies. The upside down triangles are the average
    values from COLDGASS survey. The dashed and solid blue curves are
    the average behaviour for normal galaxies and the expected
    location for starburst galaxies by \citet{Bauermeister2013},
    respectively. The light grey area is the region of the
    `main-sequence' for star-forming galaxies when adopting an average
    $\tau_{\rm M_{H2}}({\rm MS})\sim2.2$\,Gyr
    \citep[e.g.][]{Bigiel2008,Leroy2008}. The two black curves show
    semi-analytic prescriptions for galaxy formation by
    \cite{Lagos2011}, which correspond to mass halo models of $M_{\rm
      h} = 10^{11}$ and $\rm 10^{12} M_{\odot}\,{\rm h}^{-1}$, from
    thinnest to thickest, respectively.}
  \label{EGNoG}
\end{figure*}

\subsection{Star formation efficiency}
\label{Star_Formation_Efficiency}

We define the star formation efficiency as ${\rm SFE = SFR}/ \rm
M_{\rm H2}$ and the $\rm M_{H2}$ consumption time-scale ($\tau_{\rm
  M_{H2}}$) as ${\rm SFE^{-1}}$. In the left panel of
Fig.~\ref{EGNoG}, we show the SFE vs.\ $L_{\rm IR}$ for our galaxies
including other galaxy samples taken from the literature. Two
distinctive types of galaxies are evident: those galaxies that present
a long-standing mode of star formation with $\tau_{\rm
  M_{H2}}\sim1.3$\,Gyr; and those affected by a much faster starburst
processes with $\tau_{\rm M_{H2}}\sim0.2$\,Gyr.  We identify a
significant number of sources that are located in the `transition
zone' between both the sequence of disks and the sequence of
starbursts (left panel in Fig.~\ref{EGNoG}), with SFEs in the range of
$4.3 - 11.7 \, \rm Gyr^{-1}$, and with a median of $8.5 \pm 0.1 \, \rm
Gyr^{-1}$ (e.g.\ similar to those SMGs at $z = 0.22 - 0.25$ presented by \citealt{Ivison2011}).  
These sources seem to suggest the co-existence of both
modes of star-formation at intermediate efficiencies.  We note that 39
of our sources are located in the long-lasting mode,
with SFEs in the range of $0.42 - 4.32 \, \rm Gyr^{-1}$,  and with a 
median of $0.8 \pm 0.1 \, \rm Gyr^{-1}$.


We highlight the evidence for galaxies located in the 
`transition zone' between `normal' and `starburst' 
confirming a break in SFE at $\rm L_{IR} \approx 10^{11-12}\,L_{\odot}$, 
which could indicate the possibility of a single evolutionary path 
(with a large scatter) rather than a sharp bimodal behaviour in 
evolution (see \S~\ref{Correlations_between_L_FIR_&_L_CO}). This is in 
agreement with previous findings by \cite{Sargent2014}. We propose an
empirical best-fit parametrisation to describe the dependence of the
$\rm SFE$ on $\rm L_{IR}$ (based in all $z<0.4$ samples included
in left panel of Fig.~\ref{EGNoG}):

\noindent
\begin{multline}
\log [{\rm SFR/M_{H2}}\,({\rm yr}^{-1})] = 0.19 \times ({\rm \log \,[{L_{IR}}/L_\odot]}-\phi)+\alpha \\
+\beta \arctan[\rho\,(\log {\rm [L_{IR}/L_\odot]}-\phi) ],   
\label{SFR_Mh2}
\end{multline}


\noindent
where $\alpha=-8.26$, $\beta=-0.41$, $\rho=-4.84$ and $\phi=11.45$. This 
function has a scatter of $\sigma=0.5$\,dex.

In this work we highlight that our method to compute the molecular gas masses is
directly using CO($1-0$), not assuming any particular conversion for
high-$J$ transitions, facilitating the interpretation of the results.
\cite{Scoville2016}, for example, obtain different star formation
modes for normal and starburst/SMG galaxies, which are likely affected
by the different methods behind the computation of both gas masses at
high redshift (using higher-$J$ CO transitions) and different
$\alpha_{\rm CO}$ for each type of galaxy. 

In spite of the remaining uncertainties on the assumptions used to
derive $\rm M_{H2}$, the detection of galaxies in the `transition
zone', including spectrally detected/undetected with $\alpha_{\rm
  CO}=4.6$\,(K\,km\,s$^{-1}$\,pc$^{2}$)$^{-1}$ and mergers with a
smaller $\alpha_{\rm CO}$ by a factor of six, supports the scenario of
a smooth increase of SFE as a function of $\rm L_{IR}$. 
This has been hinted before in galaxies at $z=1.6$ by
\citet{Silverman2015}, where as they explored sources above the `main
sequence' they tentatively concluded a smooth increase of SFE instead
of a bimodality in star formation modes.

\subsection{Evolution of the molecular gas fraction}
\label{Evolution_Molecular_Gas_Fraction}

In this section, we explore the evolution of the molecular gas
fraction ($f_{\rm {H_2}}$) as a function of redshift. We introduce 
the molecular gas mass to the stellar mass ratio as:

\begin{equation}
\frac{M_{\rm {H_2}}}{M_{\star}} = \tau_{\rm M_{H2}} \times {\rm sSFR},
\label{gas_stellar_ratio}
\end{equation}

\noindent
thus, the gas fraction can be calculated as $f_{\rm {H2}} = M_{\rm
  H2}/(M_{\rm H2}+M_{\star})$. We find our sample covers a wide range
of values $f_{\rm {H_2}}\sim0.04 - 0.71$ (for those sources spectrally 
CO-detected above $5\sigma$), which are similar to those
shown by the Evolution of molecular Gas in Normal Galaxies
\citep[EGNoG][]{Bauermeister2013} survey in normal star-forming
galaxies. Compared to local ULIRGs, where $f_{\rm {H_2}}$ ranges at
3-5\% (e.g.\ \citealt{Solomon1997,Gao&Solomon2004,Chung2009}), our gas
fractions are typically higher than those, although if we only
consider those M--dominated galaxies we find similar values to those
seen in local ULIRGs (lying near the lower $f_{\rm {H_2}}$ envelope
defined for starburst galaxies by \citealt{Bauermeister2013}). 
We identify that our $f_{\rm H_2}$ values show a tendency to increase as a 
function of redshift (see Fig.~\ref{EGNoG}) -- probably product 
of a selection effect induced by the {\it Herschel} detectability 
(these are dusty galaxies). \cite{Bethermin2014} suggest a rapid 
increase of the average fraction of molecular gas with redshift, 
similarly to what we find in our analysis. Based on recent works
\citep[e.g.][]{Dunne2011,Hardcastle2016,Wang2016}, there is growing 
evidence for a rapid galaxy evolution at low redshifts. Particularly,
\cite{Dunne2011} find evidence for fast evolution of the dust mass
content of galaxies up to $z=0.5$, a result that suggest that the molecular gas
content also rapidly evolves in samples of {\it Herschel}-selected 
galaxies. Actually, using galaxies taken from this same work, 
\cite{Hughes2016} suggests that this rapid evolution goes together 
with a significant increment of the gas density (up to $z=0.2$), aided 
by predictions from photo-dissociation region modelling.

In the right panel of Fig.~\ref{EGNoG}, the two black curves show
semi-analytic prescriptions for galaxy formation and evolution of the
molecular ISM computed by \cite{Lagos2011}, based on an empirical star
formation law to estimate the molecular gas mass. Black solid lines
correspond to mass halo models of $M_{\rm h} = 10^{11}$ and $\rm
10^{12}\,M_{\odot}\,{\rm h}^{-1}$, from thinnest to thickest,
respectively, that trace our B-- and D--dominated galaxies. 
These models suggest that molecular gas mass content and 
SFR densities increase as a function of redshift, in rough 
agreement with what we see in our B-- and D--dominated galaxies. On
the other hand, the M--dominated galaxies are apparently associated to
more massive dark matter halos of $\sim10^{12}\,{\rm M_{\odot}}$.

\section{Conclusions}
\label{S4_Conclusion}

In this paper, we present the VALES survey -- one of the 
largest samples of CO detected normal star-forming galaxies 
up to $z=0.35$. We use the ALMA
telescope to estimate the molecular gas content via CO(1--0) emission
in a sample of 67 dusty star-forming galaxies. Sources are bright
far-IR emitters ($S_{\rm 160\mu m}\geq100\,$mJy; $L_{\rm
  IR}\approx10^{10-12}\,M_\odot$) selected from the equatorial fields
of the {\it H}-ATLAS survey (with SFRs in the range of $1.4-94.2$
$\rm M_{\odot} \, yr^{-1}$). We have spectroscopically detected 49
galaxies (72\,\% of the sample) with a $>5\sigma$ CO peak line
significance and 12 others are detected in collapsed spectra at low
signal to noise. We find that 21 galaxies are spatially resolved in CO (with
physical sizes in the range of $3.7-35.1$ kpc, allowing a multi-wavelength
exploration over a wide parameter space. We summarise our main results
as follows:

\begin{itemize}

\item  Based on a visual inspection to the optical/near-IR photometry 
  of the 49 spectrally CO-detected galaxies, we classify $36\%$ as being
  dominated by a (B)ulge morphology, $53\%$ as a (D)isk morphology,
  and $11\%$ show evidence for a (M)erger event or interaction. 
  We spatially resolve 21 galaxies which on average show optical-to-CO size ratios of 
  $\sim1.6\pm0.5$, hence the molecular gas is more concentrated
  towards the central regions than the stellar component.

\item Our sample explores the ${\rm L}'_{\rm CO}$ luminosity range of 
  $0.3\times10^{10}$\,K\,km\,s$^{-1}$\,pc$^{2}$ expanding the 
  parameter space to fainter values than previous relevant CO 
  surveys at similar redshifts. Aided by the morphological 
  classification (assuming standard $\alpha_{\rm CO}$ conversion factors 
  for disks and mergers), we estimate a range of $\rm M_{\rm
  H_2}$=\,$10^{8.9-10.9}$\,$\rm M_{\odot}$ for Bulge- and 
  Disk-dominated galaxies while $10^{9.3-9.8}$\,$\rm M_{\odot}$ for 
  Merger-dominated galaxies.

\item We explore the Schmidt-Kennicutt relation using 
  values for global $\Sigma_{\rm SFR}$ and $\Sigma_{\rm
    Mgas}$ derived from a combination of CO and optical radii. 
    Our sample perfectly complements the parameter space that
  joins both, local and high-$z$ `normal' galaxy samples. We find a
  best linear fit with a power law slope of $1.16\pm0.05$ and $1.27\pm0.05$ when
  using ${\rm M_{\rm H2}}$ and ${\rm M_{\rm H2}+H\,{\textsc i}}$,
  respectively.
  
\item The median SFE of our sample is $8.5$\,Gyr$^{-1}$ (with values
  in the range of $0.4-11.7$ Gyr$^{-1}$). Even though most of our
  galaxies follow a long standing mode of star-formation
  activity, we provide evidence for a population with efficiencies in
  the `intermediate valley' between normal star formation in disks and more rapid/violent starburst episodes. Within some galaxies there may be a mixture of star-formation modes occurring at the same time. We propose the existence of a
  continuous transition for the star formation efficiencies as a
  function of far-IR luminosities. 

\item We estimate the molecular gas fraction, finding values in the
  range of $f_{\rm H_2} = 0.06 - 0.34$. Our observations suggest 
  a strong increment of the gas fraction as a function of redshift 
  (up to $z=0.35$), faster than semi-analytical models predictions. 
  This rapid evolution might be affected by the selection criteria as 
  we are selecting {\it Herschel}-detected galaxies with preferentially 
  high dust content.

\end{itemize}
\noindent


To conclude, we note that one of the main uncertainties in this work
is produced by the different CO conversion factors between
CO luminosity and molecular gas mass, which undoubtedly impact our 
estimates. Two of the most evident drivers of these uncertainties are 
the dynamical state of the galaxies and the metallicity. We are putting 
special emphasis on tackling the uncertainty on the molecular gas mass 
estimates using: dynamical modelling of resolved galaxies
({\color{blue}Molina et al.\ in prep}), the physical conditions of the
interstellar molecular gas within them (\citealt{Hughes2016}), and the
calibration between the dust continuum luminosity and interstellar gas
content \citep{Hughes2017}.

\section*{Acknowledgements}

EI and TMH acknowledge CONICYT/ALMA funding Program in Astronomy/PCI
Project N$^\circ$:31140020. M.A. acknowledges partial support from 
FONDECYT through grant 1140099. This paper makes use of the following ALMA
data: ADS/JAO.ALMA 2012.1.01080.S \& ADS/JAO.ALMA 2013.1.00530.S. ALMA
is a partnership of ESO (representing its member states), NSF (USA)
and NINS (Japan), together with NRC (Canada), NSC and ASIAA (Taiwan),
and KASI (Republic of Korea), in cooperation with the Republic of
Chile. The Joint ALMA Observatory is operated by ESO, AUI/NRAO and
NAOJ. The {\it Herschel}-ATLAS is a project with Herschel, 
which is an ESA space observatory with science instruments provided 
by European-led Principal Investigator consortia and with important 
participation from NASA. The {\it H}-ATLAS website is \url{http://www.h-atlas.org/}. 
PACS has been developed by a consortium of institutes led by MPE (Germany) and including UVIE (Austria); KU Leuven, CSL, IMEC (Belgium); CEA, LAM (France); MPIA (Germany); INAF-IFSI/OAA/OAP/OAT, LENS, SISSA (Italy); IAC (Spain). This development has been supported by the funding agencies BMVIT (Austria), ESA-PRODEX (Belgium), CEA/CNES (France), DLR (Germany), ASI/INAF (Italy), and CICYT/MCYT (Spain). SPIRE has been developed by a consortium of institutes led by Cardiff University (UK) and including Univ. Lethbridge (Canada); NAOC (China); CEA, LAM (France); IFSI, Univ. Padua (Italy); IAC (Spain); Stockholm Observatory (Sweden); Imperial College London, RAL, UCL- MSSL, UKATC, Univ. Sussex (UK); and Caltech, JPL, NHSC, Univ. Colorado (USA). This development has been supported by national funding agencies: CSA (Canada); NAOC (China); CEA, CNES, CNRS (France); ASI (Italy); MCINN (Spain); SNSB (Sweden); STFC, UKSA (UK); and NASA (USA). GAMA is a joint European-Australasian project based around a spectroscopic campaign using the Anglo-Australian Telescope. The GAMA input catalogue is based on data taken from the Sloan Digital Sky Survey and the UKIRT Infrared Deep Sky Survey. Complementary imaging of the GAMA regions is being obtained by a number of independent survey programmes including GALEX MIS, VST KiDS, VISTA VIKING, WISE, {\it Herschel}-ATLAS, GMRT and ASKAP providing UV to radio coverage. GAMA is funded by the STFC (UK), the ARC (Australia), the AAO, and the participating institutions. The GAMA website is \url{http://www.gama-survey.org/ }. D.R. acknowledges support from the National Science Foundation under grant number AST-1614213 to Cornell University. H.D. acknowledges financial support from the Spanish Ministry of Economy and Competitiveness (MINECO) under the 2014 Ramón y Cajal program MINECO RYC-2014-15686. LD, SJM and RJI acknowledge support from European Research Council Advanced Investigator Grant COSMICISM, 321302; SJM and LD are also supported by the European Research Council Consolidator Grant {\sc CosmicDust} (ERC-2014-CoG-647939, PI H\,L\,Gomez).

\bibliographystyle{mnras}
\bibliography{paper}

\begin{landscape}
\begin{table}
\begin{center}
\resizebox{\linewidth}{!}{ 

\begin{tabular}{ccccccccccccccccccccc}
\hline\hline
GAMA ID & Source & RA (J2000) & Dec (J2000) &  $\rm z_{spec}$ & $\rm \log [L_{\rm IR}/L_{\odot}]$   & SFR  & $\rm \log[M_{\star}/M_{\odot}]$  & sSFR  & $\nu_{\rm obs}$ & v$_{\rm obs}$  &  v$_{\rm FWHM}$ &$S_{\rm CO}\, \Delta \rm v$  & $\rm {L}'_{\rm CO}/10^{10}$  & $\rm \log[M_{\rm H2}/M_{\odot}]$  & $\rm R_{\rm FWHM}$ & SFE  & $\Sigma_{\rm gas}$  & $\Sigma_{\rm SFR}$  & $\tau_{\rm gas}$ & Morphology \\
& & & & &  & [$\rm M_{\odot}$ yr$^{-1}$] &  & [Gyr$^{-1}$] & [GHz] & [km s$^{-1}$] & [km s$^{-1}$] & [Jy km s$^{-1}$] & [K km s$^{-1}$ pc$^{2}$] & &  [kpc] & [Gyr$^{-1}$] & [$\rm \log M_{\odot}$ pc$^{-2}$] & [$\log \rm M_{\odot}$ yr$^{-1}$ kpc$^{-2}$] & [Gyr]\\
\hline
214184	 & HATLASJ083601.5+002617 & 08:36:01.6 & +00:26:18.1 & 0.0332 & 10.31 $\pm$ 0.02 & 2.06 $\pm$ 0.1 & 10.59 $\pm$ 0.1 & 0.05 $\pm$ 0.01 & 111.545 & 54 $\pm$ 11 & 306 $\pm$ 27 & $^{b}$20.6 $\pm$ 0.85 & 0.104 $\pm$ 0.004 & 9.68 $\pm$ 0.018 & 10.1 $\pm$ 0.8 & 0.43 $\pm$ 0.03 & 1.3 $\pm$ 0.3 & $-$2.1 $\pm$ 0.3 & 2.318 $\pm$ 0.15 & DB	 \\
3895257	 & HATLASJ083745.1$-$005141 & 08:37:45.2 & $-$00:51:40.9 & 0.0306 & 10.13 $\pm$ 0.03 & 1.35 $\pm$ 0.09 & 10.35 $\pm$ 0.12 & 0.06 $\pm$ 0.02 & 111.843 & 18 $\pm$ 14 & 202 $\pm$ 33 & $^{b}$8.1 $\pm$ 0.62 & 0.034 $\pm$ 0.003 & 9.2 $\pm$ 0.034 & 6.4 $\pm$ 0.9 & 0.85 $\pm$ 0.09 & 1.3 $\pm$ 0.1 & $-$2.0 $\pm$ 0.1 & 1.17 $\pm$ 0.119 & D	 \\
208589	 & HATLASJ083831.9+000045 & 08:38:31.9 & +00:00:45.0 & 0.0781 & 11.15 $\pm$ 0.01 & 14.17 $\pm$ 0.4 & 10.27 $\pm$ 0.11 & 0.77 $\pm$ 0.2 & 106.911 & 38 $\pm$ 3 & 172 $\pm$ 8 & 8.8 $\pm$ 0.68 & 0.25 $\pm$ 0.019 & 10.06 $\pm$ 0.034 & 6.6 $\pm$ 0.6 & 1.23 $\pm$ 0.1 & 1.2 $\pm$ 0.3 & $-$2.1 $\pm$ 0.3 & 0.81 $\pm$ 0.067 & DBC	 \\
345647	 & HATLASJ084139.5+015346 & 08:41:39.5 & +01:53:46.7 & 0.0736 & 10.98 $\pm$ 0.01 & 9.48 $\pm$ 0.25 & 10.29 $\pm$ 0.11 & 0.49 $\pm$ 0.13 & 107.561 &  &  & <1.3   & <0.032 & & & & & & & \\  
417395	 & HATLASJ084217.7+021222 & 08:42:17.9 & +02:12:23.4 & 0.096 & 10.93 $\pm$ 0.04 & 8.51 $\pm$ 0.79 & 10.53 $\pm$ 0.11 & 0.25 $\pm$ 0.07 & 105.185 & $-$35 $\pm$ 5 & 186 $\pm$ 12 & $^{b}$5.7 $\pm$ 0.45 & 0.249 $\pm$ 0.02 & 10.059 $\pm$ 0.034 & 17.3 $\pm$ 1.4 & 0.74 $\pm$ 0.09 & 1.1 $\pm$ 0.2 & $-$2.3 $\pm$ 0.2 & 1.346 $\pm$ 0.163 & DBC	 \\
300757	 & HATLASJ084305.0+010858 & 08:43:05.1 & +01:08:56.0 & 0.0777 & 11.05 $\pm$ 0.03 & 11.31 $\pm$ 0.78 & 10.41 $\pm$ 0.17 & 0.44 $\pm$ 0.17 & 106.975 & $-$40 $\pm$ 4 & 187 $\pm$ 11 & 5.9 $\pm$ 0.6 & 0.166 $\pm$ 0.017 & 9.883 $\pm$ 0.044 &  & 1.48 $\pm$ 0.18 &  &  & 0.676 $\pm$ 0.083 & DBC	 \\
371334	 & HATLASJ084350.7+005535 & 08:43:50.8 & +00:55:34.8 & 0.0729 & 11.03 $\pm$ 0.01 & 10.61 $\pm$ 0.27 & 10.64 $\pm$ 0.12 & 0.24 $\pm$ 0.07 & 107.478 & 30 $\pm$ 17 & 283 $\pm$ 47 & 7.7 $\pm$ 0.64 & 0.191 $\pm$ 0.016 & 9.943 $\pm$ 0.036 &  & 1.21 $\pm$ 0.11 &  &  & 0.826 $\pm$ 0.072 & DBC	 \\
345754	 & HATLASJ084428.3+020349 & 08:44:28.4 & +02:03:49.8 & 0.0254 & 10.25 $\pm$ 0.01 & 1.8 $\pm$ 0.04 & 10.29 $\pm$ 0.12 & 0.09 $\pm$ 0.03 & 112.401 & 46 $\pm$ 4 & 203 $\pm$ 20 & 14.0 $\pm$ 1.18 & 0.041 $\pm$ 0.003 & 9.276 $\pm$ 0.037 &  & 0.95 $\pm$ 0.08 &  &  & 1.049 $\pm$ 0.091 & DBC	 \\
386263	 & HATLASJ084428.3+020657 & 08:44:28.4 & +02:06:57.4 & 0.0786 & 11.01 $\pm$ 0.03 & 10.33 $\pm$ 0.64 & 10.78 $\pm$ 0.11 & 0.17 $\pm$ 0.05 & 106.865 & 5 $\pm$ 10 & 349 $\pm$ 25 & 13.6 $\pm$ 1.78 & 0.392 $\pm$ 0.051 & 10.255 $\pm$ 0.057 &  & 0.57 $\pm$ 0.08 &  &  & 1.744 $\pm$ 0.253 & DC	 \\
278475	 & HATLASJ084630.7+005055 & 08:46:30.9 & +00:50:53.3 & 0.1323 & 11.51 $\pm$ 0.02 & 32.13 $\pm$ 1.69 & 10.36 $\pm$ 0.12 & 1.42 $\pm$ 0.39 & 101.802 & $-$4 $\pm$ 9 & 324 $\pm$ 23 & 5.5 $\pm$ 0.5 & 0.463 $\pm$ 0.042 & 9.569 $\pm$ 0.039 &  & 8.67 $\pm$ 0.91 &  &  & 0.115 $\pm$ 0.012 & M	 \\
3624571	 & HATLASJ084907.0$-$005139 & 08:49:07.1 & $-$00:51:37.7 & 0.0698 & 11.18 $\pm$ 0.01 & 15.1 $\pm$ 0.33 & 10.48 $\pm$ 0.11 & 0.5 $\pm$ 0.13 & 107.752 & $-$3 $\pm$ 4 & 244 $\pm$ 9 & 12.3 $\pm$ 0.98 & 0.279 $\pm$ 0.022 & 10.108 $\pm$ 0.035 &  & 1.18 $\pm$ 0.1 &  &  & 0.85 $\pm$ 0.07 & BC	 \\
376293	 & HATLASJ085111.5+013006 & 08:51:11.4 & +01:30:06.9 & 0.0594 & 10.72 $\pm$ 0.02 & 5.29 $\pm$ 0.26 & 10.56 $\pm$ 0.1 & 0.15 $\pm$ 0.04 & 108.805 & 16 $\pm$ 15 & 372 $\pm$ 35 & $^{b}$12.2 $\pm$ 0.46 & 0.198 $\pm$ 0.007 & 9.96 $\pm$ 0.016 & 13.9 $\pm$ 0.7 & 0.58 $\pm$ 0.04 & 1.0 $\pm$ 0.3 & $-$2.1 $\pm$ 0.3 & 1.724 $\pm$ 0.107 & DB	 \\
371789	 & HATLASJ085112.9+010342 & 08:51:12.8 & +01:03:43.7 & 0.0267 & 10.2 $\pm$ 0.01 & 1.58 $\pm$ 0.04 & 10.14 $\pm$ 0.12 & 0.11 $\pm$ 0.03 & 112.28 & $-$13 $\pm$ 4 & 143 $\pm$ 11 & 6.3 $\pm$ 0.87 & 0.02 $\pm$ 0.003 & 8.972 $\pm$ 0.06 &  & 1.68 $\pm$ 0.24 &  &  & 0.595 $\pm$ 0.083 & DB	 \\
323772	 & HATLASJ085234.4+013419 & 08:52:33.9 & +01:34:22.7 & 0.195 & 11.92 $\pm$ 0.01 & 83.43 $\pm$ 2.18 & 10.57 $\pm$ 0.12 & 2.25 $\pm$ 0.62 & 96.417 & 134 $\pm$ 16 & 340 $\pm$ 38 & 10.7 $\pm$ 0.07 & 1.999 $\pm$ 0.012 & 10.963 $\pm$ 0.003 &  & 0.91 $\pm$ 0.02 &  &  & 1.102 $\pm$ 0.03 & BC	 \\
323855	 & HATLASJ085340.7+013348 & 08:53:40.7 & +01:33:47.9 & 0.041 & 10.28 $\pm$ 0.03 & 1.92 $\pm$ 0.14 & 10.36 $\pm$ 0.12 & 0.08 $\pm$ 0.02 & 110.722 & 26 $\pm$ 14 & 140 $\pm$ 32 & $^{b}$8.0 $\pm$ 0.37 & 0.061 $\pm$ 0.003 & 9.451 $\pm$ 0.02 & 11.2 $\pm$ 2.3 & 0.68 $\pm$ 0.06 & 1.9 $\pm$ 0.3 & $-$1.0 $\pm$ 0.3 & 1.471 $\pm$ 0.125 & DC	 \\
600024	 & HATLASJ085346.4+001252 & 08:53:46.3 & +00:12:52.4 & 0.0504 & 10.71 $\pm$ 0.01 & 5.11 $\pm$ 0.15 & 10.31 $\pm$ 0.12 & 0.25 $\pm$ 0.07 & 109.73 & 8 $\pm$ 12 & 259 $\pm$ 29 & 6.5 $\pm$ 0.13 & 0.076 $\pm$ 0.002 & 9.543 $\pm$ 0.009 & 5.5 $\pm$ 0.7 & 1.46 $\pm$ 0.05 & 2.0 $\pm$ 0.3 & $-$1.2 $\pm$ 0.3 & 0.684 $\pm$ 0.025 & D	 \\
600026	 & HATLASJ085356.5+001256 & 08:53:56.3 & +00:12:56.3 & 0.0508 & 10.33 $\pm$ 0.03 & 2.14 $\pm$ 0.16 & 10.01 $\pm$ 0.12 & 0.21 $\pm$ 0.06 & 109.683 & 31 $\pm$ 5 & 138 $\pm$ 13 & $^{b}$5.7 $\pm$ 0.14 & 0.068 $\pm$ 0.002 & 9.493 $\pm$ 0.011 & 9.5 $\pm$ 2.5 & 0.69 $\pm$ 0.06 & 1.1 $\pm$ 0.2 & $-$2.1 $\pm$ 0.2 & 1.45 $\pm$ 0.116 & DB	 \\
301346	 & HATLASJ085406.0+011129 & 08:54:05.9 & +01:11:30.4 & 0.0441 & 10.54 $\pm$ 0.02 & 3.5 $\pm$ 0.14 & 9.79 $\pm$ 0.13 & 0.57 $\pm$ 0.17 & 110.405 & $-$5 $\pm$ 17 & 255 $\pm$ 41 & 4.8 $\pm$ 1.25 & 0.043 $\pm$ 0.011 & 9.295 $\pm$ 0.113 & 4.5 $\pm$ 1.2 & 1.77 $\pm$ 0.47 & 0.8 $\pm$ 0.2 & $-$2.6 $\pm$ 0.2 & 0.564 $\pm$ 0.149 & DBC	 \\
386720	 & HATLASJ085450.2+021207 & 08:54:50.2 & +02:12:08.3 & 0.0583 & 10.7 $\pm$ 0.02 & 5.05 $\pm$ 0.27 & 10.66 $\pm$ 0.1 & 0.11 $\pm$ 0.03 & 108.909 & 29 $\pm$ 15 & 387 $\pm$ 36 & 12.9 $\pm$ 1.23 & 0.202 $\pm$ 0.019 & 9.969 $\pm$ 0.042 & 8.6 $\pm$ 0.8 & 0.54 $\pm$ 0.06 & 1.0 $\pm$ 0.1 & $-$2.1 $\pm$ 0.1 & 1.842 $\pm$ 0.201 & DB	 \\
278874	 & HATLASJ085615.9+005237 & 08:56:16.0 & +00:52:36.2 & 0.1692 & 10.94 $\pm$ 0.01 & 8.62 $\pm$ 0.22 & 10.96 $\pm$ 0.1 & 0.09 $\pm$ 0.02 & 98.588 & 15 $\pm$ 13 & 252 $\pm$ 32 & 3.2 $\pm$ 0.55 & 0.443 $\pm$ 0.076 & 10.31 $\pm$ 0.075 & 17.1 $\pm$ 2.9 & 0.42 $\pm$ 0.07 & 1.1 $\pm$ 0.2 & $-$2.5 $\pm$ 0.2 & 2.365 $\pm$ 0.411 & B	 \\
600164	 & HATLASJ085623.6+001352 & 08:56:23.7 & +00:13:51.7 & 0.1774 & 11.14 $\pm$ 0.01 & 13.86 $\pm$ 0.36 & 10.7 $\pm$ 0.12 & 0.27 $\pm$ 0.07 & 97.305 &  &  & <0.8   & <0.119 & & & & & & & \\  
622662	 & HATLASJ085748.0+004641 & 08:57:48.0 & +00:46:38.7 & 0.0718 & 11.27 $\pm$ 0.01 & 18.68 $\pm$ 0.33 & 10.37 $\pm$ 0.11 & 0.79 $\pm$ 0.2 & 107.552 & 0 $\pm$ 2 & 182 $\pm$ 6 & 11.5 $\pm$ 0.58 & 0.276 $\pm$ 0.014 & 9.344 $\pm$ 0.022 &  & 8.46 $\pm$ 0.45 &  &  & 0.118 $\pm$ 0.006 & M	 \\
376631	 & HATLASJ085750.0+012808 & 08:57:50.0 & +01:28:06.7 & 0.1993 & 10.92 $\pm$ 0.01 & 8.38 $\pm$ 0.21 & 10.49 $\pm$ 0.12 & 0.27 $\pm$ 0.08 & 96.117 &  &  & $^{a}$0.3 $\pm$ 0.17   & 0.056 & 9.409 & & 3.26 & & & 0.306 & DB	 \\  
301648	 & HATLASJ085828.4+012211 & 08:58:28.5 & +01:22:11.5 & 0.1992 & 11.46 $\pm$ 0.01 & 29.15 $\pm$ 0.77 & 10.7 $\pm$ 0.12 & 0.58 $\pm$ 0.16 & 96.14 &  &  & $^{a}$0.4 $\pm$ 0.18   & 0.079 & 9.563 & & 7.98 & & & 0.125 & BD	 \\  
622694	 & HATLASJ085828.5+003815 & 08:58:28.6 & +00:38:14.8 & 0.0524 & 10.44 $\pm$ 0.02 & 2.72 $\pm$ 0.14 & 10.43 $\pm$ 0.11 & 0.1 $\pm$ 0.03 & 109.538 & $-$6 $\pm$ 8 & 223 $\pm$ 19 & 3.4 $\pm$ 0.42 & 0.043 $\pm$ 0.005 & 9.301 $\pm$ 0.053 & 8.2 $\pm$ 0.8 & 1.36 $\pm$ 0.18 & 1.5 $\pm$ 0.3 & $-$1.4 $\pm$ 0.3 & 0.734 $\pm$ 0.097 & DBC	 \\
376679	 & HATLASJ085836.0+013149 & 08:58:36.0 & +01:31:49.0 & 0.1068 & 11.22 $\pm$ 0.01 & 16.51 $\pm$ 0.43 & 10.9 $\pm$ 0.1 & 0.21 $\pm$ 0.05 & 104.131 & 57 $\pm$ 6 & 217 $\pm$ 15 & $^{b}$10.3 $\pm$ 0.2 & 0.554 $\pm$ 0.011 & 10.406 $\pm$ 0.008 & 24.9 $\pm$ 1.2 & 0.65 $\pm$ 0.02 & 1.4 $\pm$ 0.2 & $-$1.6 $\pm$ 0.2 & 1.544 $\pm$ 0.051 & DBC	 \\
382034	 & HATLASJ085957.9+015632 & 08:59:57.9 & +01:56:34.2 & 0.1943 & 11.34 $\pm$ 0.01 & 21.69 $\pm$ 0.57 & 10.81 $\pm$ 0.2 & 0.33 $\pm$ 0.15 & 96.531 &  &  & $^{a}$0.5 $\pm$ 0.18   & 0.087 & 8.843 & & 31.15 & & & 0.032 & M	 \\  
209807	 & HATLASJ090004.9+000447 & 09:00:05.0 & +00:04:46.8 & 0.0539 & 10.57 $\pm$ 0.02 & 3.68 $\pm$ 0.15 & 10.7 $\pm$ 0.1 & 0.07 $\pm$ 0.02 & 109.375 & 14 $\pm$ 10 & 308 $\pm$ 24 & 11.4 $\pm$ 1.64 & 0.153 $\pm$ 0.022 & 9.848 $\pm$ 0.062 &  & 0.52 $\pm$ 0.08 &  &  & 1.916 $\pm$ 0.286 & DBC	 \\
346900	 & HATLASJ090120.7+020224 & 09:01:20.7 & +02:02:24.9 & 0.2009 & 11.28 $\pm$ 0.01 & 18.95 $\pm$ 0.49 & 11.06 $\pm$ 0.15 & 0.16 $\pm$ 0.06 & 95.954 & 95 $\pm$ 26 & 393 $\pm$ 62 & 4.0 $\pm$ 0.51 & 0.789 $\pm$ 0.101 & 10.56 $\pm$ 0.056 & 21.2 $\pm$ 5.0 & 0.52 $\pm$ 0.07 & 1.9 $\pm$ 0.2 & $-$1.6 $\pm$ 0.2 & 1.916 $\pm$ 0.25 & DBC	 \\
203879	 & HATLASJ090223.8$-$001639 & 09:02:23.8 & $-$00:16:39.6 & 0.1963 & 10.94 $\pm$ 0.01 & 8.72 $\pm$ 0.22 & 11.03 $\pm$ 0.1 & 0.08 $\pm$ 0.02 & 96.338 &  &  & $^{a}$0.2 $\pm$ 0.17   & 0.032 & 9.166 & & 5.95 & & & 0.168 & BD	 \\  
382362	 & HATLASJ090532.6+020220 & 09:05:32.6 & +02:02:21.9 & 0.0519 & 10.69 $\pm$ 0.02 & 4.89 $\pm$ 0.17 & 10.8 $\pm$ 0.1 & 0.08 $\pm$ 0.02 & 109.59 & $-$13 $\pm$ 16 & 256 $\pm$ 37 & 19.2 $\pm$ 4.08 & 0.238 $\pm$ 0.051 & 10.039 $\pm$ 0.092 & 6.8 $\pm$ 1.4 & 0.45 $\pm$ 0.1 & 1.6 $\pm$ 0.2 & $-$1.3 $\pm$ 0.2 & 2.24 $\pm$ 0.483 & DB	 \\
600656	 & HATLASJ090633.6+001526 & 09:06:33.6 & +00:15:27.9 & 0.165 & 10.88 $\pm$ 0.01 & 7.66 $\pm$ 0.19 & 10.98 $\pm$ 0.11 & 0.08 $\pm$ 0.02 & 98.913 & 95 $\pm$ 22 & 334 $\pm$ 53 & 2.3 $\pm$ 0.1 & 0.301 $\pm$ 0.013 & 10.141 $\pm$ 0.019 &  & 0.55 $\pm$ 0.03 &  &  & 1.807 $\pm$ 0.092 & DBC	 \\
279387	 & HATLASJ090750.0+010141 & 09:07:50.1 & +01:01:41.8 & 0.1281 & 11.7 $\pm$ 0.01 & 50.07 $\pm$ 1.05 & 10.14 $\pm$ 0.12 & 3.62 $\pm$ 0.97 & 102.191 & $-$22 $\pm$ 3 & 150 $\pm$ 7 & 6.8 $\pm$ 0.58 & 0.535 $\pm$ 0.045 & 9.631 $\pm$ 0.037 &  & 11.71 $\pm$ 1.02 &  &  & 0.085 $\pm$ 0.007 & M	 \\
324842	 & HATLASJ090949.6+014847 & 09:09:49.6 & +01:48:46.0 & 0.1819 & 11.84 $\pm$ 0.02 & 68.47 $\pm$ 2.7 & 10.89 $\pm$ 0.12 & 0.88 $\pm$ 0.24 & 97.531 & 6 $\pm$ 1 & 73 $\pm$ 3 & 8.5 $\pm$ 0.58 & 1.364 $\pm$ 0.093 & 10.798 $\pm$ 0.03 &  & 1.09 $\pm$ 0.09 &  &  & 0.917 $\pm$ 0.072 & BC	 \\
324931	 & HATLASJ091157.2+014454 & 09:11:57.2 & +01:44:53.9 & 0.1694 & 11.39 $\pm$ 0.01 & 24.49 $\pm$ 0.64 & 10.9 $\pm$ 0.16 & 0.31 $\pm$ 0.11 & 98.569 & $-$3 $\pm$ 5 & 168 $\pm$ 13 & 5.3 $\pm$ 0.52 & 0.737 $\pm$ 0.072 & 10.53 $\pm$ 0.043 &  & 0.72 $\pm$ 0.07 &  &  & 1.384 $\pm$ 0.141 & DBC	 \\
216401	 & HATLASJ091205.8+002655 & 09:12:05.8 & +00:26:55.6 & 0.0545 & 11.09 $\pm$ 0.01 & 12.32 $\pm$ 0.2 & 10.33 $\pm$ 0.11 & 0.58 $\pm$ 0.14 & 109.314 & 10 $\pm$ 2 & 171 $\pm$ 6 & 13.6 $\pm$ 0.84 & 0.187 $\pm$ 0.011 & 9.934 $\pm$ 0.027 &  & 1.44 $\pm$ 0.09 &  &  & 0.697 $\pm$ 0.044 & BC	 \\
210543	 & HATLASJ091420.1+000509 & 09:14:20.0 & +00:05:10.0 & 0.2022 & 11.55 $\pm$ 0.01 & 35.48 $\pm$ 0.94 & 10.62 $\pm$ 0.13 & 0.85 $\pm$ 0.25 & 95.874 & 37 $\pm$ 9 & 302 $\pm$ 22 & 3.3 $\pm$ 0.57 & 0.667 $\pm$ 0.114 & 10.487 $\pm$ 0.074 &  & 1.16 $\pm$ 0.2 &  &  & 0.864 $\pm$ 0.15 & B	 \\
378002	 & HATLASJ091956.8+013851 & 09:19:57.0 & +01:38:51.6 & 0.1764 & 11.13 $\pm$ 0.01 & 13.48 $\pm$ 0.35 & 10.45 $\pm$ 0.12 & 0.48 $\pm$ 0.13 & 97.994 & $-$10 $\pm$ 7 & 151 $\pm$ 16 & 2.4 $\pm$ 0.32 & 0.365 $\pm$ 0.048 & 10.225 $\pm$ 0.057 &  & 0.8 $\pm$ 0.11 &  &  & 1.246 $\pm$ 0.168 & BD	 \\
216973	 & HATLASJ092232.2+002707 & 09:22:32.3 & +00:27:08.4 & 0.173 & 11.19 $\pm$ 0.01 & 15.6 $\pm$ 0.4 & 10.36 $\pm$ 0.12 & 0.68 $\pm$ 0.18 & 98.305 & $-$96 $\pm$ 24 & 518 $\pm$ 57 & 3.5 $\pm$ 0.64 & 0.504 $\pm$ 0.093 & 10.366 $\pm$ 0.08 &  & 0.67 $\pm$ 0.12 &  &  & 1.487 $\pm$ 0.276 & DC	 \\
30313	 & HATLASJ113740.6$-$010454 & 11:37:40.7 & $-$01:04:54.1 & 0.1517 & 10.93 $\pm$ 0.01 & 8.49 $\pm$ 0.21 & 10.71 $\pm$ 0.12 & 0.17 $\pm$ 0.05 & 100.123 &  &  & $^{a}$0.9 $\pm$ 0.11   & 0.096 & 9.646 & & 1.92 & & & 0.521 & DB	 \\  
53724	 & HATLASJ113858.4$-$001629 & 11:38:58.5 & $-$00:16:30.2 & 0.1637 & 11.21 $\pm$ 0.01 & 16.12 $\pm$ 0.42 & 10.84 $\pm$ 0.12 & 0.23 $\pm$ 0.06 & 99.075 & $-$58 $\pm$ 19 & 261 $\pm$ 45 & 4.2 $\pm$ 1.0 & 0.546 $\pm$ 0.129 & 10.4 $\pm$ 0.103 & 18.8 $\pm$ 4.2 & 0.64 $\pm$ 0.15 & 1.4 $\pm$ 0.1 & $-$2.1 $\pm$ 0.1 & 1.559 $\pm$ 0.372 & DBC	 \\
69661	 & HATLASJ114343.9+000203 & 11:43:44.1 & +00:02:02.5 & 0.1872 & 11.05 $\pm$ 0.01 & 11.18 $\pm$ 0.28 & 10.1 $\pm$ 0.12 & 0.89 $\pm$ 0.25 & 97.106 & $-$24 $\pm$ 4 & 67 $\pm$ 10 & 2.8 $\pm$ 0.52 & 0.485 $\pm$ 0.089 & 10.348 $\pm$ 0.08 & 18.0 $\pm$ 4.1 & 0.5 $\pm$ 0.09 & 1.5 $\pm$ 0.1 & $-$1.8 $\pm$ 0.1 & 1.994 $\pm$ 0.369 & BC	 \\
84183	 & HATLASJ114540.7+002554 & 11:45:41.0 & +00:25:55.3 & 0.3429 & 11.56 $\pm$ 0.01 & 36.44 $\pm$ 0.97 & 11.04 $\pm$ 0.11 & 0.33 $\pm$ 0.09 & 85.827 &  &  & $^{a}$0.2 $\pm$ 0.18   & 0.147 & 9.831 & & 5.38 & & & 0.186 & DB	 \\  
136959	 & HATLASJ114625.1$-$014511 & 11:46:25.0 & $-$01:45:13.0 & 0.1645 & 11.72 $\pm$ 0.01 & 52.8 $\pm$ 1.42 & 10.71 $\pm$ 0.1 & 1.03 $\pm$ 0.24 & 98.956 & 95 $\pm$ 11 & 381 $\pm$ 28 & 6.6 $\pm$ 0.64 & 0.861 $\pm$ 0.084 & 10.598 $\pm$ 0.043 &  & 1.33 $\pm$ 0.14 &  &  & 0.75 $\pm$ 0.076 & D	 \\
758786	 & HATLASJ114702.7+001207 & 11:47:02.8 & +00:12:07.0 & 0.328 & 11.67 $\pm$ 0.01 & 46.3 $\pm$ 1.24 & 10.81 $\pm$ 0.11 & 0.71 $\pm$ 0.19 & 86.779 & 61 $\pm$ 32 & 335 $\pm$ 76 & 2.3 $\pm$ 0.72 & 1.279 $\pm$ 0.393 & 10.769 $\pm$ 0.134 & 35.0 $\pm$ 9.5 & 0.79 $\pm$ 0.24 & 2.5 $\pm$ 0.4 & $-$0.5 $\pm$ 0.4 & 1.27 $\pm$ 0.392 & D	 \\
30979	 & HATLASJ115039.6$-$010639 & 11:50:39.7 & $-$01:06:40.7 & 0.343 & 11.93 $\pm$ 0.01 & 84.51 $\pm$ 2.32 & 11.3 $\pm$ 0.14 & 0.42 $\pm$ 0.14 & 85.84 &  &  & $^{a}$0.5 $\pm$ 0.18   & 0.284 & 10.117 & & 6.46 & & & 0.155 & B	 \\  
39703	 & HATLASJ115141.3$-$004239 & 11:51:41.5 & $-$00:42:39.5 & 0.321 & 11.77 $\pm$ 0.01 & 58.93 $\pm$ 1.6 & 10.94 $\pm$ 0.11 & 0.68 $\pm$ 0.17 & 87.265 &  &  & $^{a}$1.1 $\pm$ 0.16   & 0.548 & 10.402 & & 2.34 & & & 0.428 & B	 \\  
535507	 & HATLASJ115317.4$-$010123 & 11:53:17.4 & $-$01:01:21.3 & 0.1804 & 11.18 $\pm$ 0.01 & 15.13 $\pm$ 0.39 & 10.73 $\pm$ 0.19 & 0.28 $\pm$ 0.12 & 97.661 & $-$21 $\pm$ 45 & 506 $\pm$ 107 & 3.3 $\pm$ 0.76 & 0.526 $\pm$ 0.12 & 10.384 $\pm$ 0.099 &  & 0.62 $\pm$ 0.14 &  &  & 1.6 $\pm$ 0.368 & BD	 \\
179053	 & HATLASJ121141.8$-$015731 & 12:11:41.8 & $-$01:57:29.7 & 0.317 & 11.8 $\pm$ 0.01 & 63.65 $\pm$ 1.73 & 11.18 $\pm$ 0.1 & 0.42 $\pm$ 0.09 & 87.599 &  &  & $^{a}$0.4 $\pm$ 0.17   & 0.21 & 9.984 & & 6.6 & & & 0.152 & B	 \\  
186110	 & HATLASJ121206.2$-$013425 & 12:12:06.2 & $-$01:34:24.3 & 0.1588 & 10.77 $\pm$ 0.01 & 5.84 $\pm$ 0.14 & 10.86 $\pm$ 0.11 & 0.08 $\pm$ 0.02 & 99.604 &  &  & <0.9   & <0.108 & & & & & & & \\  
55717	 & HATLASJ121253.5$-$002203 & 12:12:53.5 & $-$00:22:04.4 & 0.1855 & 11.11 $\pm$ 0.01 & 13.01 $\pm$ 0.33 & 10.79 $\pm$ 0.11 & 0.21 $\pm$ 0.05 & 97.238 & $-$8 $\pm$ 12 & 202 $\pm$ 28 & 2.7 $\pm$ 0.39 & 0.447 $\pm$ 0.065 & 10.313 $\pm$ 0.063 &  & 0.63 $\pm$ 0.09 &  &  & 1.581 $\pm$ 0.234 & DB	 \\
99611	 & HATLASJ121427.4+005818 & 12:14:27.4 & +00:58:18.3 & 0.184 & 11.27 $\pm$ 0.01 & 18.44 $\pm$ 0.48 & 10.93 $\pm$ 0.13 & 0.22 $\pm$ 0.06 & 97.34 & 41 $\pm$ 9 & 215 $\pm$ 22 & 2.8 $\pm$ 0.42 & 0.46 $\pm$ 0.069 & 9.566 $\pm$ 0.065 &  & 5.01 $\pm$ 0.77 &  &  & 0.2 $\pm$ 0.03 & MBC	 \\
32169	 & HATLASJ121446.4$-$011155 & 12:14:46.5 & $-$01:11:55.6 & 0.1797 & 11.55 $\pm$ 0.01 & 35.36 $\pm$ 0.94 & 10.82 $\pm$ 0.11 & 0.54 $\pm$ 0.14 & 97.707 & 13 $\pm$ 19 & 402 $\pm$ 47 & 4.6 $\pm$ 0.6 & 0.725 $\pm$ 0.094 & 10.523 $\pm$ 0.056 &  & 1.06 $\pm$ 0.14 &  &  & 0.944 $\pm$ 0.125 & BC	 \\
537340	 & HATLASJ121908.8$-$010201 & 12:19:08.8 & $-$01:02:00.5 & 0.1575 & 10.89 $\pm$ 0.01 & 7.76 $\pm$ 0.19 & 11.12 $\pm$ 0.11 & 0.06 $\pm$ 0.01 & 99.635 &  &  & $^{a}$0.5 $\pm$ 0.18   & 0.061 & 9.447 & & 2.77 & & & 0.361 & B	 \\  
543473	 & HATLASJ140649.0$-$005647 & 14:06:49.0 & $-$00:56:47.9 & 0.2715 & 11.66 $\pm$ 0.01 & 45.51 $\pm$ 1.22 & 11.25 $\pm$ 0.12 & 0.25 $\pm$ 0.07 & 90.656 & 12 $\pm$ 36 & 390 $\pm$ 86 & 2.3 $\pm$ 0.52 & 0.834 $\pm$ 0.192 & 10.584 $\pm$ 0.1 &  & 1.19 $\pm$ 0.28 &  &  & 0.843 $\pm$ 0.196 & BD	 \\
491545	 & HATLASJ140912.3$-$013454 & 14:09:12.5 & $-$01:34:54.9 & 0.2649 & 11.89 $\pm$ 0.01 & 78.16 $\pm$ 2.14 & 10.97 $\pm$ 0.13 & 0.84 $\pm$ 0.26 & 91.098 & 100 $\pm$ 11 & 271 $\pm$ 27 & 4.3 $\pm$ 0.66 & 1.494 $\pm$ 0.231 & 10.837 $\pm$ 0.067 &  & 1.14 $\pm$ 0.18 &  &  & 0.879 $\pm$ 0.138 & BC	 \\
105572	 & HATLASJ141008.0+005106 & 14:10:08.0 & +00:51:06.9 & 0.2564 & 11.83 $\pm$ 0.01 & 67.9 $\pm$ 1.85 & 11.1 $\pm$ 0.12 & 0.54 $\pm$ 0.15 & 91.702 & 144 $\pm$ 29 & 411 $\pm$ 70 & 4.0 $\pm$ 0.9 & 1.311 $\pm$ 0.295 & 10.78 $\pm$ 0.098 & 13.7 $\pm$ 3.6 & 1.13 $\pm$ 0.26 & 1.8 $\pm$ 0.2 & $-$1.5 $\pm$ 0.2 & 0.888 $\pm$ 0.201 & BD	 \\
15049	 & HATLASJ141522.0+004413 & 14:15:22.1 & +00:44:14.9 & 0.2682 & 11.37 $\pm$ 0.01 & 23.69 $\pm$ 0.62 & 10.84 $\pm$ 0.14 & 0.34 $\pm$ 0.11 & 90.928 & $-$48 $\pm$ 90 & 805 $\pm$ 213 & 1.7 $\pm$ 0.45 & 0.614 $\pm$ 0.162 & 10.451 $\pm$ 0.114 & 22.7 $\pm$ 3.3 & 0.84 $\pm$ 0.22 & 1.4 $\pm$ 0.1 & $-$1.9 $\pm$ 0.1 & 1.193 $\pm$ 0.315 & BDC	 \\
227768	 & HATLASJ141908.4+011313 & 14:19:08.6 & +01:13:10.7 & 0.2801 & 11.52 $\pm$ 0.01 & 33.23 $\pm$ 0.88 & 11.0 $\pm$ 0.11 & 0.33 $\pm$ 0.09 & 90.11 &  &  & <0.8   & <0.321 & & & & & & & \\  
511560	 & HATLASJ141925.3$-$011130 & 14:19:25.4 & $-$01:11:30.5 & 0.2533 & 11.75 $\pm$ 0.01 & 56.6 $\pm$ 1.53 & 10.15 $\pm$ 0.12 & 4.03 $\pm$ 1.16 & 91.991 &  &  & $^{a}$0.3 $\pm$ 0.16   & 0.091 & 9.62 & & 13.58 & & & 0.074 & BDC	 \\  
319660	 & HATLASJ142057.9+015233 & 14:20:58.0 & +01:52:32.1 & 0.2646 & 11.64 $\pm$ 0.01 & 43.85 $\pm$ 1.18 & 10.86 $\pm$ 0.12 & 0.6 $\pm$ 0.17 & 91.145 & 19 $\pm$ 15 & 244 $\pm$ 37 & 3.5 $\pm$ 0.66 & 1.238 $\pm$ 0.231 & 10.755 $\pm$ 0.081 & 15.4 $\pm$ 0.8 & 0.77 $\pm$ 0.15 & 2.0 $\pm$ 0.1 & $-$1.5 $\pm$ 0.1 & 1.298 $\pm$ 0.245 & BDC	 \\
106336	 & HATLASJ142208.8+005428 & 14:22:09.0 & +00:54:27.4 & 0.2541 & 11.46 $\pm$ 0.01 & 28.77 $\pm$ 0.76 & 11.17 $\pm$ 0.13 & 0.19 $\pm$ 0.06 & 91.953 & $-$117 $\pm$ 26 & 261 $\pm$ 63 & 2.6 $\pm$ 0.16 & 0.822 $\pm$ 0.051 & 9.818 $\pm$ 0.027 &  & 4.38 $\pm$ 0.3 &  &  & 0.229 $\pm$ 0.016 & MC	 \\
228157	 & HATLASJ142517.1+010546 & 14:25:17.1 & +01:05:46.6 & 0.2807 & 11.84 $\pm$ 0.01 & 68.52 $\pm$ 1.87 & 11.07 $\pm$ 0.12 & 0.58 $\pm$ 0.16 & 90.023 & $-$52 $\pm$ 12 & 308 $\pm$ 29 & 4.3 $\pm$ 0.6 & 1.714 $\pm$ 0.237 & 10.897 $\pm$ 0.06 &  & 0.87 $\pm$ 0.12 &  &  & 1.151 $\pm$ 0.162 & BD	 \\
771527	 & HATLASJ144116.3+002724 & 14:41:16.4 & +00:27:26.0 & 0.2441 & 11.08 $\pm$ 0.01 & 12.16 $\pm$ 0.31 & 10.83 $\pm$ 0.11 & 0.18 $\pm$ 0.05 & 92.647 &  &  & <0.9   & <0.262 & & & & & & & \\  
594509	 & HATLASJ144129.5$-$000902 & 14:41:29.7 & $-$00:09:01.2 & 0.2432 & 11.25 $\pm$ 0.01 & 17.84 $\pm$ 0.46 & 10.98 $\pm$ 0.13 & 0.19 $\pm$ 0.06 & 92.688 &  &  & <0.9   & <0.261 & & & & & & & \\  
93478	 & HATLASJ144218.7+003615 & 14:42:18.7 & +00:36:15.5 & 0.2409 & 11.55 $\pm$ 0.01 & 35.27 $\pm$ 0.94 & 10.58 $\pm$ 0.12 & 0.92 $\pm$ 0.25 & 92.859 & 112 $\pm$ 18 & 363 $\pm$ 42 & 3.5 $\pm$ 0.96 & 1.014 $\pm$ 0.276 & 10.669 $\pm$ 0.118 &  & 0.76 $\pm$ 0.21 &  &  & 1.322 $\pm$ 0.362 & B	 \\
16985	 & HATLASJ144515.0+003907 & 14:45:15.1 & +00:39:06.4 & 0.2818 & 11.85 $\pm$ 0.01 & 70.93 $\pm$ 1.94 & 10.85 $\pm$ 0.15 & 1.01 $\pm$ 0.36 & 89.886 &  &  & $^{a}$0.6 $\pm$ 0.17   & 0.239 & 9.282 & & 37.04 & & & 0.027 & M	 \\
\hline
\end{tabular}}
\caption[Table]{\small{Observed CO(1--0) line parameters. The $\rm z_{spec}$ and $L_{\rm IR}$ are the optical redshift and total IR luminosity ($8-1000_{\rm \mu m}$). The $\nu_{\rm obs}$ is the observed frequency of the line. The v$_{\rm obs}$ and v$_{\rm FWHM}$  are the velocity where the line is centred (with respect to the GAMA redshift) and the {\sc fwhm} computed by a single Gaussian fit (see Fig.~\ref{Spectra_1}). $S_{\rm CO} \Delta \nu$ is the velocity integrated flux density from the data cubes collapsed between v$_{\rm obs}$ - v$_{\rm FWHM}$ and v$_{\rm obs}$ + v$_{\rm FWHM}$. $\rm {L}'_{\rm CO}$ is the CO(1--0) line luminosity using Eqn.~\ref{Solomon}. $M_{\rm H2}$ is the mass of the molecular gas using the morphological criteria from \S~\ref{Morphological_Properties}. $\rm R_{\rm FWHM}$ is the size of the deconvolved semi--major axis of the sources that can be spatially resolved by ALMA (see \S~\ref{Data_Reduction}). SFE is calculated as $\rm SFR / M_{\rm H2}$. $\Sigma_{\rm gas}$ and $\Sigma_{\rm SFR}$ are the surface density of the gas (molecular and atomic) and SFR, respectively. $\tau_{\rm gas}$ is the gas consumption timescale, and is calculated as $\rm SFE^{-1}$. Fluxes for sources that are not spectrally detected in CO are calculated collapsing the data cubes between v$_{\rm obs}$ - $\left \langle \rm v_{\rm FWHM}\right \rangle$ and v$_{\rm obs}$ + $\left \langle \rm v_{\rm FWHM}\right \rangle$, where $\left \langle \rm v_{\rm FWHM}\right \rangle$ = 250 km s$^{-1}$, the average {\sc fwhm} found for the whole sample (see \S~\ref{Source_Properties}). $^{(a)}$ These sources were spectroscopically undetected, and their fluxes were computed from collapsed cubes (between v$_{\rm obs}$ $\pm$ $\left \langle \rm v_{\rm FWHM}\right \rangle$) using a 2D Gaussian fit with task {\sc gaussfit}}. $^{(b)}$ These are extended sources. Velocity integrated flux densities have been calculated over an irregular region covering the whole extension of the emission.}\label{Table}
\end{center}
\end{table}
\end{landscape}

{\color{red}


}

\begin{figure*}
$
\begin{array}{ccc}
\includegraphics[width=8.4cm]{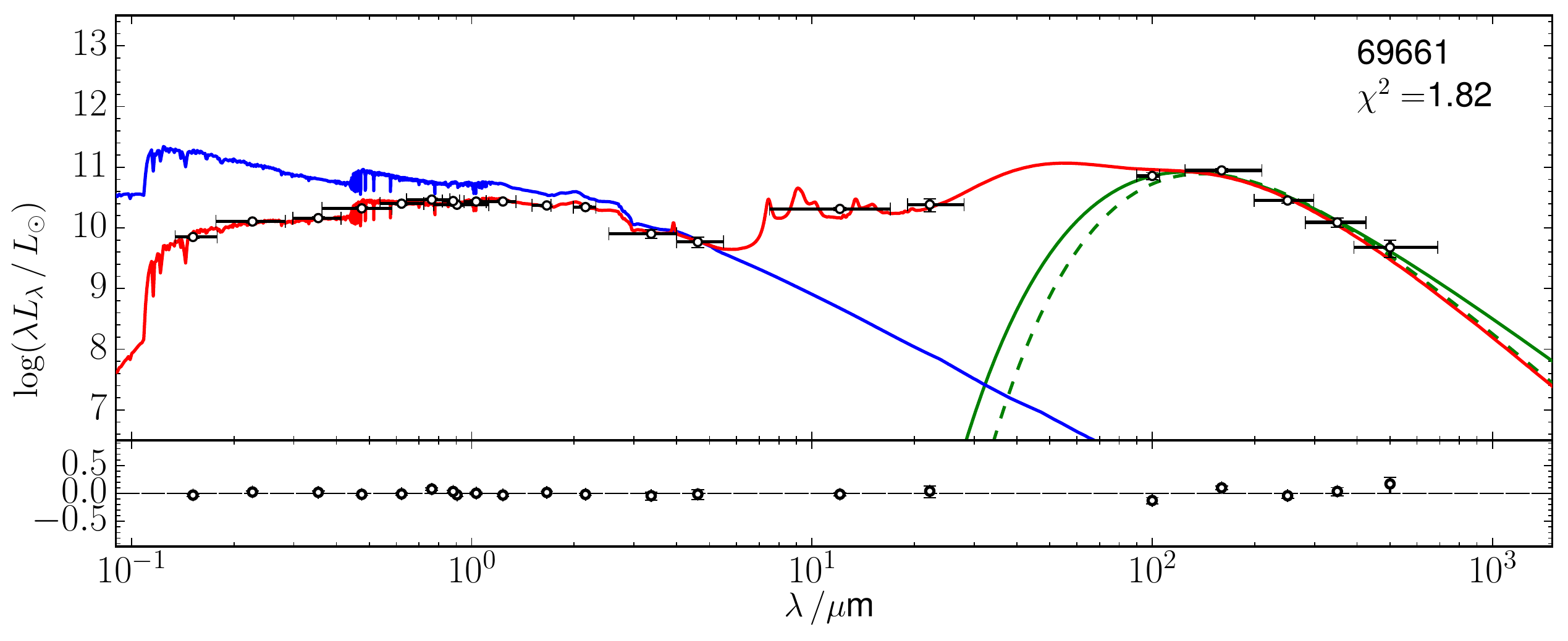} &
\includegraphics[width=5.0cm]{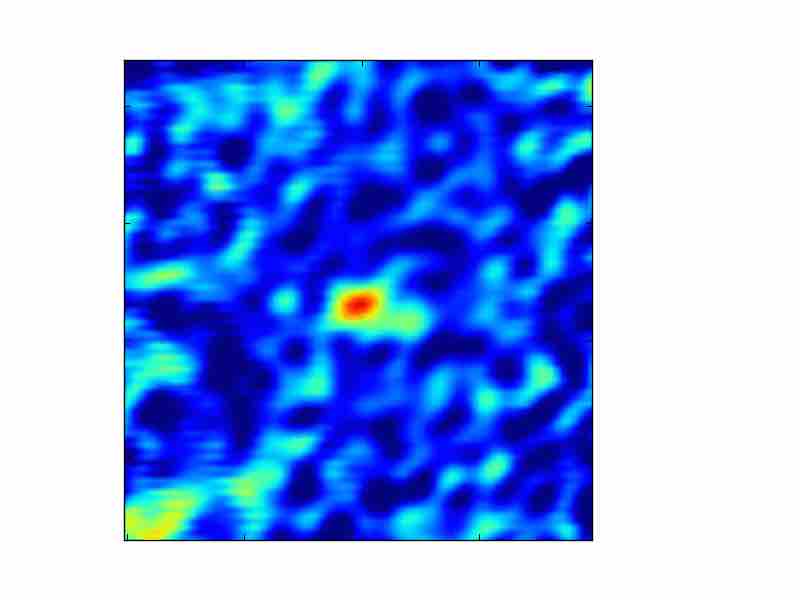} &
\hspace*{-1.2cm}\begin{overpic}[width=3.4cm]{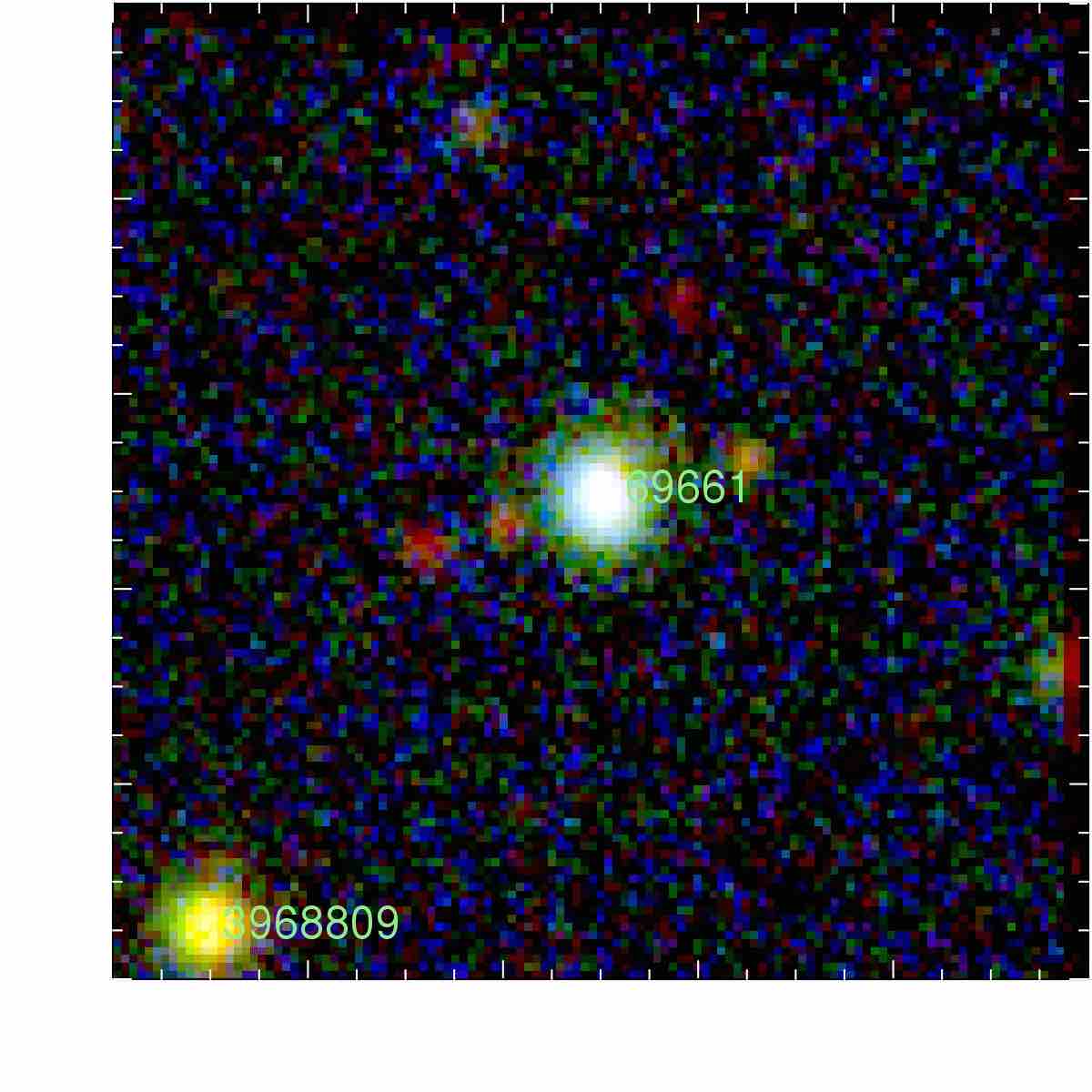} \put (9,85) { \begin{fitbox}{2.25cm}{0.2cm} \color{white}$\bf BC$ \end{fitbox}} \end{overpic} \\ 	
	
\includegraphics[width=8.4cm]{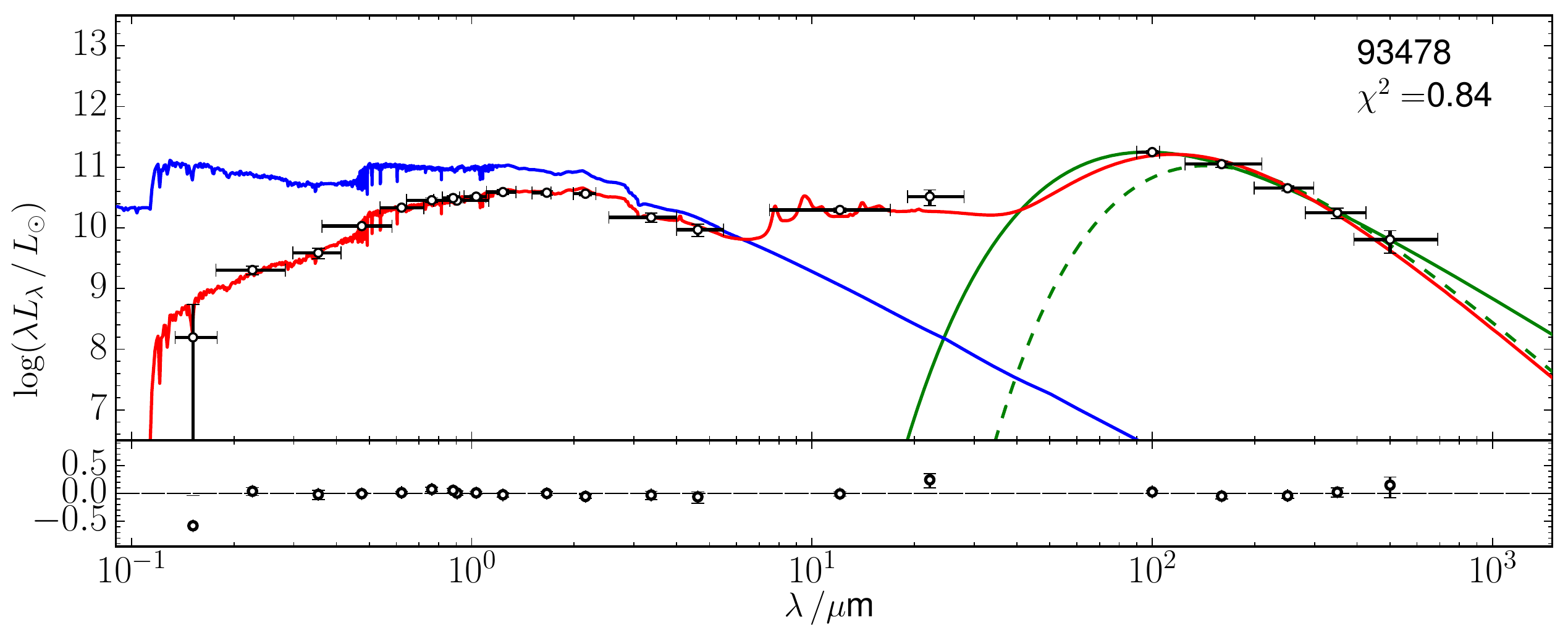} &
\includegraphics[width=5.0cm]{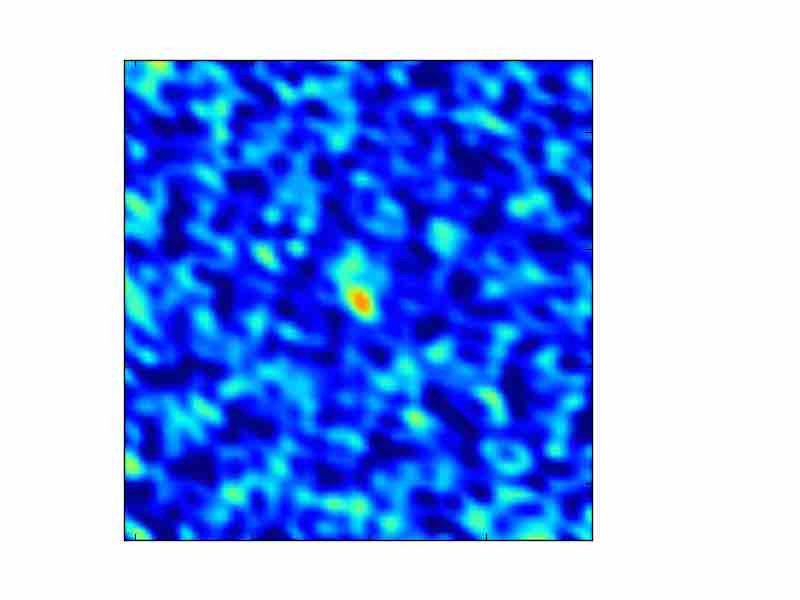} &
\hspace*{-1.2cm}\begin{overpic}[width=3.4cm]{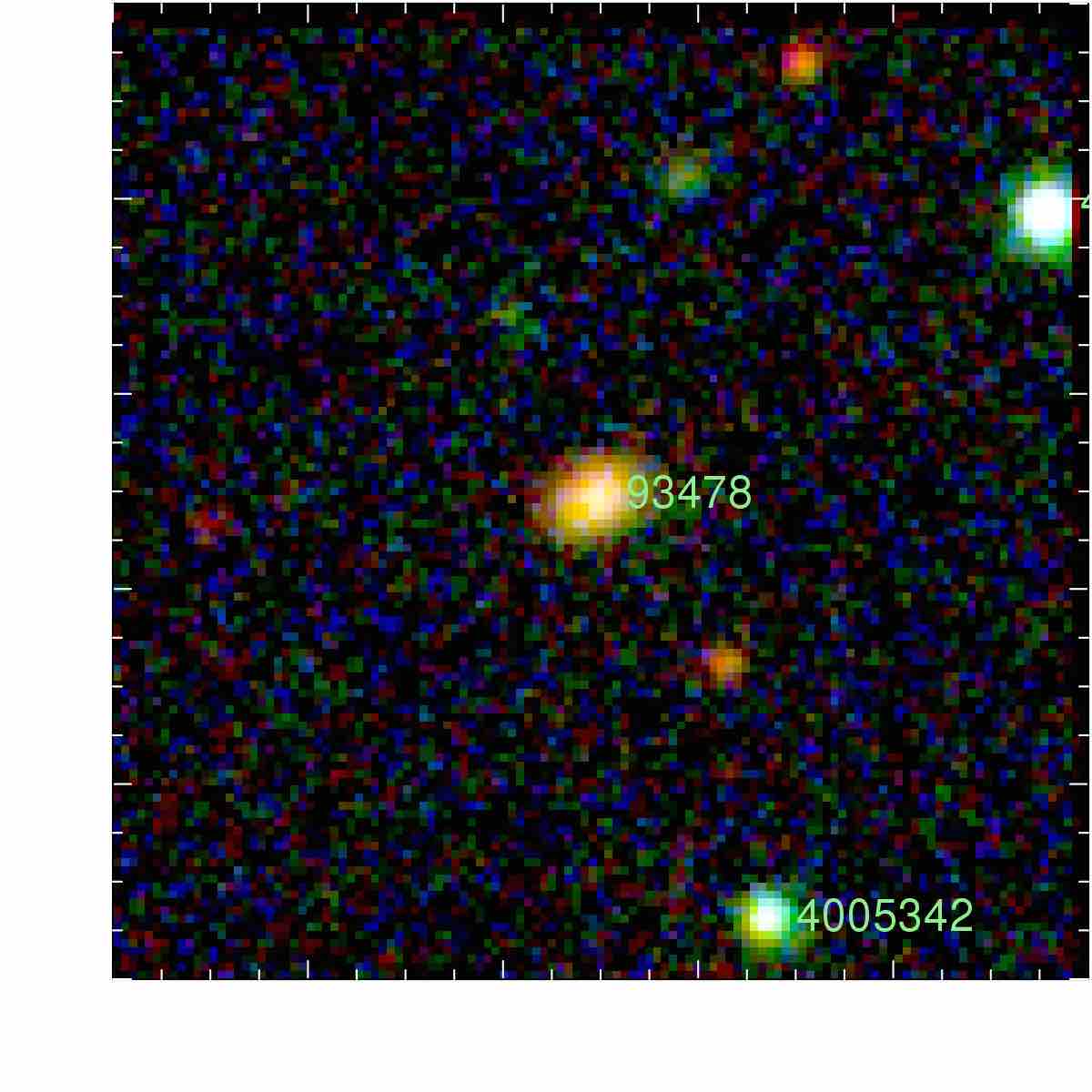} \put (9,85) { \begin{fitbox}{2.25cm}{0.2cm} \color{white}$\bf B$ \end{fitbox}} \end{overpic} \\ 	

\includegraphics[width=8.4cm]{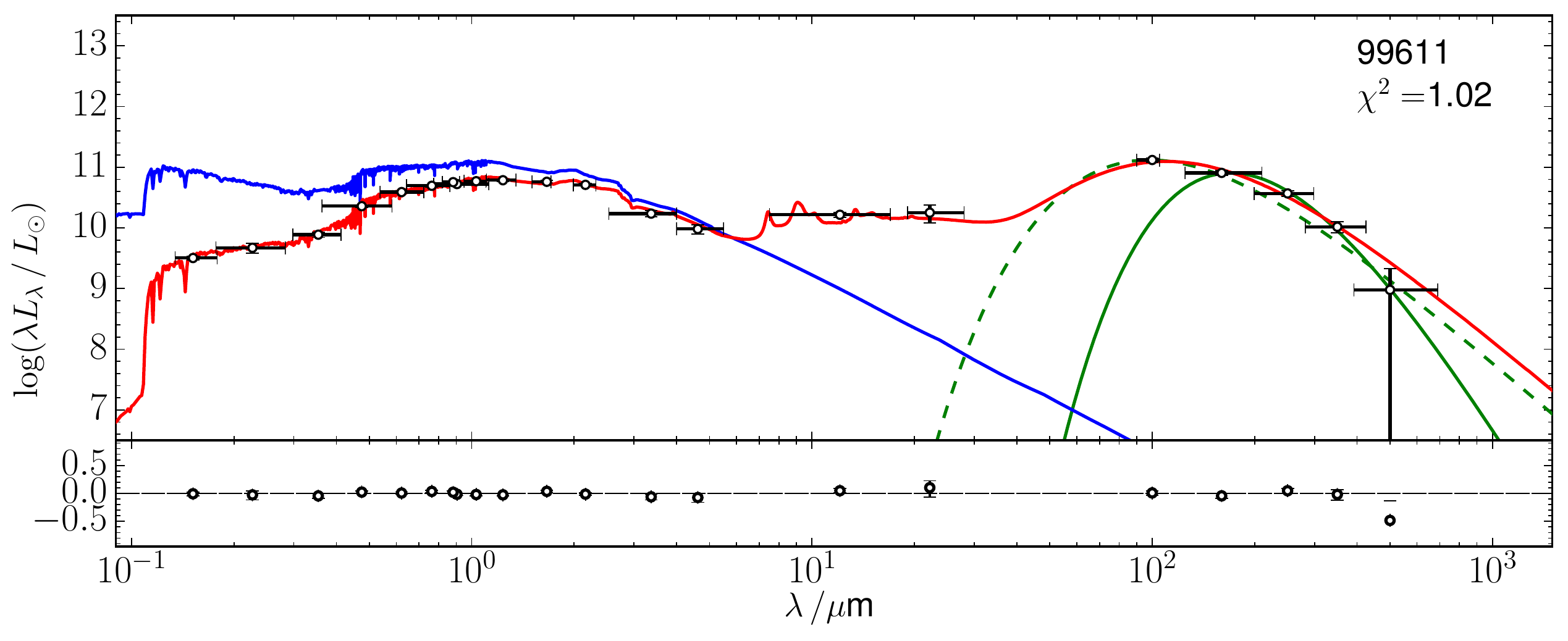} &
\includegraphics[width=5.0cm]{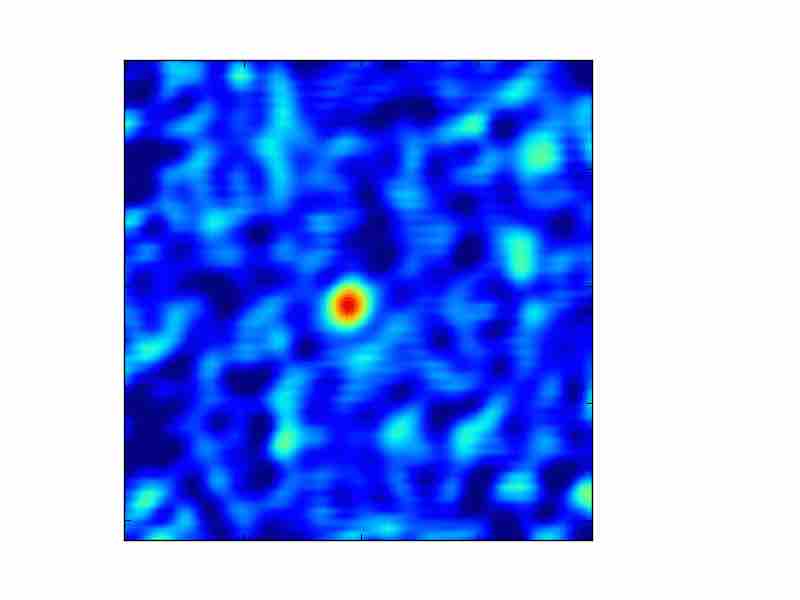} &
\hspace*{-1.2cm}\begin{overpic}[width=3.4cm]{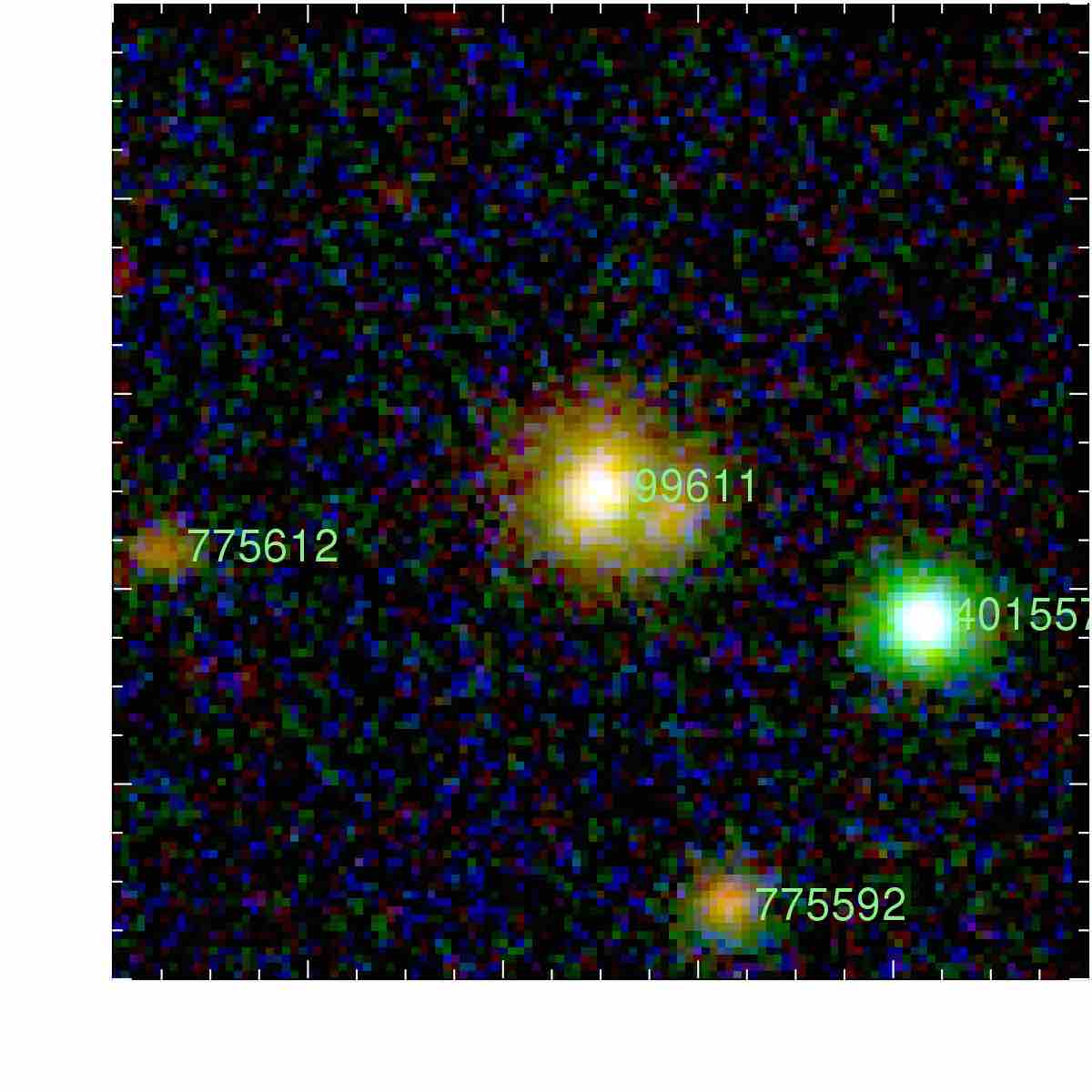} \put (9,85) { \begin{fitbox}{2.25cm}{0.2cm} \color{white}$\bf MBC$ \end{fitbox}} \end{overpic} \\ 

\includegraphics[width=8.4cm]{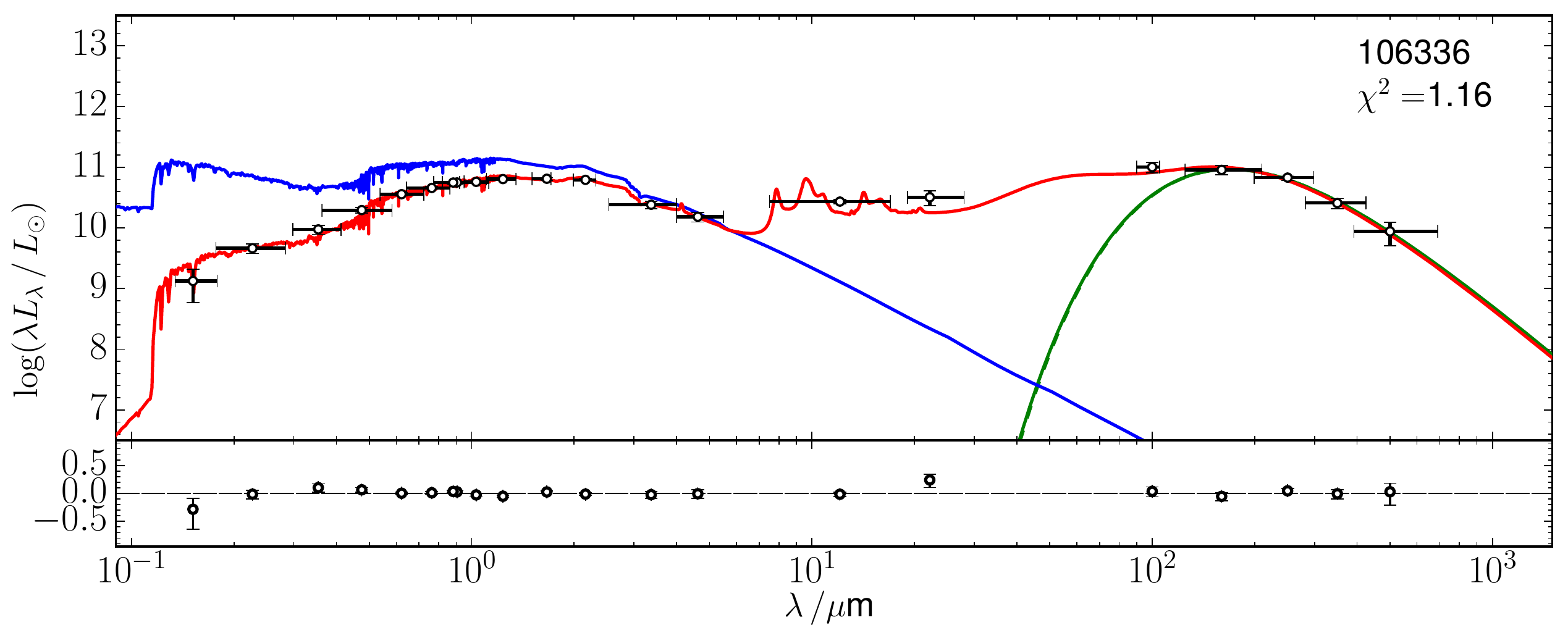} &
\includegraphics[width=5.0cm]{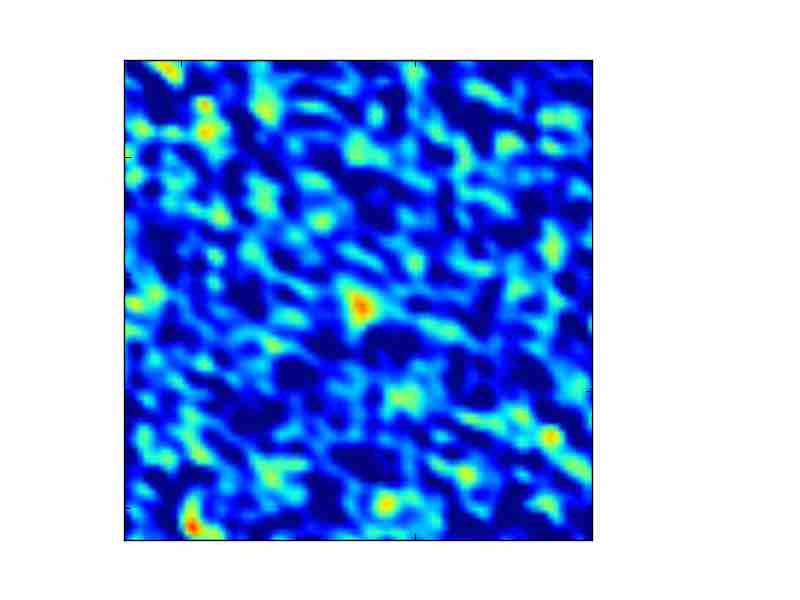} &
\hspace*{-1.2cm}\begin{overpic}[width=3.4cm]{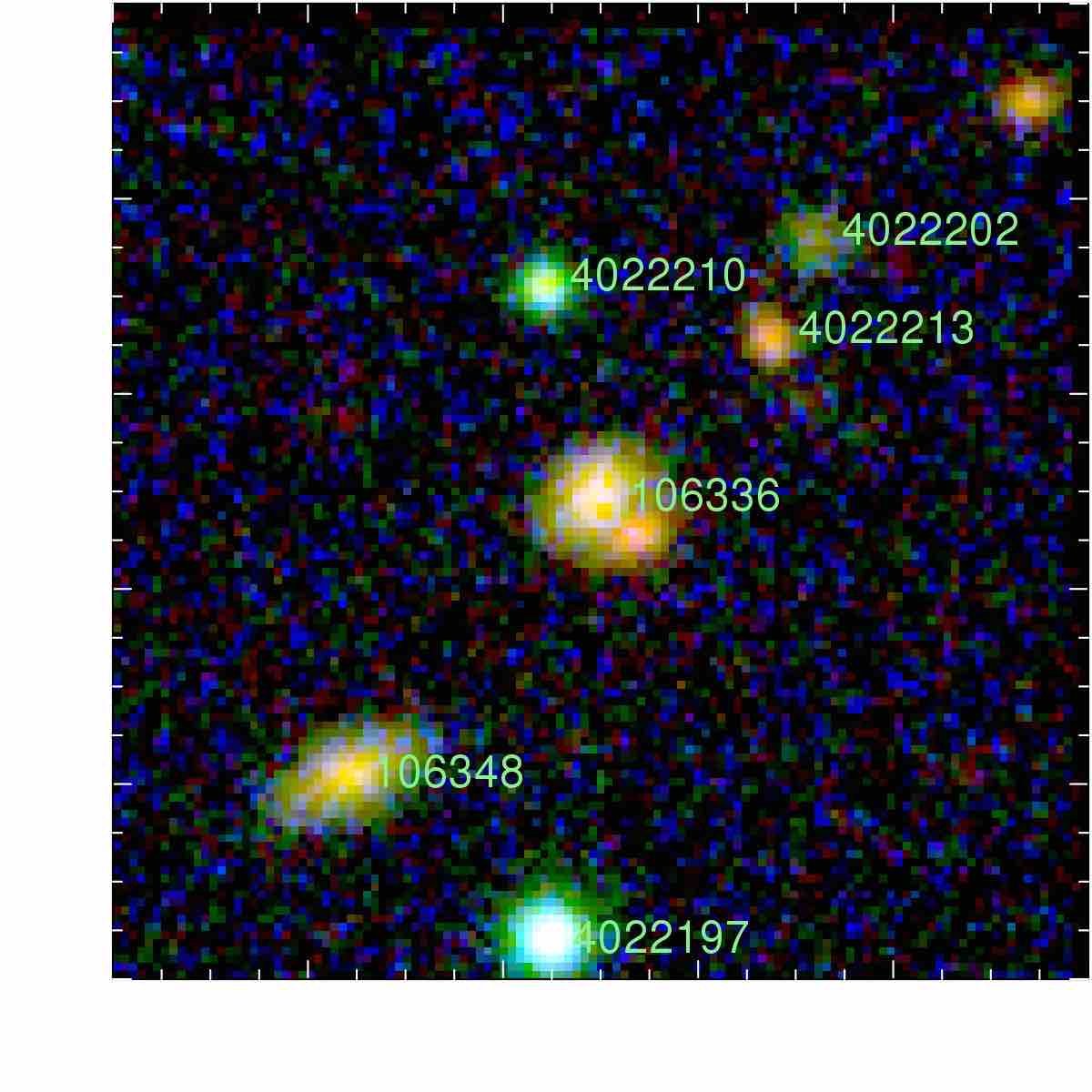} \put (9,85) { \begin{fitbox}{2.25cm}{0.2cm} \color{white}$\bf MC$ \end{fitbox}} \end{overpic} \\ 

\includegraphics[width=8.4cm]{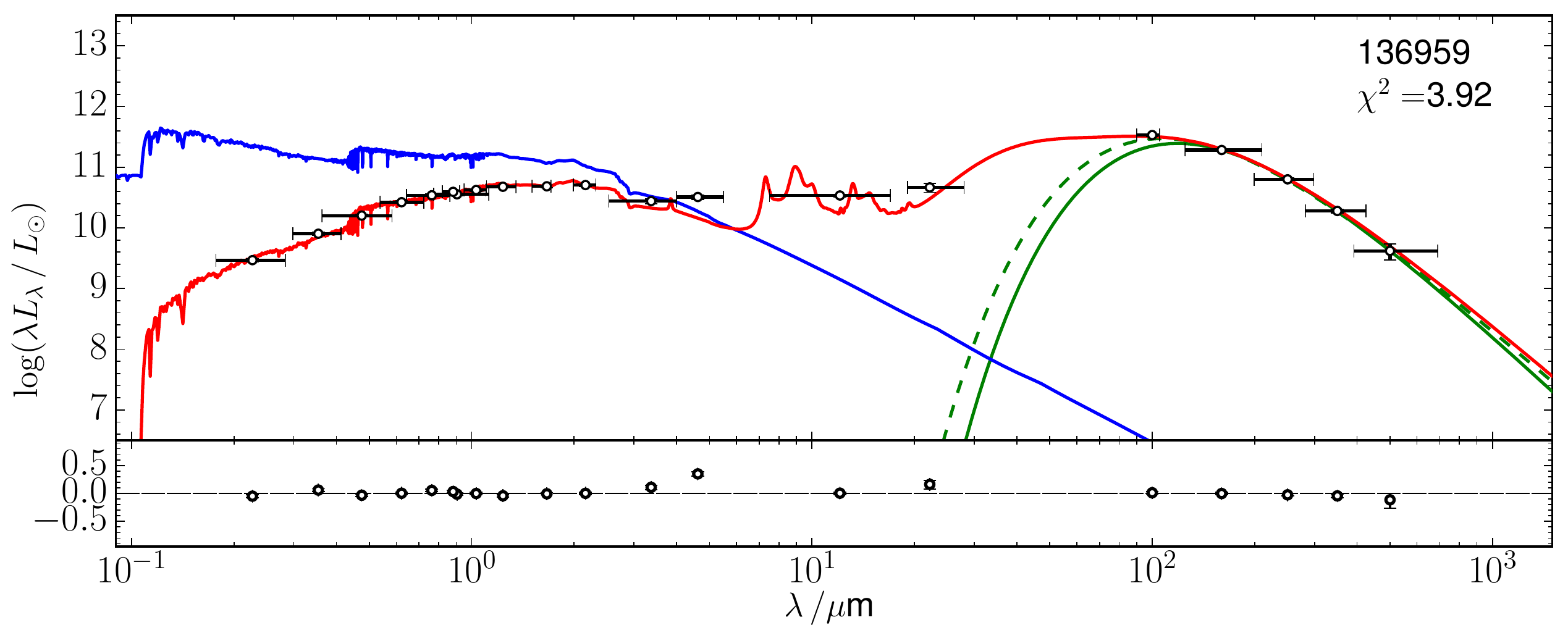} &
\includegraphics[width=5.0cm]{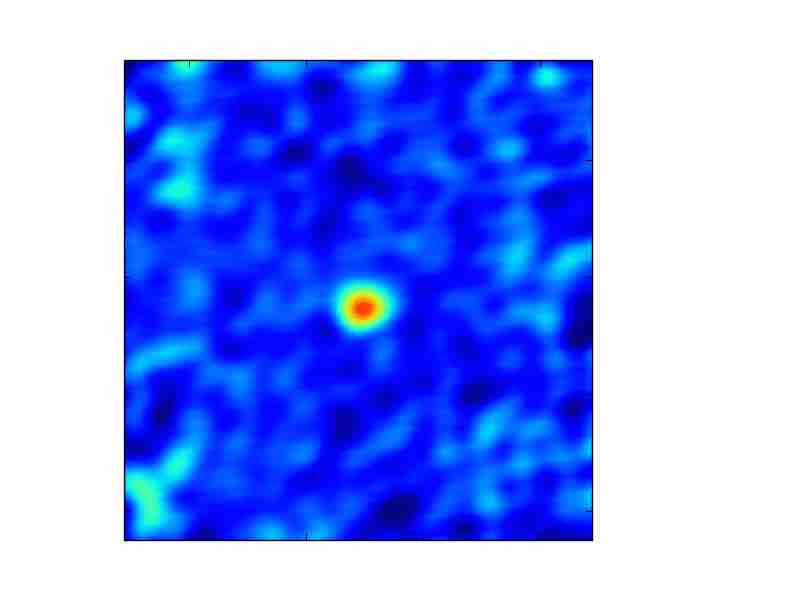} &
\hspace*{-1.2cm}\begin{overpic}[width=3.4cm]{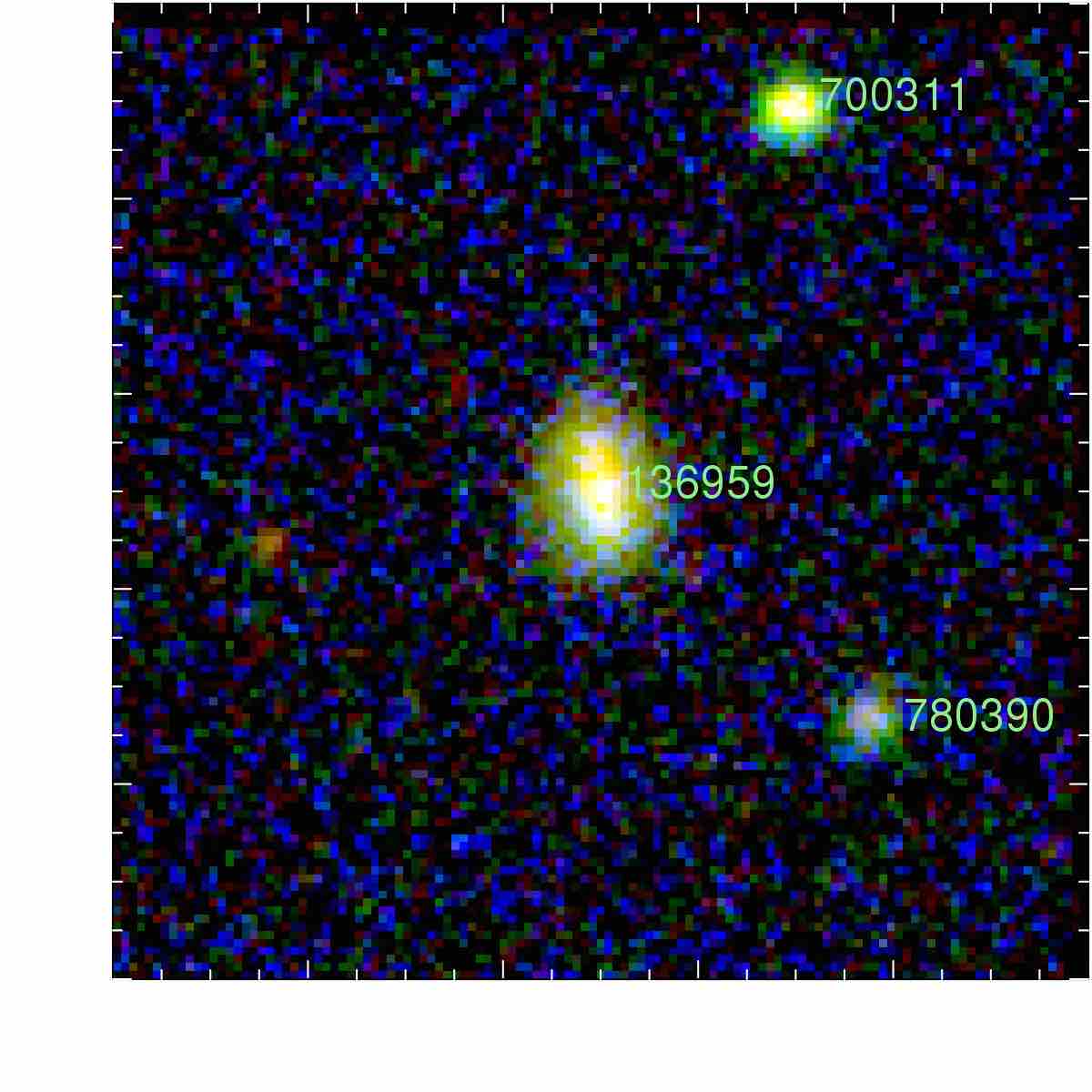} \put (9,85) { \begin{fitbox}{2.25cm}{0.2cm} \color{white}$\bf D$ \end{fitbox}} \end{overpic} \\ 

\end{array}
$
{\textbf{Figure~\ref{pdrdiaglit}.} continued}

\end{figure*}


\begin{figure*}
$
\begin{array}{ccc}
\includegraphics[width=8.4cm]{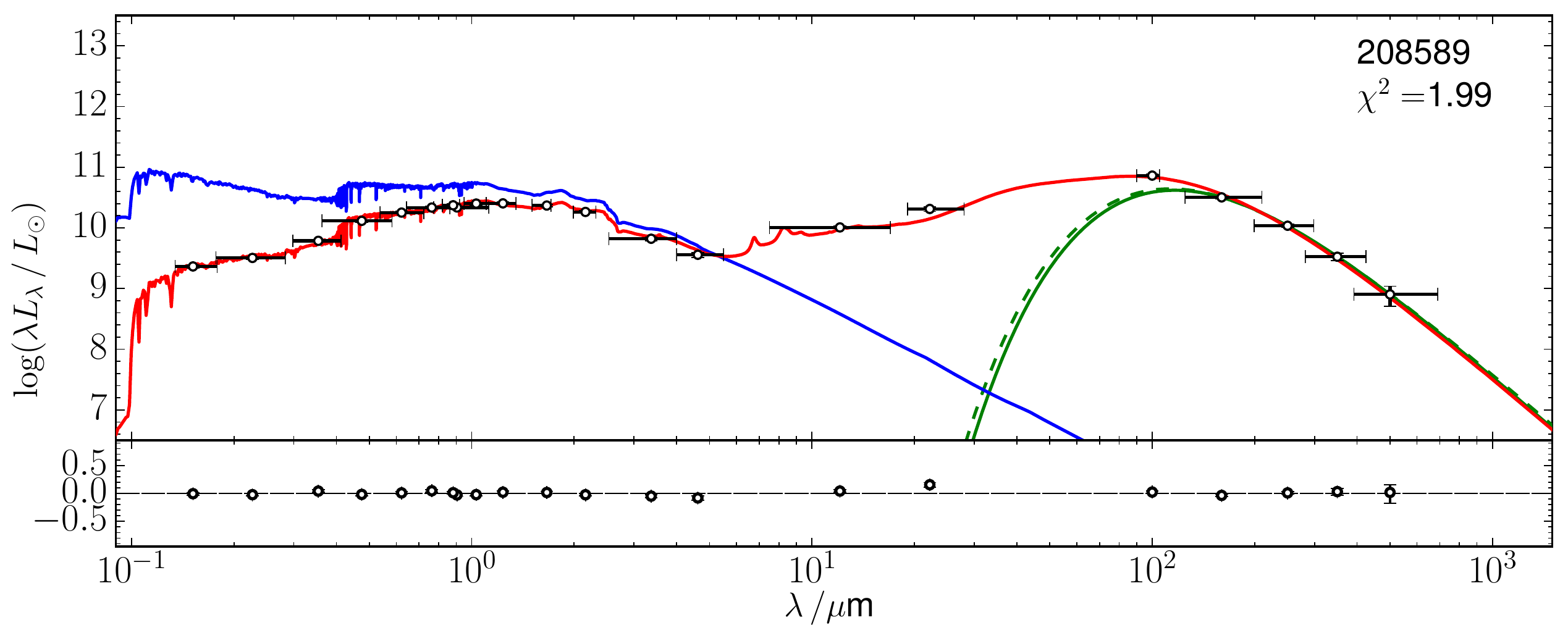} &
\includegraphics[width=5.0cm]{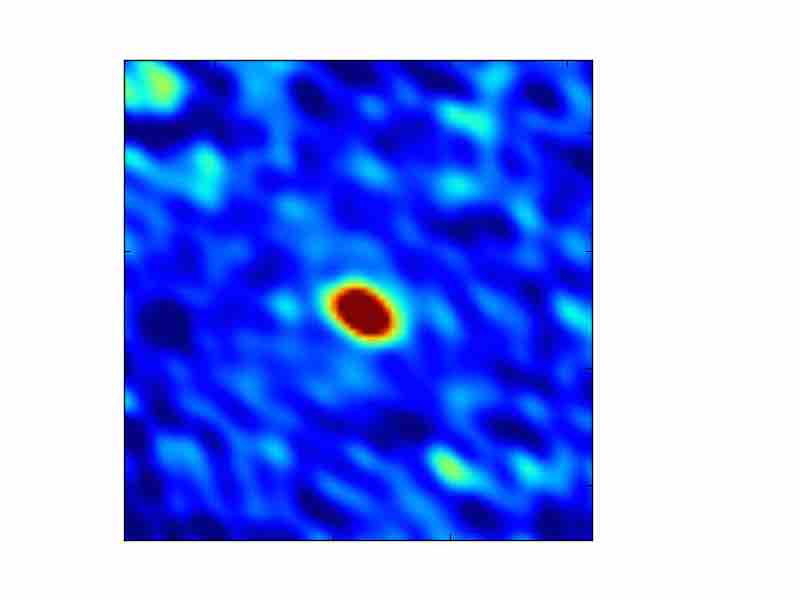} &
\hspace*{-1.2cm}\begin{overpic}[width=3.4cm]{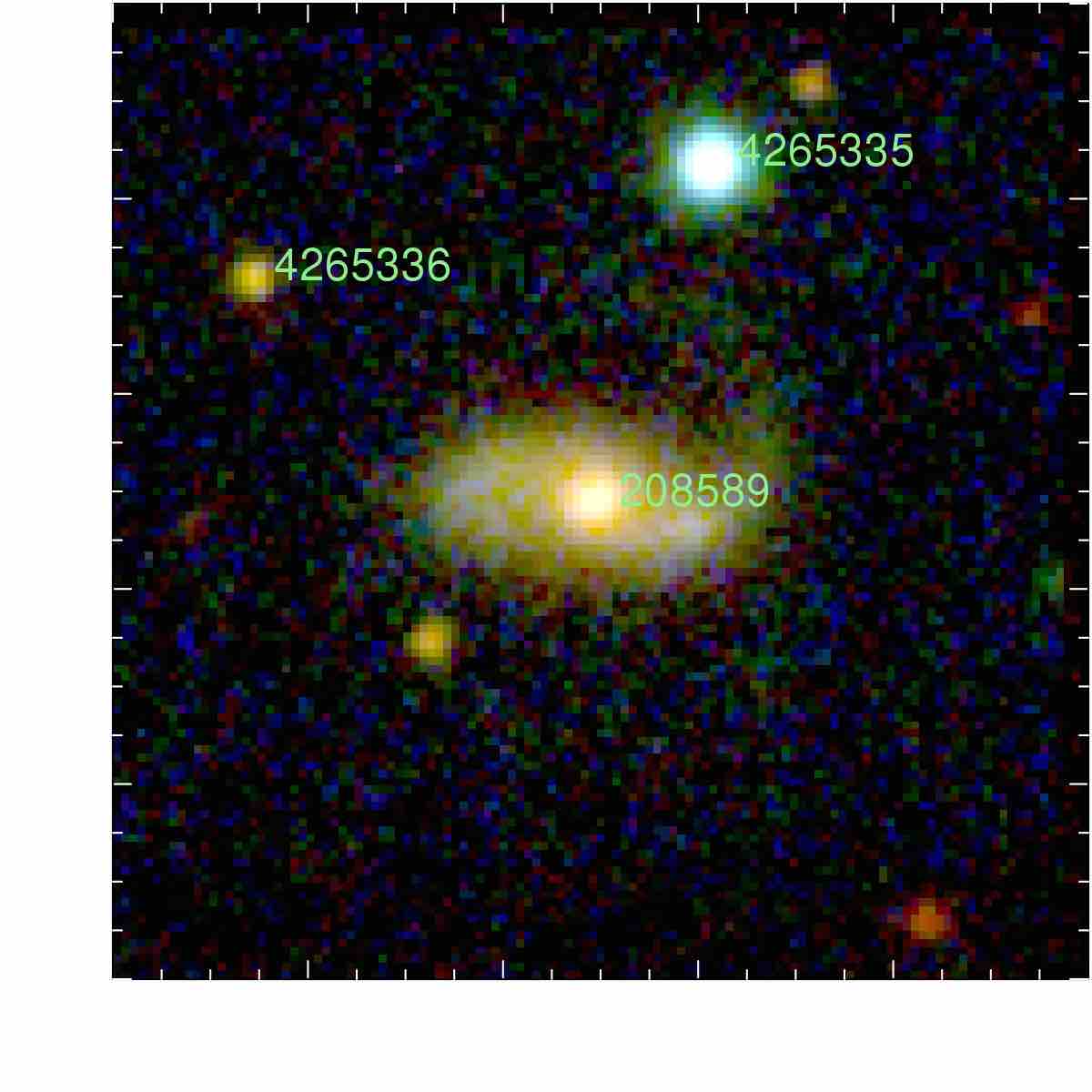} \put (9,85) { \begin{fitbox}{2.25cm}{0.2cm} \color{white}$\bf DBC$ \end{fitbox}} \end{overpic} \\ 	
	
\includegraphics[width=8.4cm]{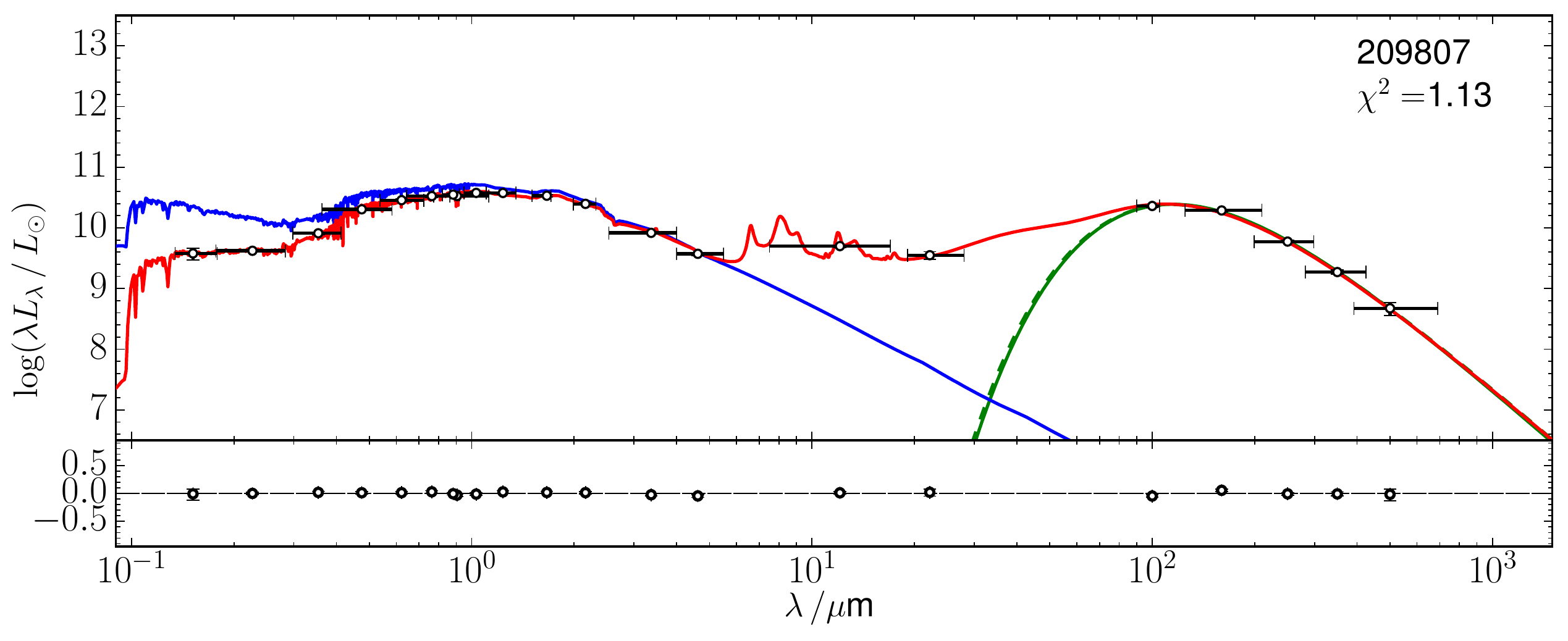} &
\includegraphics[width=5.0cm]{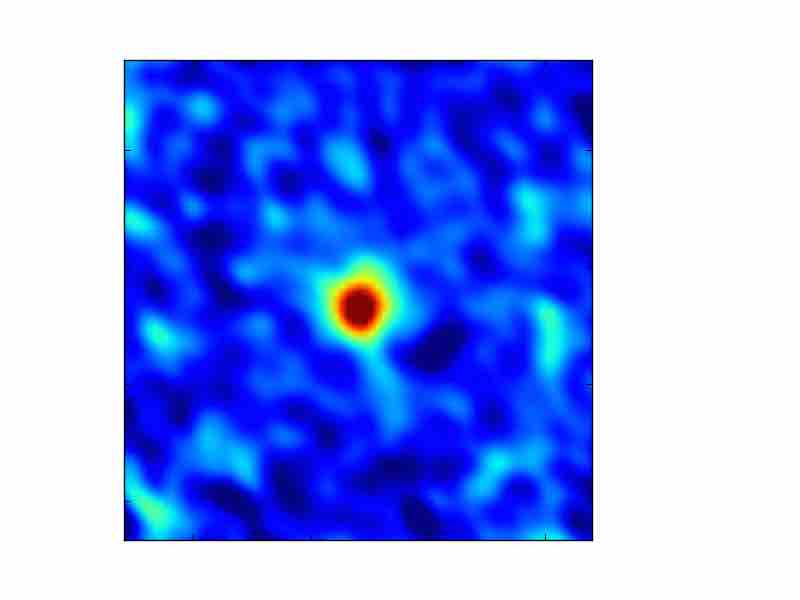} &
\hspace*{-1.2cm}\begin{overpic}[width=3.4cm]{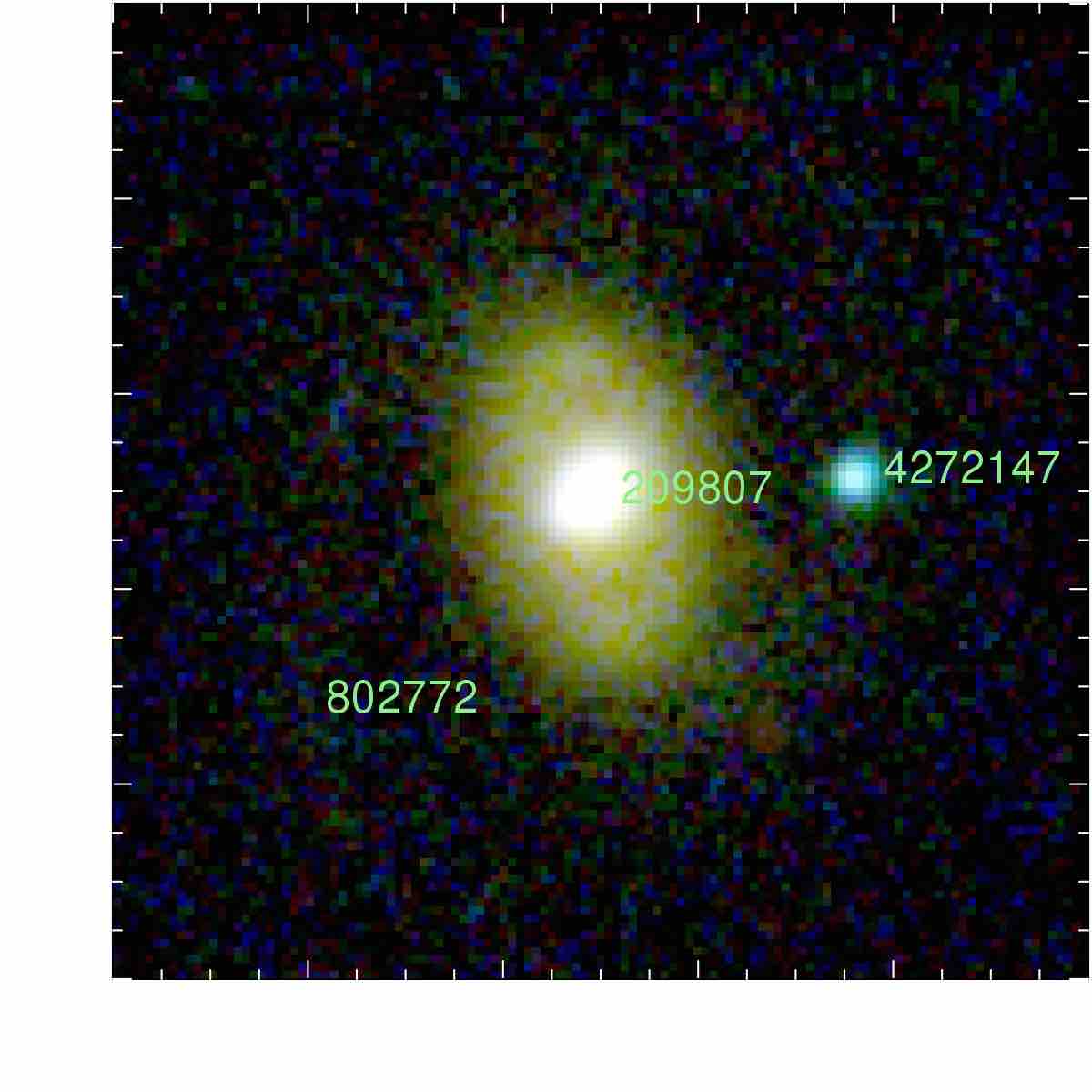} \put (9,85) { \begin{fitbox}{2.25cm}{0.2cm} \color{white}$\bf DBC$ \end{fitbox}} \end{overpic} \\ 	

\includegraphics[width=8.4cm]{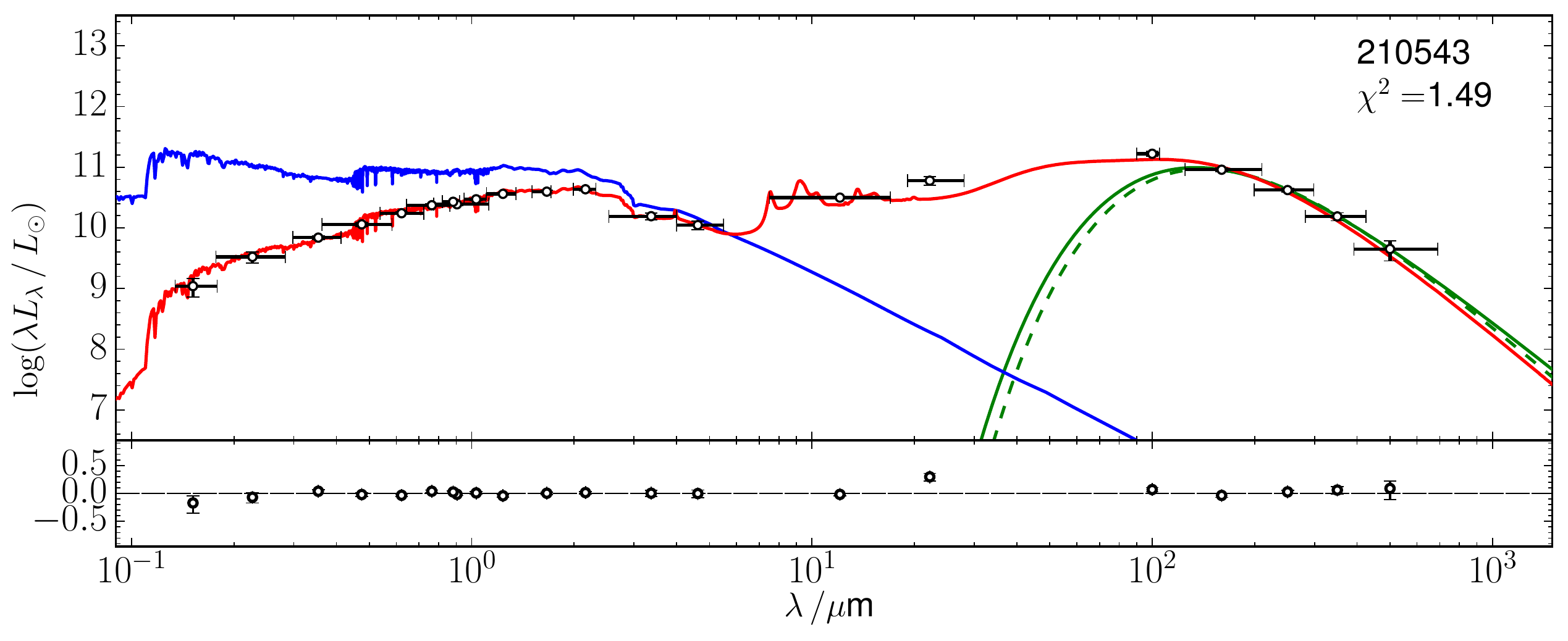} &
\includegraphics[width=5.0cm]{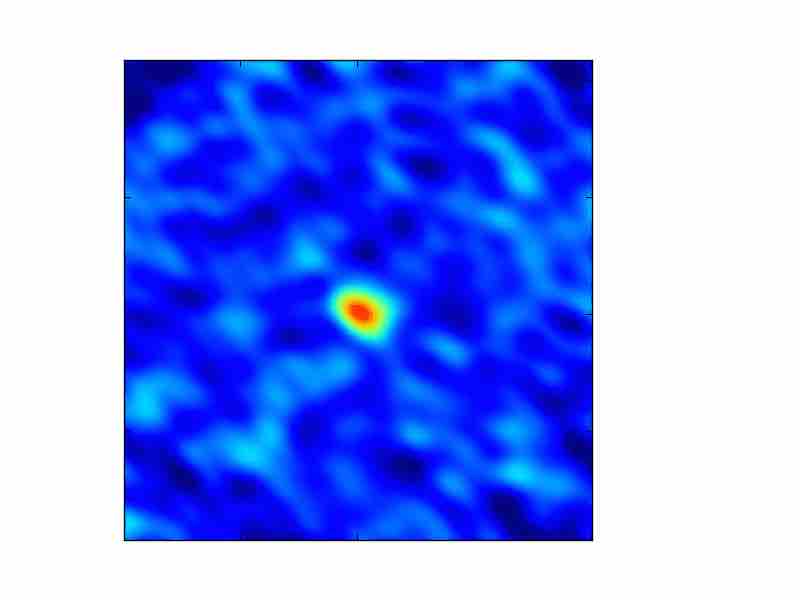} &
\hspace*{-1.2cm}\begin{overpic}[width=3.4cm]{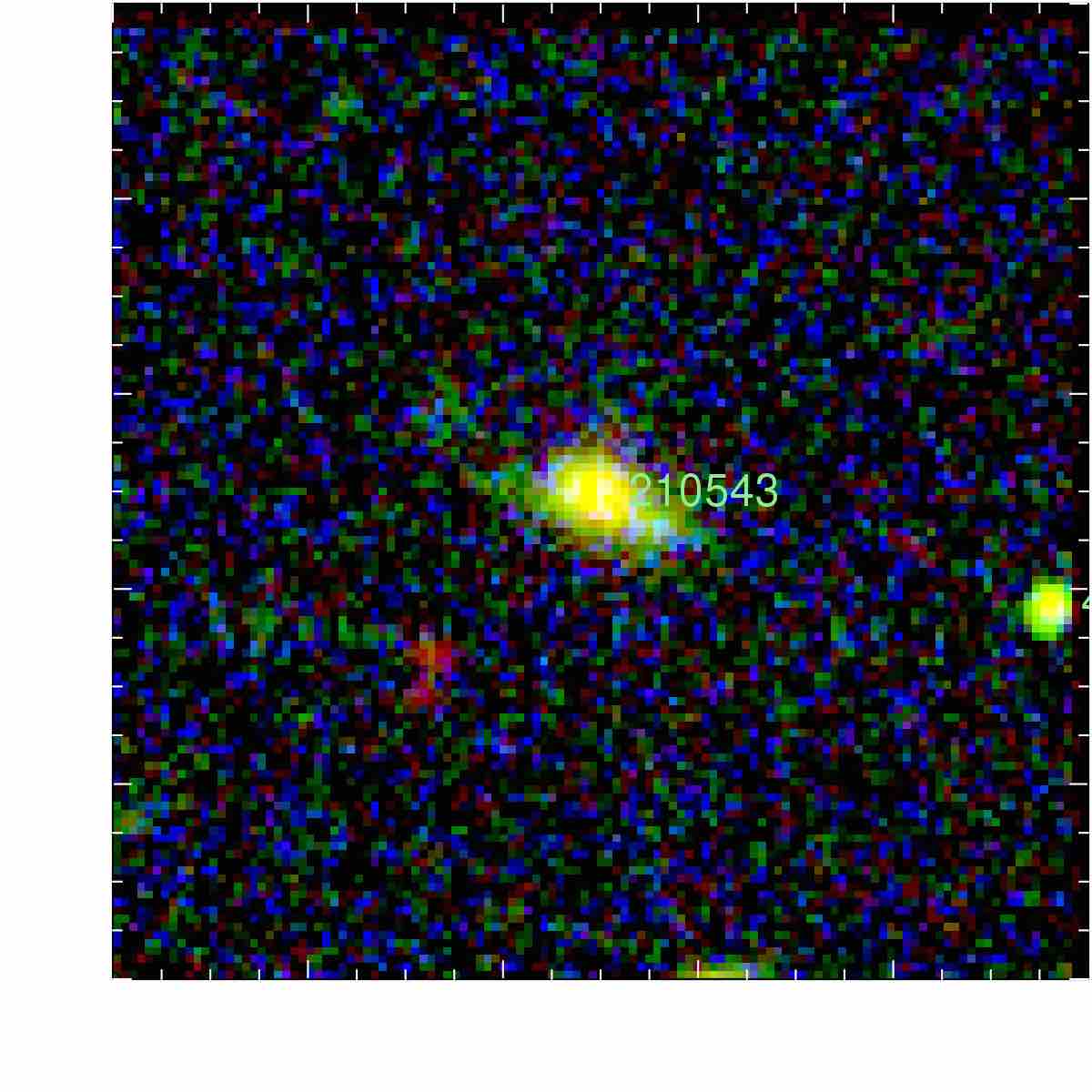} \put (9,85) { \begin{fitbox}{2.25cm}{0.2cm} \color{white}$\bf B$ \end{fitbox}} \end{overpic} \\ 	

\includegraphics[width=8.4cm]{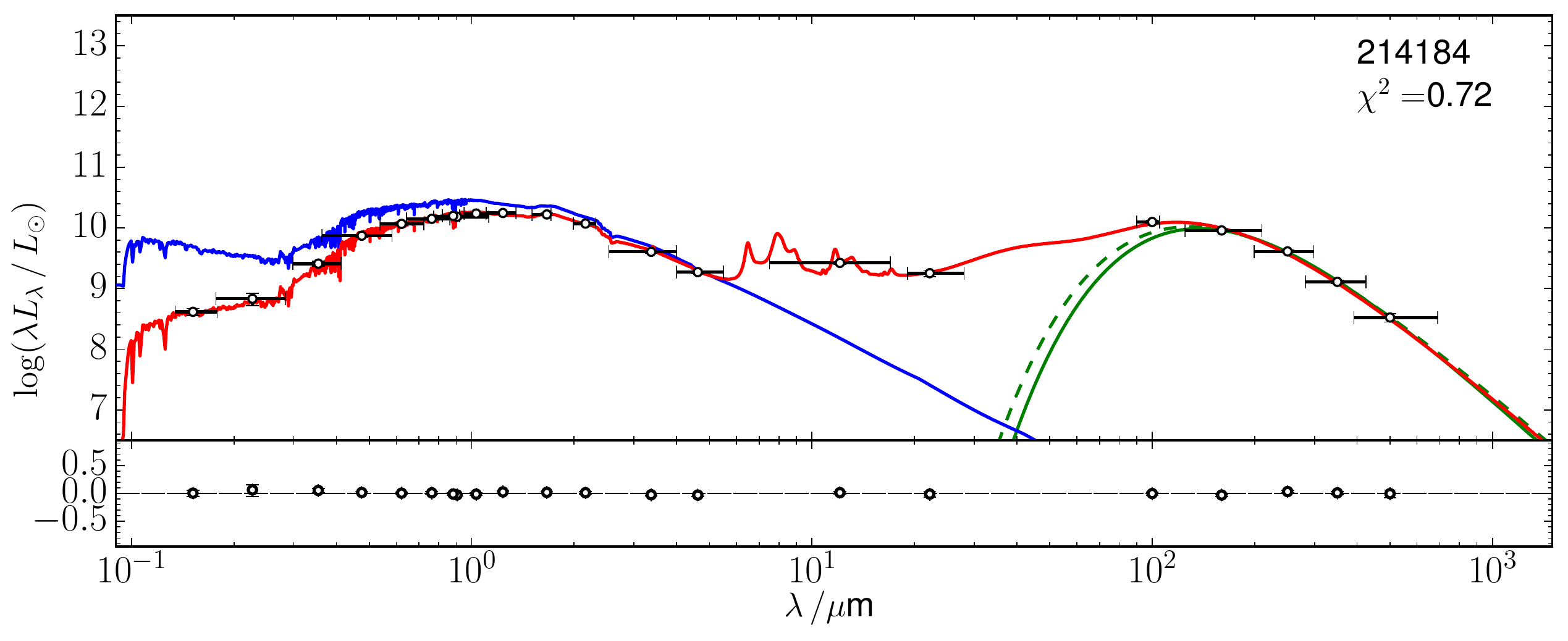} &
\includegraphics[width=5.0cm]{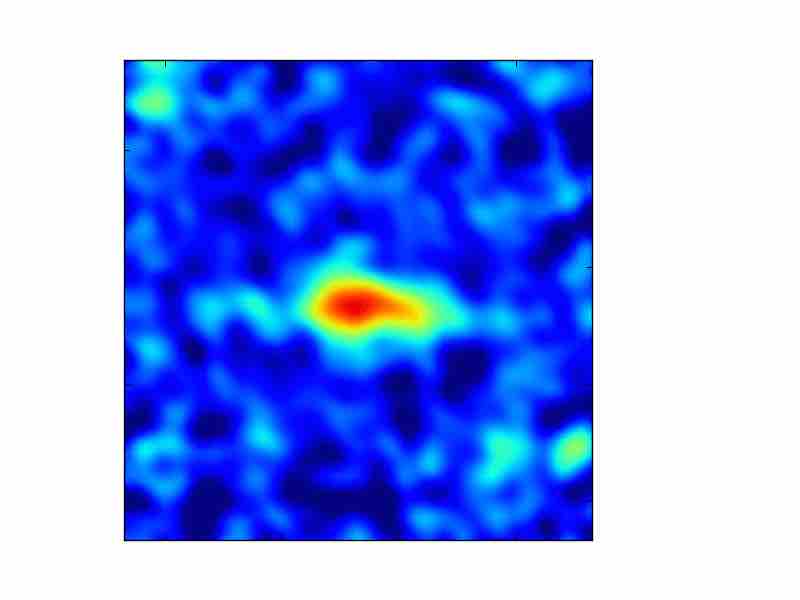} &
\hspace*{-1.2cm}\begin{overpic}[width=3.4cm]{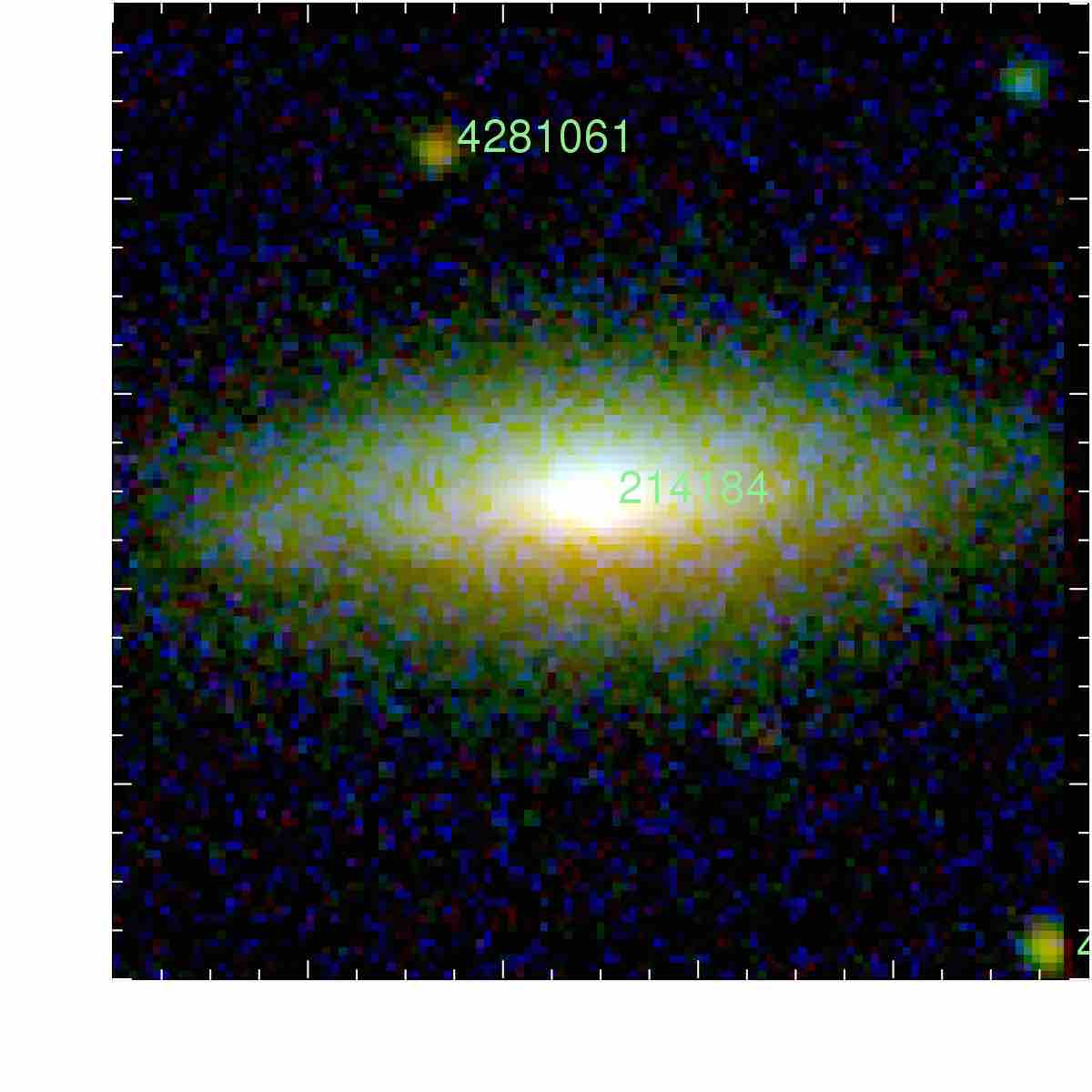} \put (9,85) { \begin{fitbox}{2.25cm}{0.2cm} \color{white}$\bf DB$ \end{fitbox}} \end{overpic} \\ 

\includegraphics[width=8.4cm]{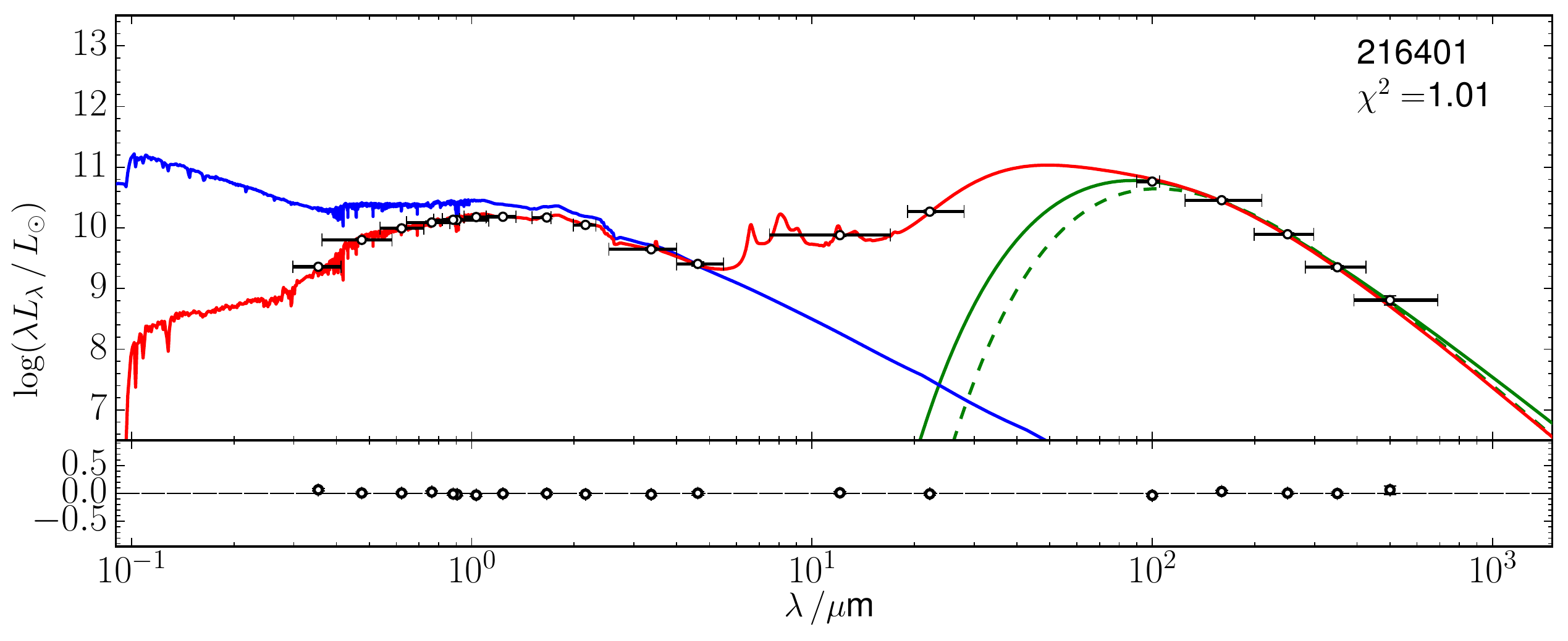} &
\includegraphics[width=5.0cm]{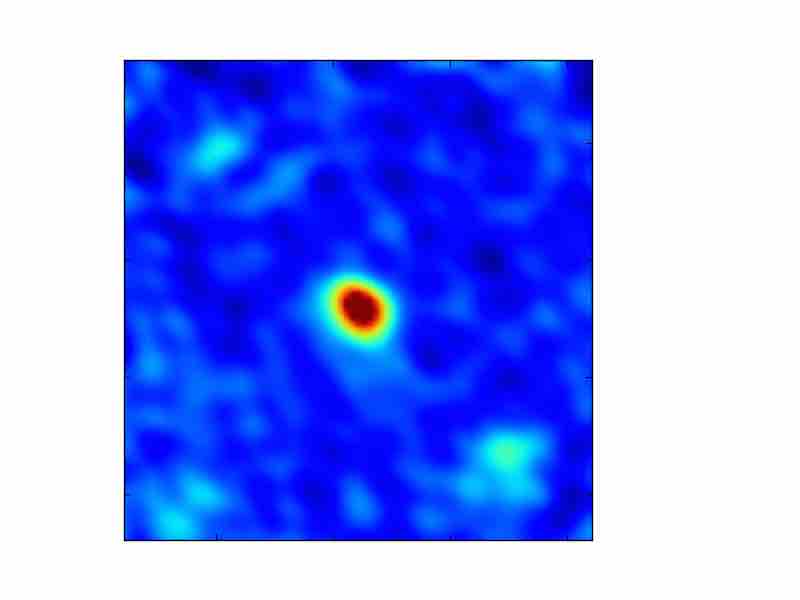} &
\hspace*{-1.2cm}\begin{overpic}[width=3.4cm]{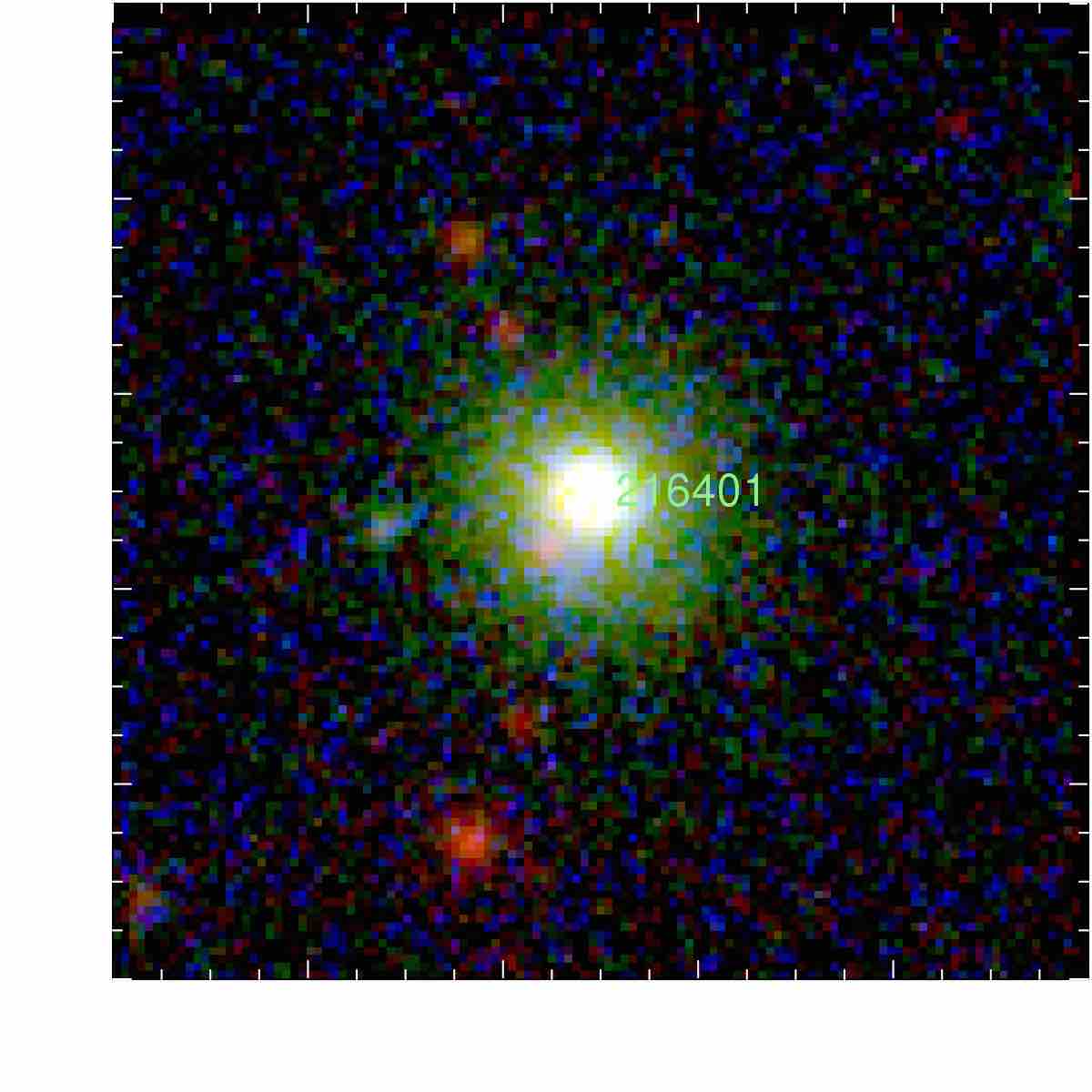} \put (9,85) { \begin{fitbox}{2.25cm}{0.2cm} \color{white}$\bf BC$ \end{fitbox}} \end{overpic} \\ 

\end{array}
$
{\textbf{Figure~\ref{pdrdiaglit}.} continued}

\end{figure*}


\begin{figure*}
$
\begin{array}{ccc}
\includegraphics[width=8.4cm]{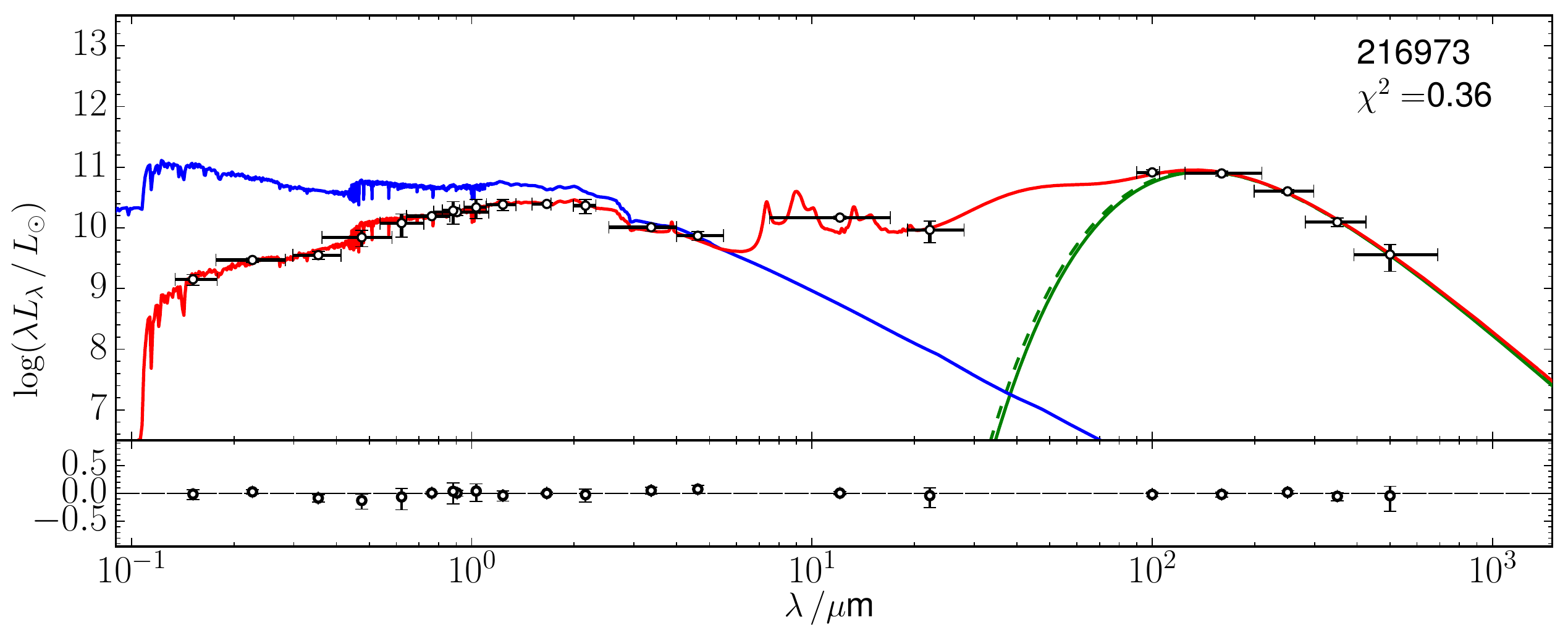} &
\includegraphics[width=5.0cm]{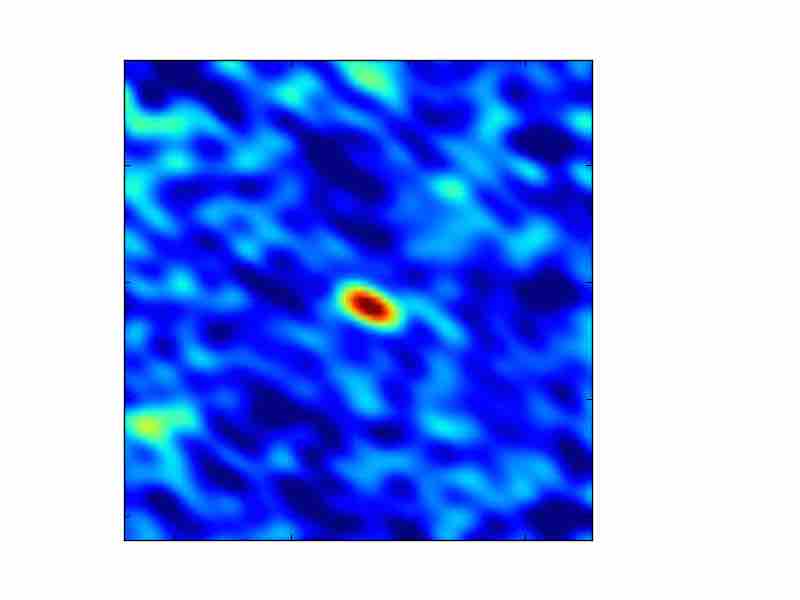} &
\hspace*{-1.2cm}\begin{overpic}[width=3.4cm]{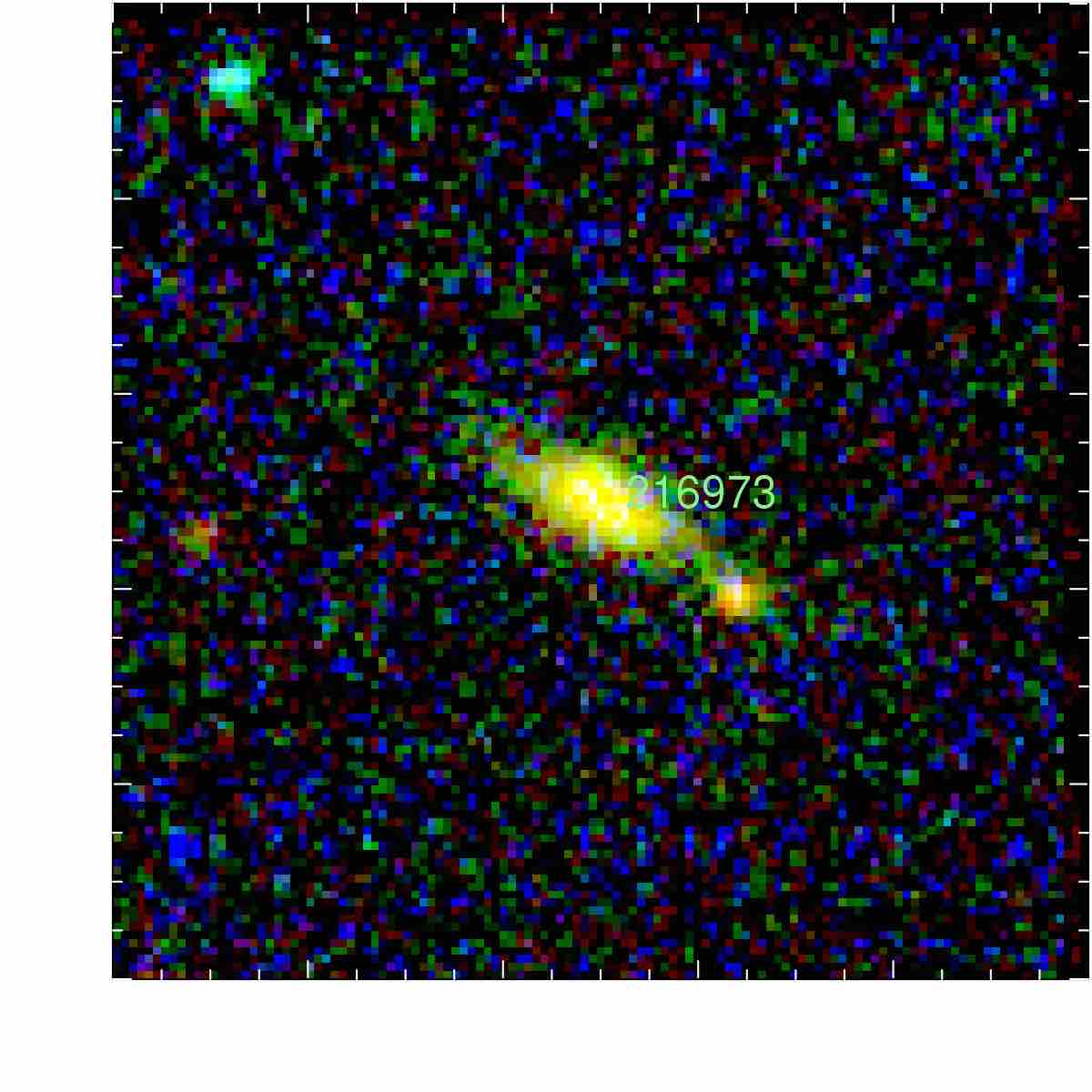} \put (9,85) { \begin{fitbox}{2.25cm}{0.2cm} \color{white}$\bf DC$ \end{fitbox}} \end{overpic} \\ 	
	
\includegraphics[width=8.4cm]{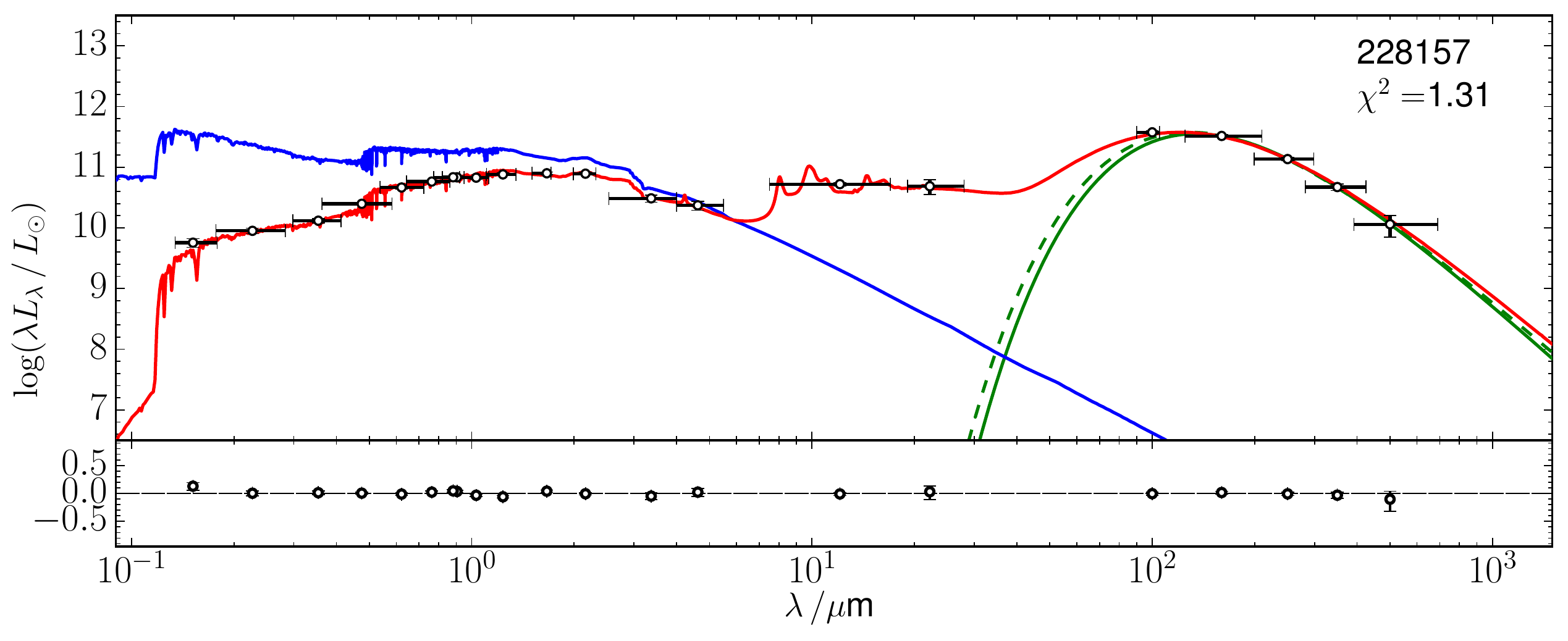} &
\includegraphics[width=5.0cm]{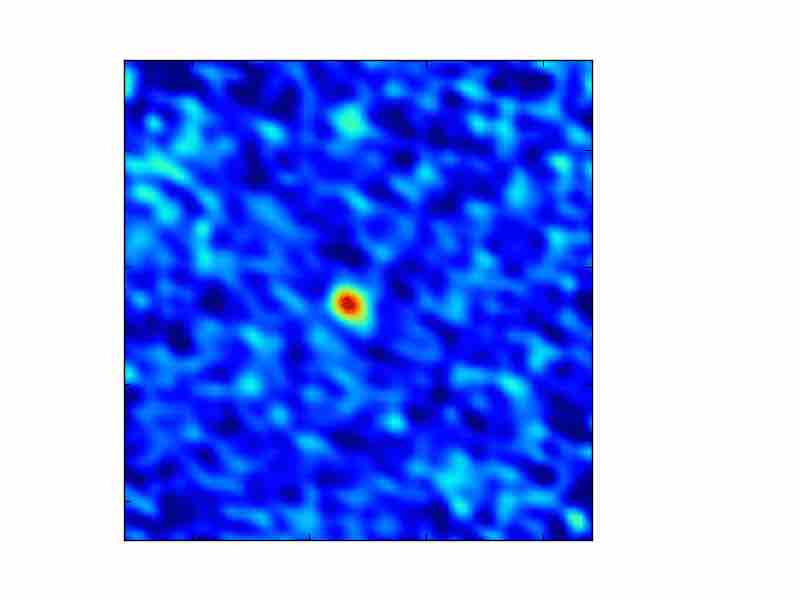} &
\hspace*{-1.2cm}\begin{overpic}[width=3.4cm]{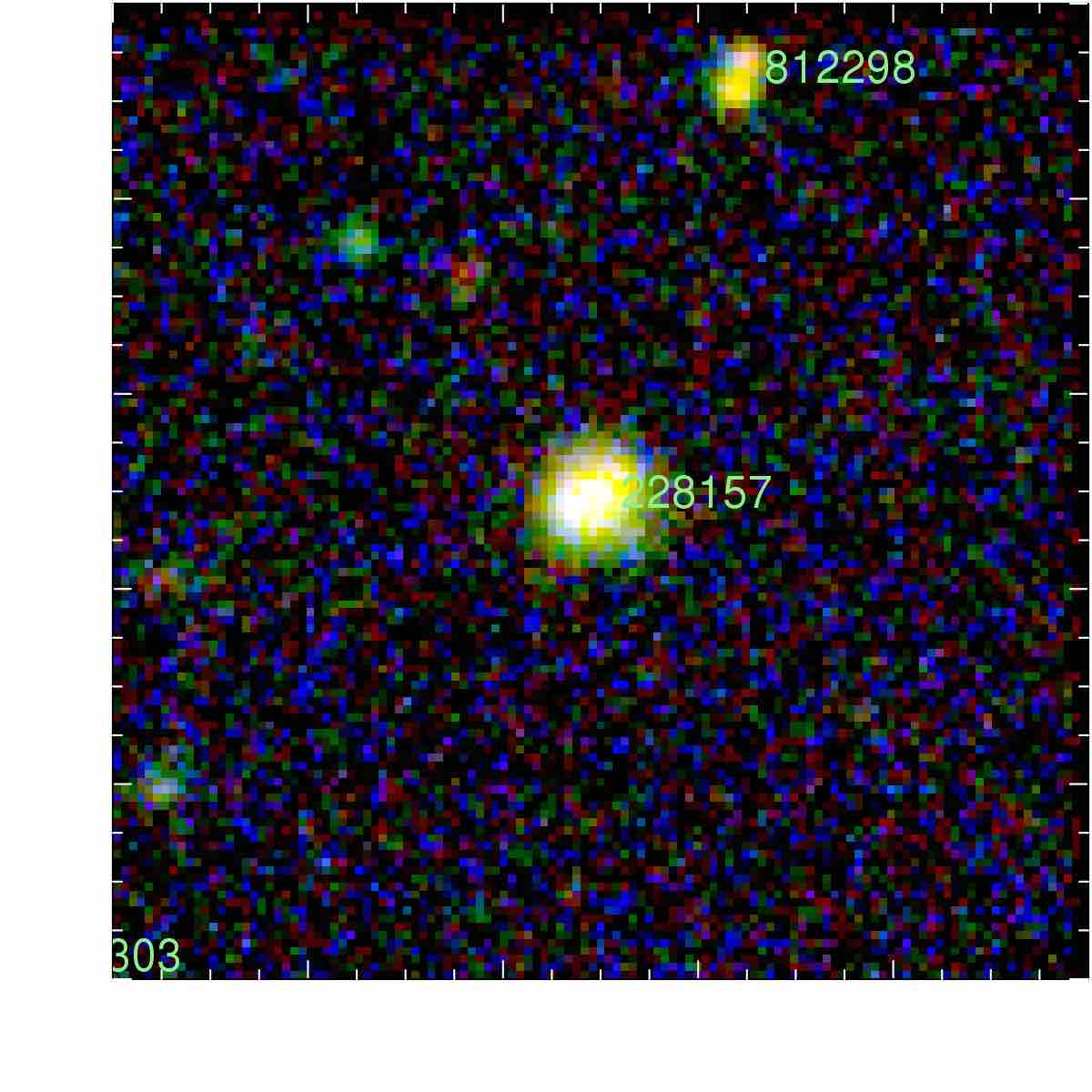} \put (9,85) { \begin{fitbox}{2.25cm}{0.2cm} \color{white}$\bf BD$ \end{fitbox}} \end{overpic} \\ 		

\includegraphics[width=8.4cm]{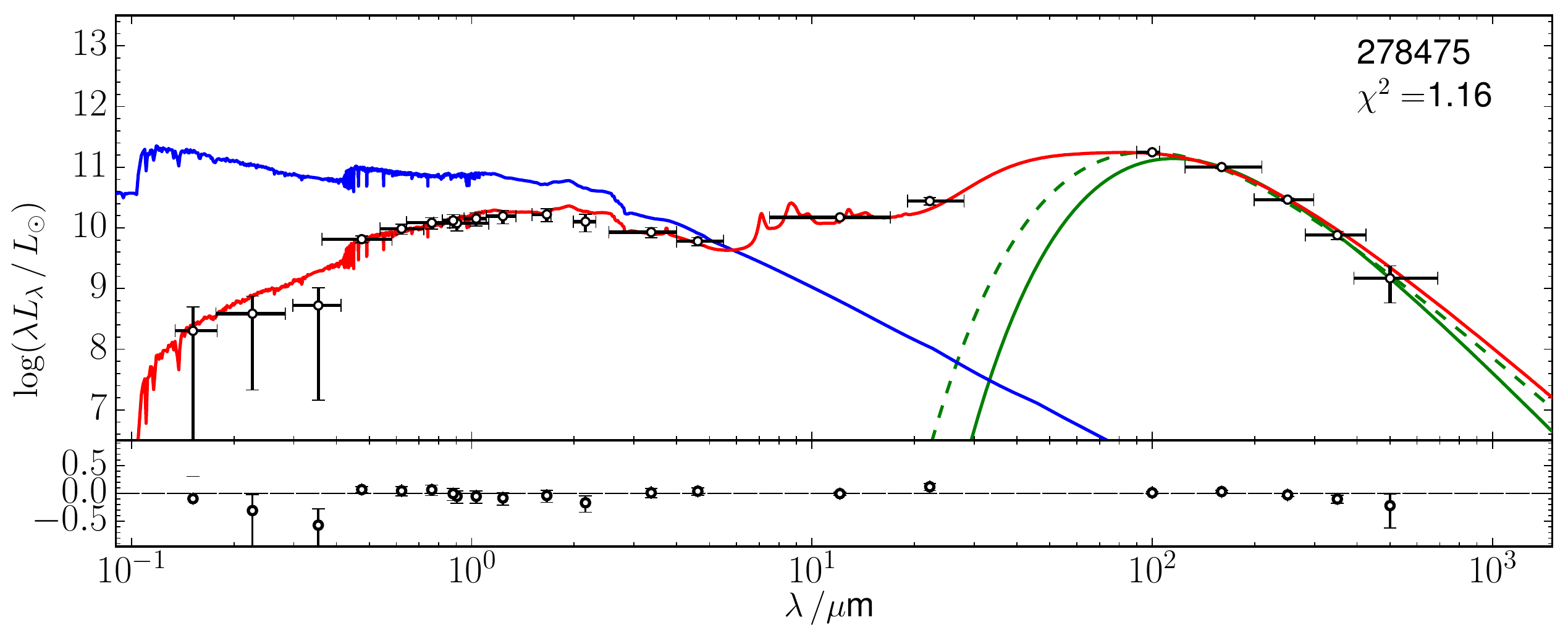} &
\includegraphics[width=5.0cm]{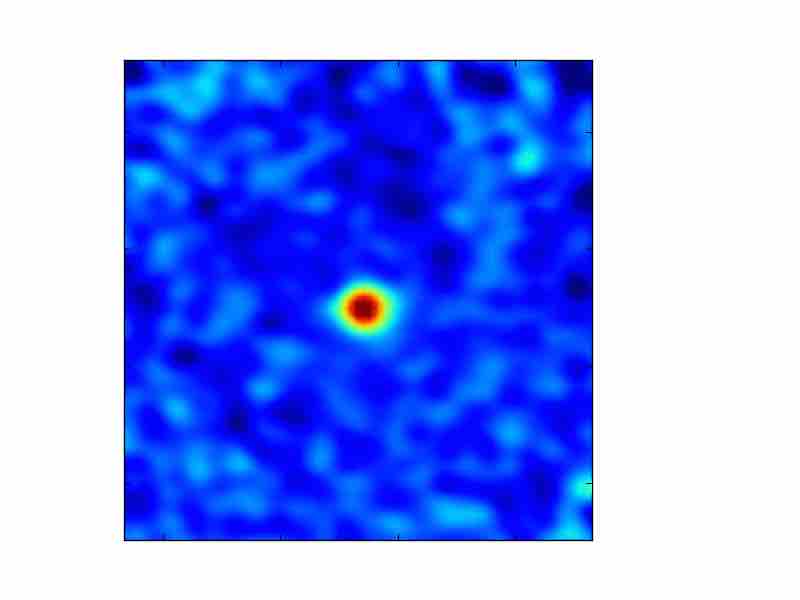} &
\hspace*{-1.2cm}\begin{overpic}[width=3.4cm]{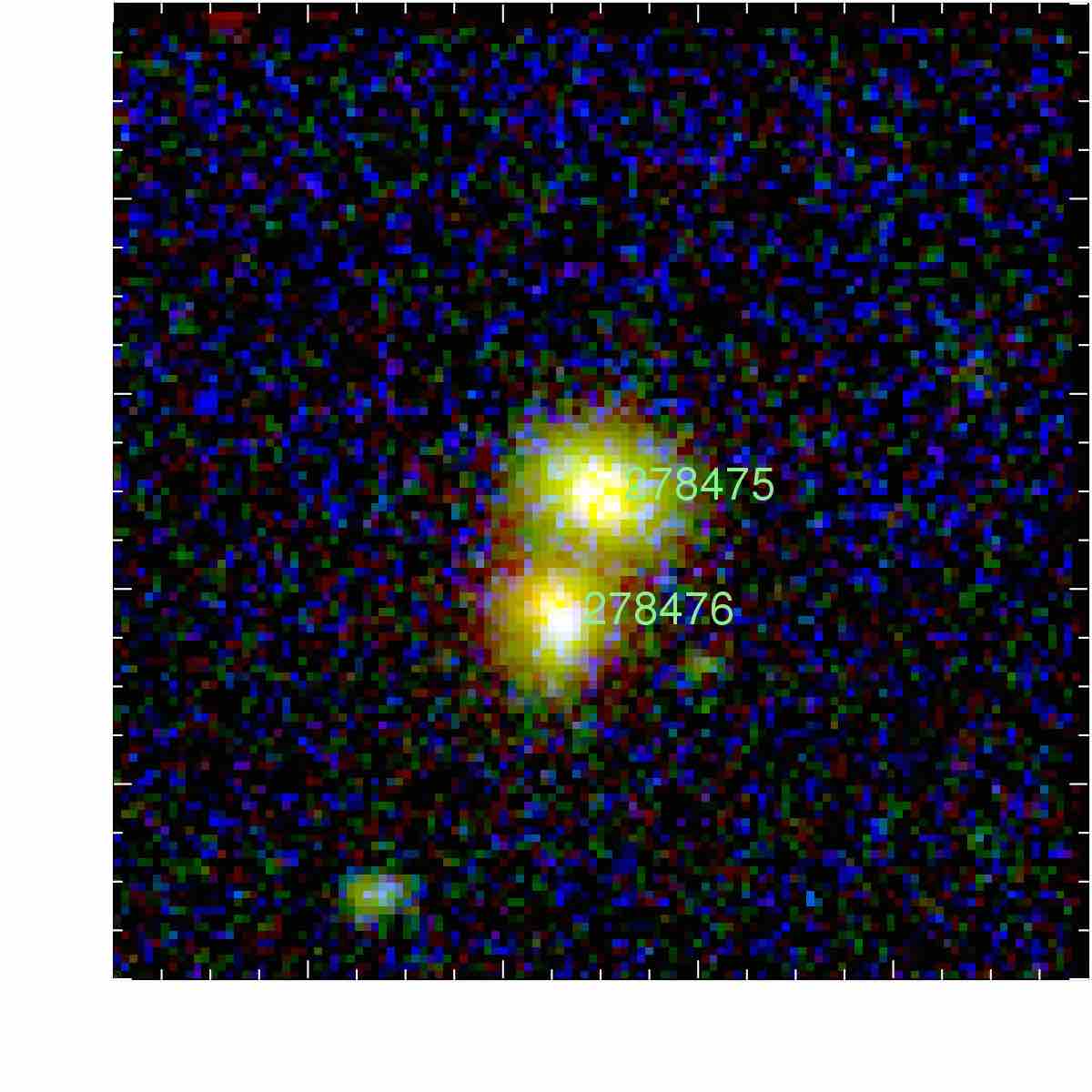} \put (9,85) { \begin{fitbox}{2.25cm}{0.2cm} \color{white}$\bf M$ \end{fitbox}} \end{overpic} \\ 

\includegraphics[width=8.4cm]{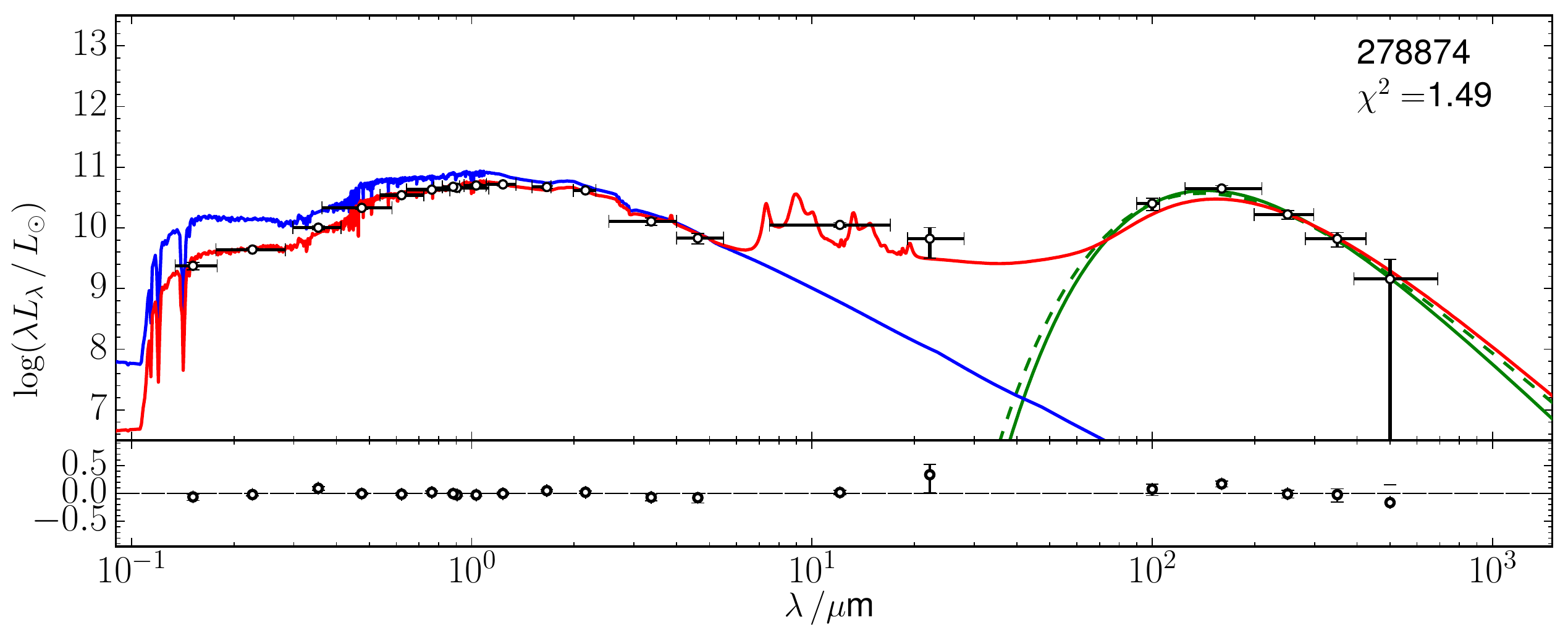} &
\includegraphics[width=5.0cm]{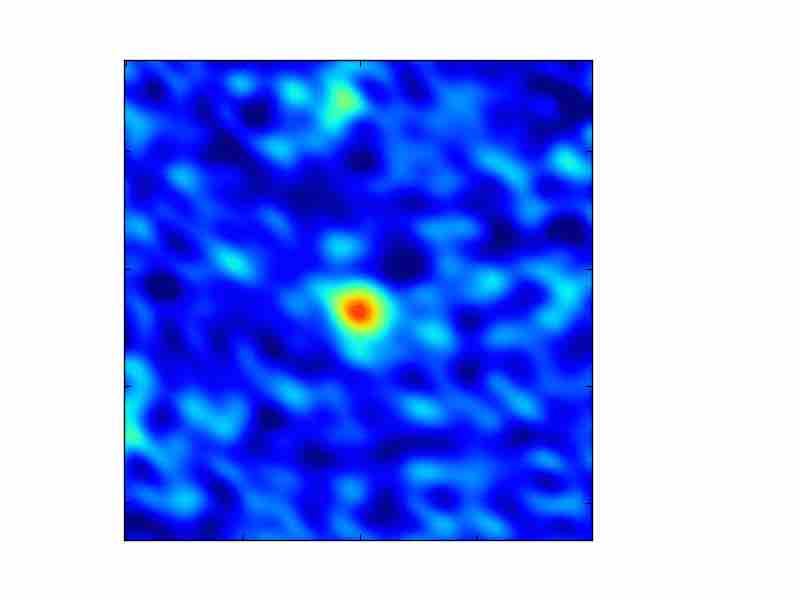} &
\hspace*{-1.2cm}\begin{overpic}[width=3.4cm]{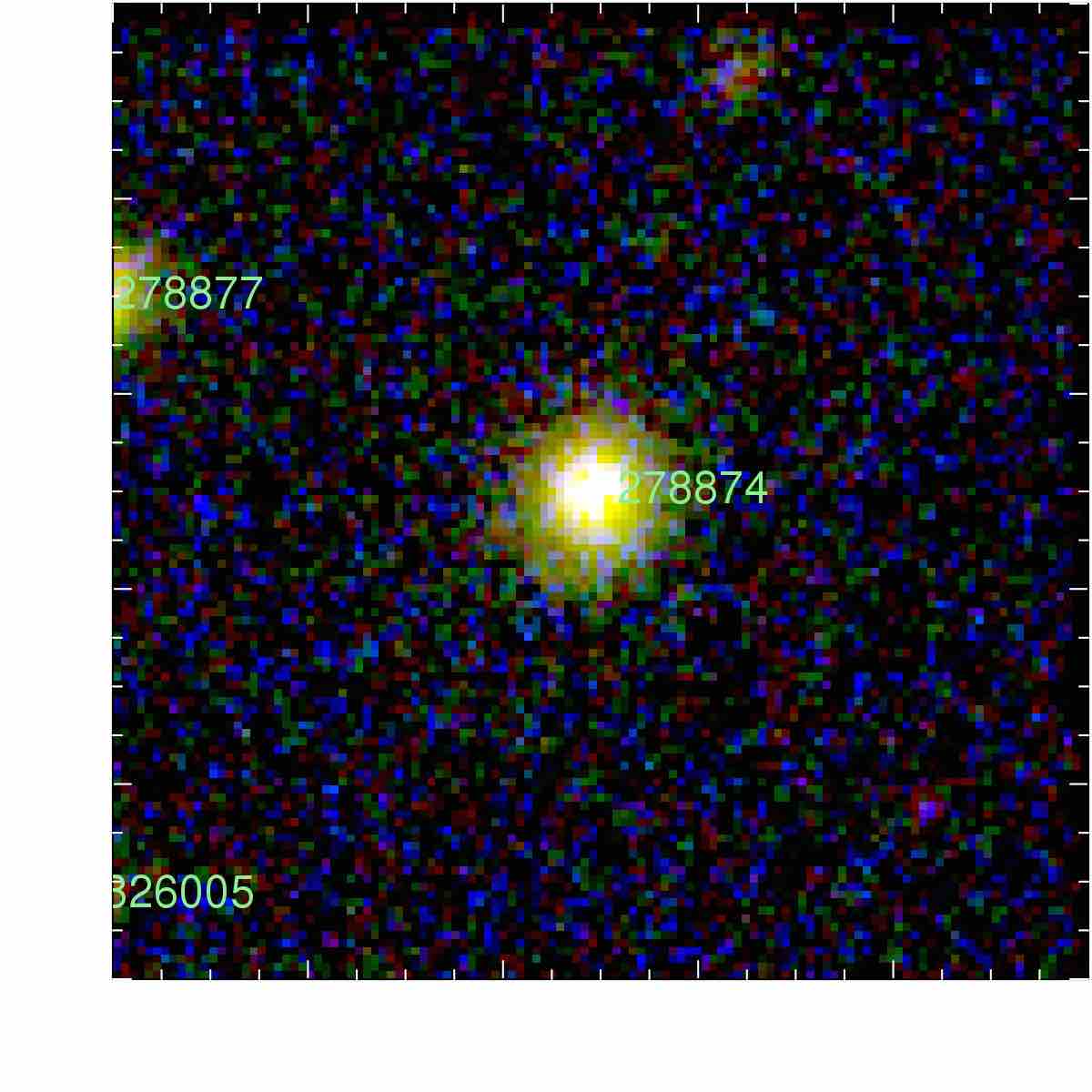} \put (9,85) { \begin{fitbox}{2.25cm}{0.2cm} \color{white}$\bf B$ \end{fitbox}} \end{overpic} \\ 

\includegraphics[width=8.4cm]{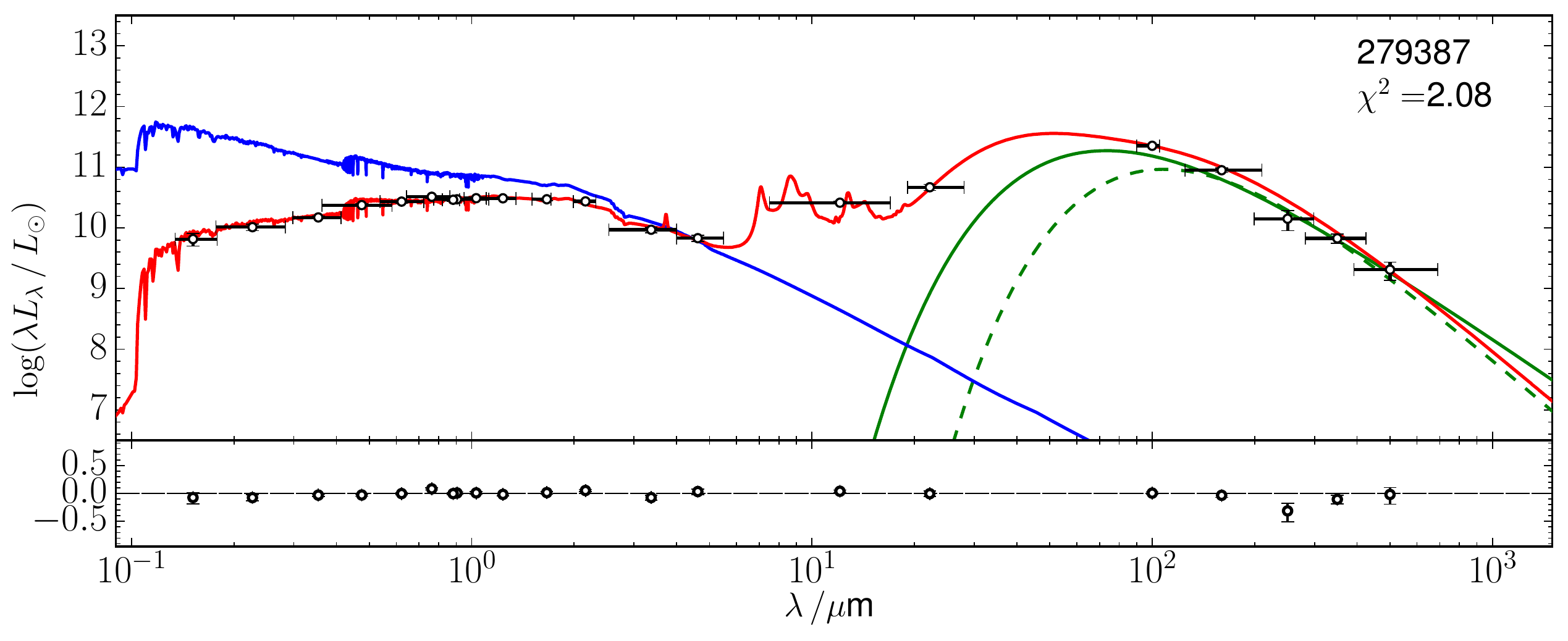} &
\includegraphics[width=5.0cm]{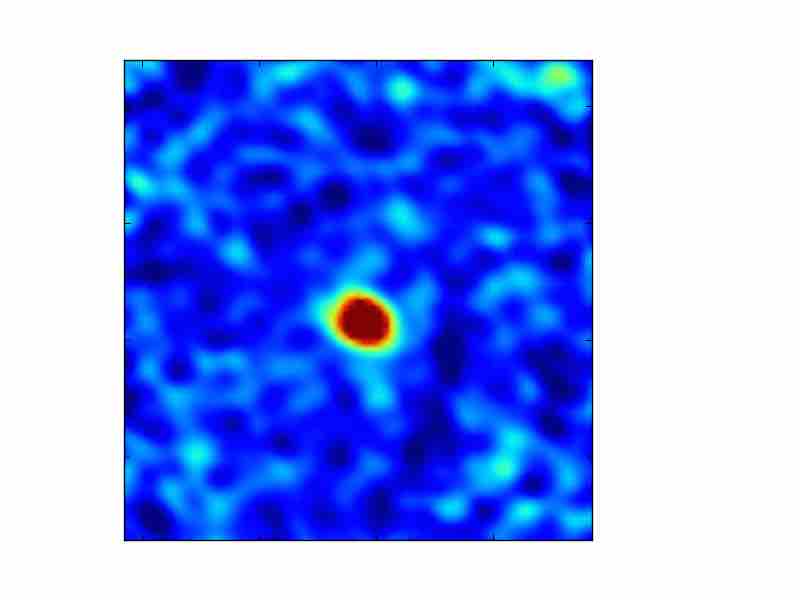} &
\hspace*{-1.2cm}\begin{overpic}[width=3.4cm]{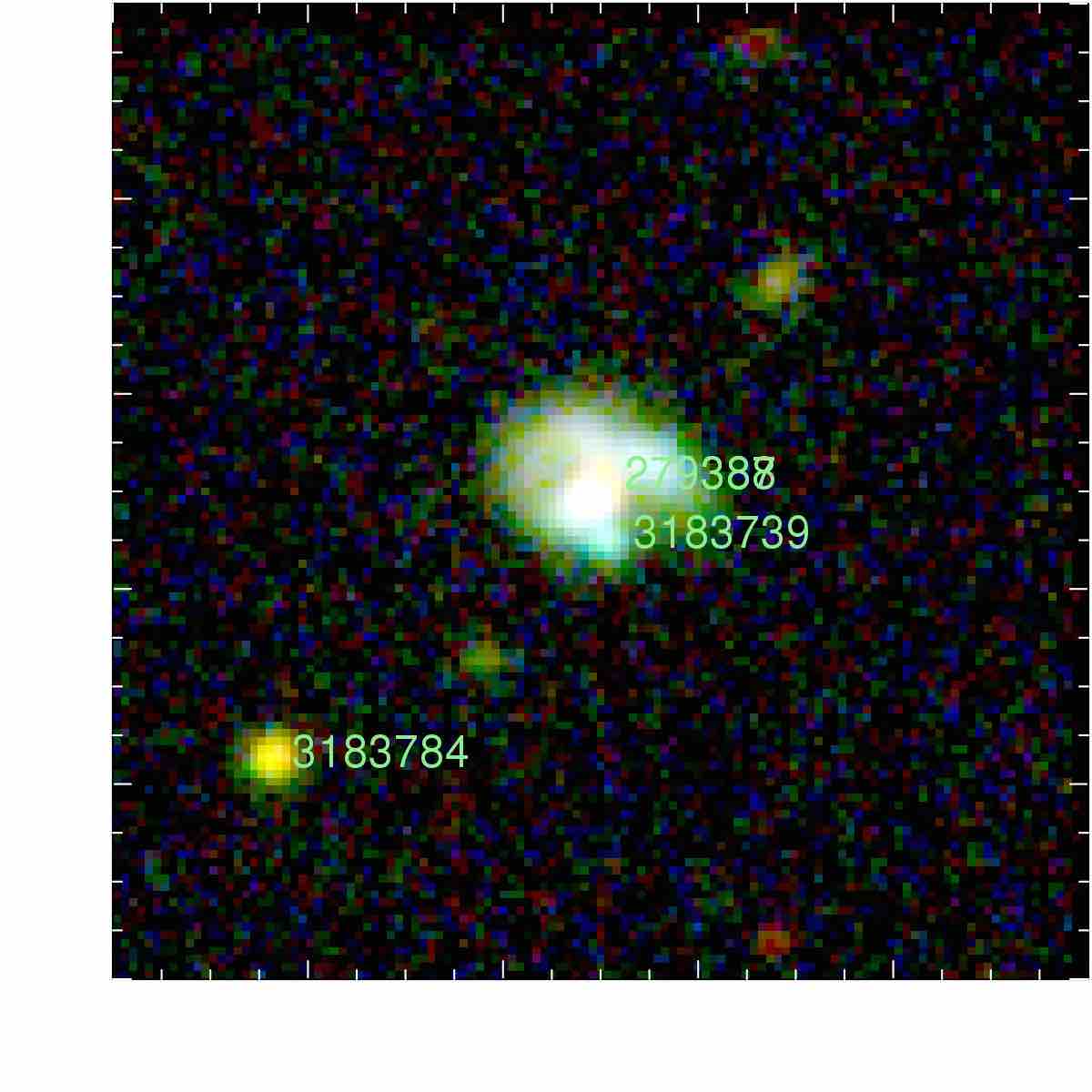} \put (9,85) { \begin{fitbox}{2.25cm}{0.2cm} \color{white}$\bf M$ \end{fitbox}} \end{overpic} \\ 

\end{array}
$
{\textbf{Figure~\ref{pdrdiaglit}.} continued}

\end{figure*}


\begin{figure*}
$
\begin{array}{ccc}
\includegraphics[width=8.4cm]{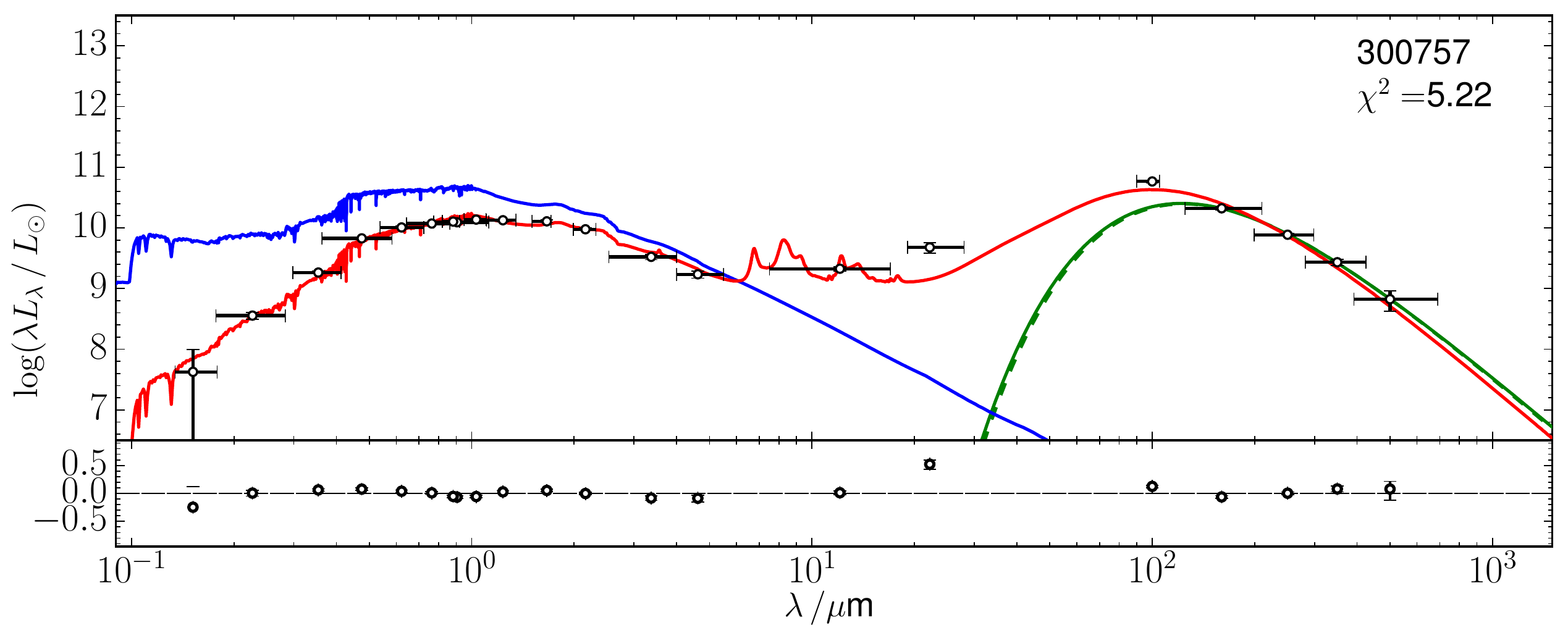} &
\includegraphics[width=5.0cm]{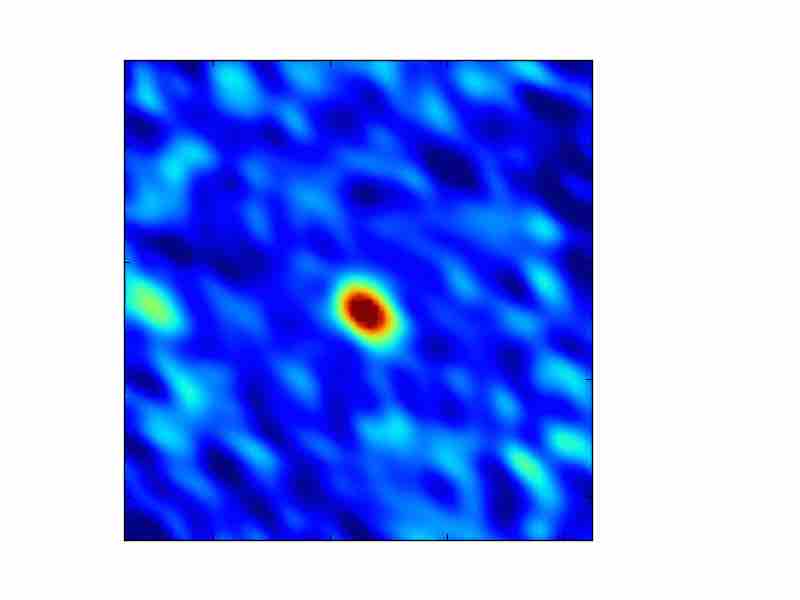} &
\hspace*{-1.2cm}\begin{overpic}[width=3.4cm]{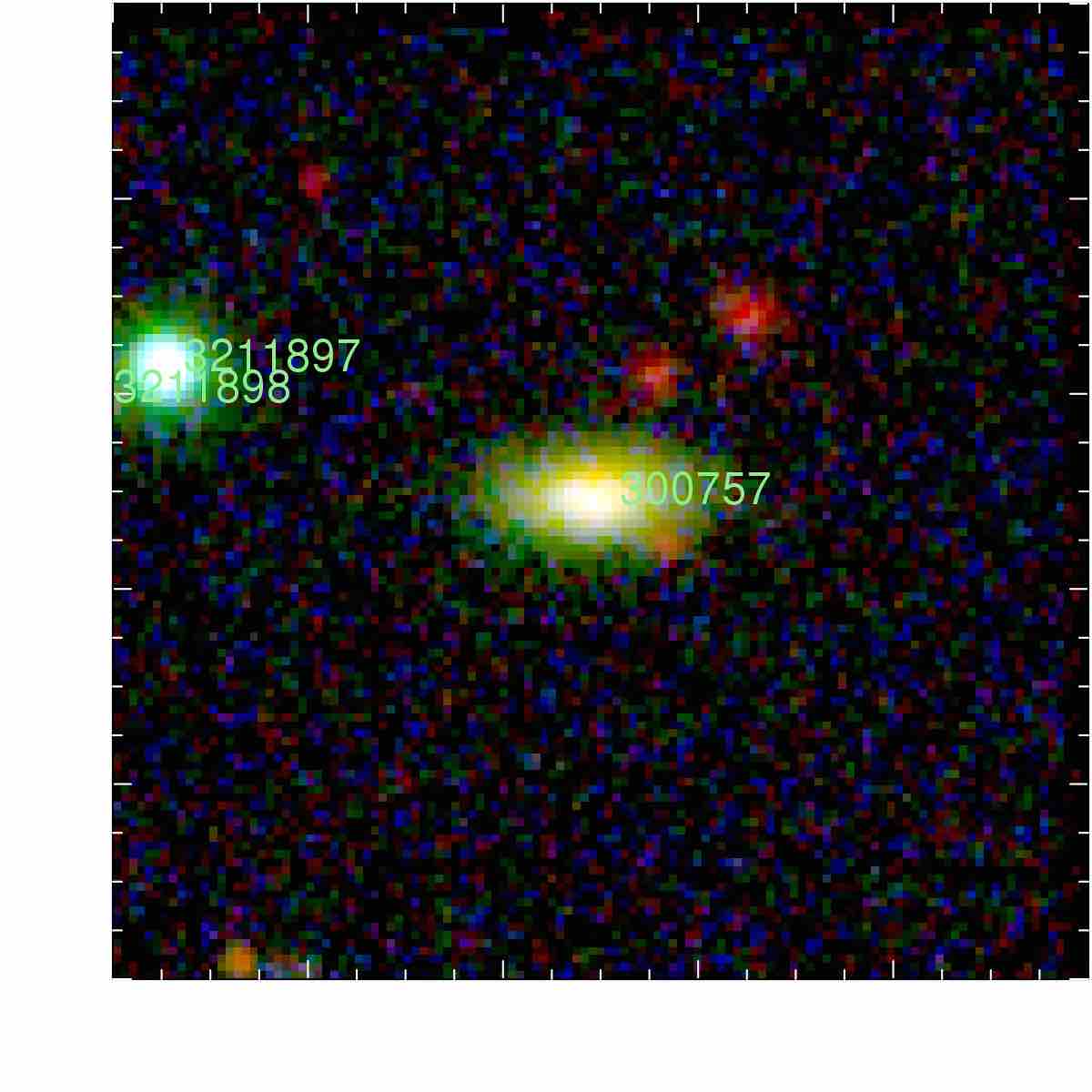} \put (9,85) { \begin{fitbox}{2.25cm}{0.2cm} \color{white}$\bf DBC$ \end{fitbox}} \end{overpic} \\ 	
	
\includegraphics[width=8.4cm]{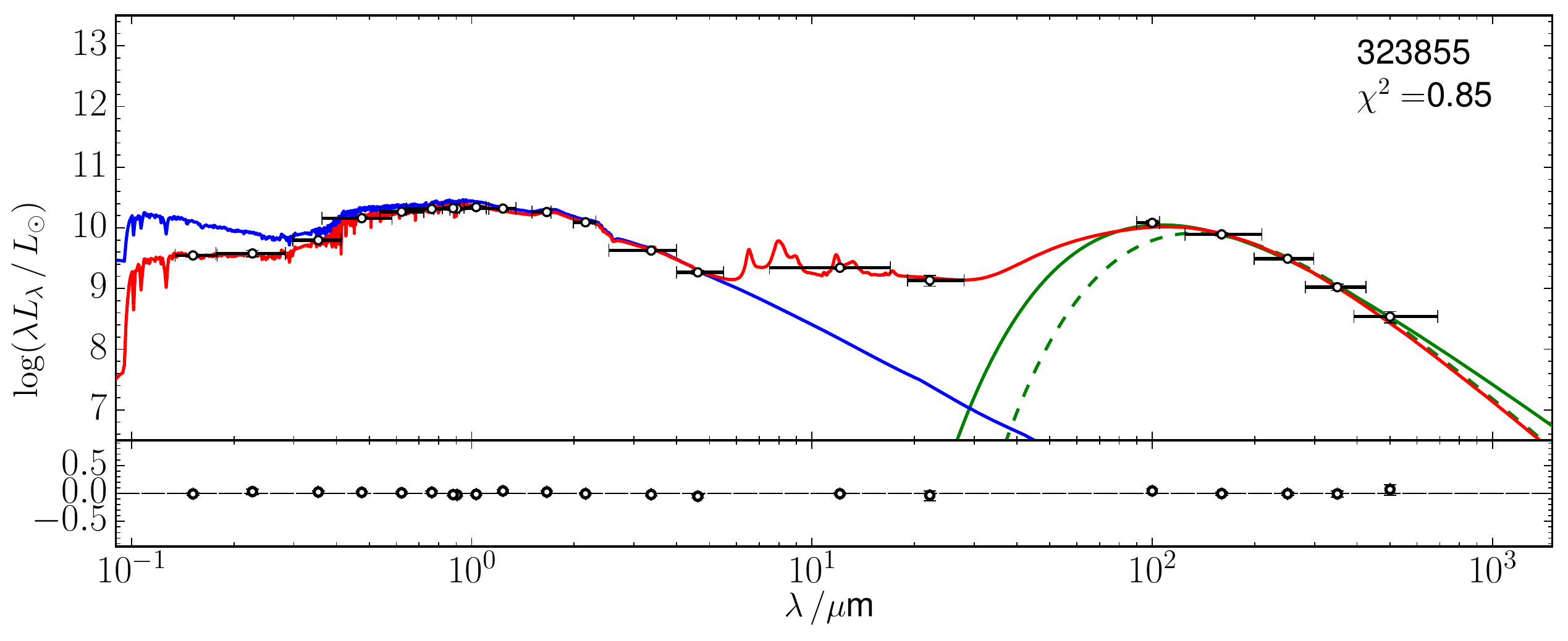} &
\includegraphics[width=5.0cm]{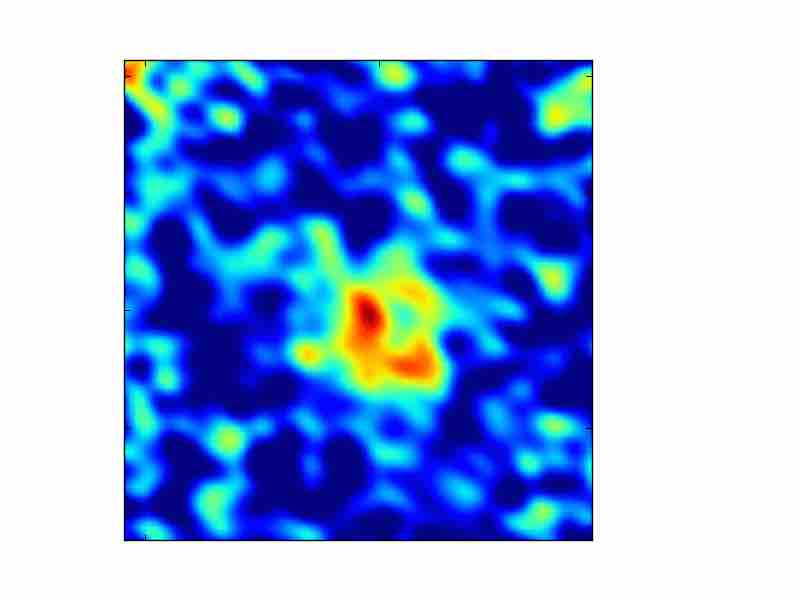} &
\hspace*{-1.2cm}\begin{overpic}[width=3.4cm]{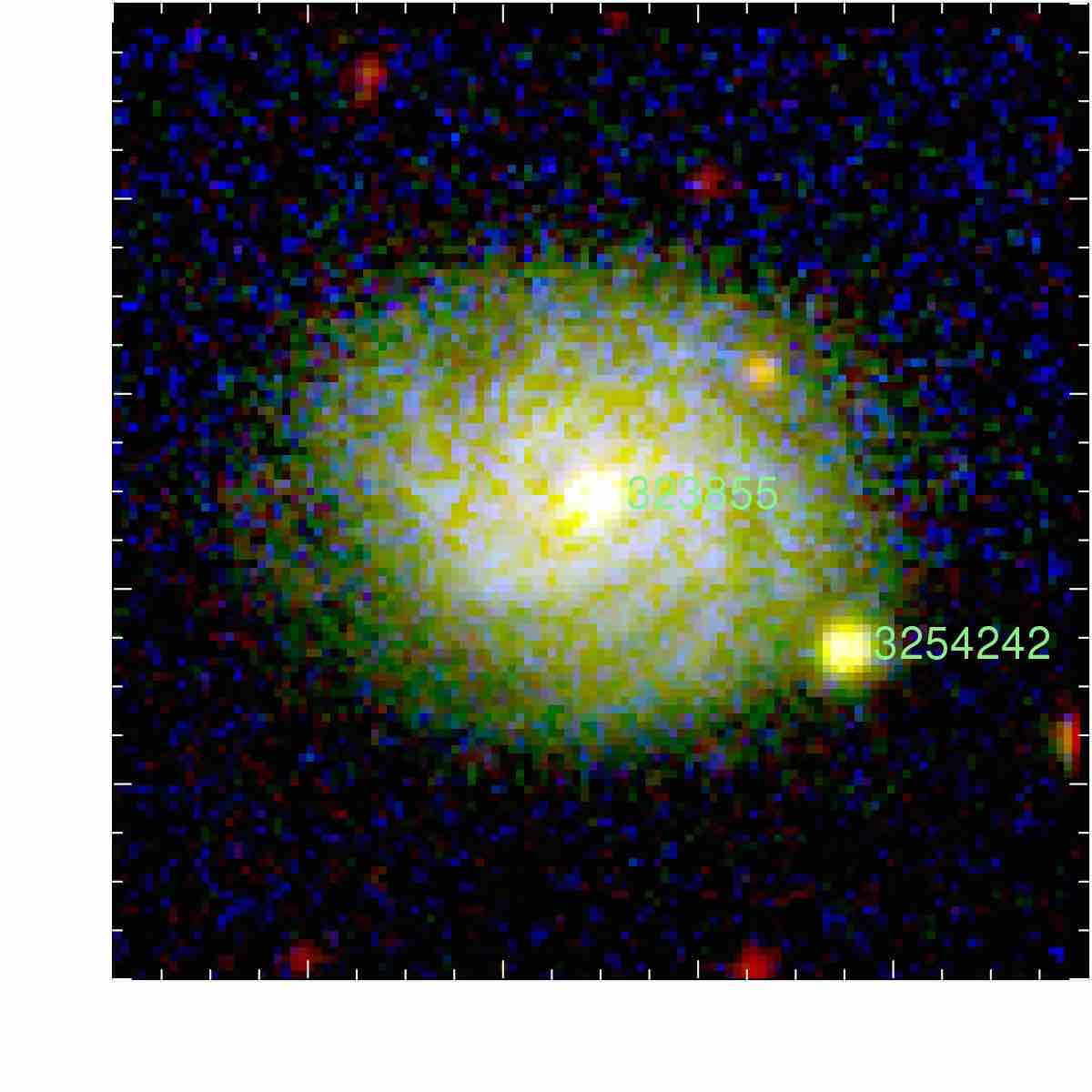} \put (9,85) { \begin{fitbox}{2.25cm}{0.2cm} \color{white}$\bf DC$ \end{fitbox}} \end{overpic} \\ 		

\includegraphics[width=8.4cm]{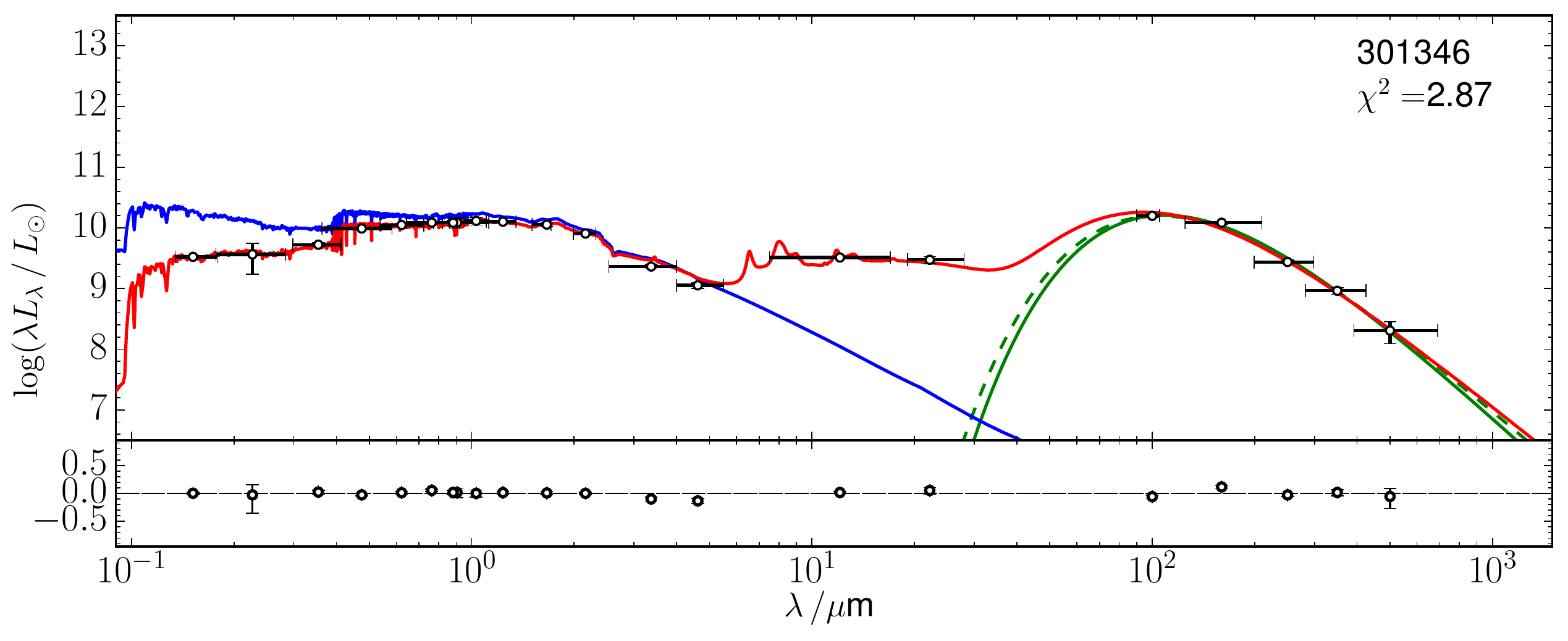} &
\includegraphics[width=5.0cm]{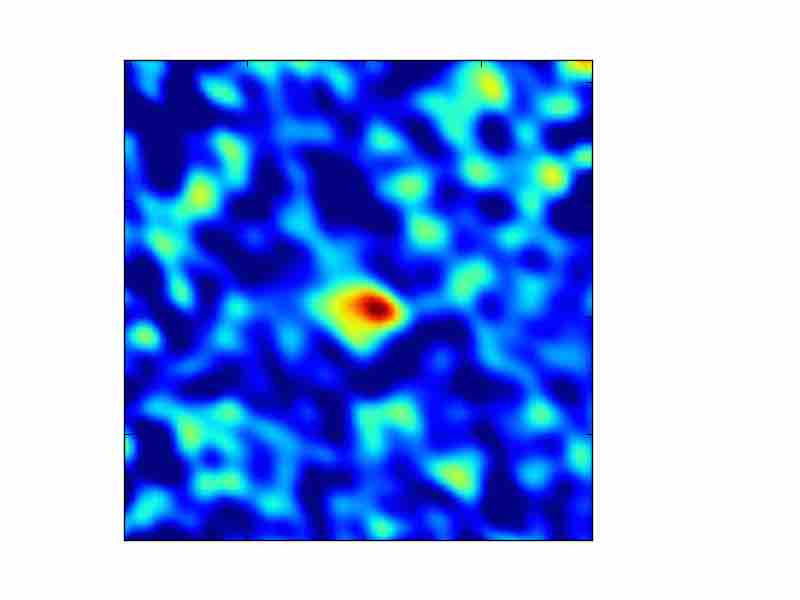} &
\hspace*{-1.2cm}\begin{overpic}[width=3.4cm]{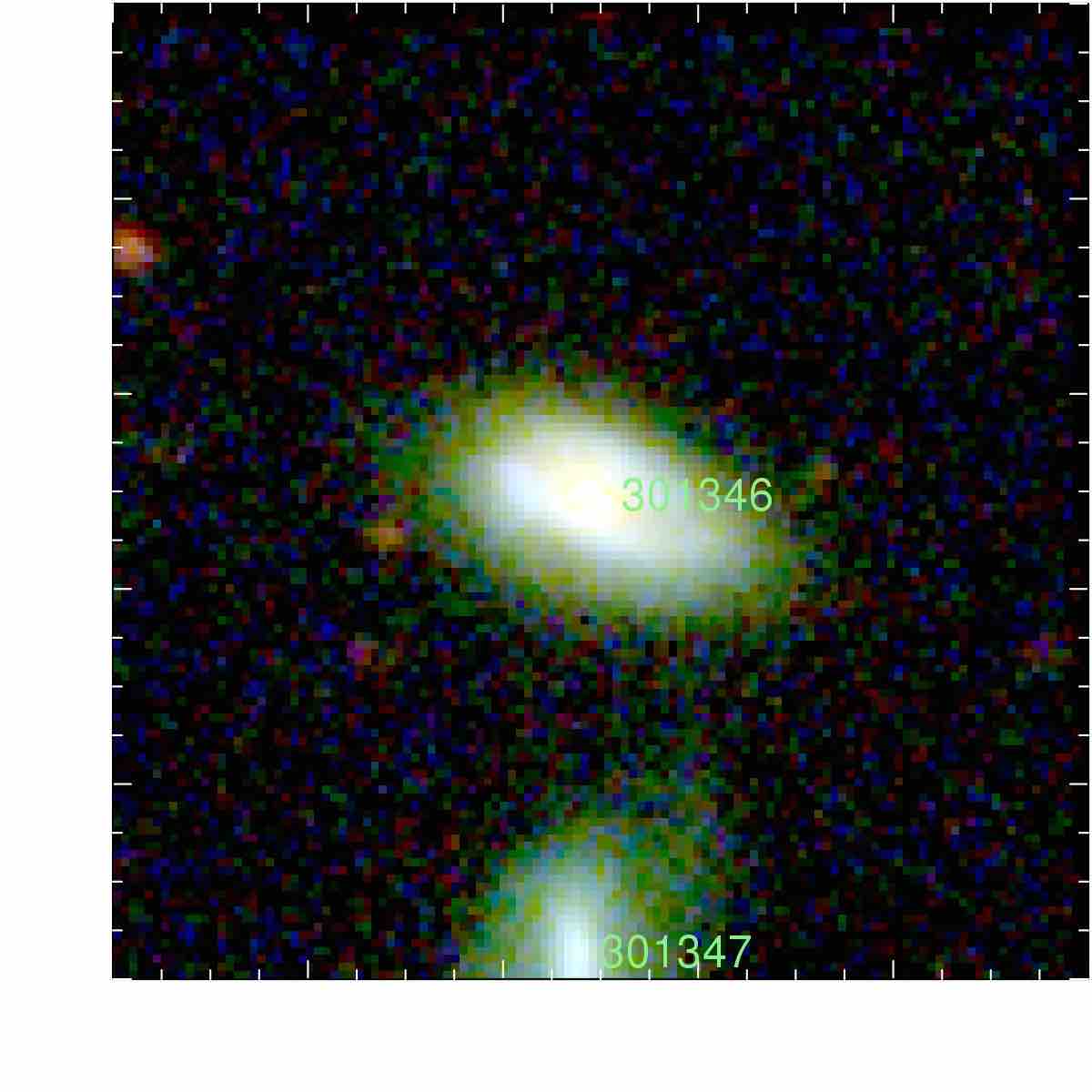} \put (9,85) { \begin{fitbox}{2.25cm}{0.2cm} \color{white}$\bf DBC$ \end{fitbox}} \end{overpic} \\ 

\includegraphics[width=8.4cm]{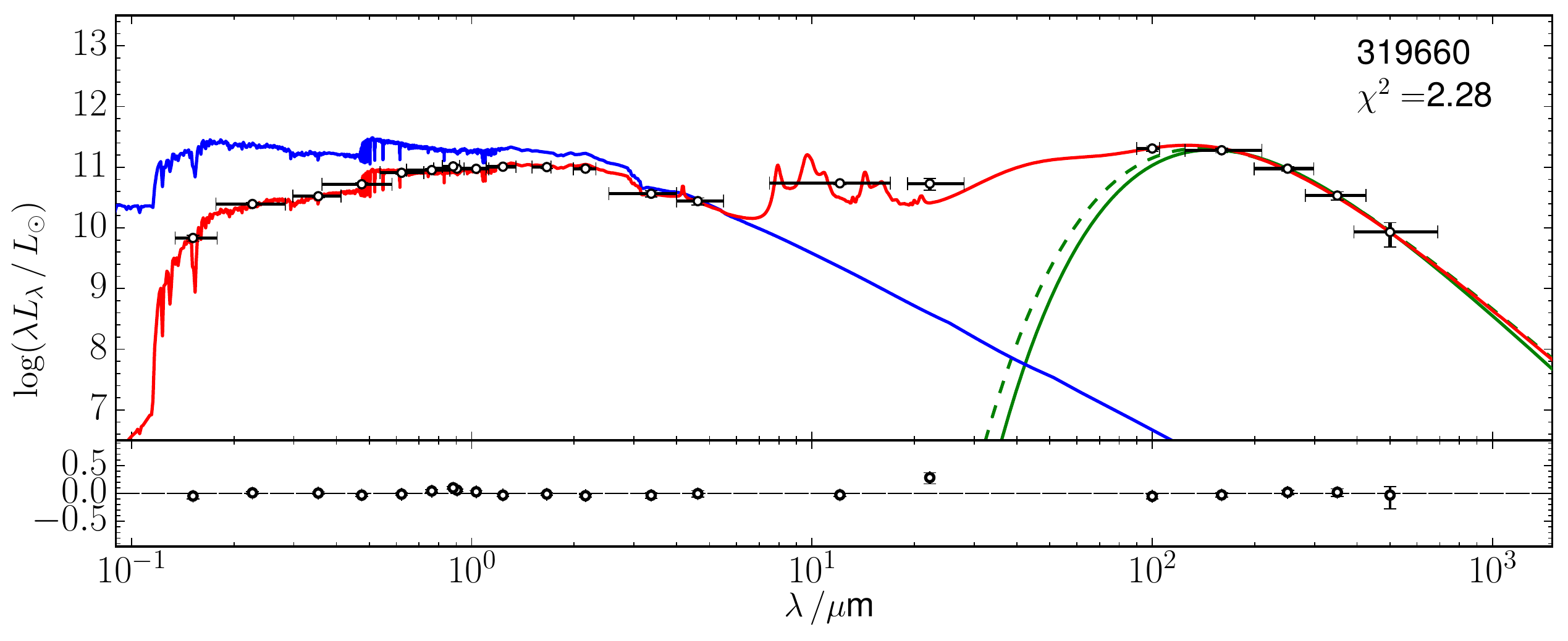} &
\includegraphics[width=5.0cm]{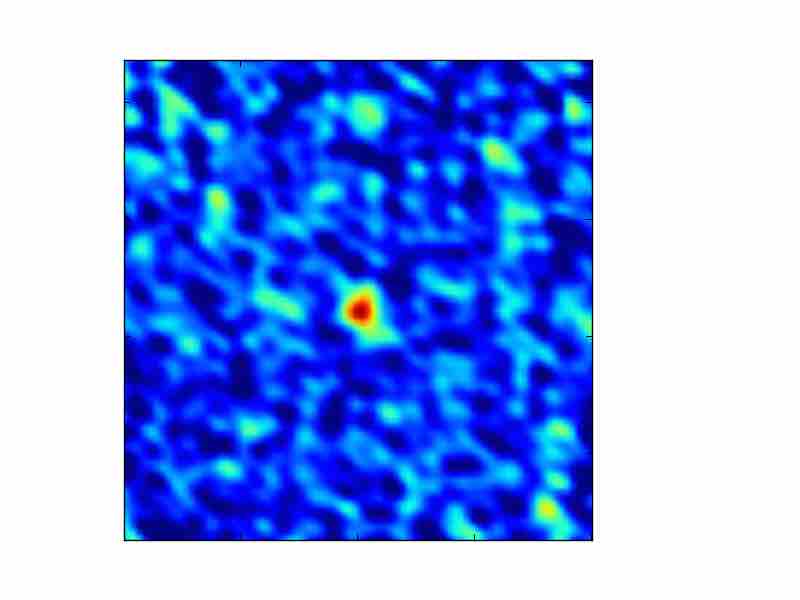} &
\hspace*{-1.2cm}\begin{overpic}[width=3.4cm]{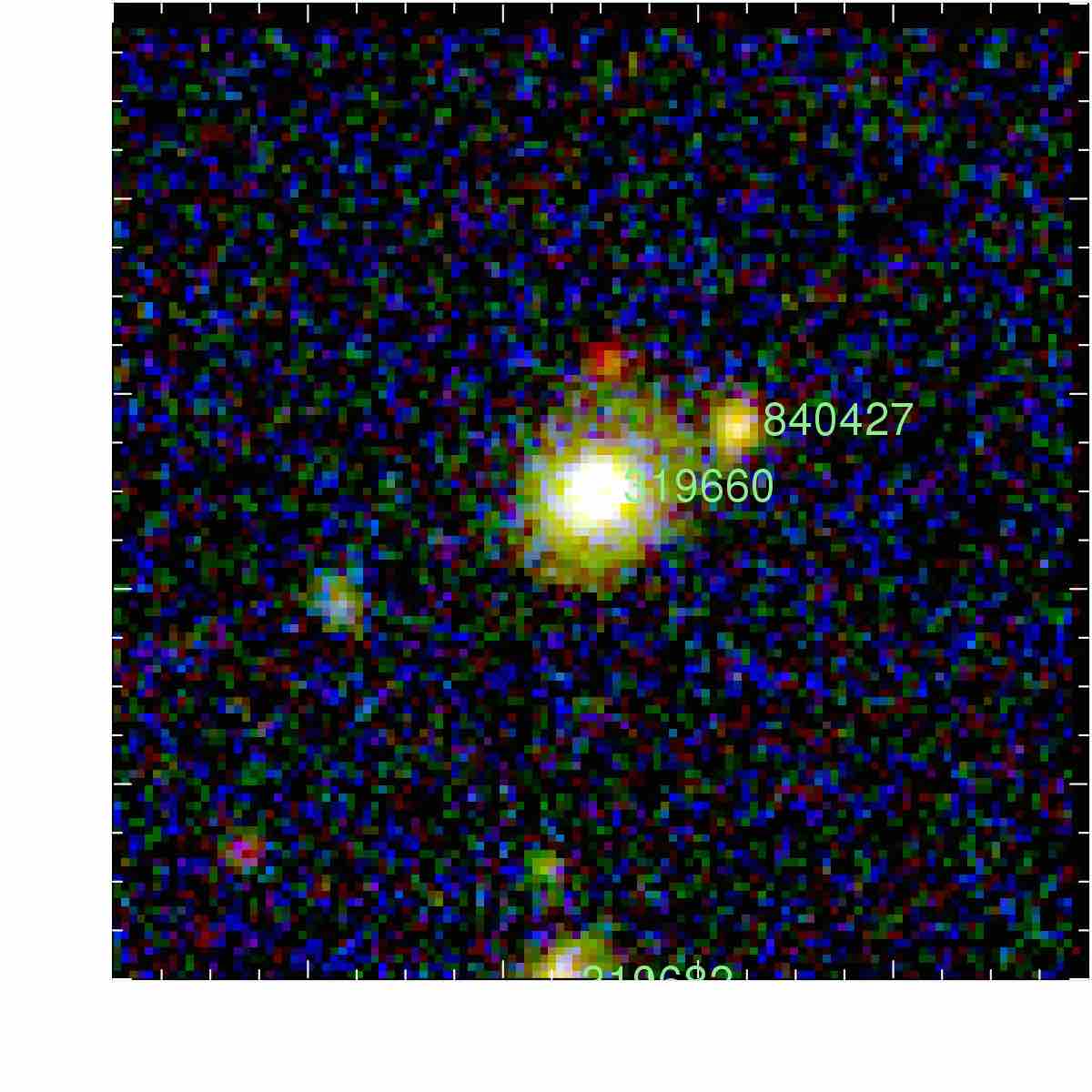} \put (9,85) { \begin{fitbox}{2.25cm}{0.2cm} \color{white}$\bf BDC$ \end{fitbox}} \end{overpic} \\ 

\includegraphics[width=8.4cm]{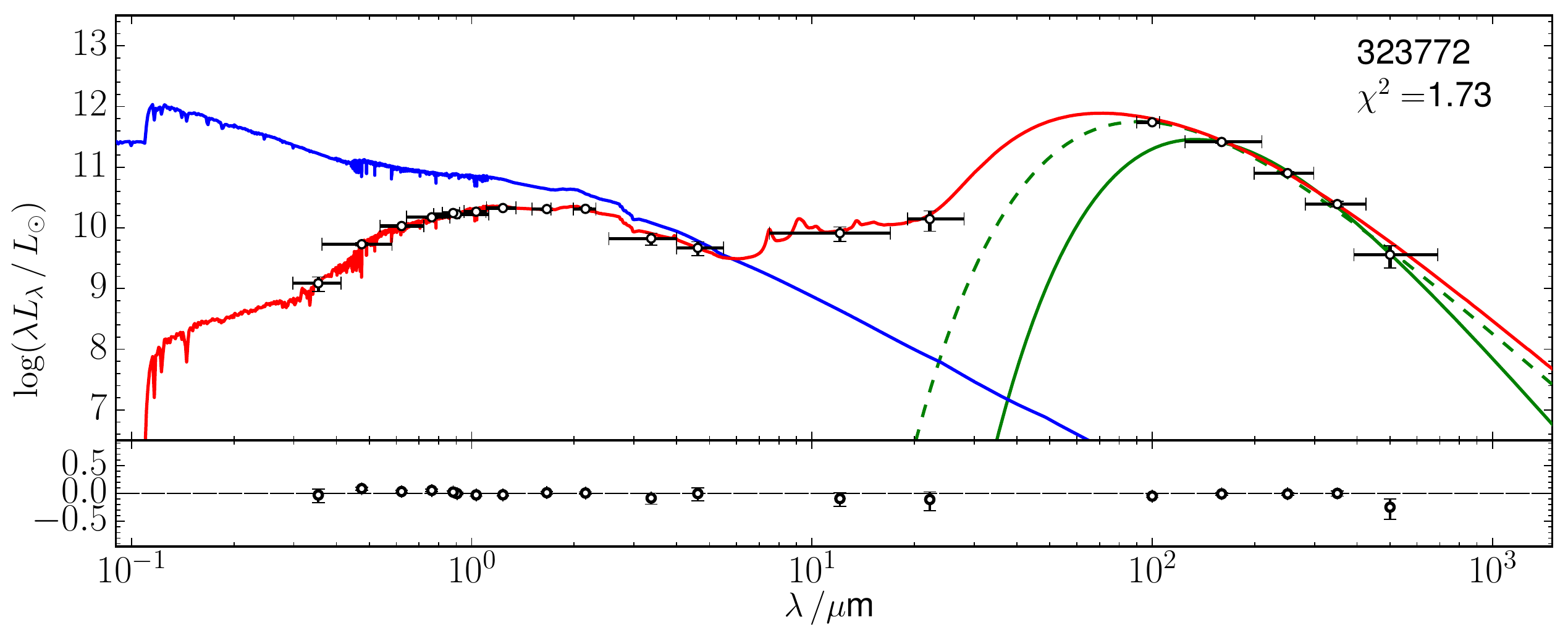} &
\includegraphics[width=5.0cm]{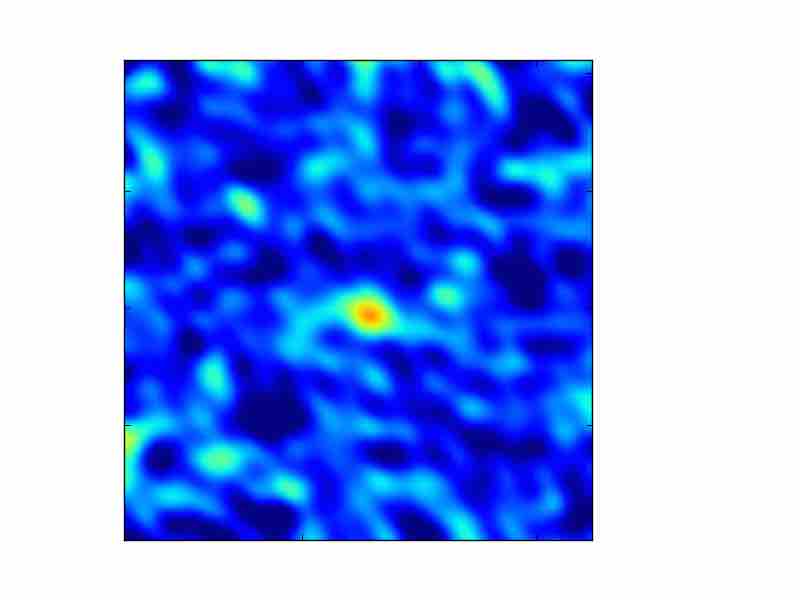} &
\hspace*{-1.2cm}\begin{overpic}[width=3.4cm]{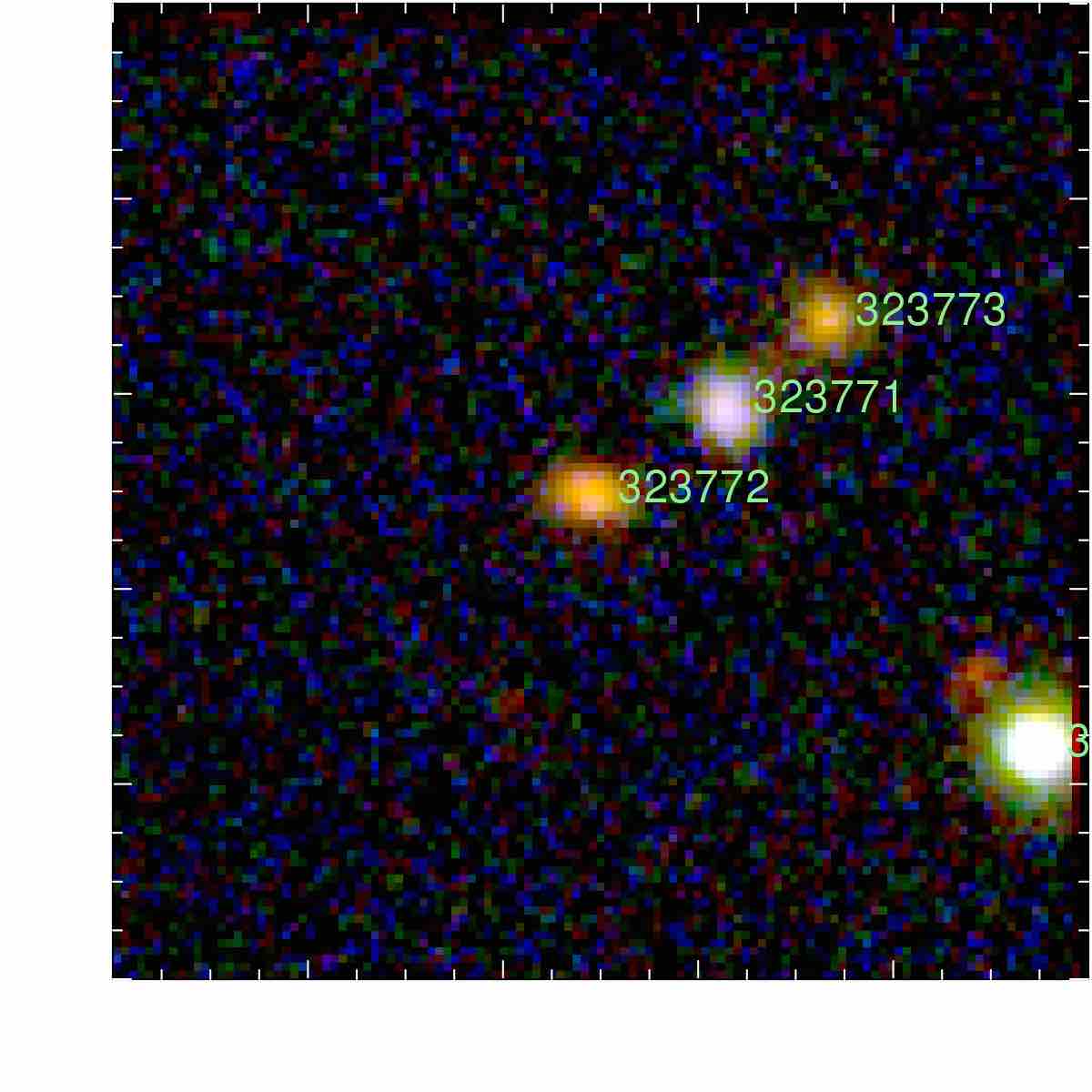} \put (9,85) { \begin{fitbox}{2.25cm}{0.2cm} \color{white}$\bf BC$ \end{fitbox}} \end{overpic} \\ 

\end{array}
$
{\textbf{Figure~\ref{pdrdiaglit}.} continued}

\end{figure*}


\begin{figure*}
$
\begin{array}{ccc}
\includegraphics[width=8.4cm]{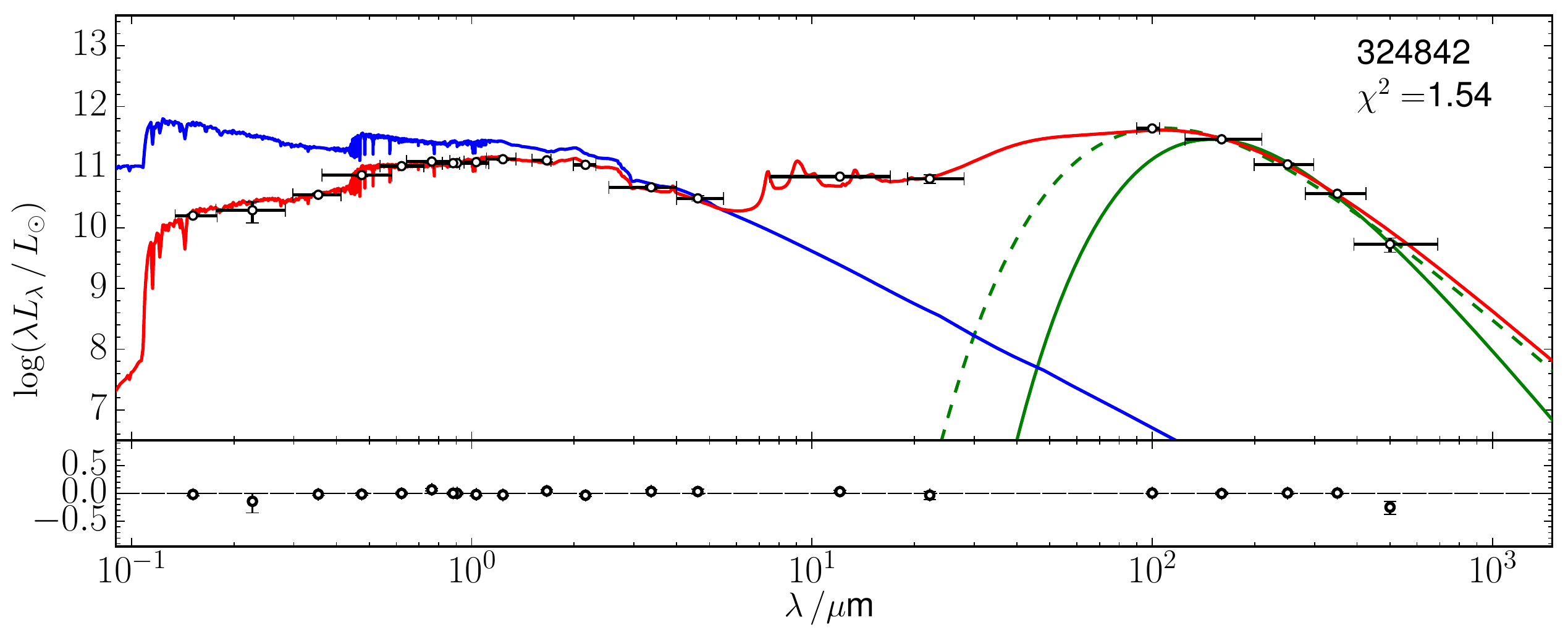} &
\includegraphics[width=5.0cm]{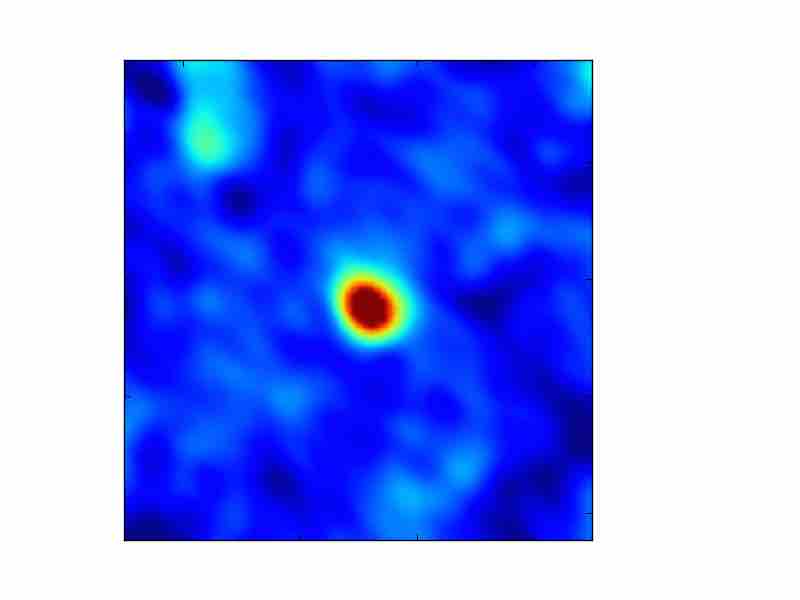} &
\hspace*{-1.2cm}\begin{overpic}[width=3.4cm]{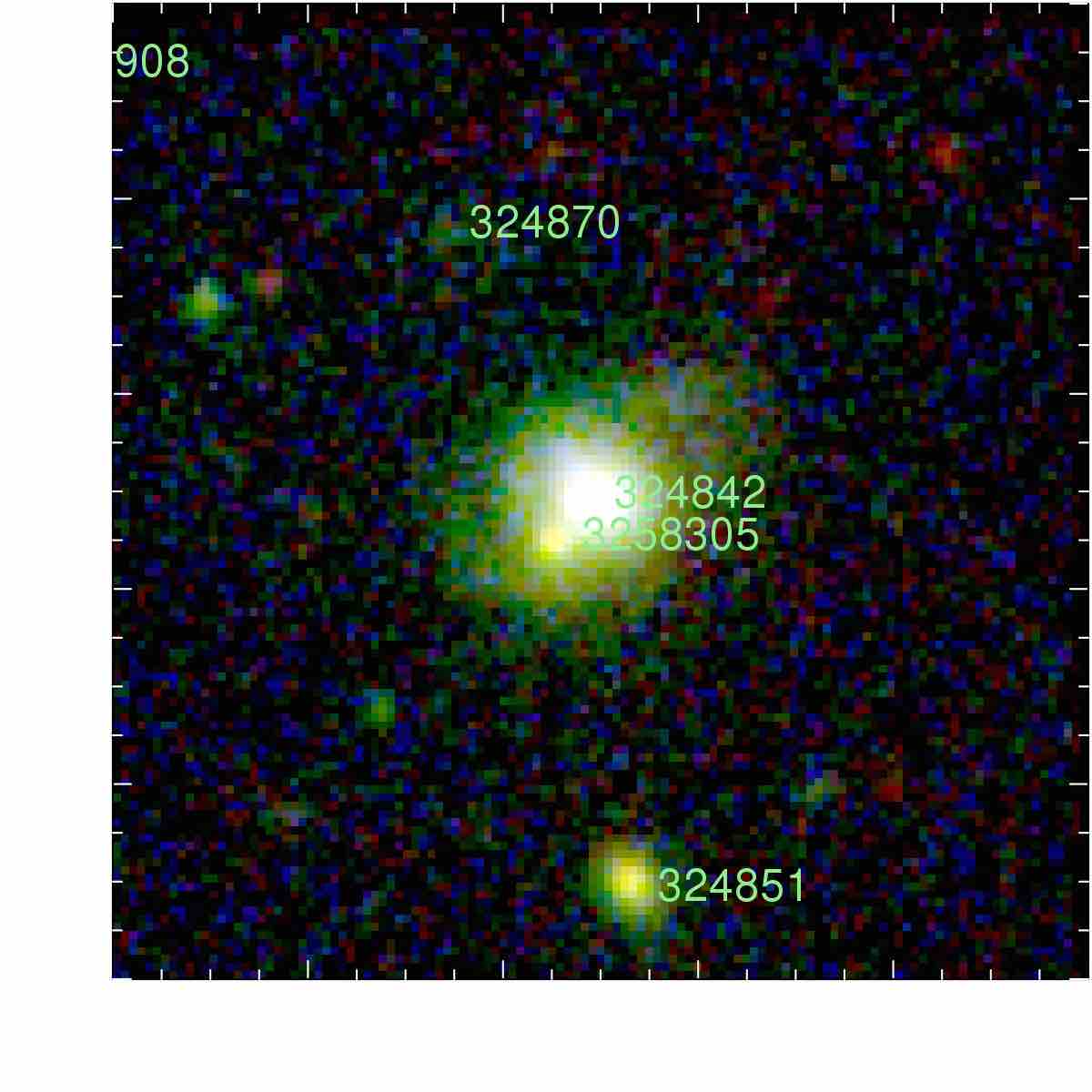} \put (9,85) { \begin{fitbox}{2.25cm}{0.2cm} \color{white}$\bf BC$ \end{fitbox}} \end{overpic} \\ 	
	
\includegraphics[width=8.4cm]{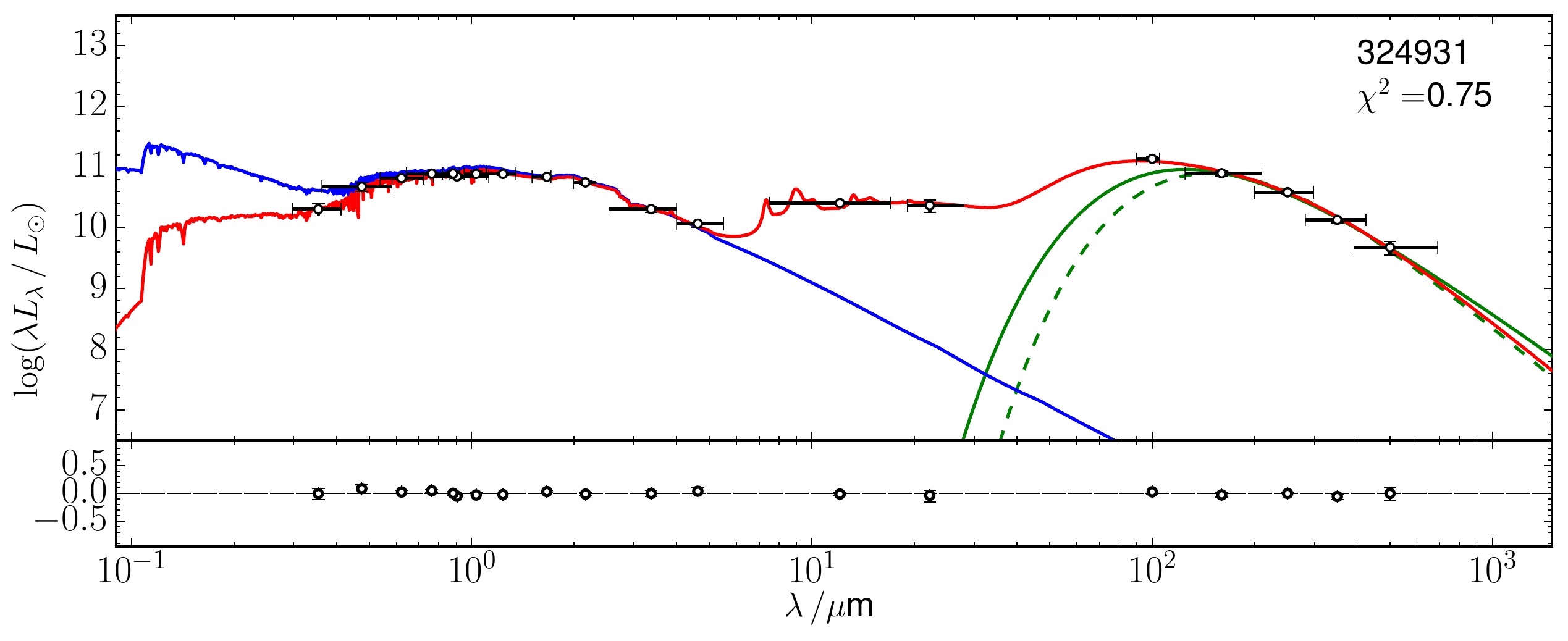} &
\includegraphics[width=5.0cm]{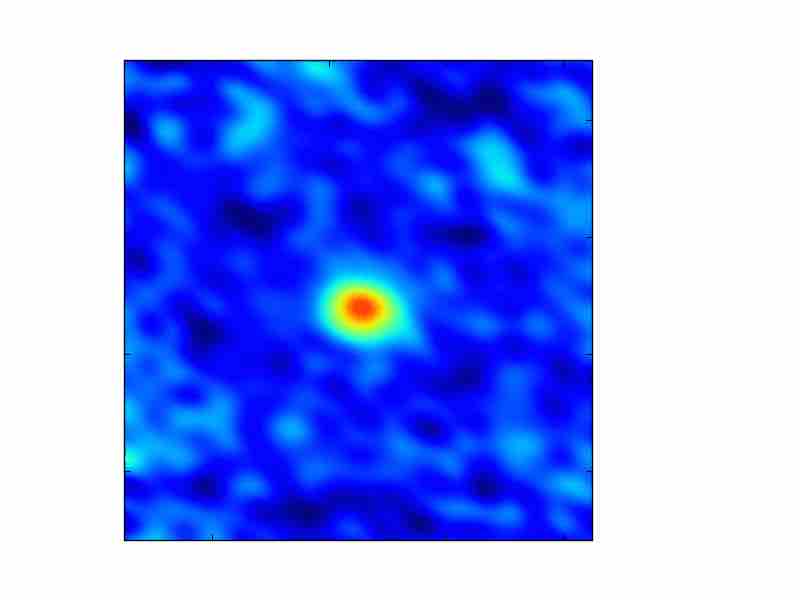} &
\hspace*{-1.2cm}\begin{overpic}[width=3.4cm]{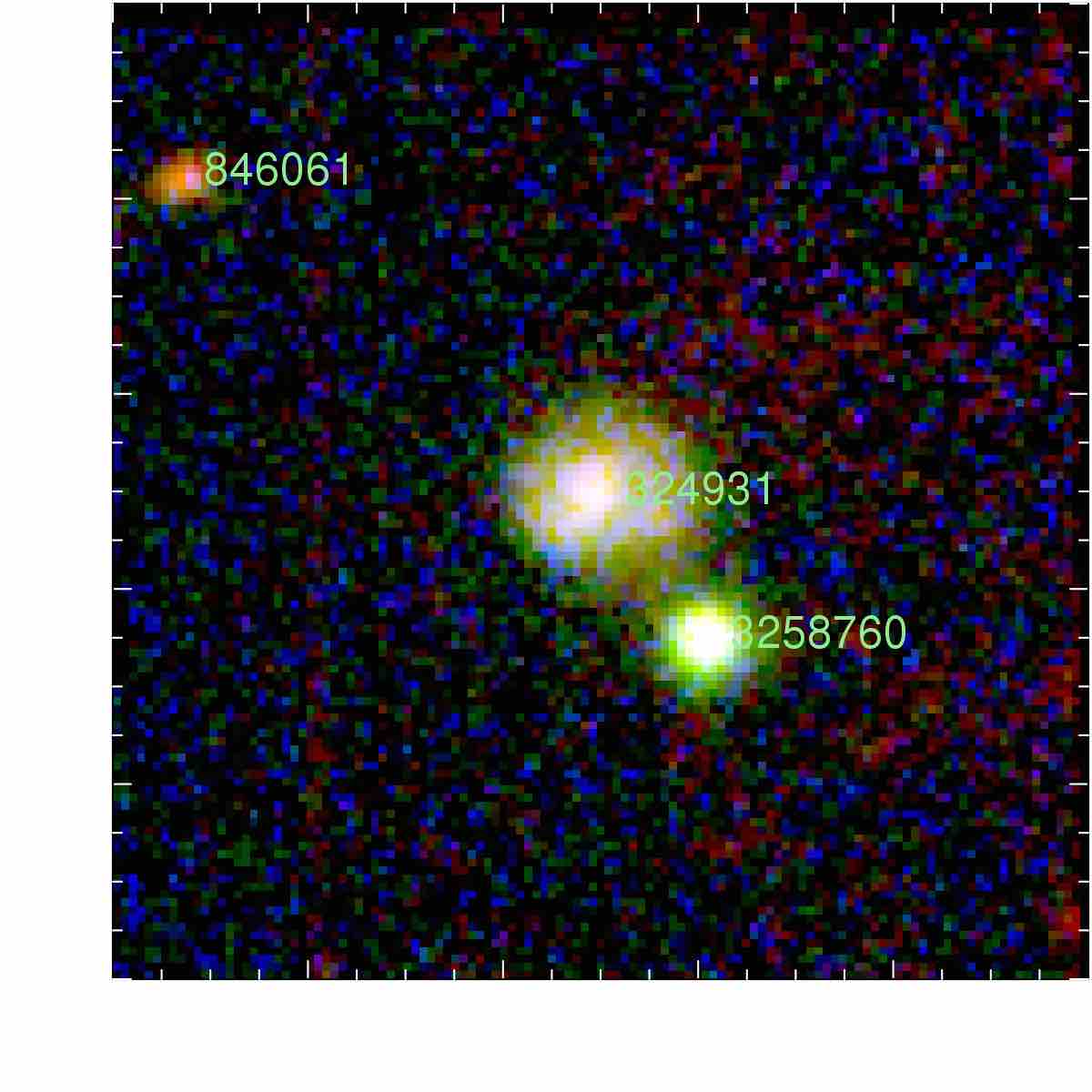} \put (9,85) { \begin{fitbox}{2.25cm}{0.2cm} \color{white}$\bf DBC$ \end{fitbox}} \end{overpic} \\ 	

\includegraphics[width=8.4cm]{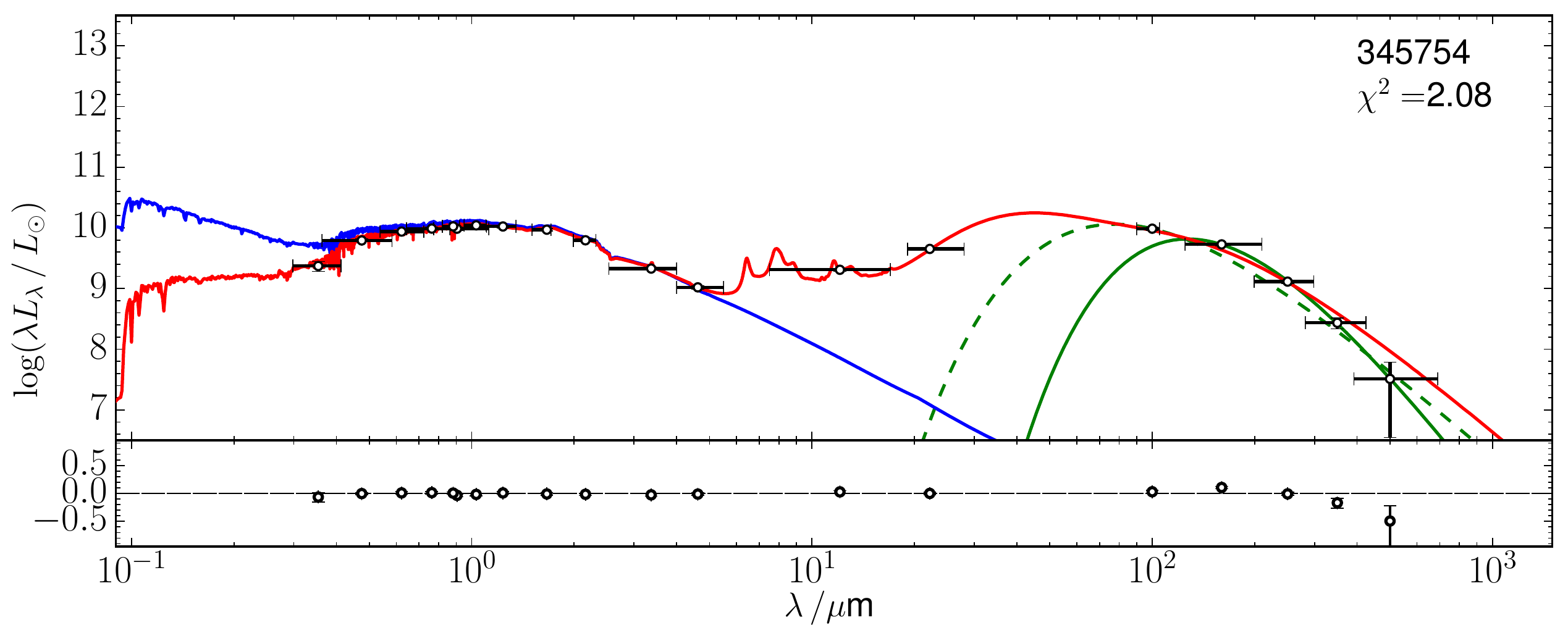} &
\includegraphics[width=5.0cm]{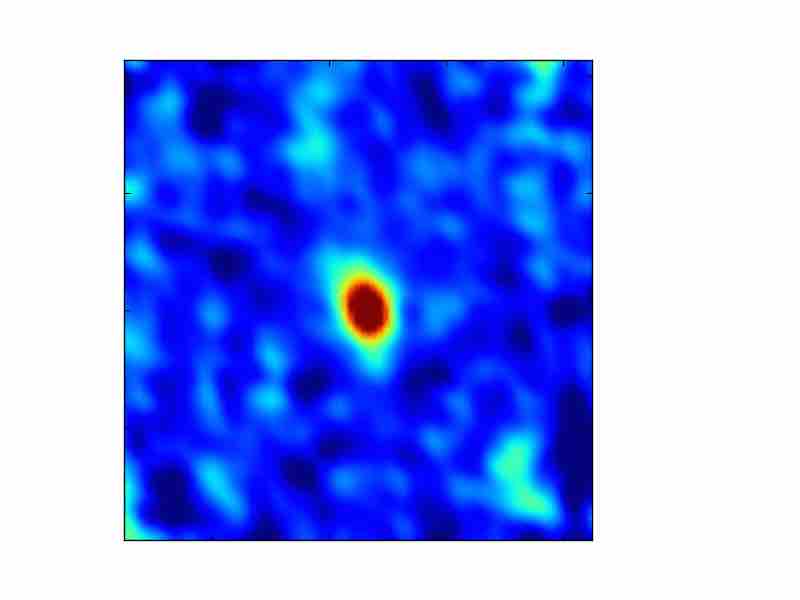} &
\hspace*{-1.2cm}\begin{overpic}[width=3.4cm]{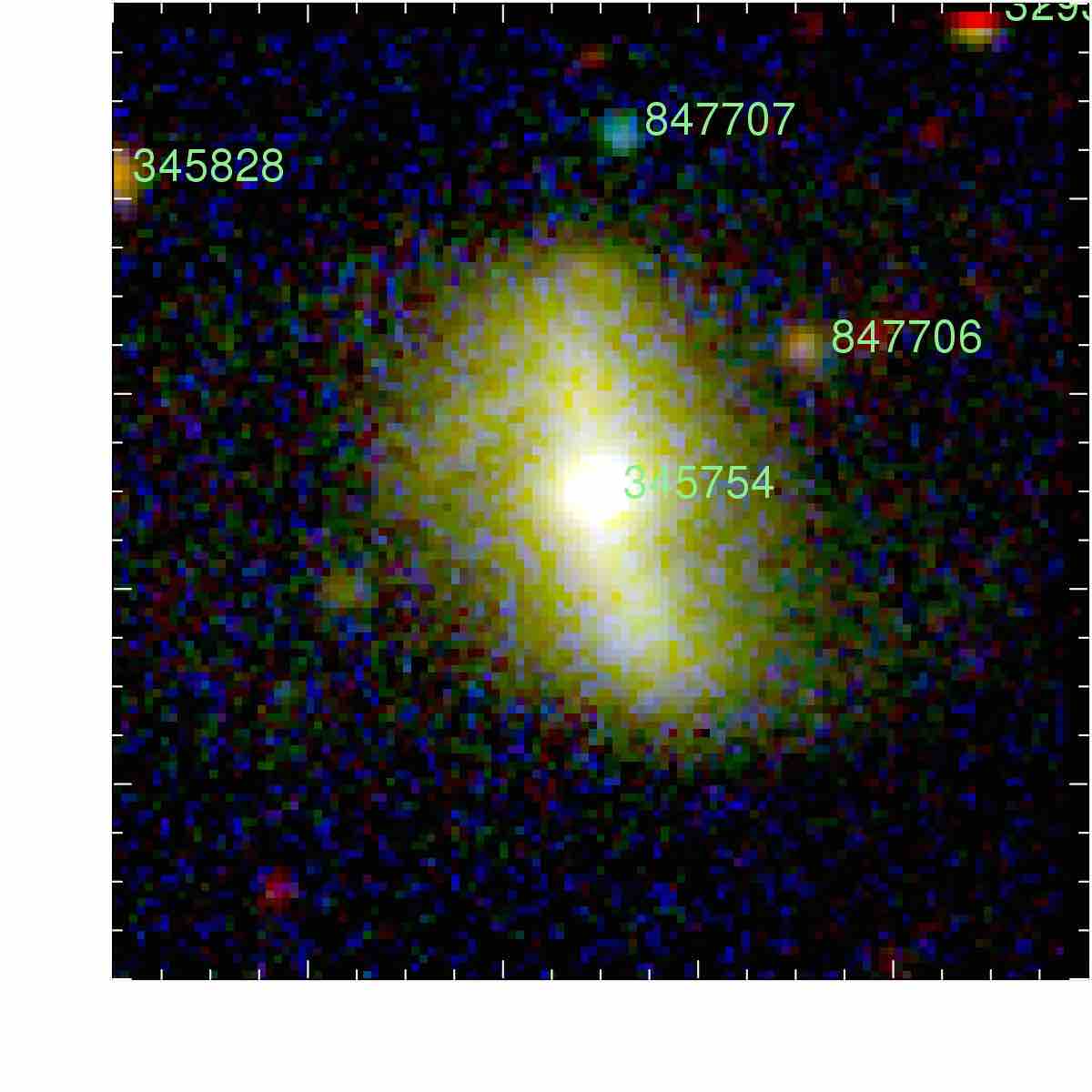} \put (9,85) { \begin{fitbox}{2.25cm}{0.2cm} \color{white}$\bf DBC$ \end{fitbox}} \end{overpic} \\ 

\includegraphics[width=8.4cm]{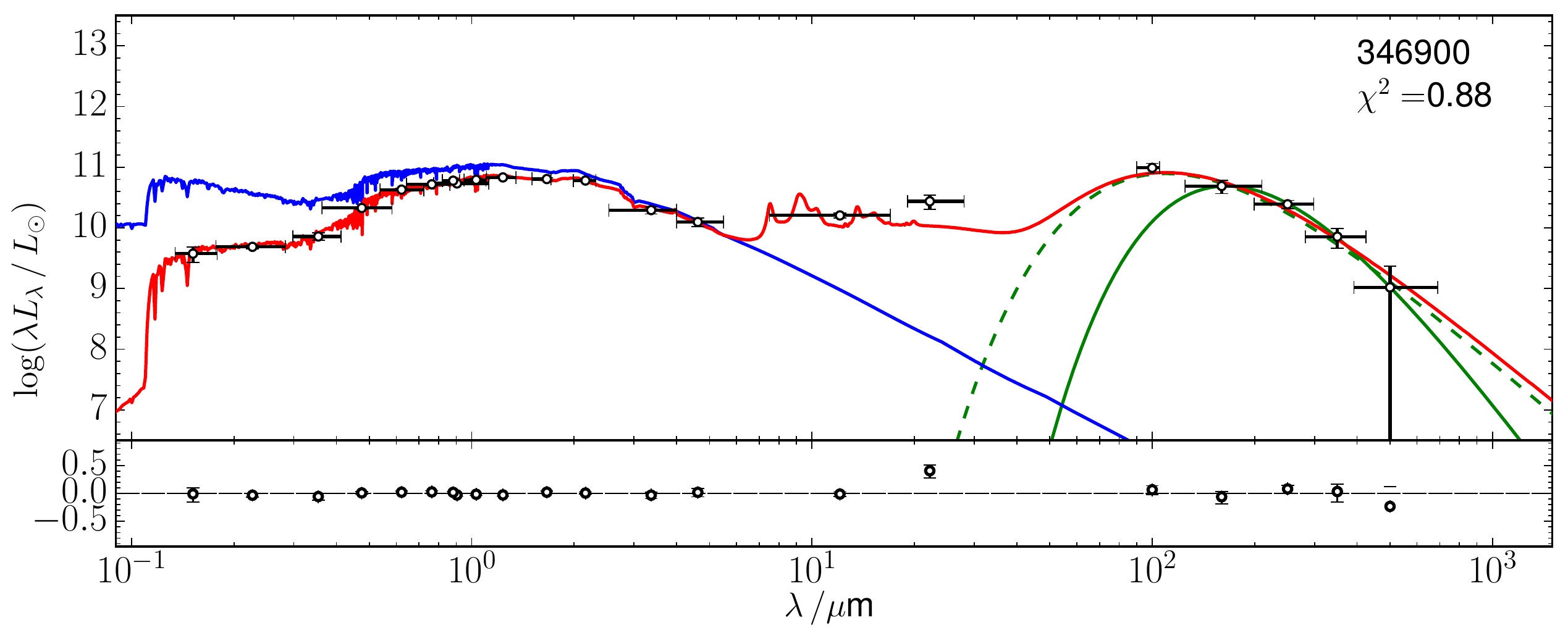} &
\includegraphics[width=5.0cm]{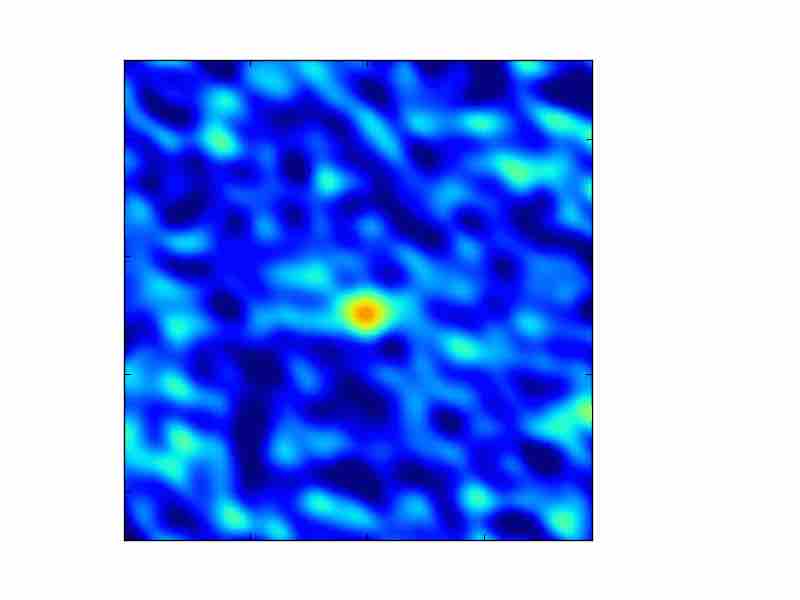} &
\hspace*{-1.2cm}\begin{overpic}[width=3.4cm]{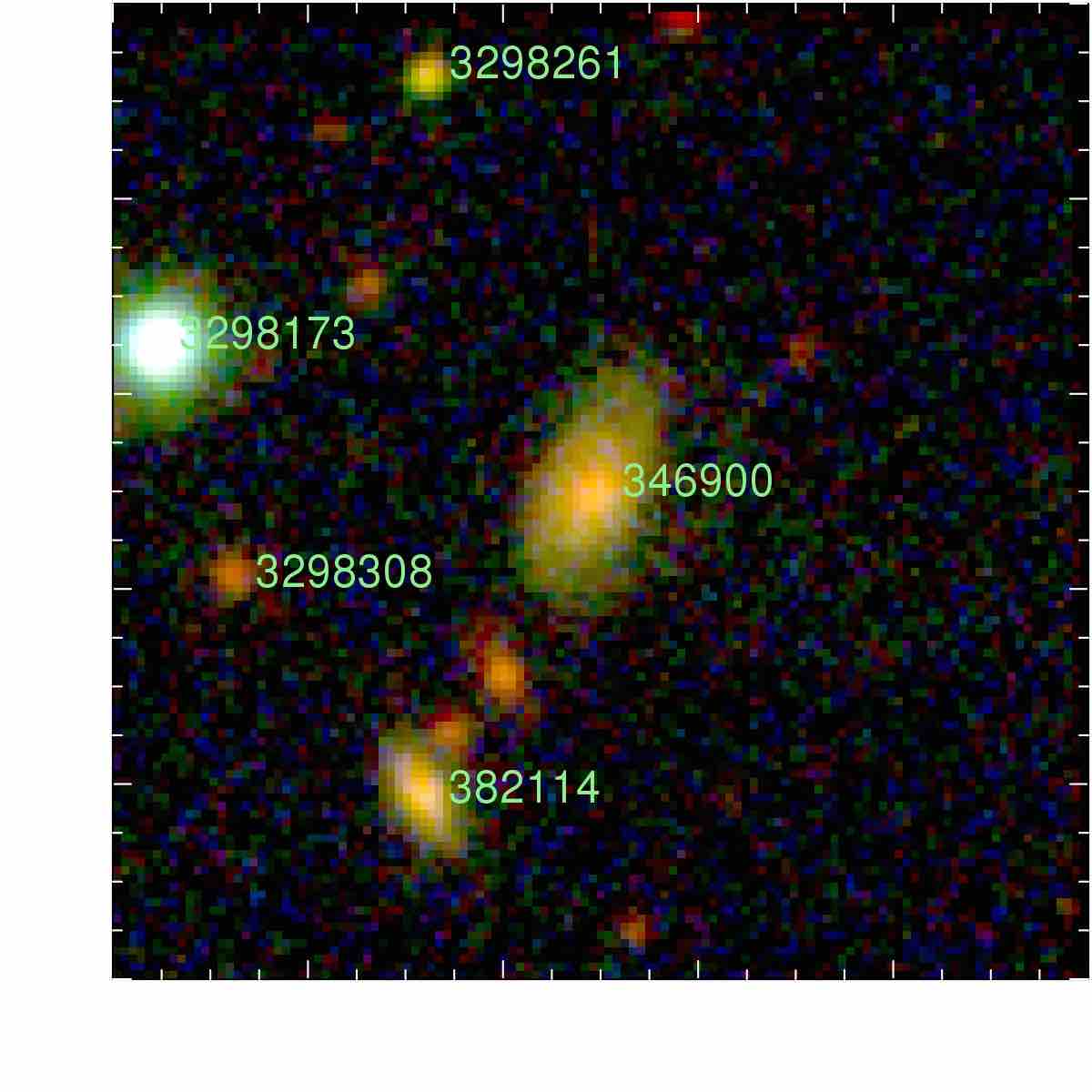} \put (9,85) { \begin{fitbox}{2.25cm}{0.2cm} \color{white}$\bf DBC$ \end{fitbox}} \end{overpic} \\ 

\includegraphics[width=8.4cm]{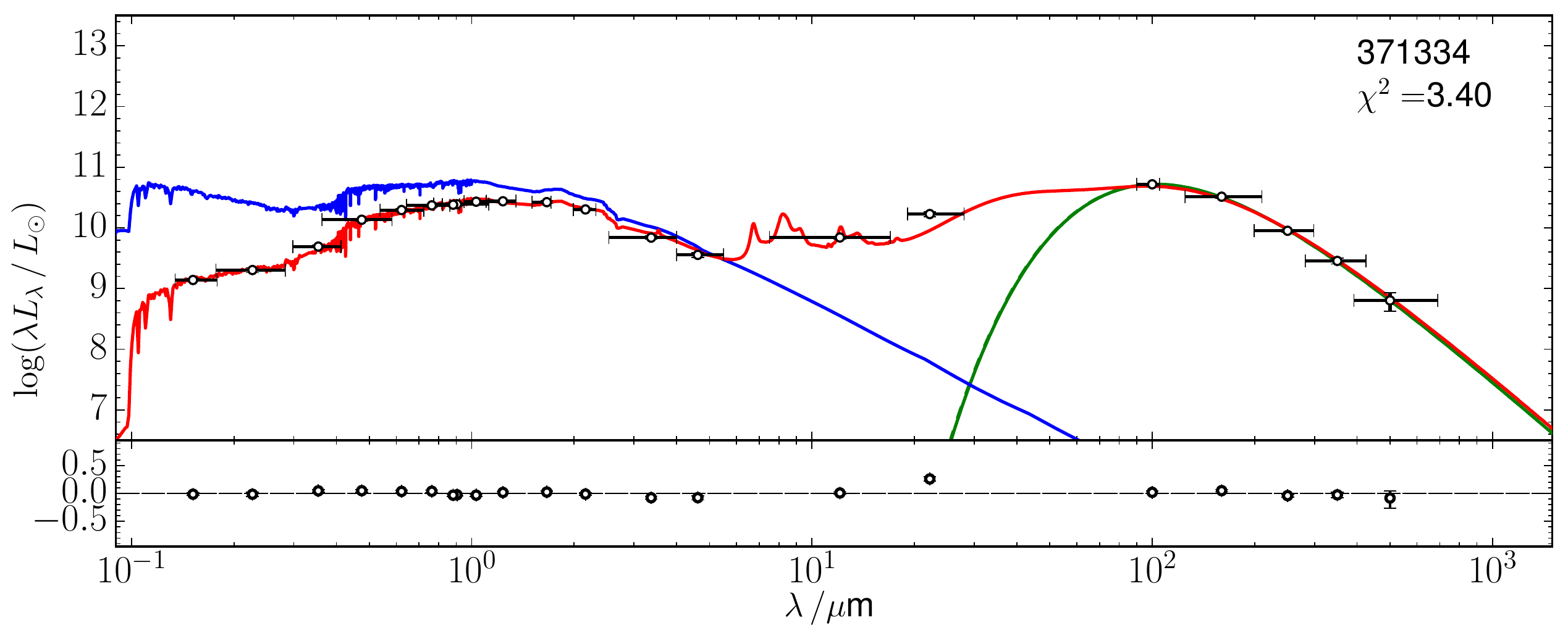} &
\includegraphics[width=5.0cm]{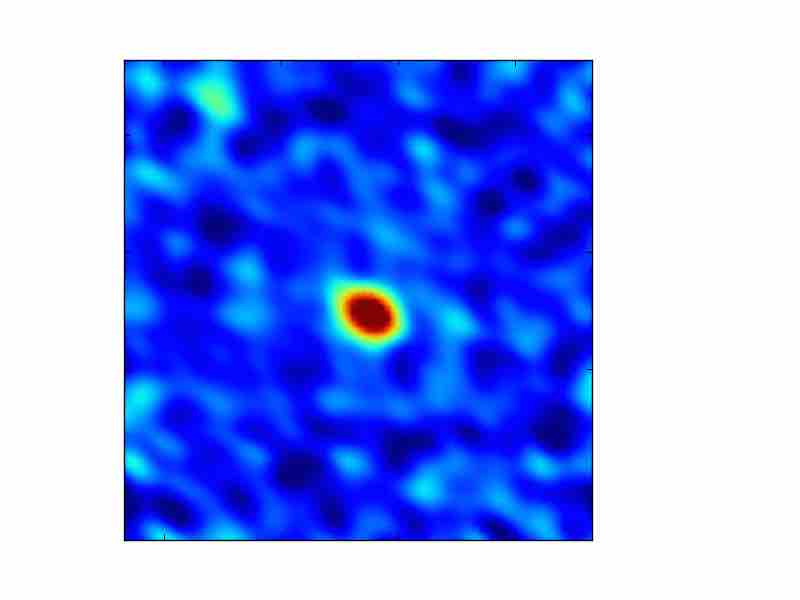} &
\hspace*{-1.2cm}\begin{overpic}[width=3.4cm]{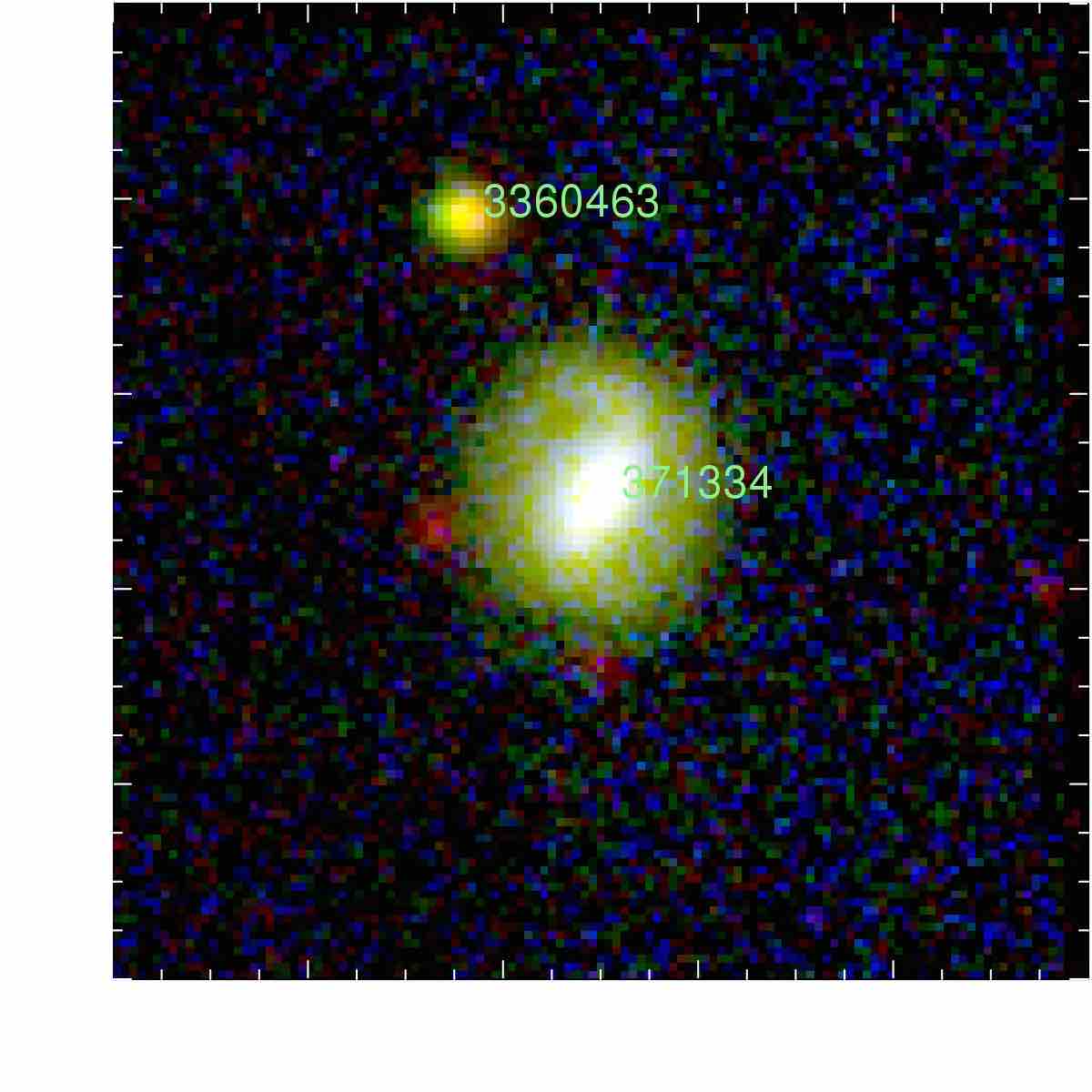} \put (9,85) { \begin{fitbox}{2.25cm}{0.2cm} \color{white}$\bf DBC$ \end{fitbox}} \end{overpic} \\ 

\end{array}
$
{\textbf{Figure~\ref{pdrdiaglit}.} continued}

\end{figure*}


\begin{figure*}
$
\begin{array}{ccc}
\includegraphics[width=8.4cm]{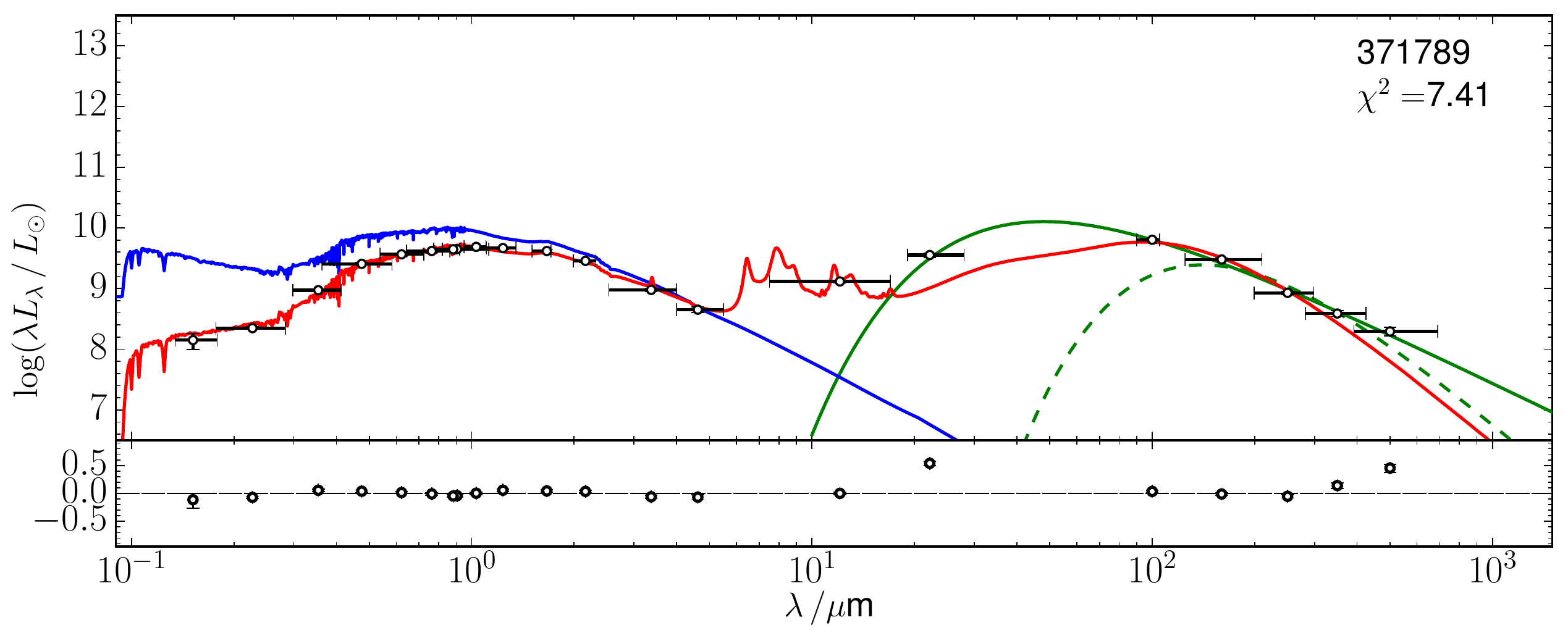} &
\includegraphics[width=5.0cm]{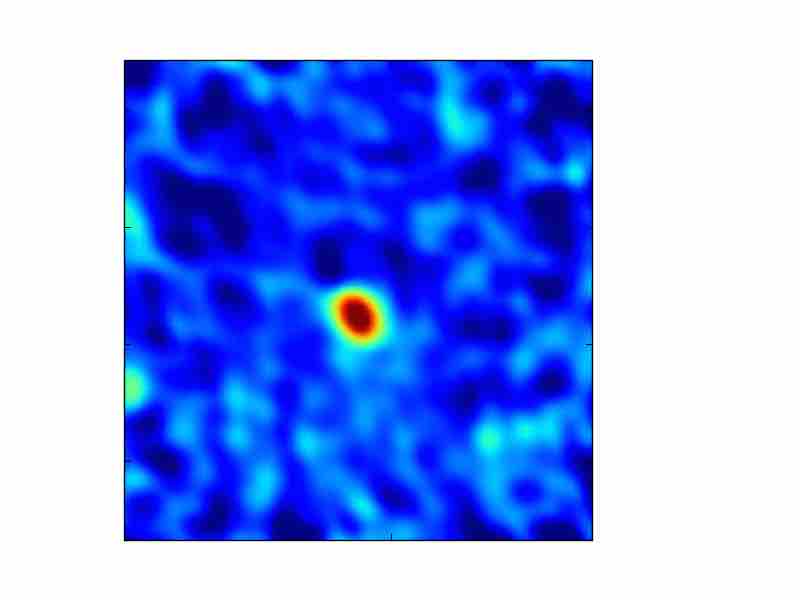} &
\hspace*{-1.2cm}\begin{overpic}[width=3.4cm]{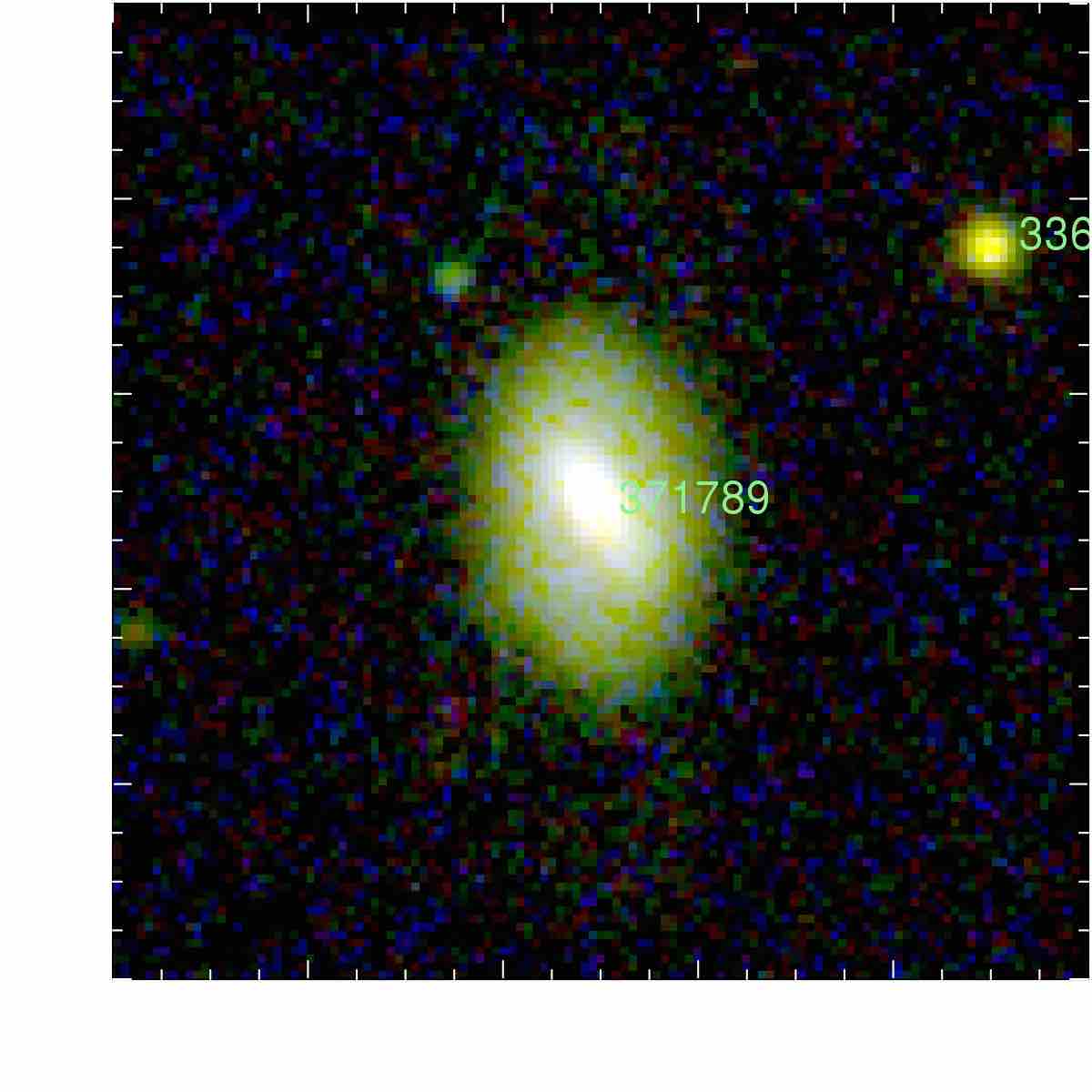} \put (9,85) { \begin{fitbox}{2.25cm}{0.2cm} \color{white}$\bf DB$ \end{fitbox}} \end{overpic} \\ 	
	
\includegraphics[width=8.4cm]{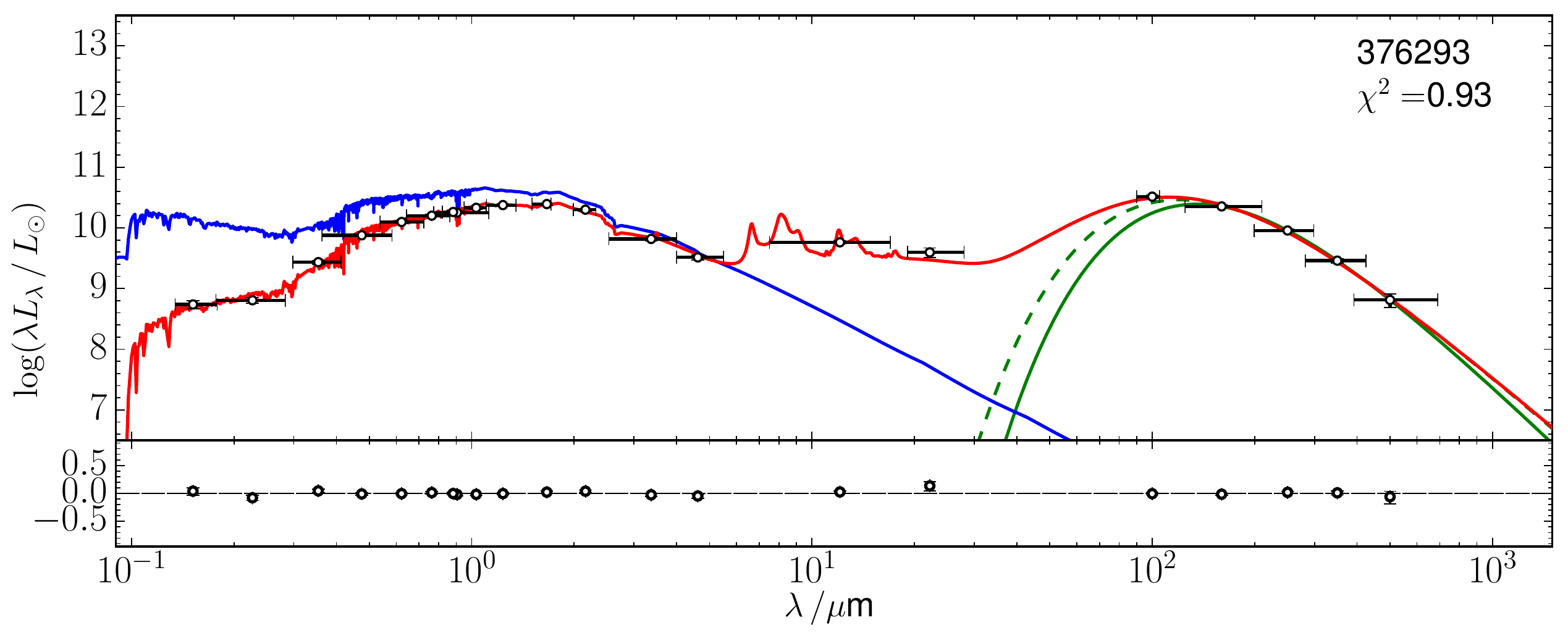} &
\includegraphics[width=5.0cm]{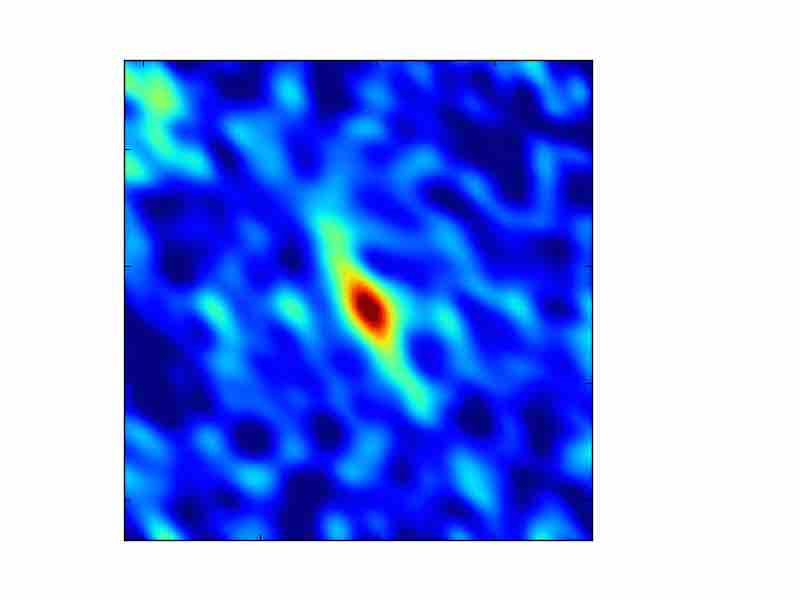} &
\hspace*{-1.2cm}\begin{overpic}[width=3.4cm]{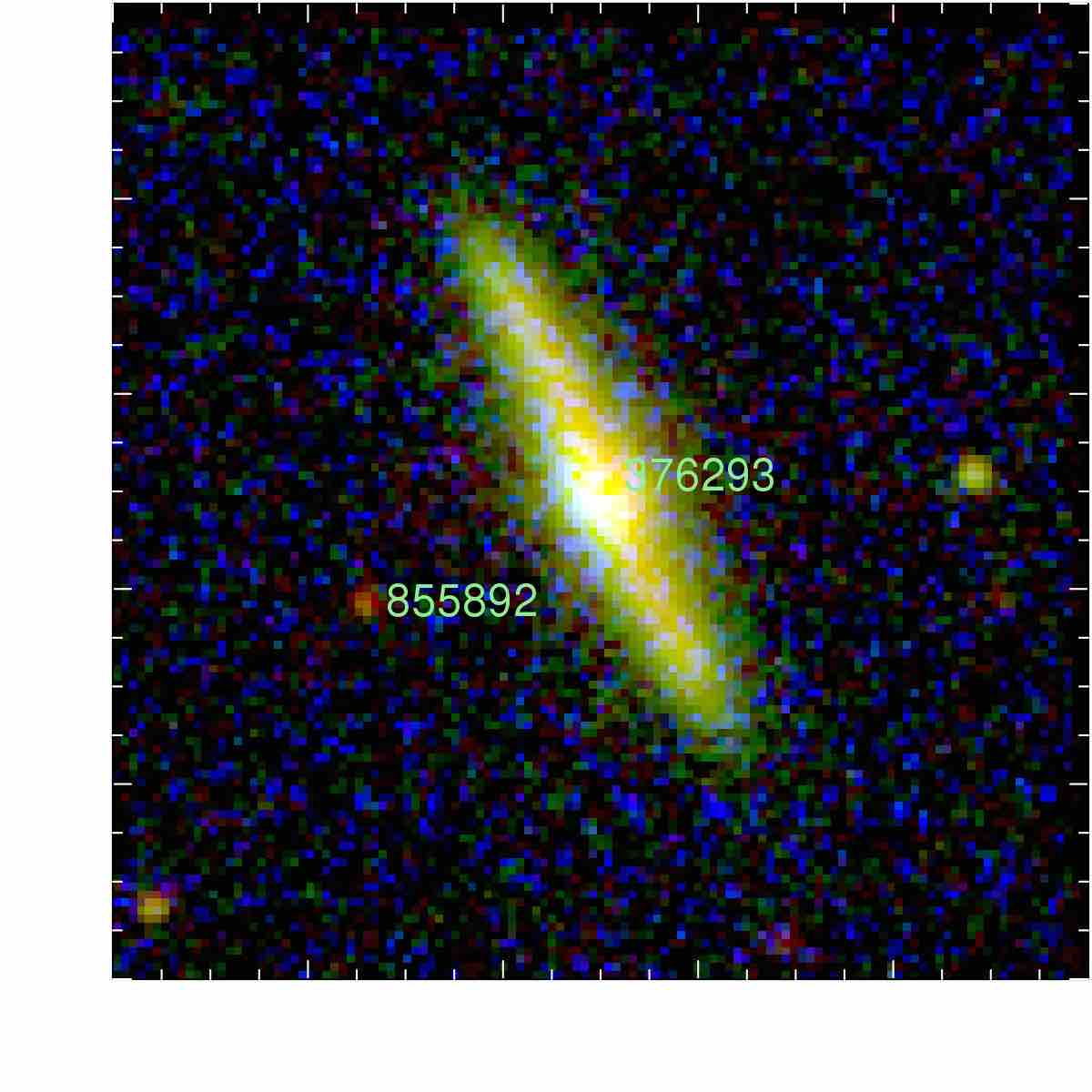} \put (9,85) { \begin{fitbox}{2.25cm}{0.2cm} \color{white}$\bf DB$ \end{fitbox}} \end{overpic} \\ 

\includegraphics[width=8.4cm]{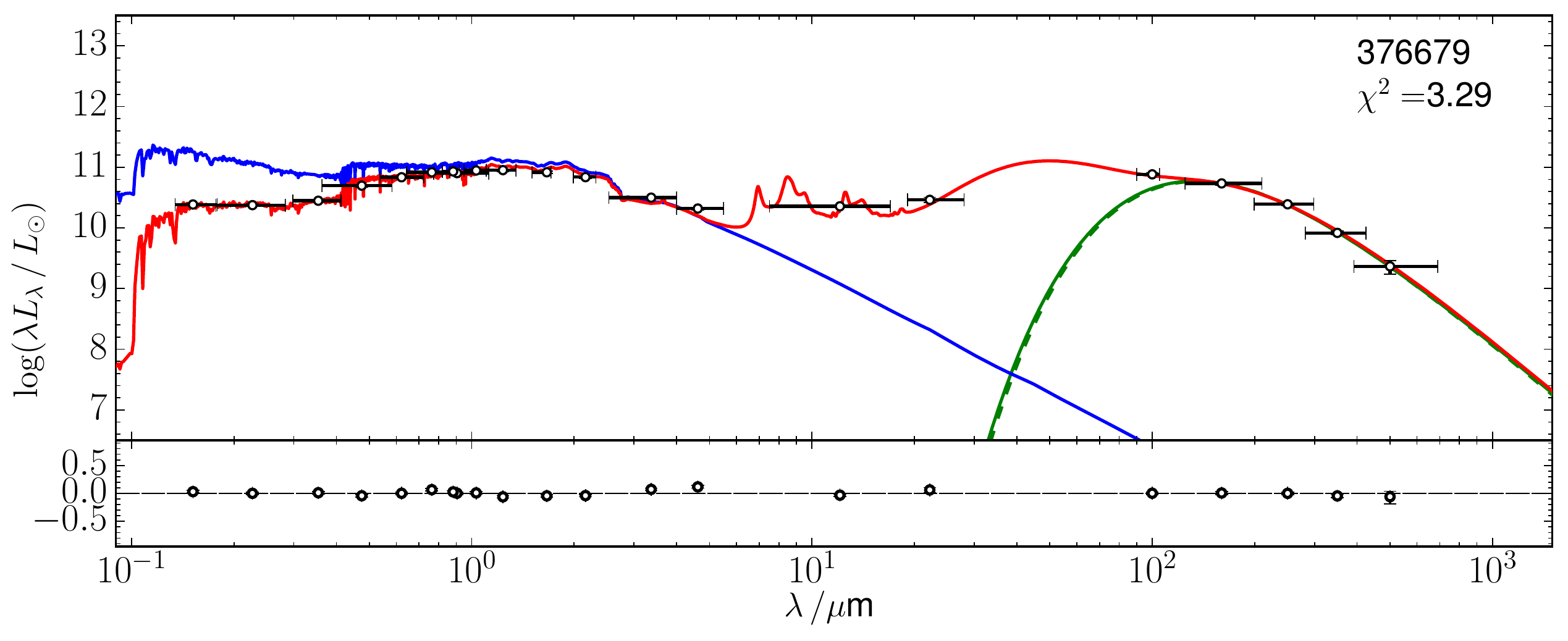} &
\includegraphics[width=5.0cm]{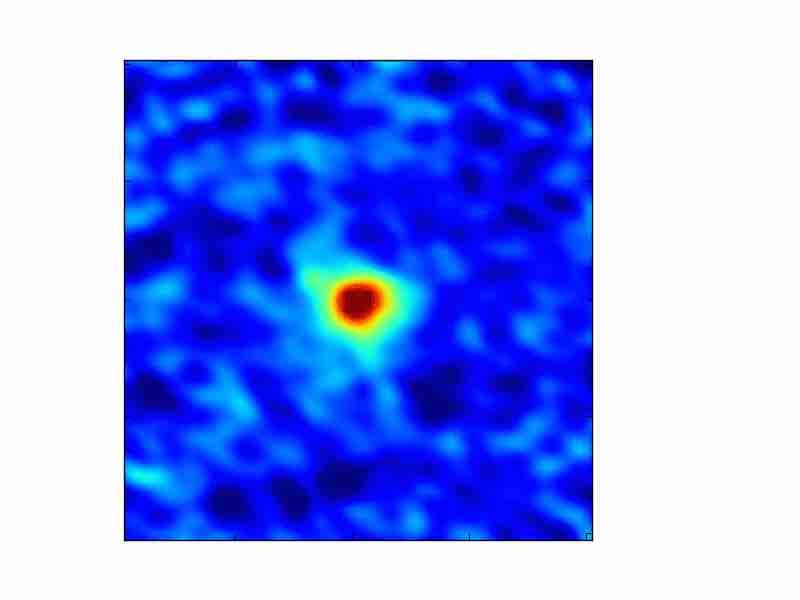} &
\hspace*{-1.2cm}\begin{overpic}[width=3.4cm]{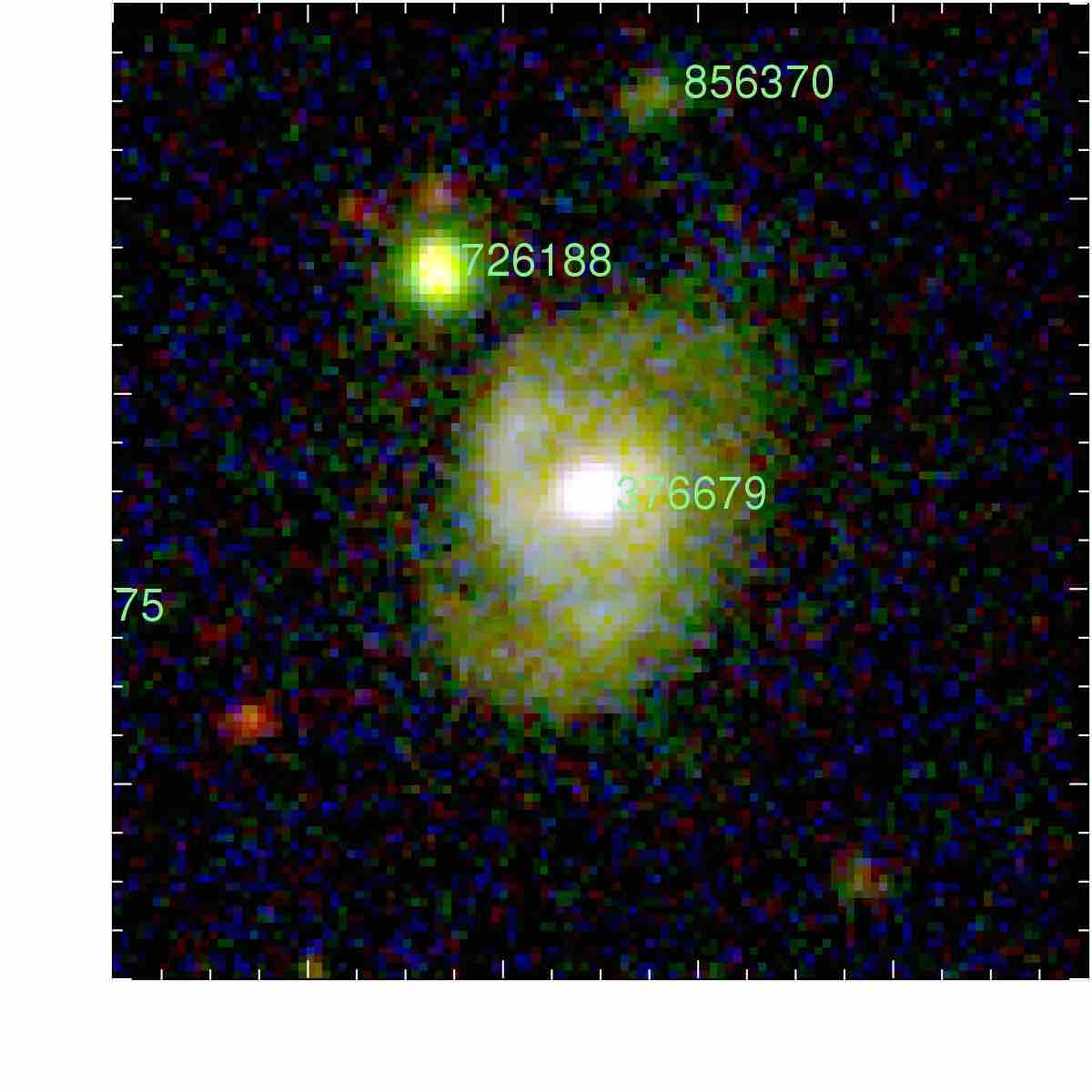} \put (9,85) { \begin{fitbox}{2.25cm}{0.2cm} \color{white}$\bf DBC$ \end{fitbox}} \end{overpic} \\ 

\includegraphics[width=8.4cm]{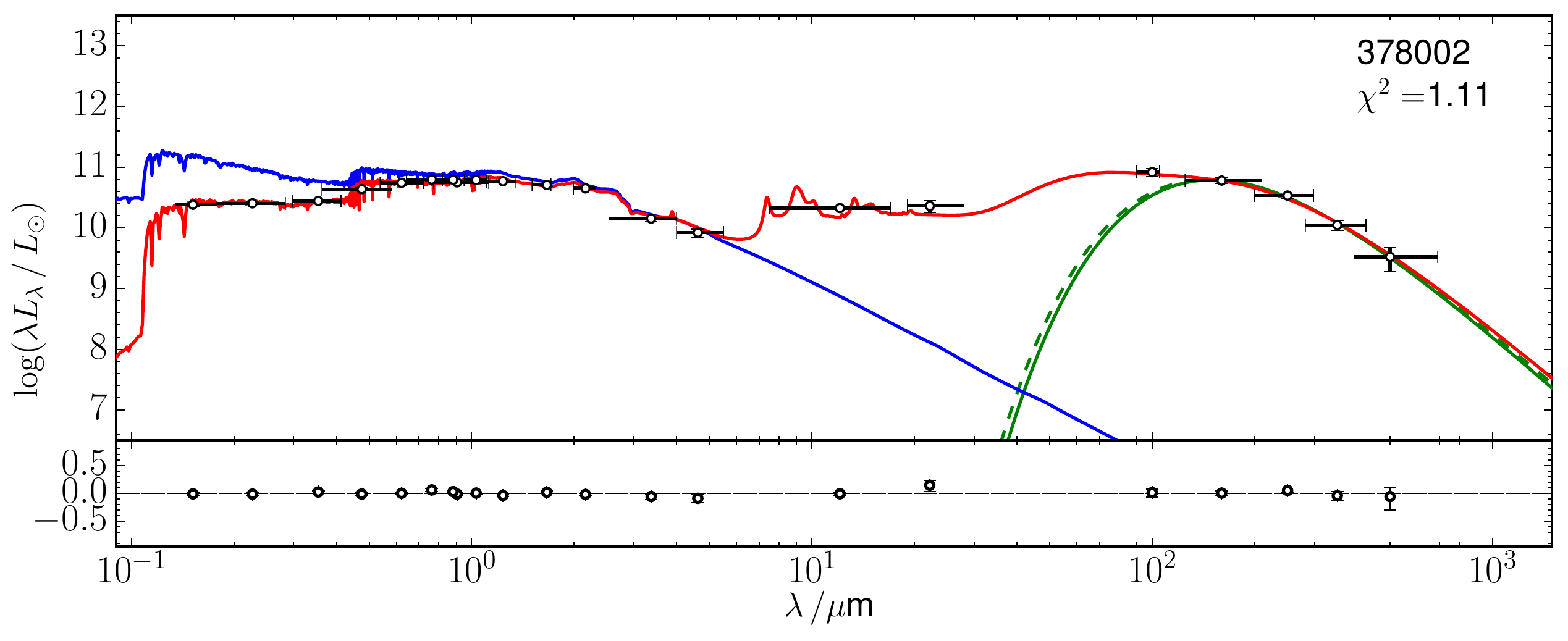} &
\includegraphics[width=5.0cm]{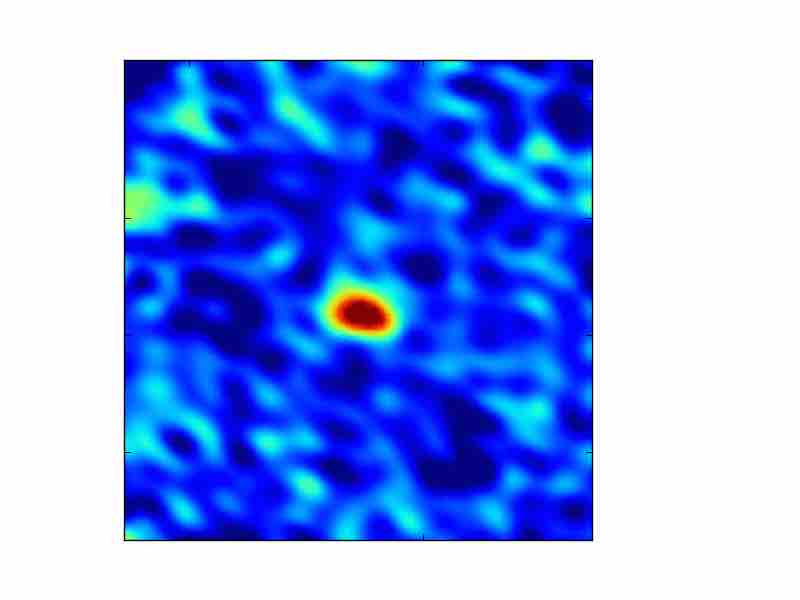} &
\hspace*{-1.2cm}\begin{overpic}[width=3.4cm]{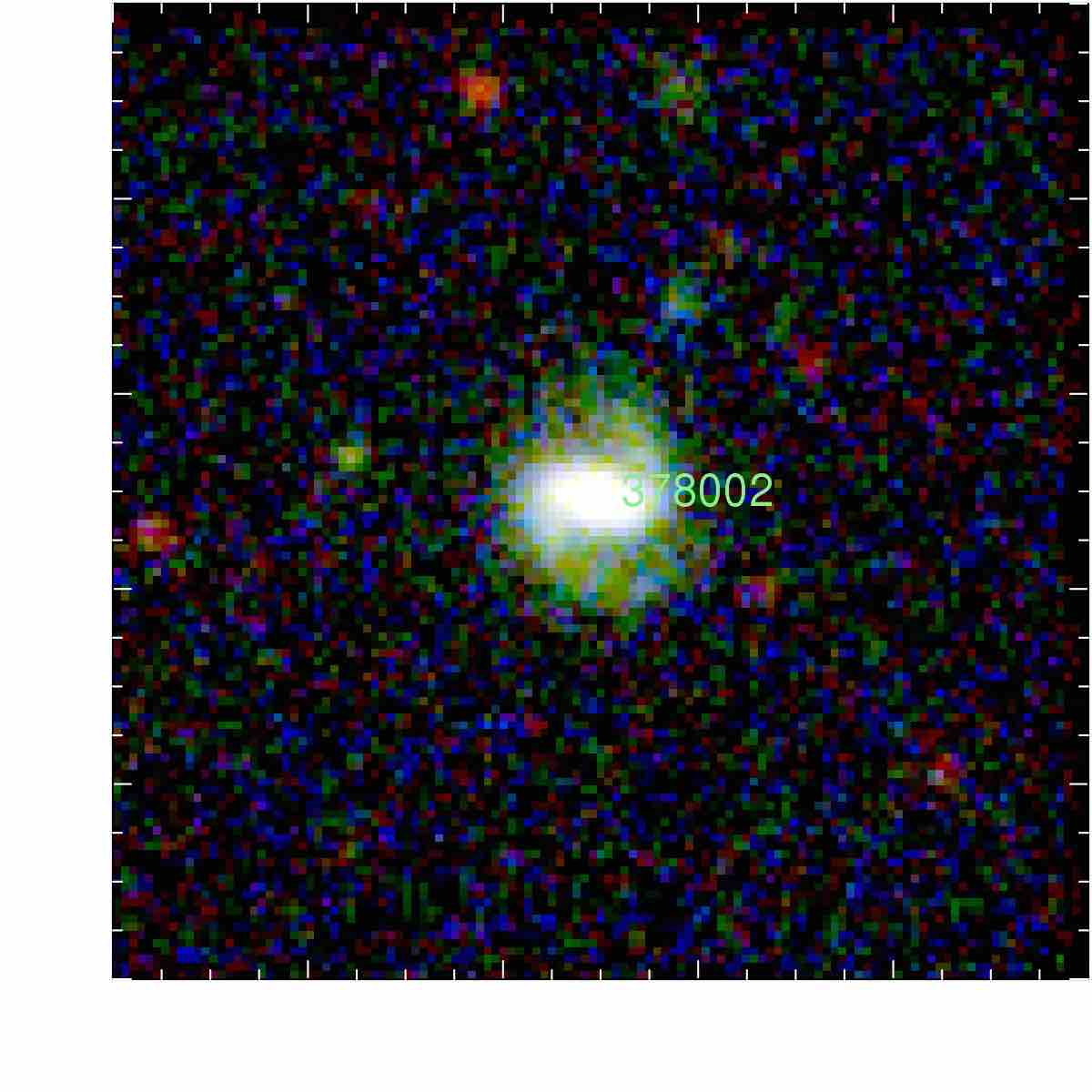} \put (9,85) { \begin{fitbox}{2.25cm}{0.2cm} \color{white}$\bf BD$ \end{fitbox}} \end{overpic} \\ 

\includegraphics[width=8.4cm]{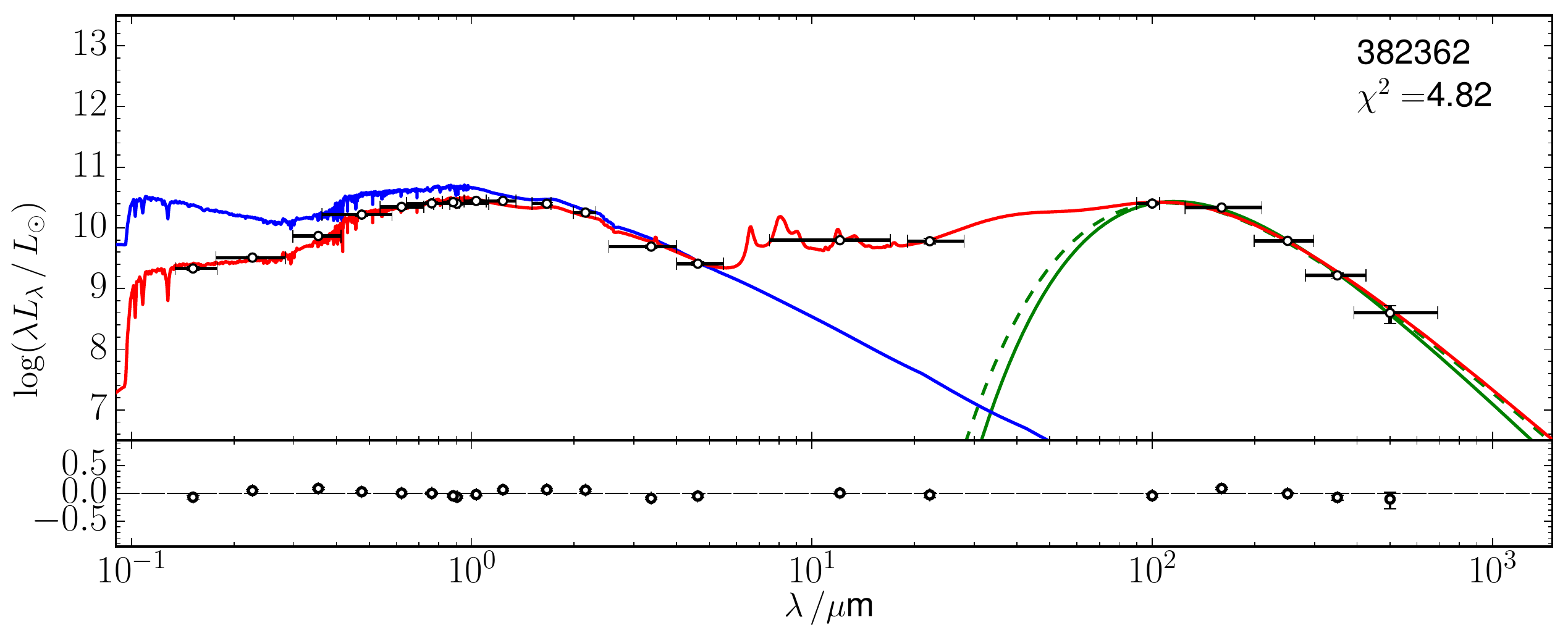} &
\includegraphics[width=5.0cm]{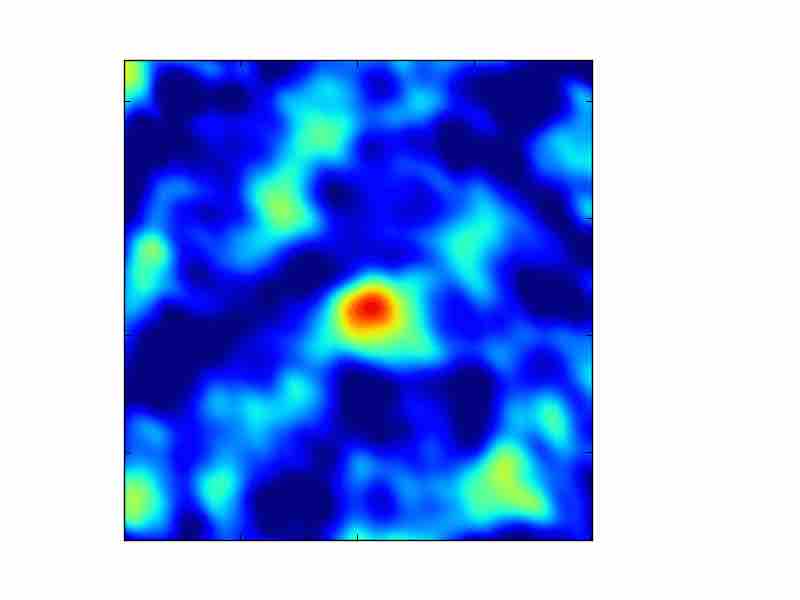} &
\hspace*{-1.2cm}\begin{overpic}[width=3.4cm]{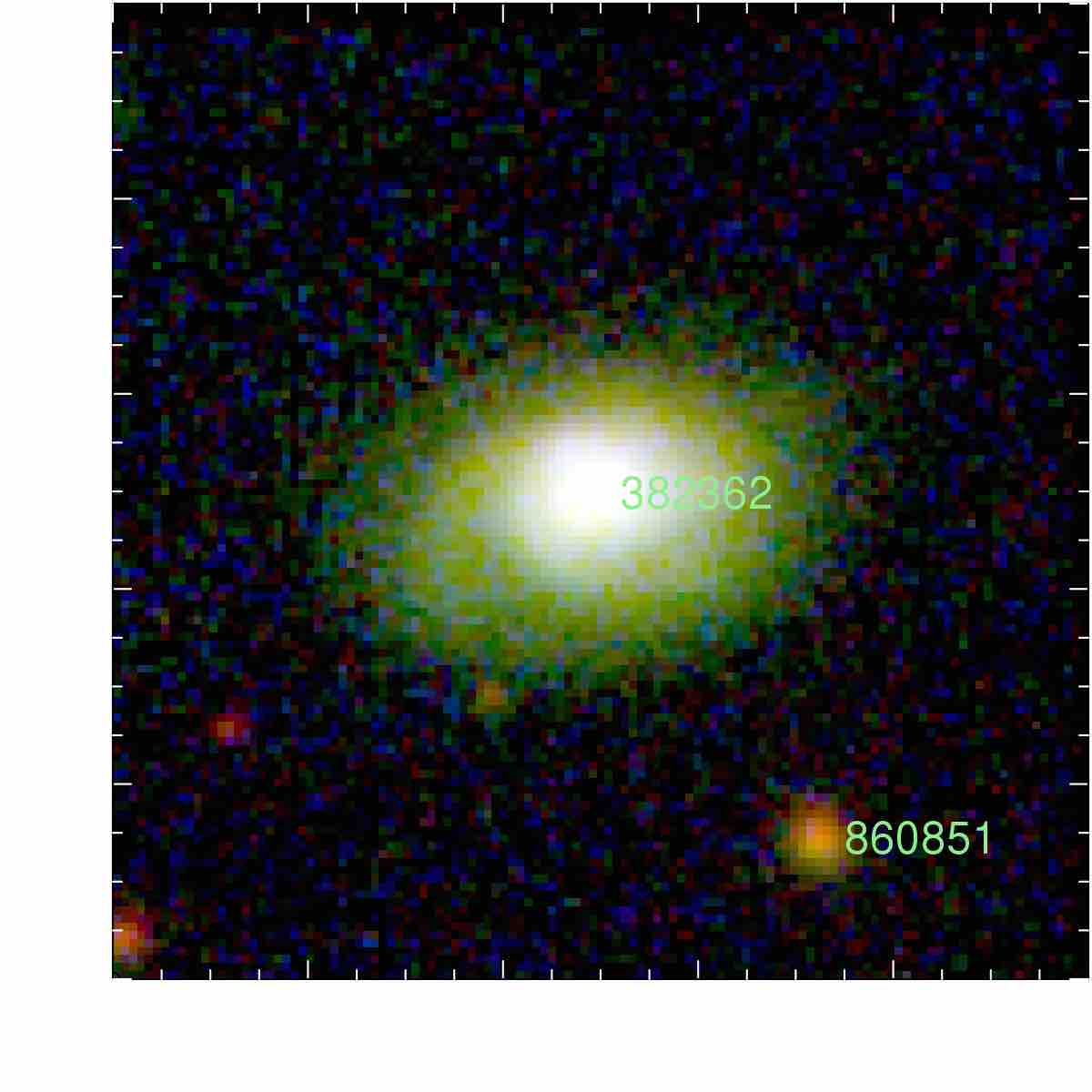} \put (9,85) { \begin{fitbox}{2.25cm}{0.2cm} \color{white}$\bf DB$ \end{fitbox}} \end{overpic} \\ 

\end{array}
$
{\textbf{Figure~\ref{pdrdiaglit}.} continued}

\end{figure*}


\begin{figure*}
$
\begin{array}{ccc}
\includegraphics[width=8.4cm]{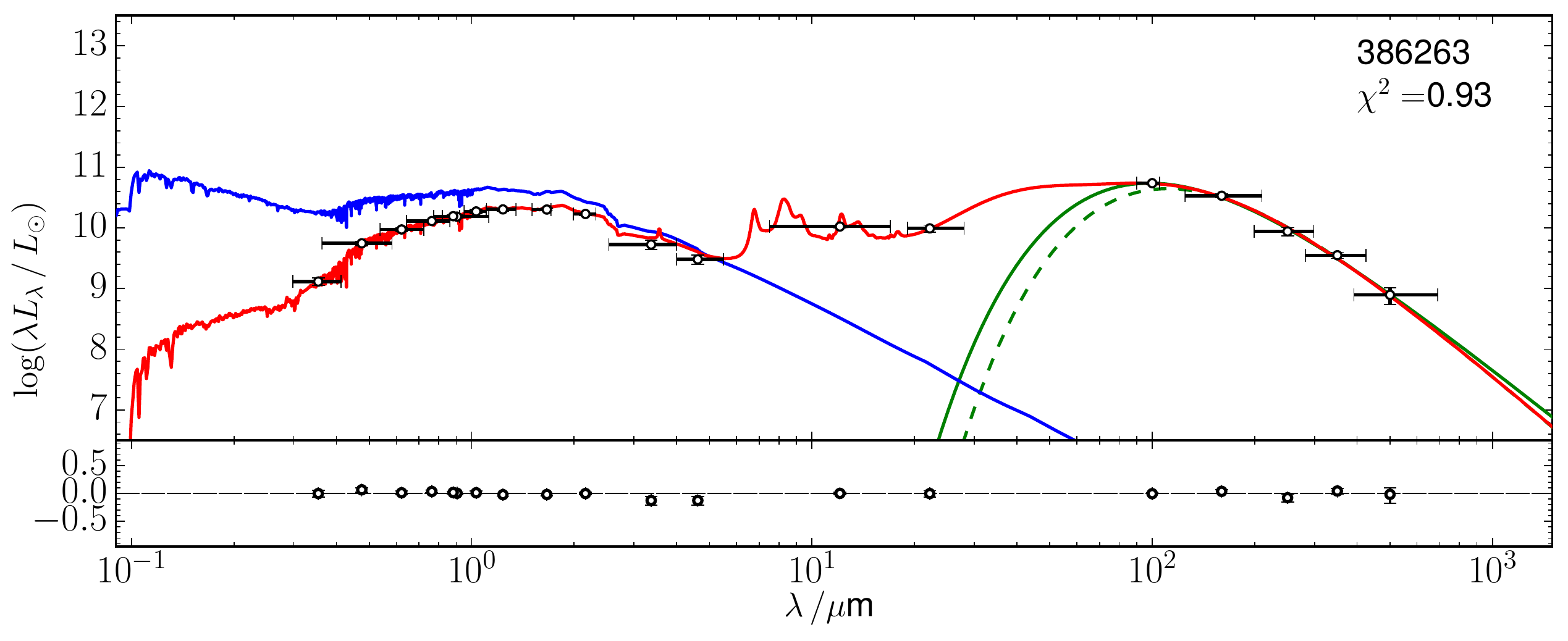} &
\includegraphics[width=5.0cm]{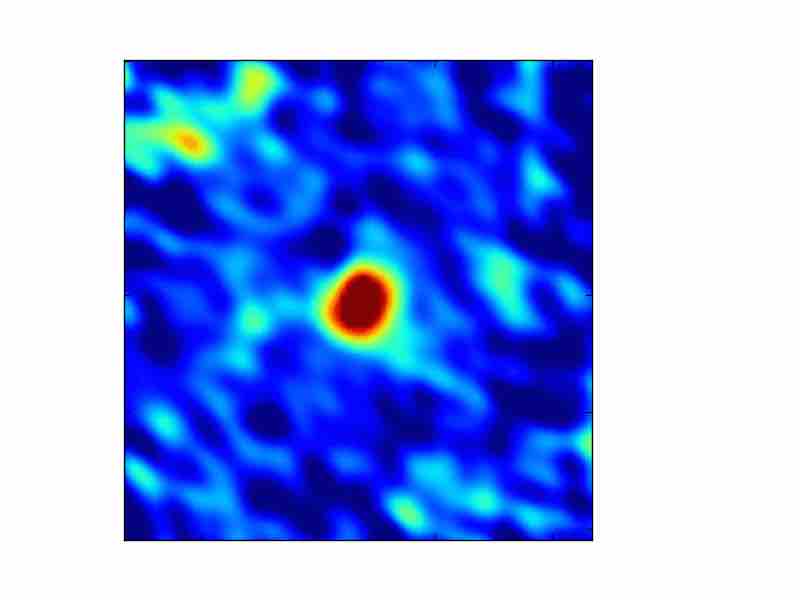} &
\hspace*{-1.2cm}\begin{overpic}[width=3.4cm]{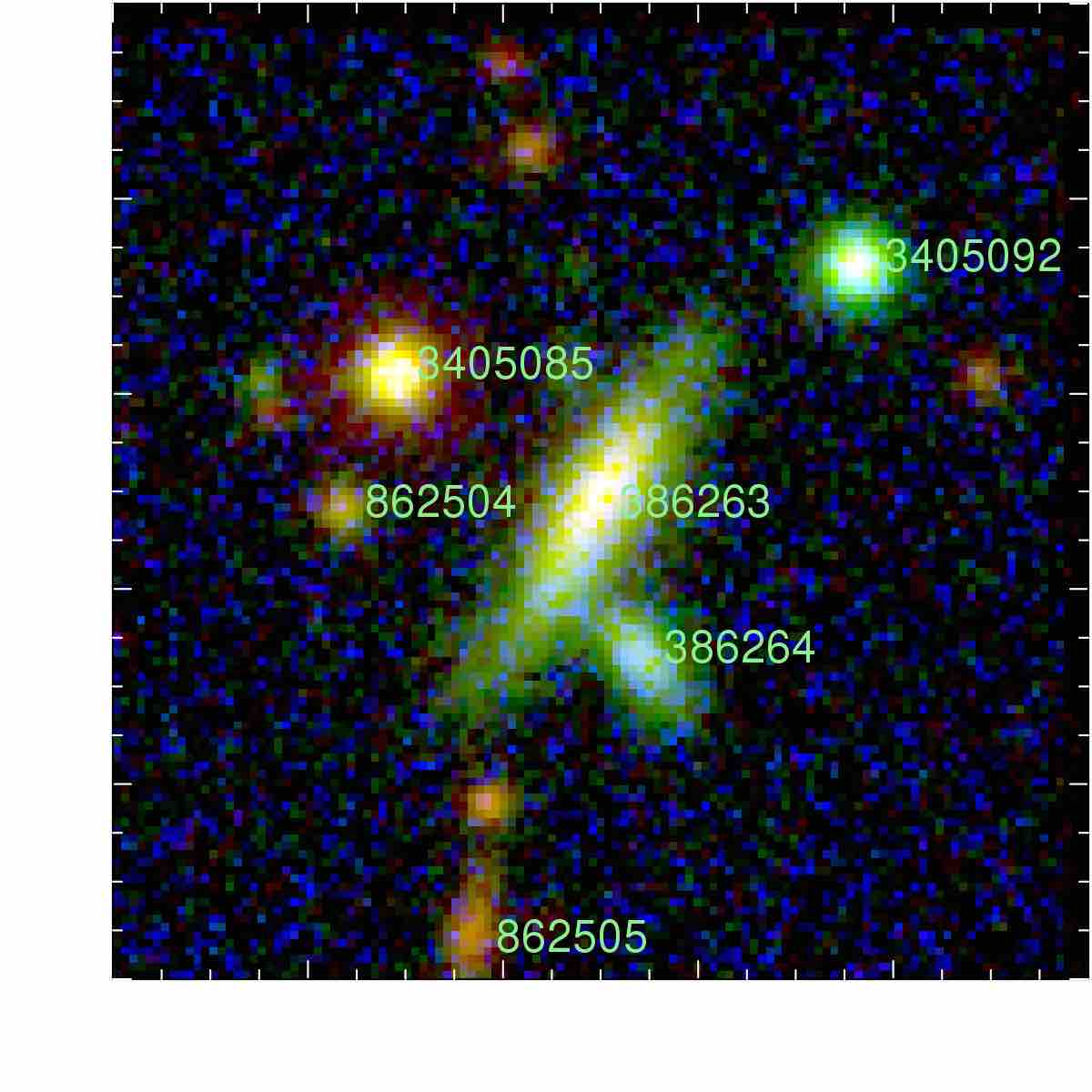} \put (9,85) { \begin{fitbox}{2.25cm}{0.2cm} \color{white}$\bf DC$ \end{fitbox}} \end{overpic} \\ 	
	
\includegraphics[width=8.4cm]{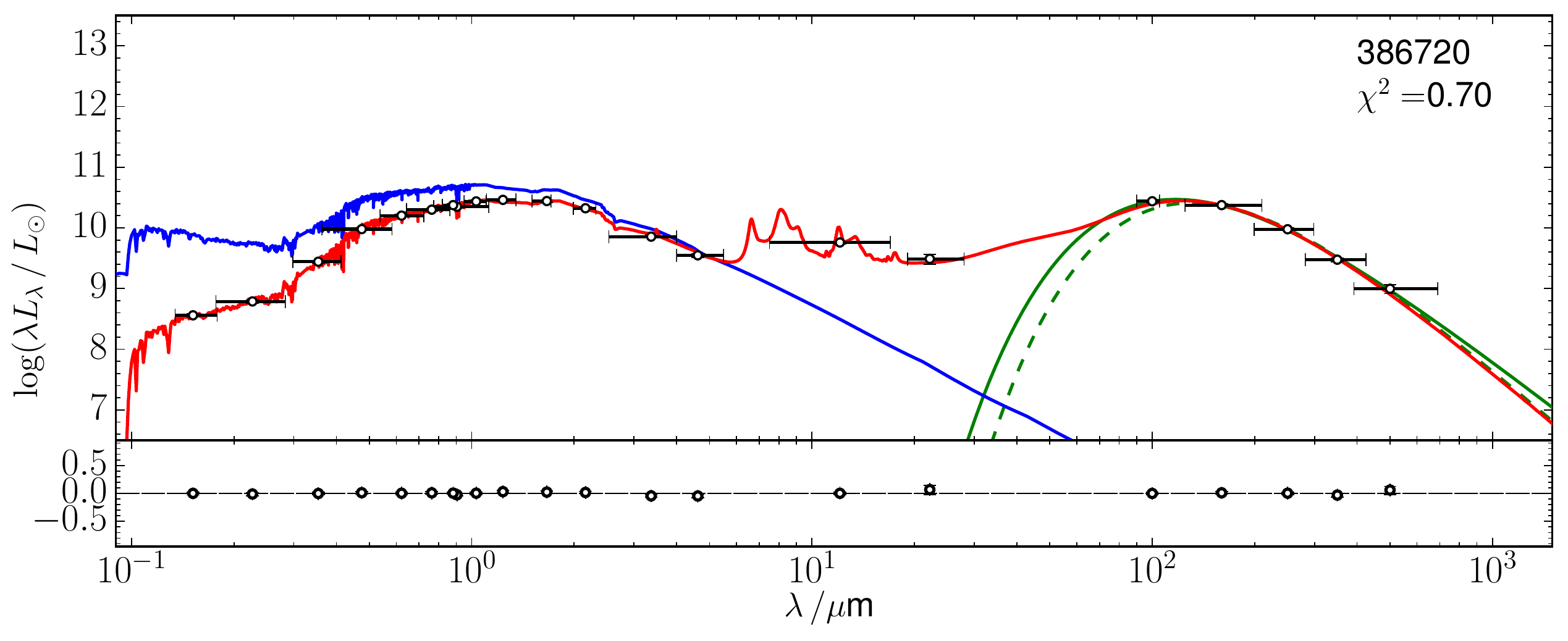} &
\includegraphics[width=5.0cm]{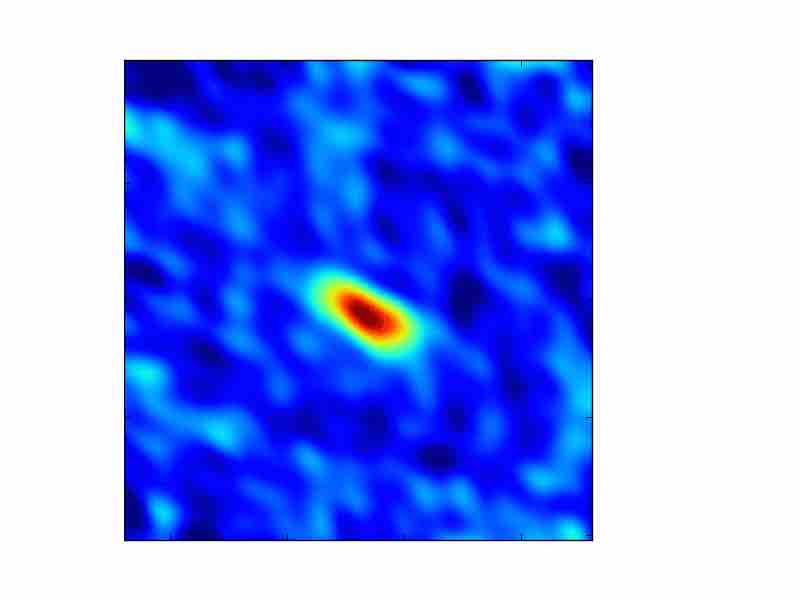} &
\hspace*{-1.2cm}\begin{overpic}[width=3.4cm]{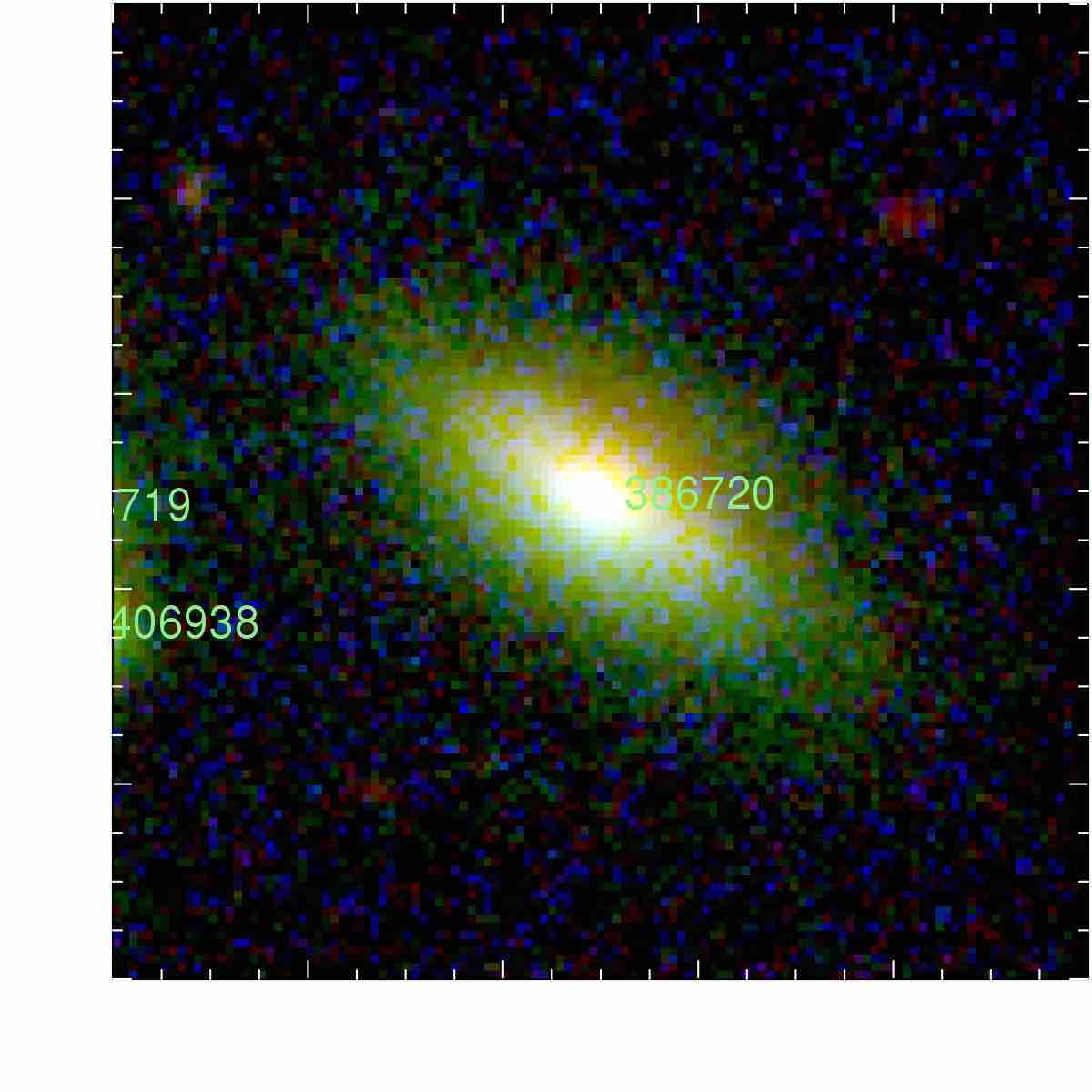} \put (9,85) { \begin{fitbox}{2.25cm}{0.2cm} \color{white}$\bf DB$ \end{fitbox}} \end{overpic} \\ 	

\includegraphics[width=8.4cm]{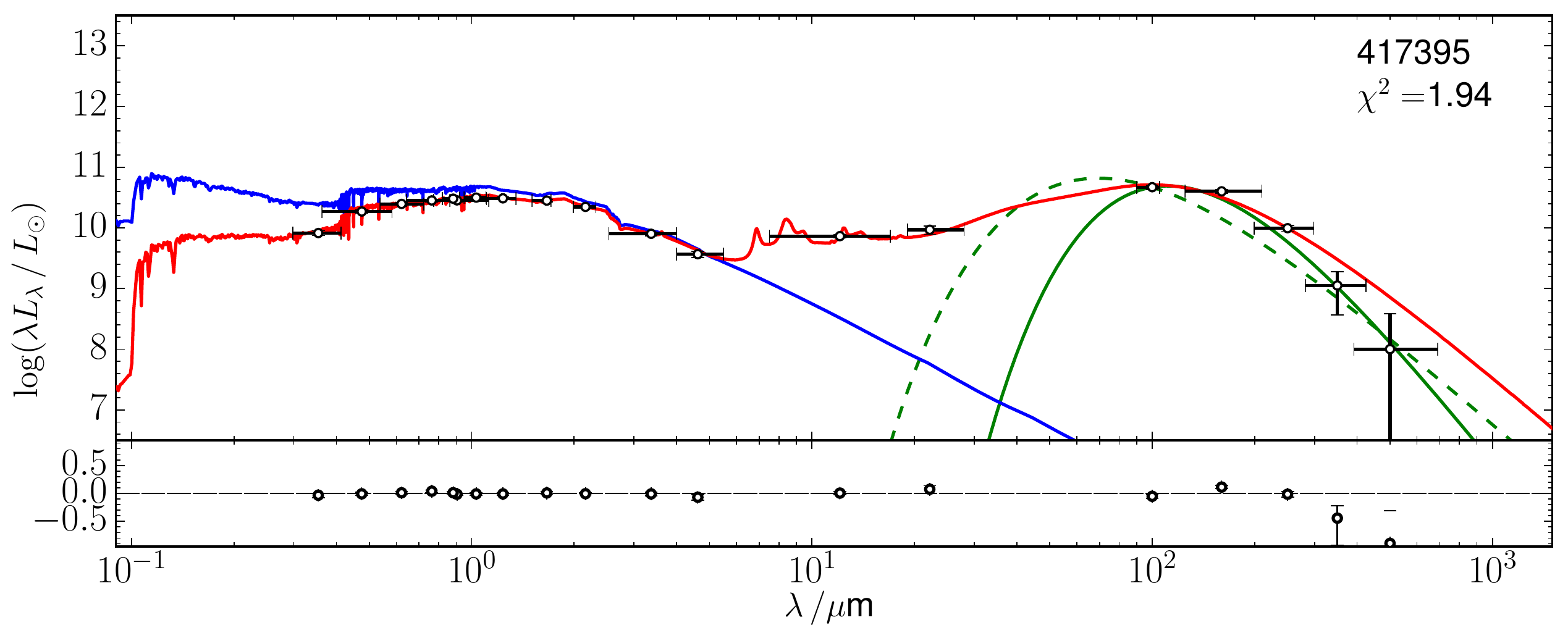} &
\includegraphics[width=5.0cm]{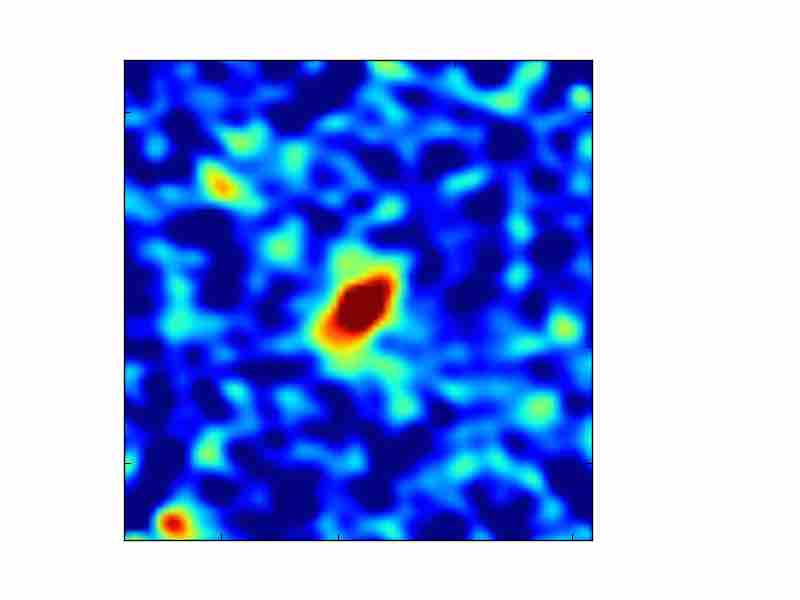} &
\hspace*{-1.2cm}\begin{overpic}[width=3.4cm]{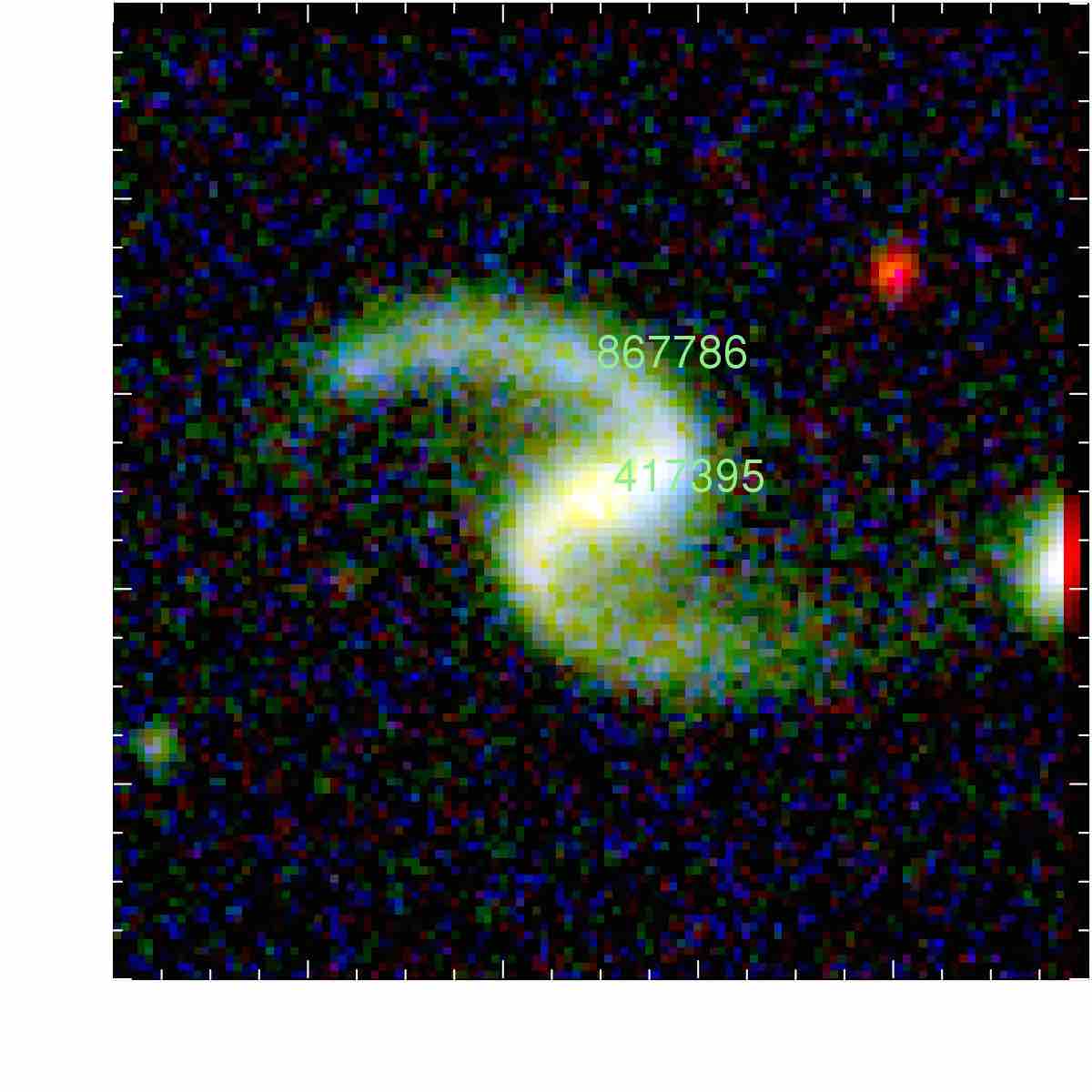} \put (9,85) { \begin{fitbox}{2.25cm}{0.2cm} \color{white}$\bf DBC$ \end{fitbox}} \end{overpic} \\ 	

\includegraphics[width=8.4cm]{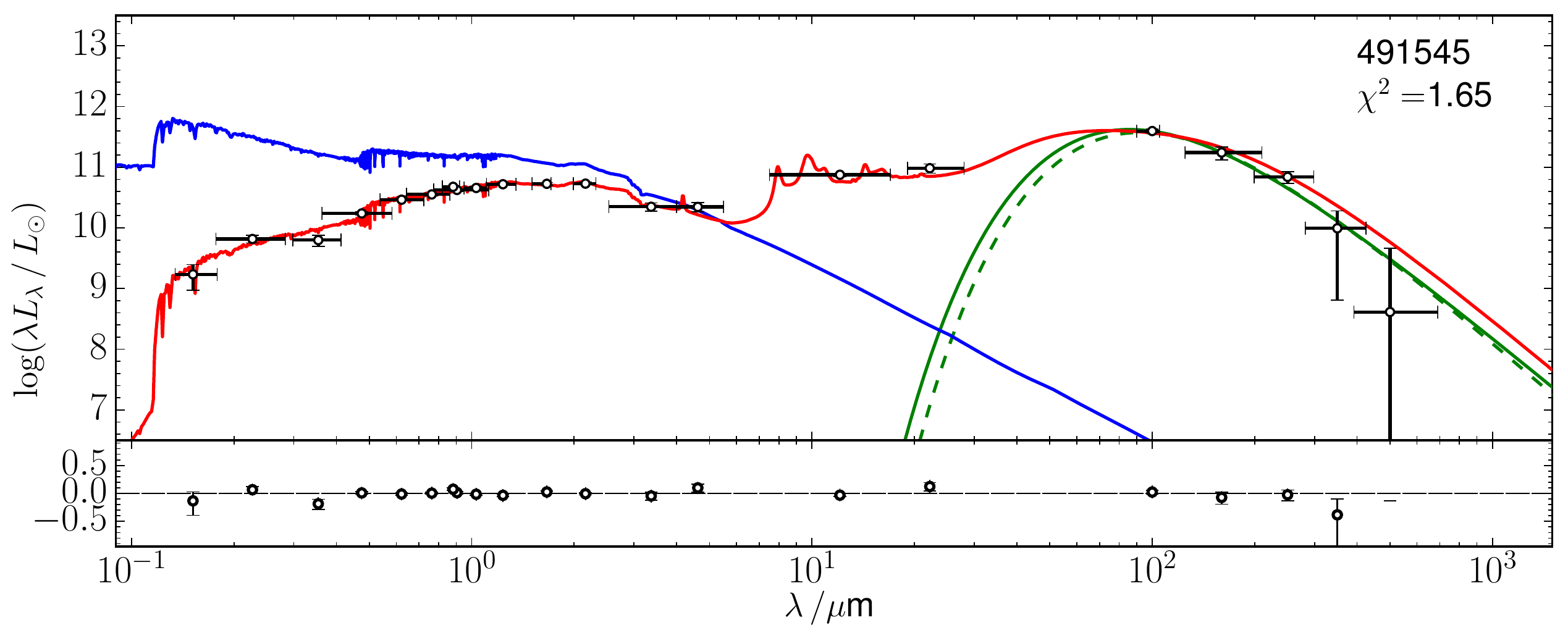} &
\includegraphics[width=5.0cm]{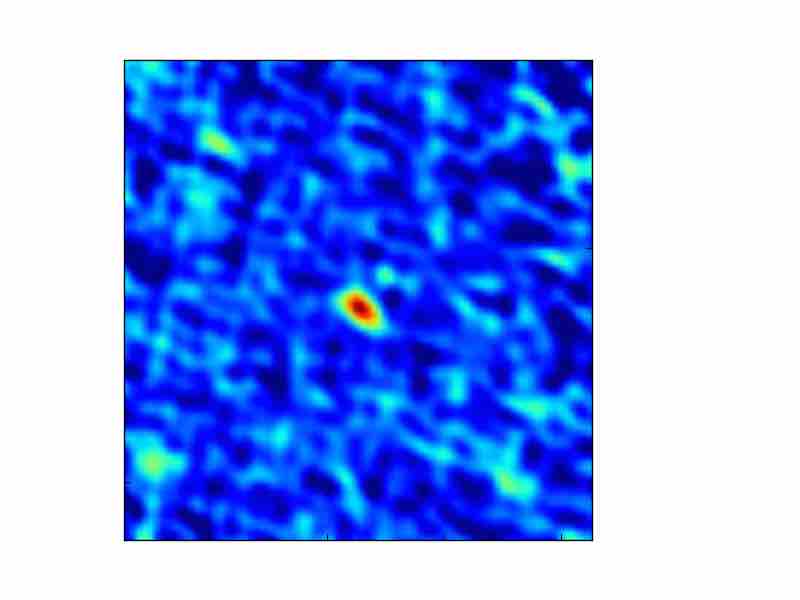} &
\hspace*{-1.2cm}\begin{overpic}[width=3.4cm]{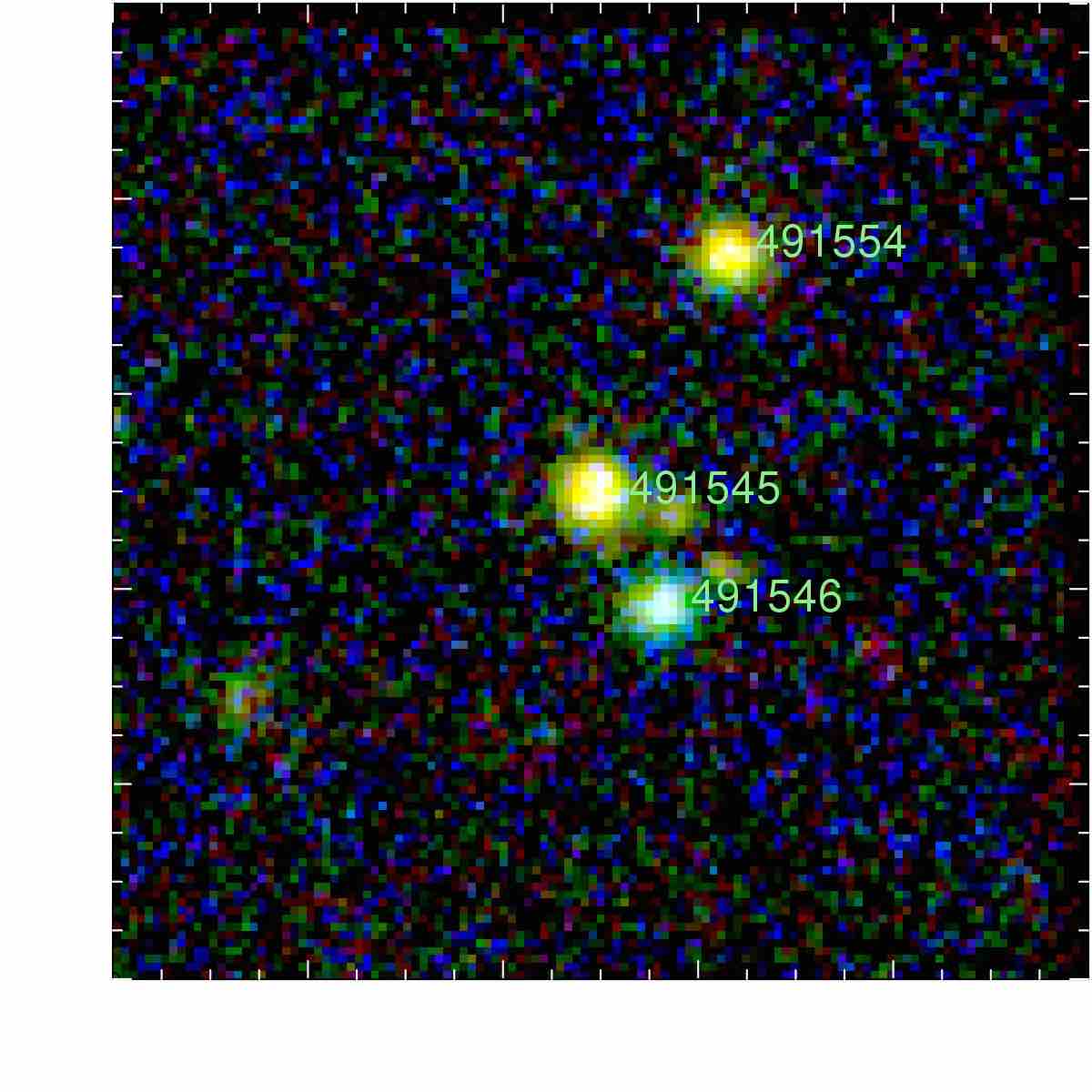} \put (9,85) { \begin{fitbox}{2.25cm}{0.2cm} \color{white}$\bf BC$ \end{fitbox}} \end{overpic} \\ 

\includegraphics[width=8.4cm]{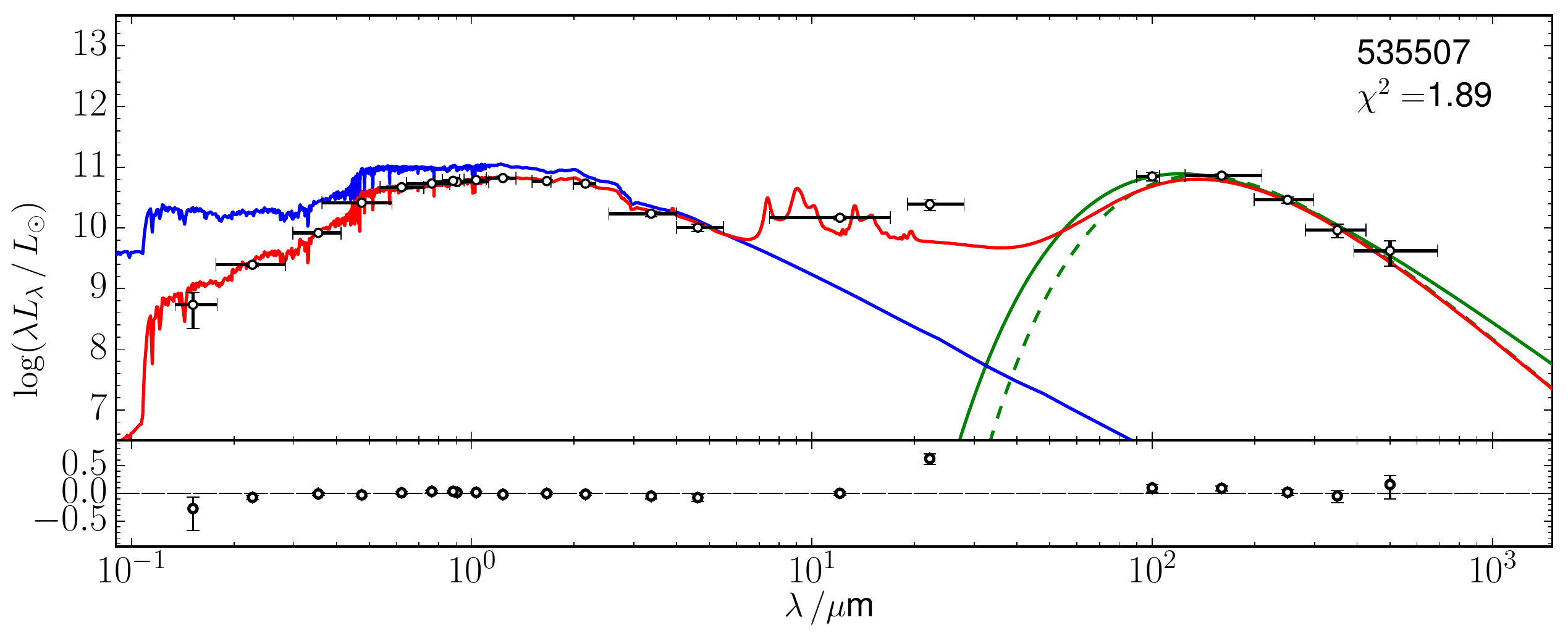} &
\includegraphics[width=5.0cm]{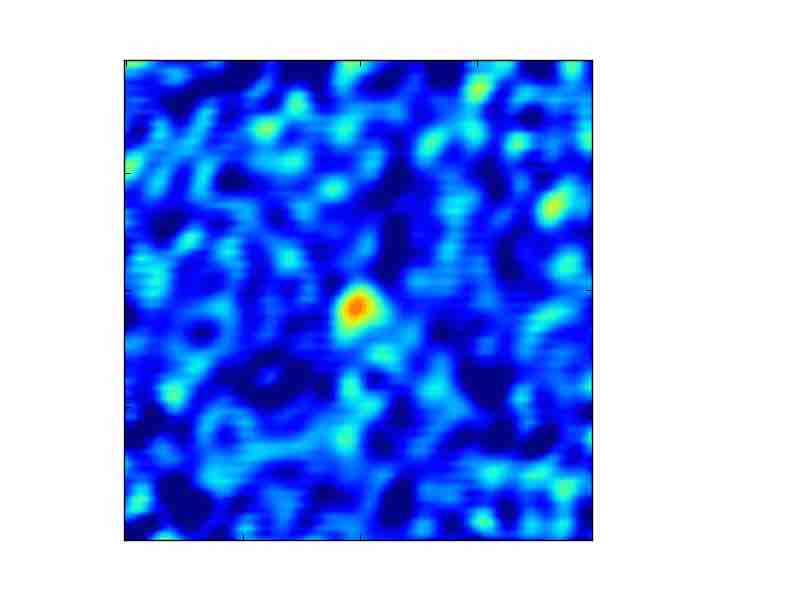} &
\hspace*{-1.2cm}\begin{overpic}[width=3.4cm]{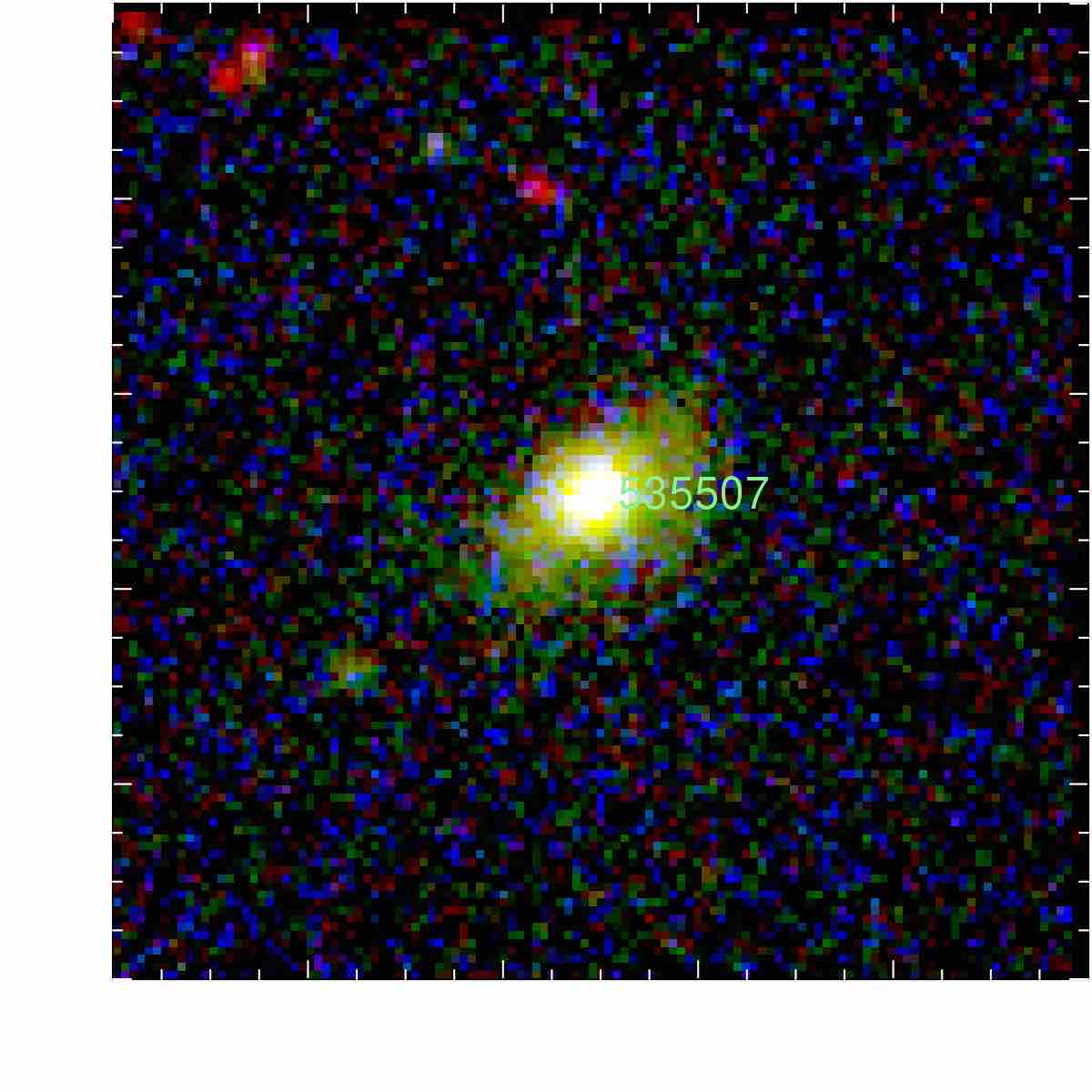} \put (9,85) { \begin{fitbox}{2.25cm}{0.2cm} \color{white}$\bf BD$ \end{fitbox}} \end{overpic} \\ 

\end{array}
$
{\textbf{Figure~\ref{pdrdiaglit}.} continued}

\end{figure*}


\begin{figure*}
$
\begin{array}{ccc}
\includegraphics[width=8.4cm]{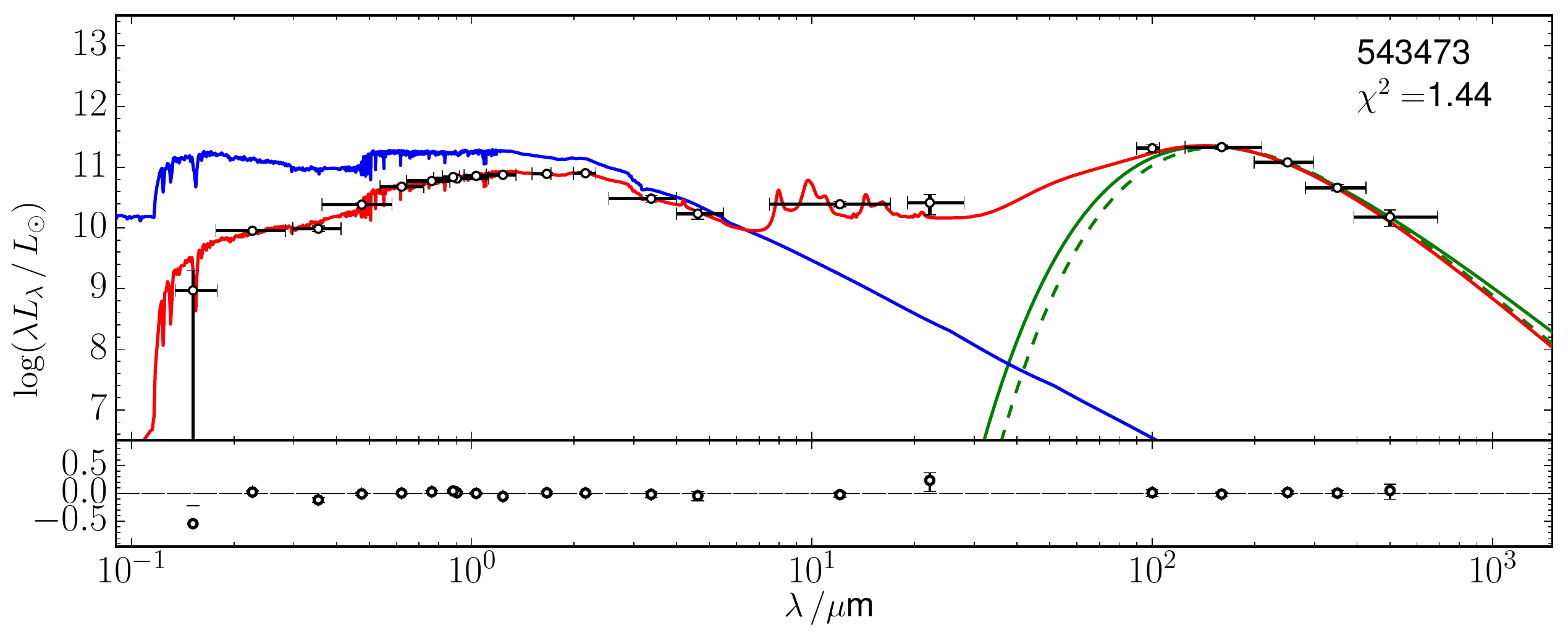} &
\includegraphics[width=5.0cm]{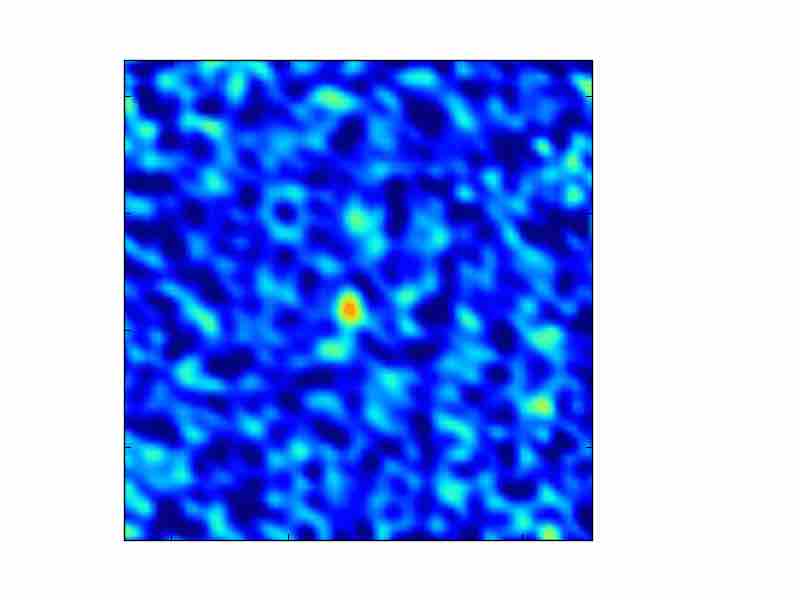} &
\hspace*{-1.2cm}\begin{overpic}[width=3.4cm]{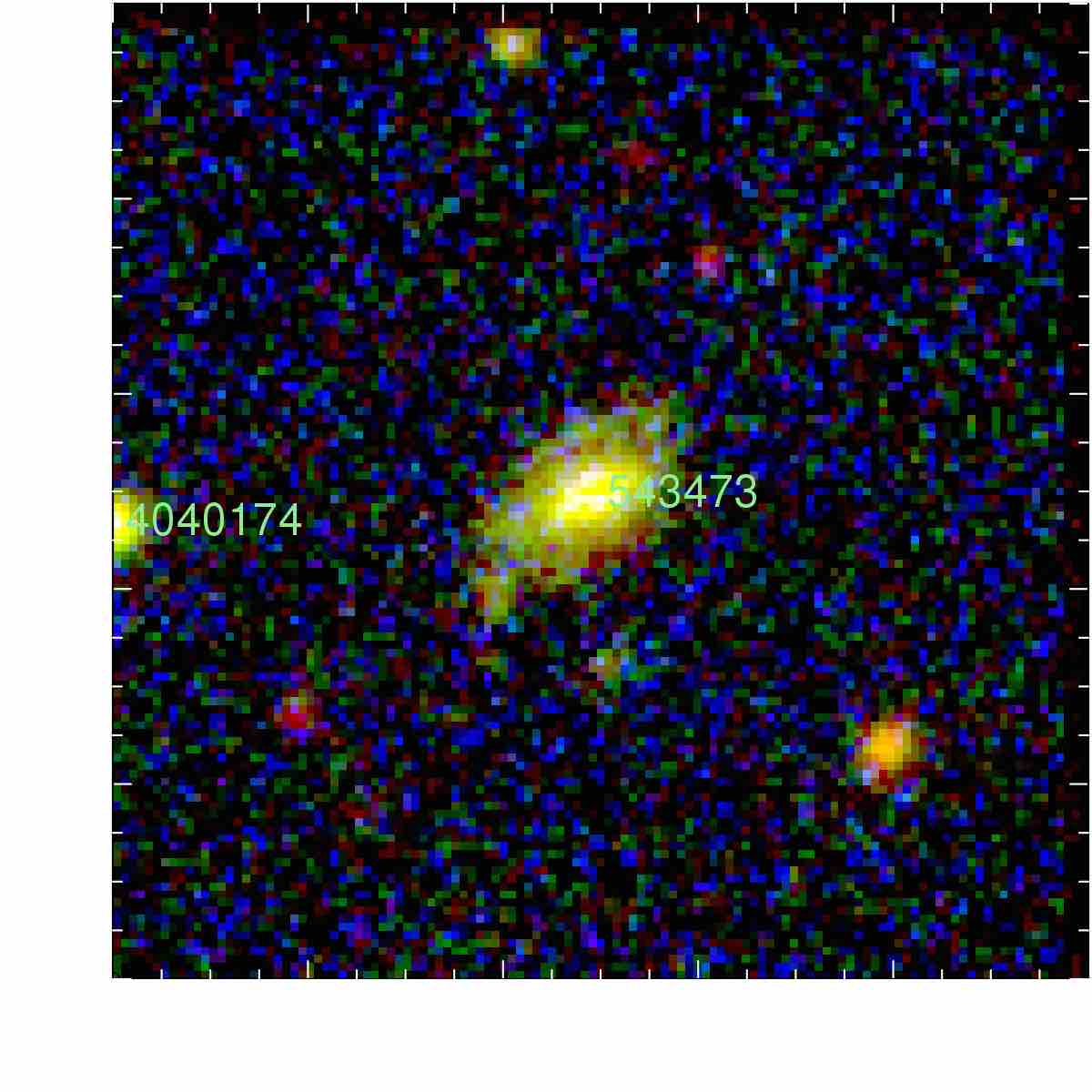} \put (9,85) { \begin{fitbox}{2.25cm}{0.2cm} \color{white}$\bf BD$ \end{fitbox}} \end{overpic} \\ 	
	
\includegraphics[width=8.4cm]{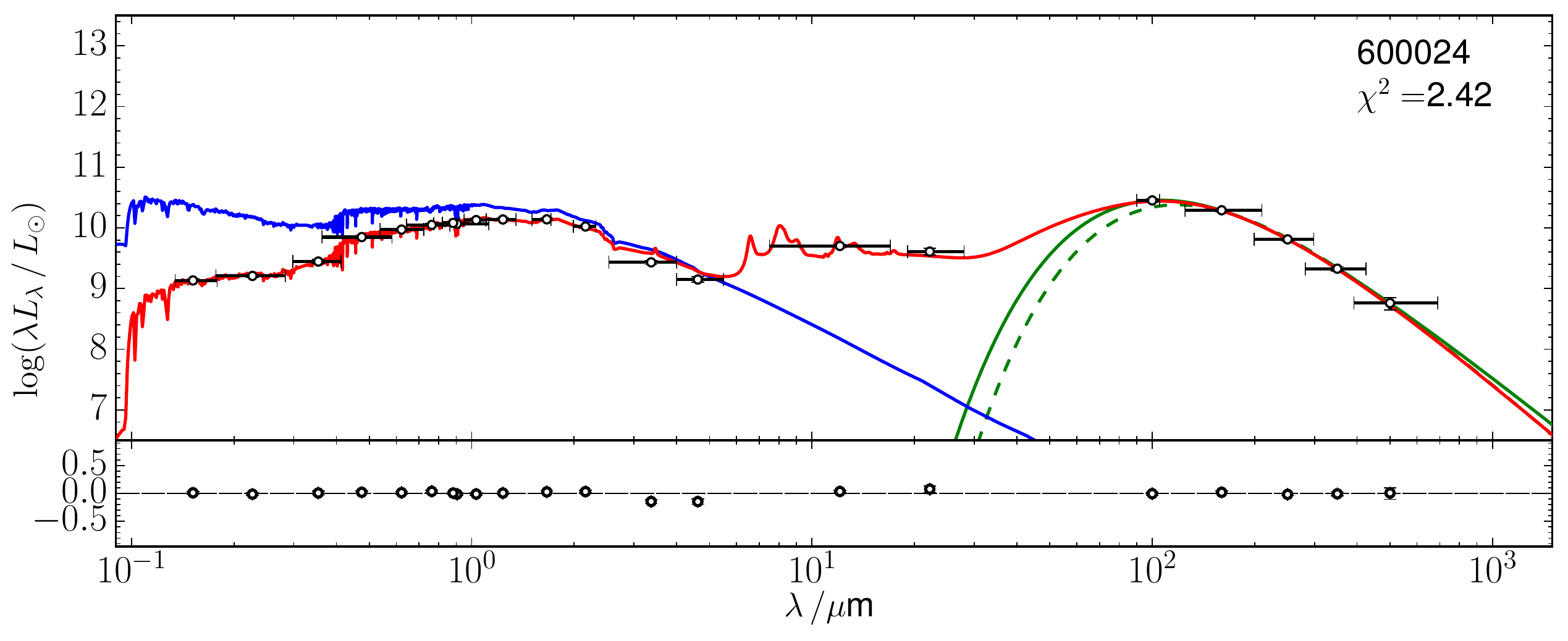} &
\includegraphics[width=5.0cm]{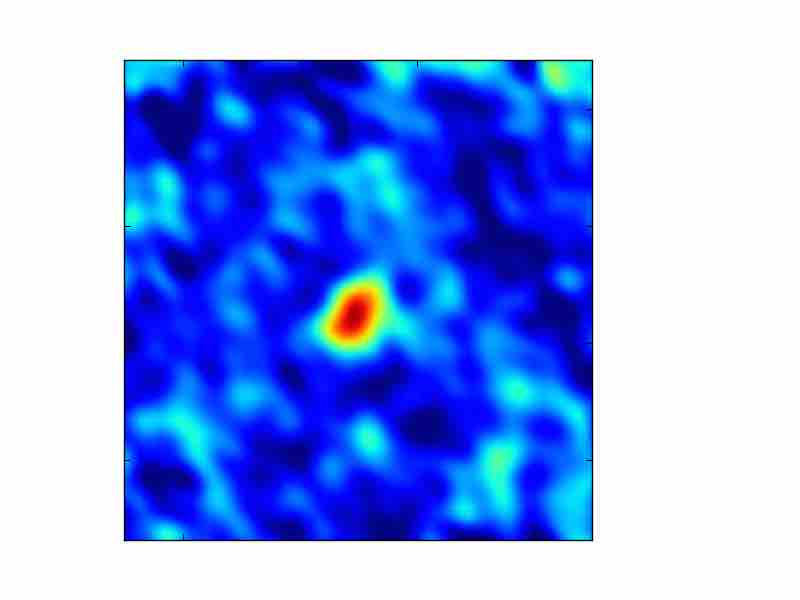} &
\hspace*{-1.2cm}\begin{overpic}[width=3.4cm]{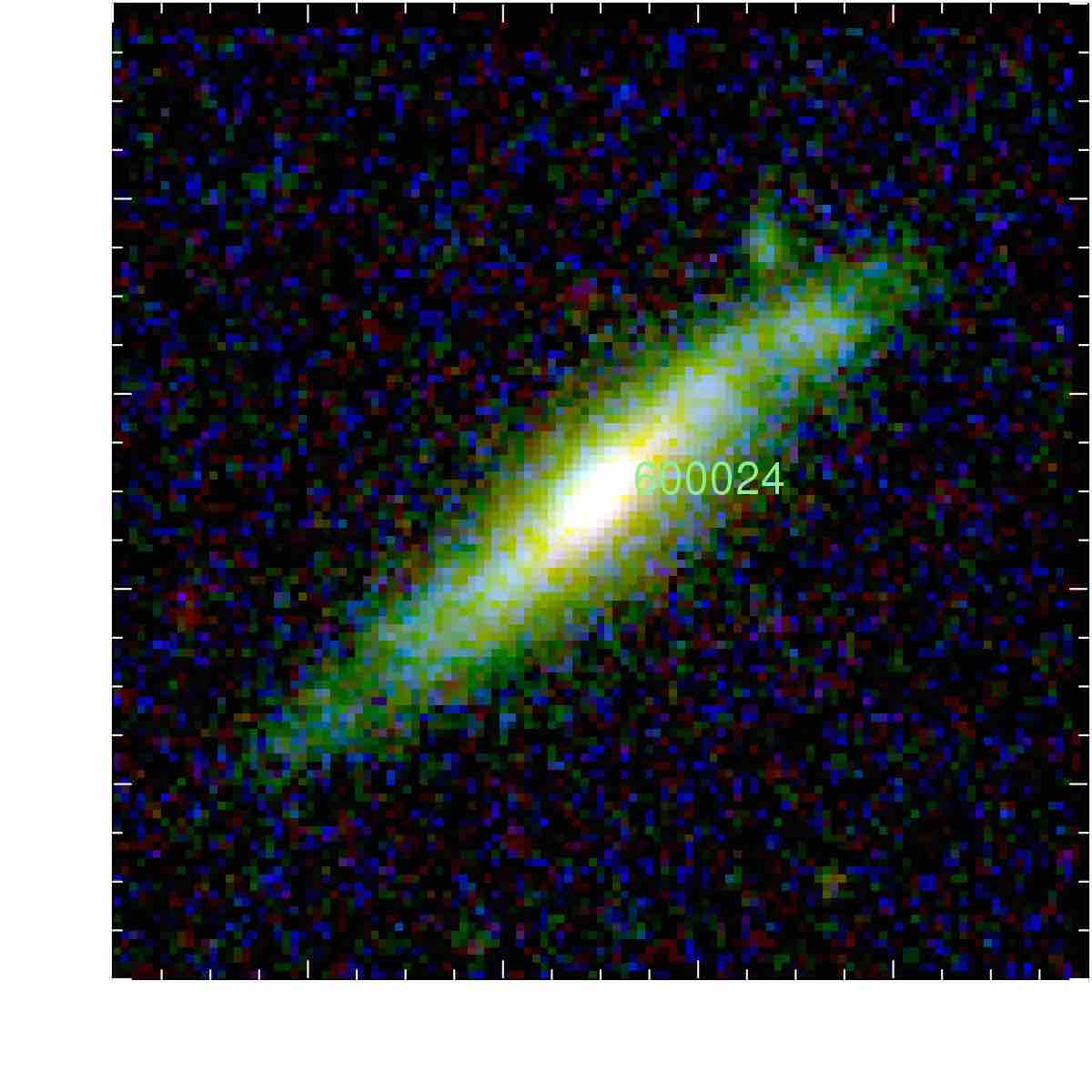} \put (9,85) { \begin{fitbox}{2.25cm}{0.2cm} \color{white}$\bf D$ \end{fitbox}} \end{overpic} \\ 	

\includegraphics[width=8.4cm]{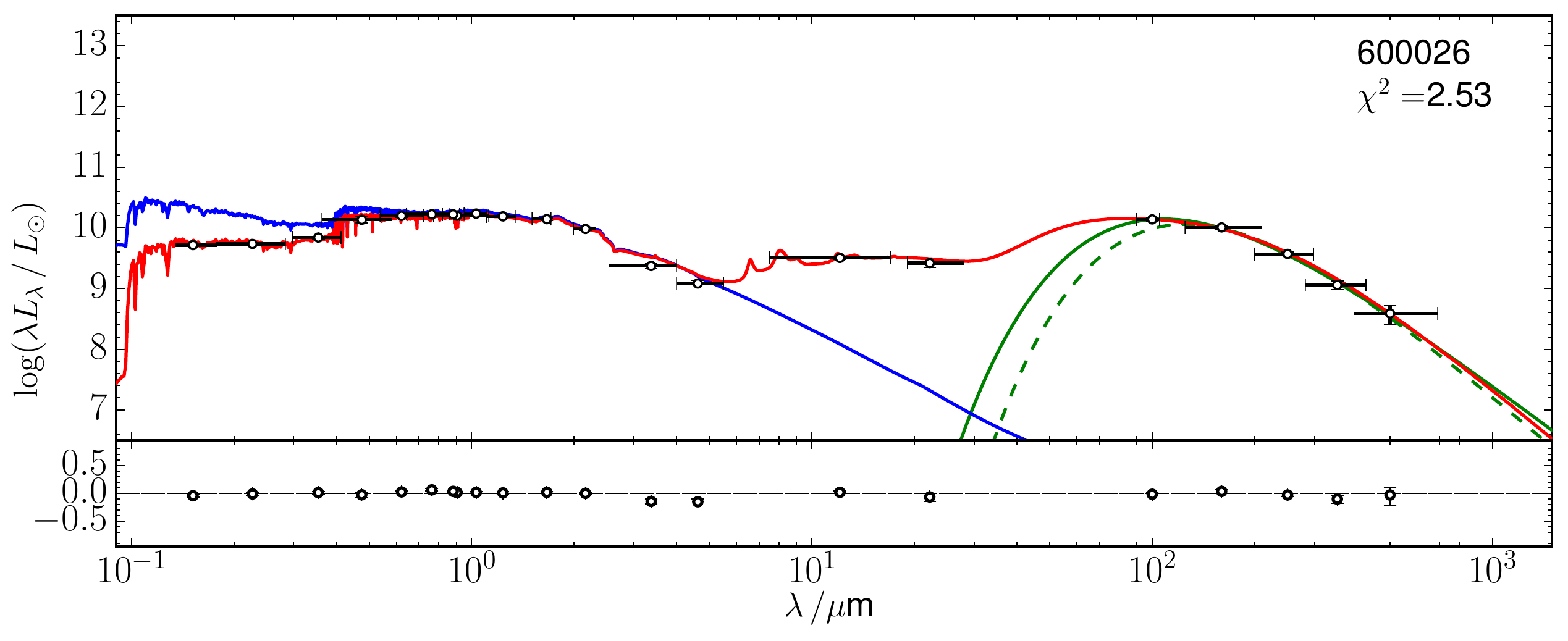} &
\includegraphics[width=5.0cm]{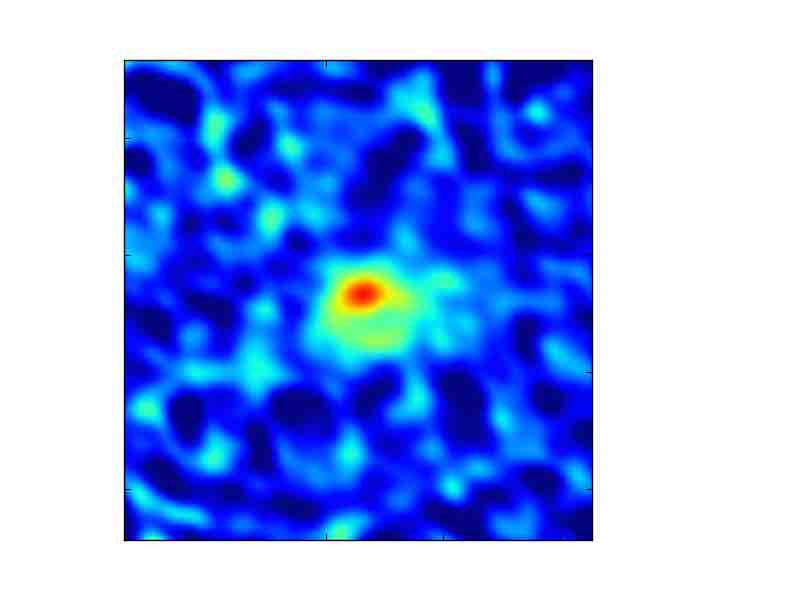} &
\hspace*{-1.2cm}\begin{overpic}[width=3.4cm]{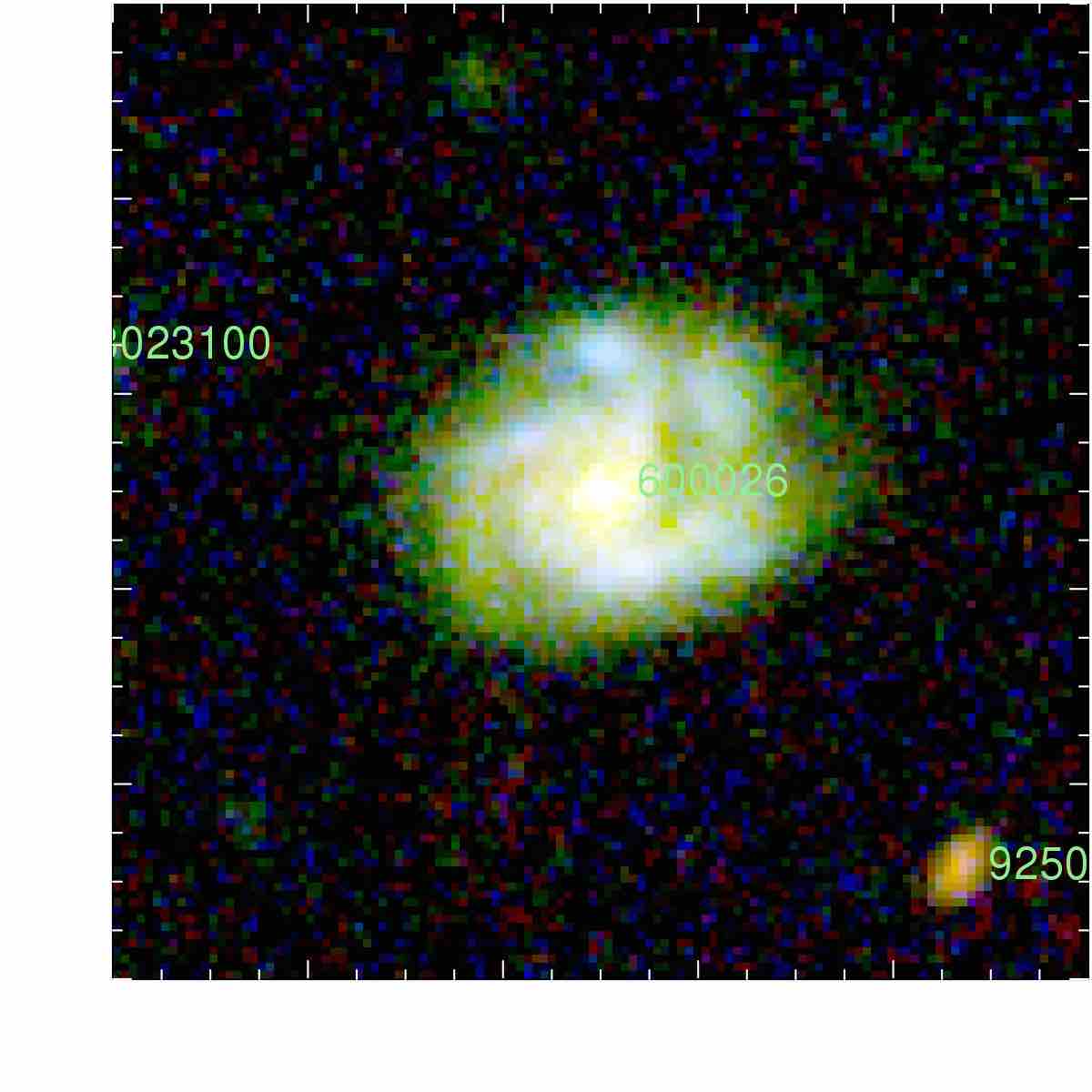} \put (9,85) { \begin{fitbox}{2.25cm}{0.2cm} \color{white}$\bf DB$ \end{fitbox}} \end{overpic} \\ 

\includegraphics[width=8.4cm]{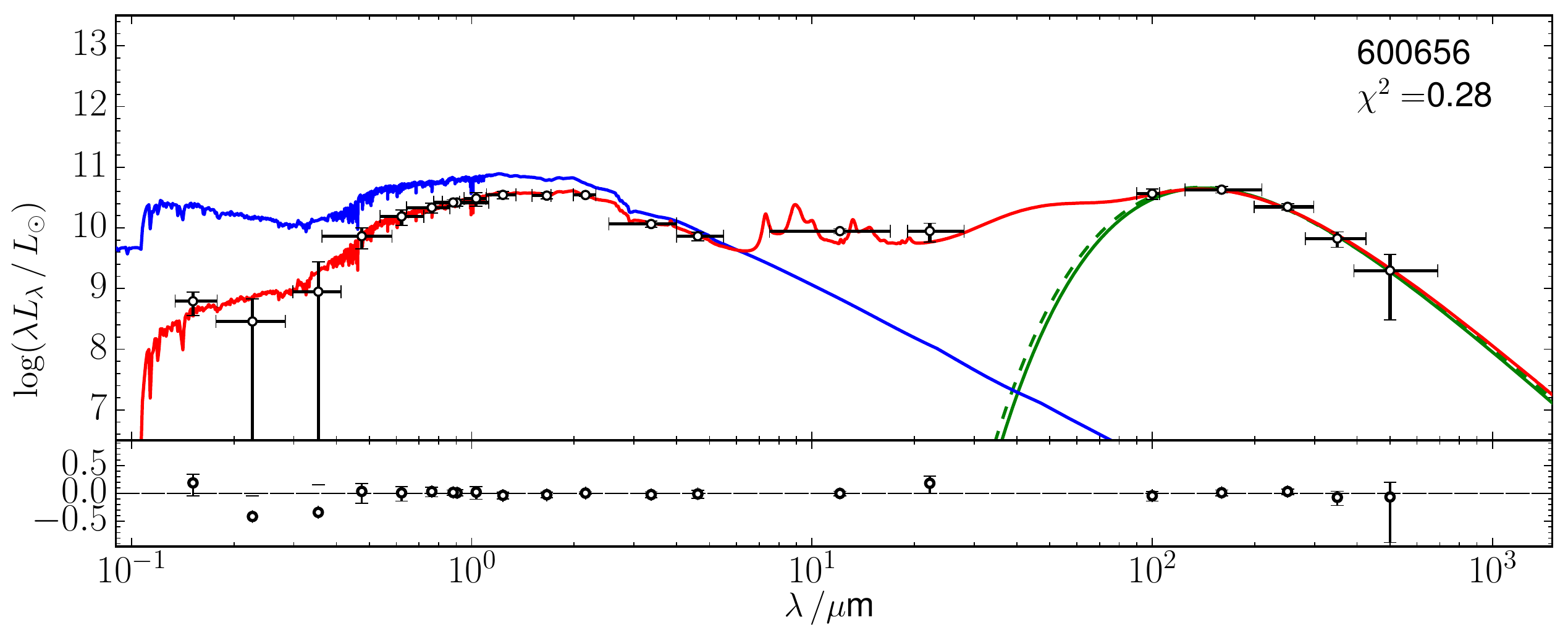} &
\includegraphics[width=5.0cm]{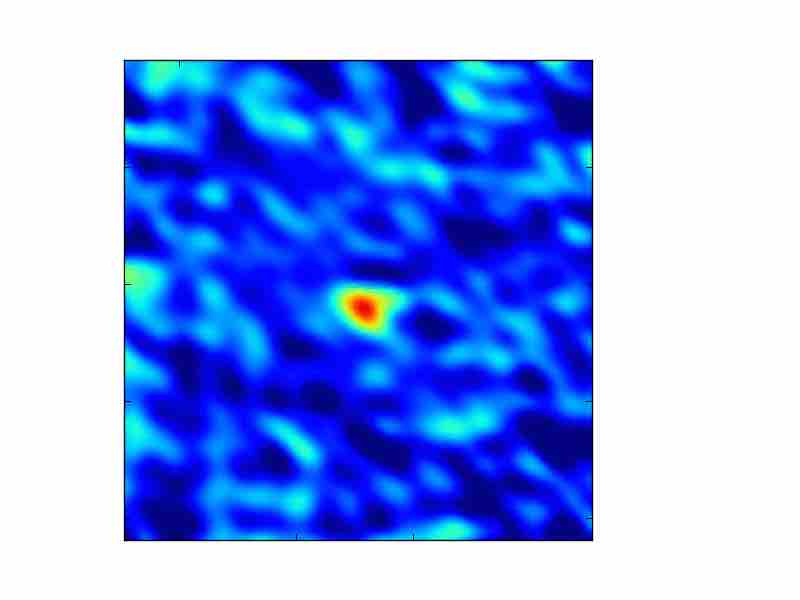} &
\hspace*{-1.2cm}\begin{overpic}[width=3.4cm]{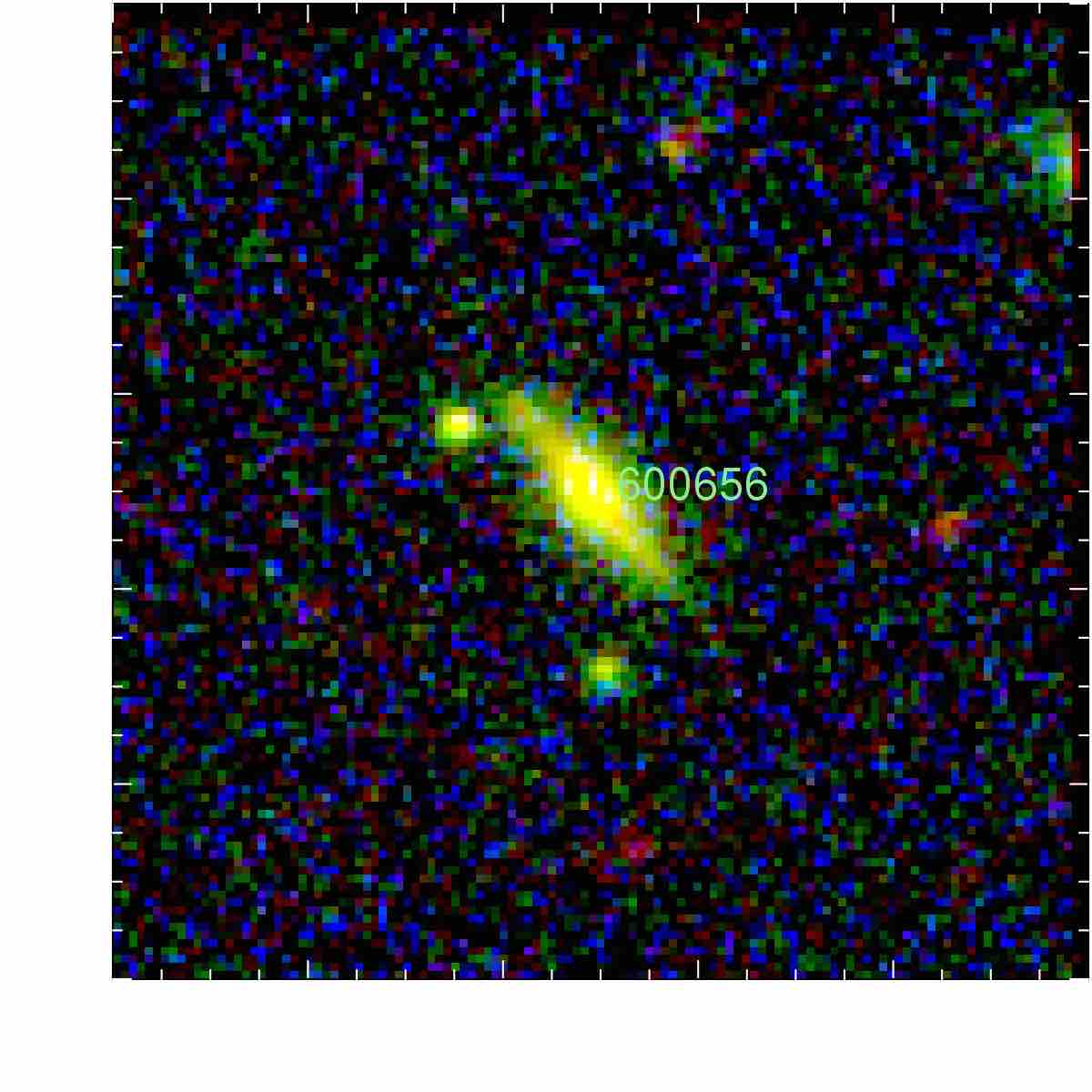} \put (9,85) { \begin{fitbox}{2.25cm}{0.2cm} \color{white}$\bf DBC$ \end{fitbox}} \end{overpic} \\ 

\includegraphics[width=8.4cm]{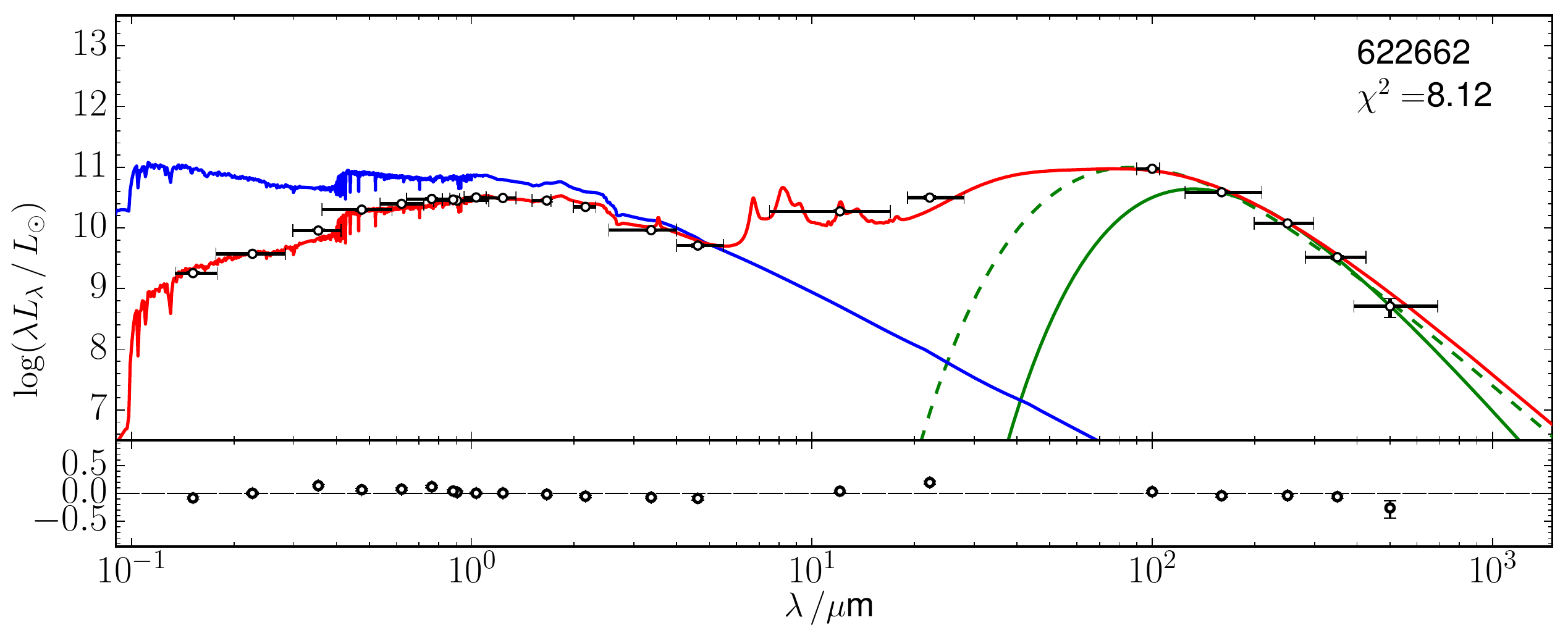} &
\includegraphics[width=5.0cm]{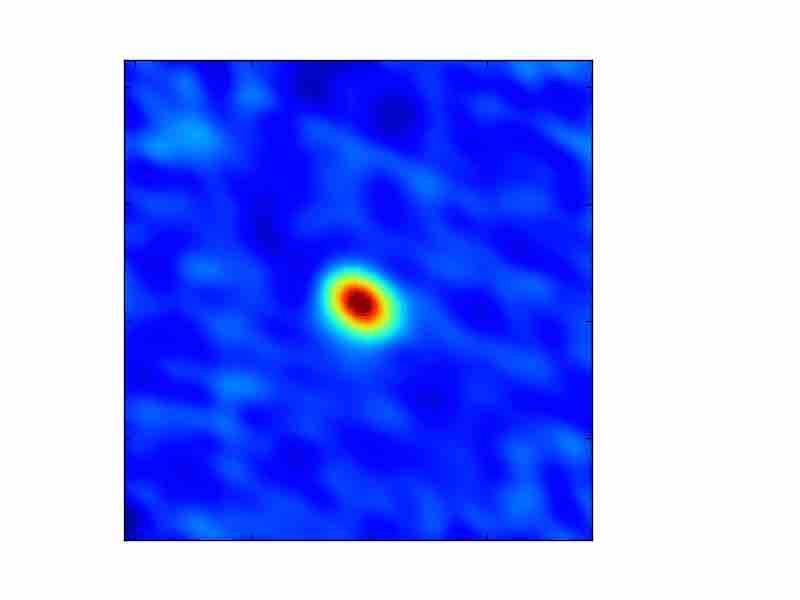} &
\hspace*{-1.2cm}\begin{overpic}[width=3.4cm]{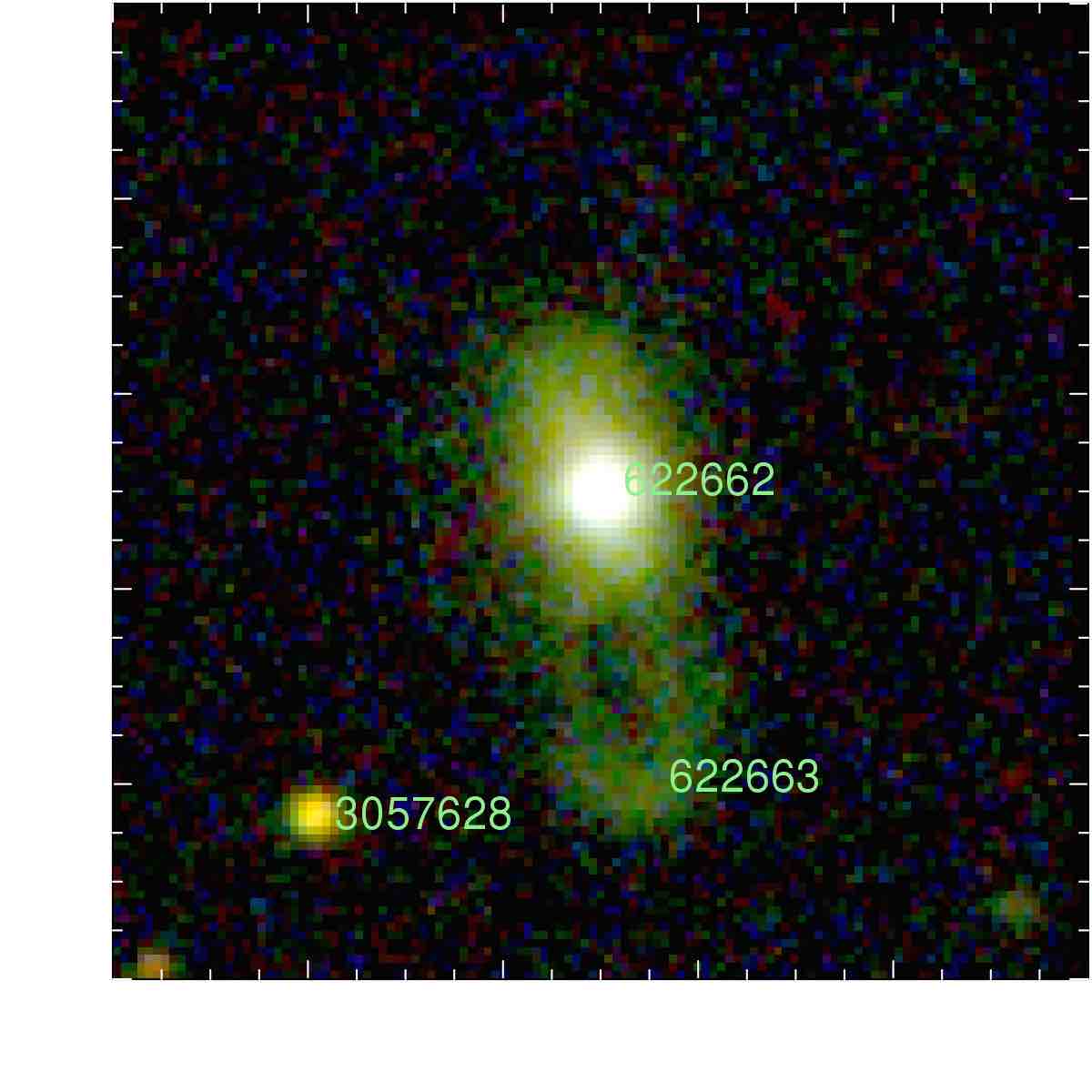} \put (9,85) { \begin{fitbox}{2.25cm}{0.2cm} \color{white}$\bf M$ \end{fitbox}} \end{overpic} \\ 

\end{array}
$
{\textbf{Figure~\ref{pdrdiaglit}.} continued}

\end{figure*}


\begin{figure*}
$
\begin{array}{ccc}
\includegraphics[width=8.4cm]{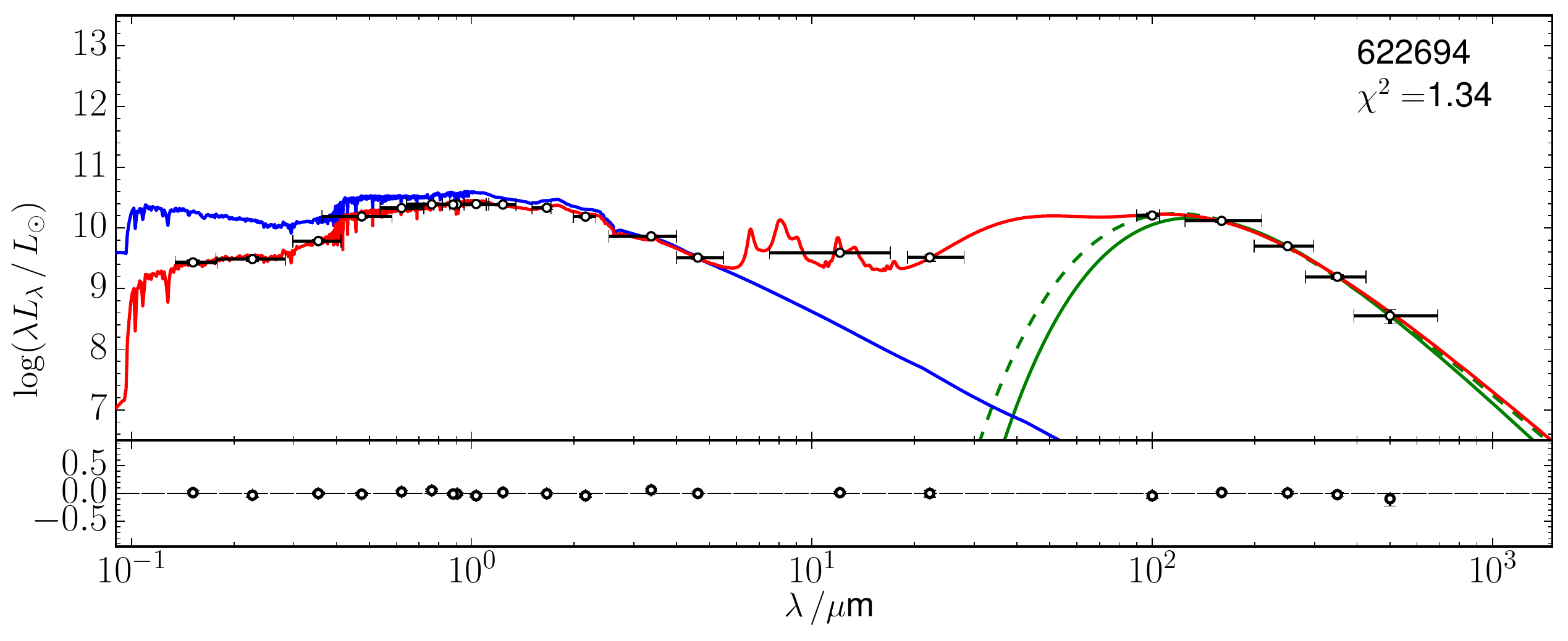} &
\includegraphics[width=5.0cm]{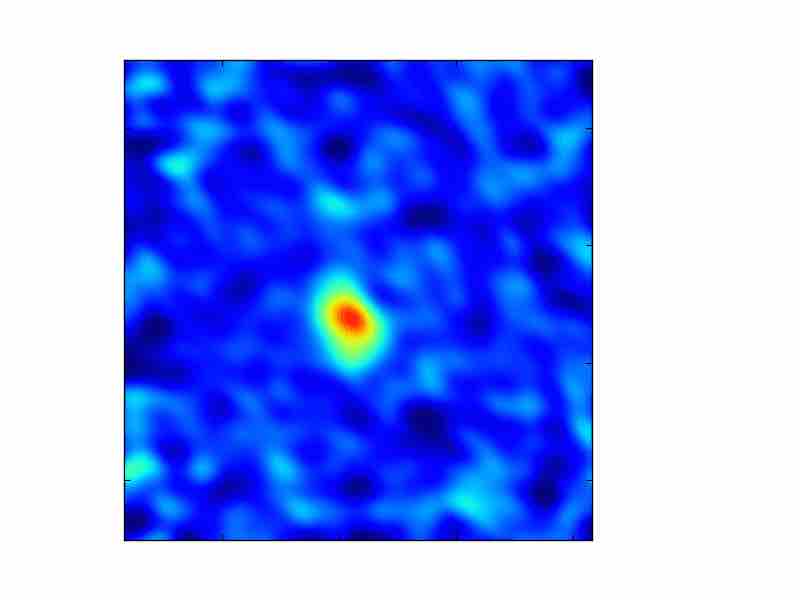} &
\hspace*{-1.2cm}\begin{overpic}[width=3.4cm]{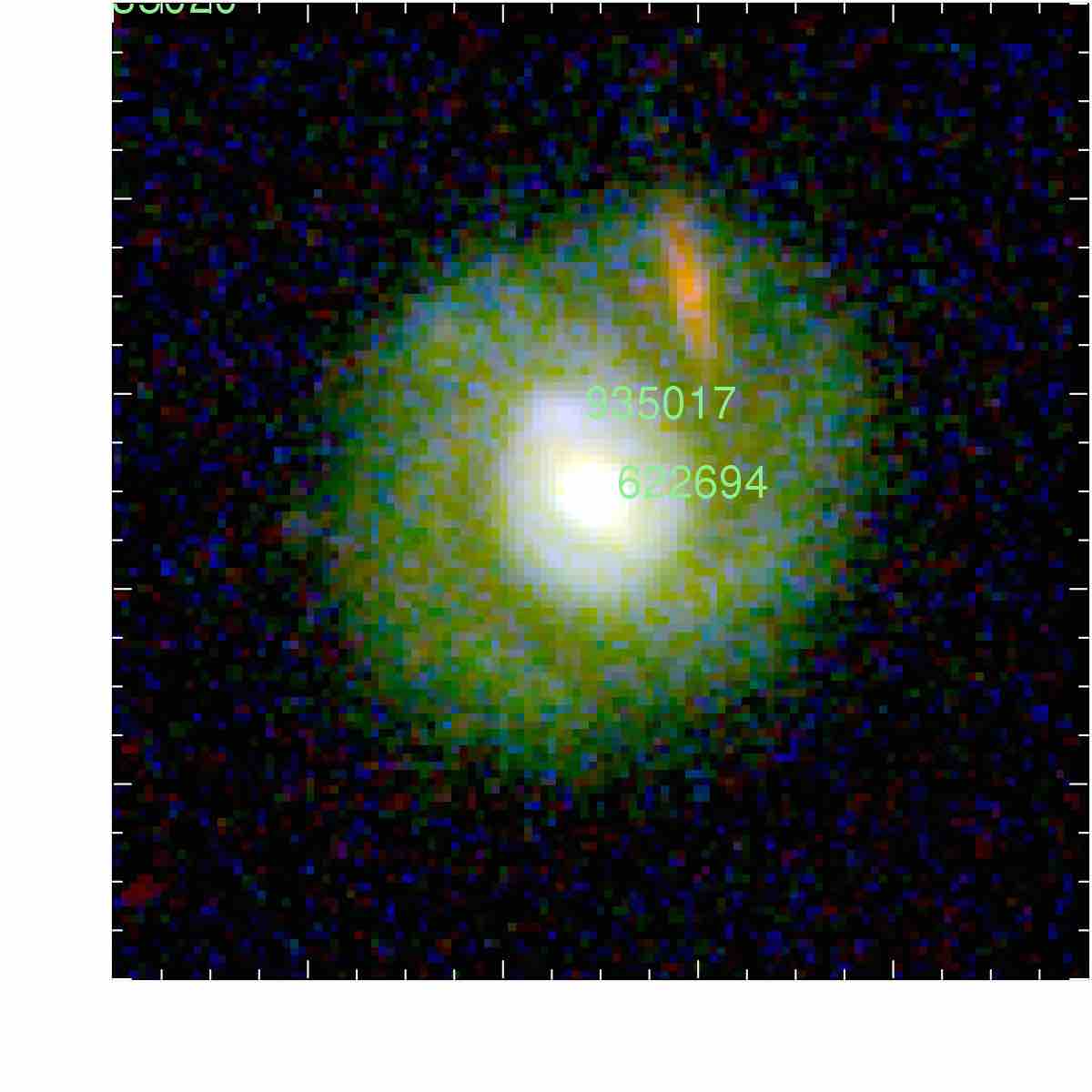} \put (9,85) { \begin{fitbox}{2.25cm}{0.2cm} \color{white}$\bf DBC$ \end{fitbox}} \end{overpic} \\ 	
	
\includegraphics[width=8.4cm]{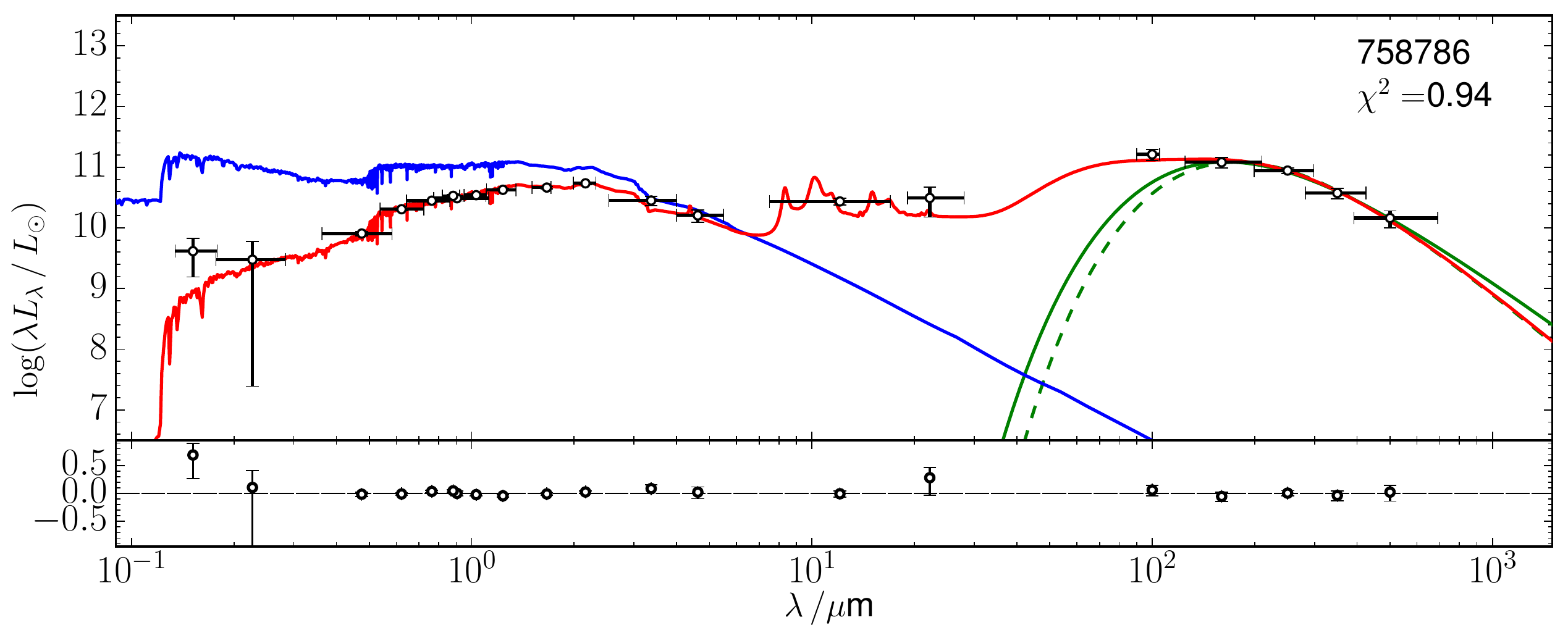} &
\includegraphics[width=5.0cm]{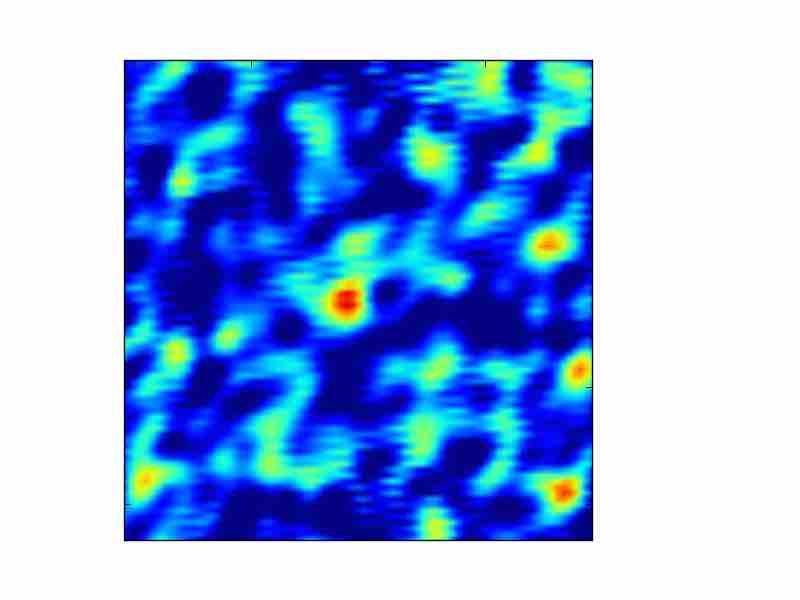} &
\hspace*{-1.2cm}\begin{overpic}[width=3.4cm]{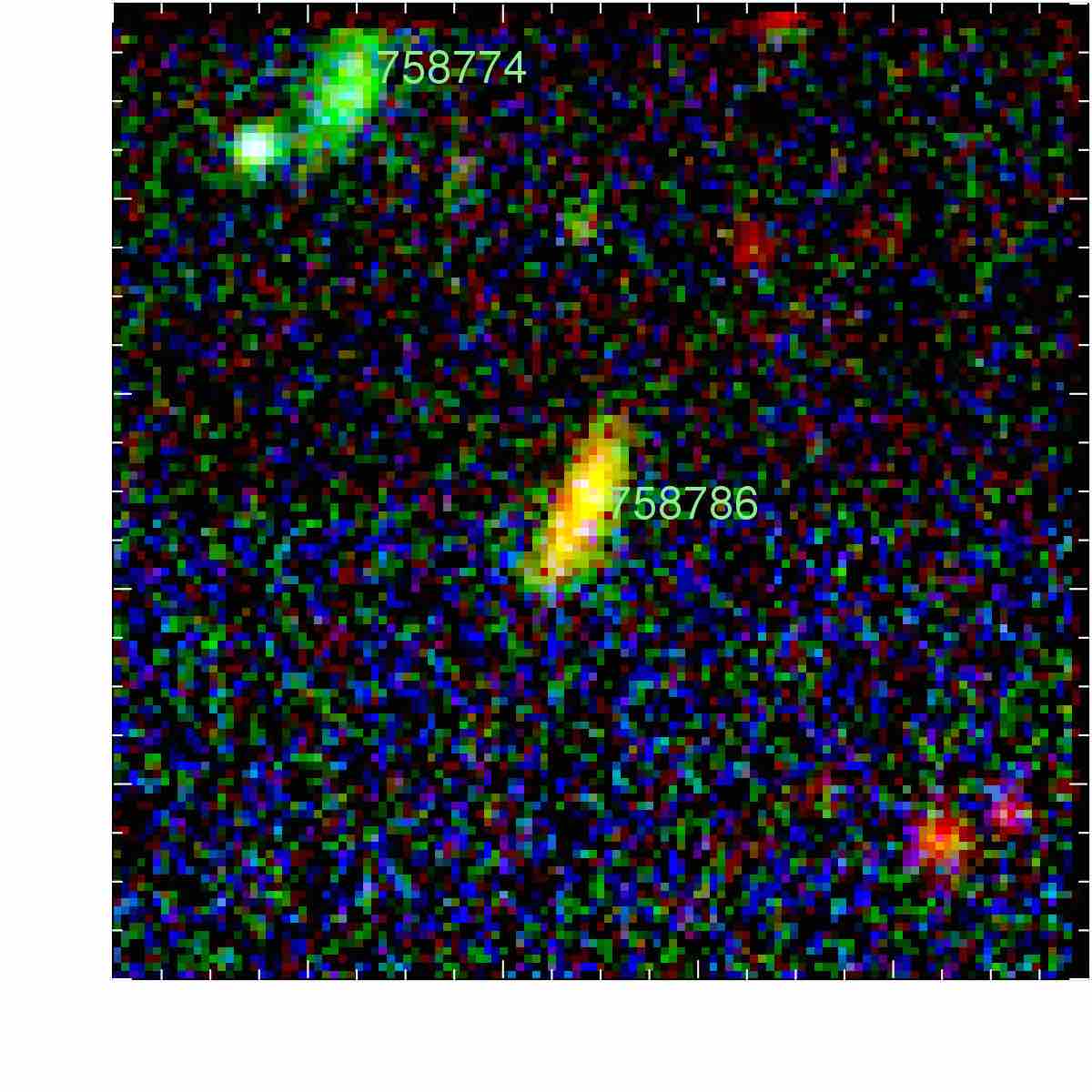} \put (9,85) { \begin{fitbox}{2.25cm}{0.2cm} \color{white}$\bf D$ \end{fitbox}} \end{overpic} \\ 		

\includegraphics[width=8.4cm]{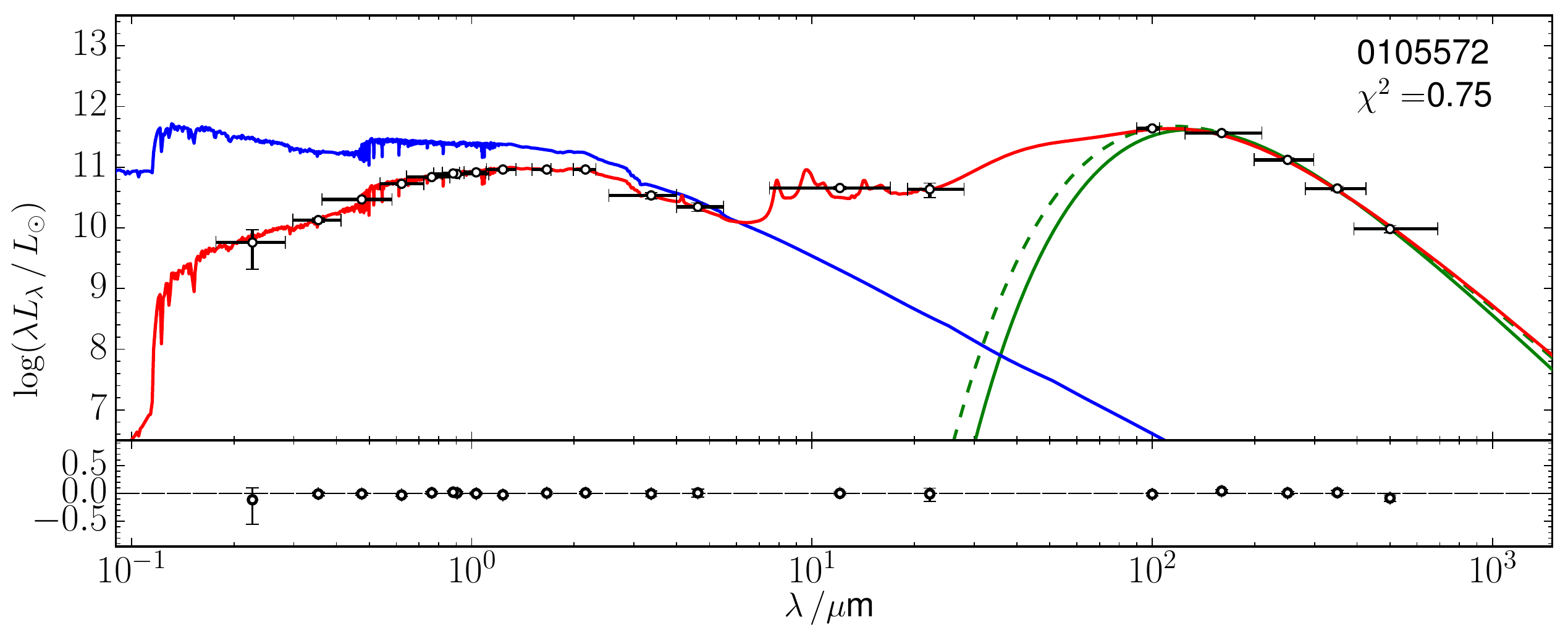} &
\includegraphics[width=5.0cm]{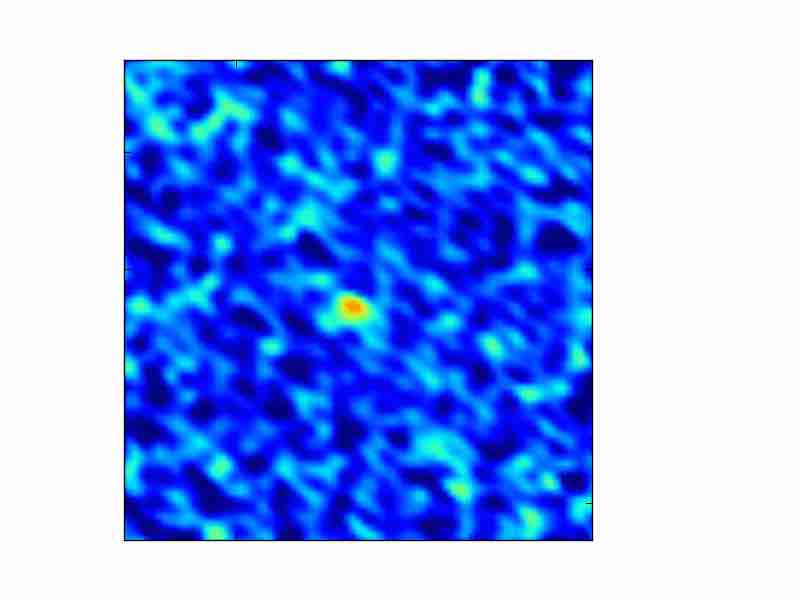} &
\hspace*{-1.2cm}\begin{overpic}[width=3.4cm]{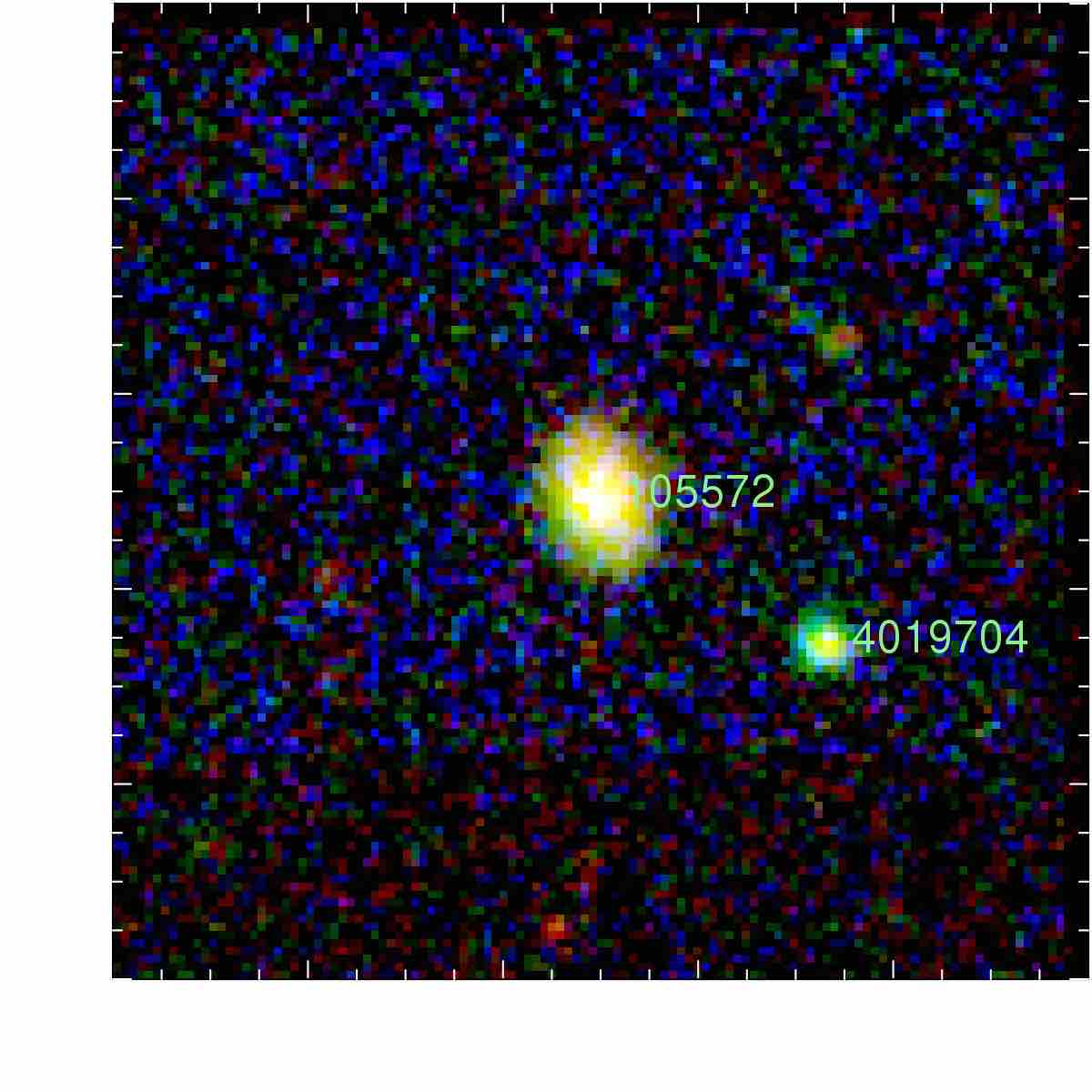} \put (9,85) { \begin{fitbox}{2.25cm}{0.2cm} \color{white}$\bf BD$ \end{fitbox}} \end{overpic} \\ 	

\includegraphics[width=8.4cm]{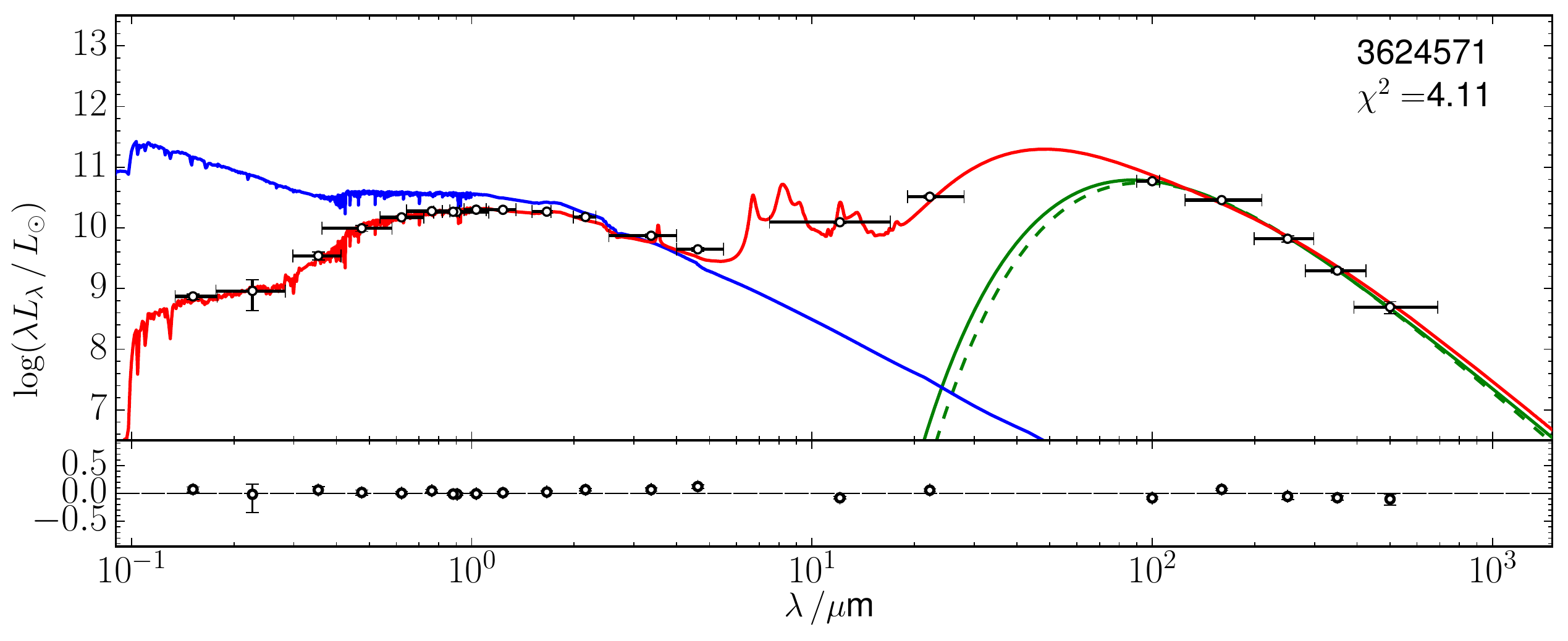} &
\includegraphics[width=5.0cm]{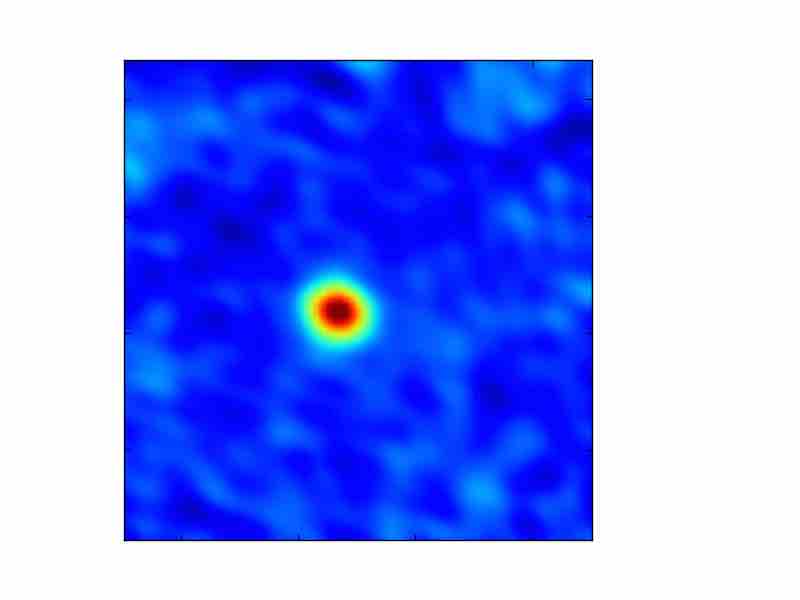} &
\hspace*{-1.2cm}\begin{overpic}[width=3.4cm]{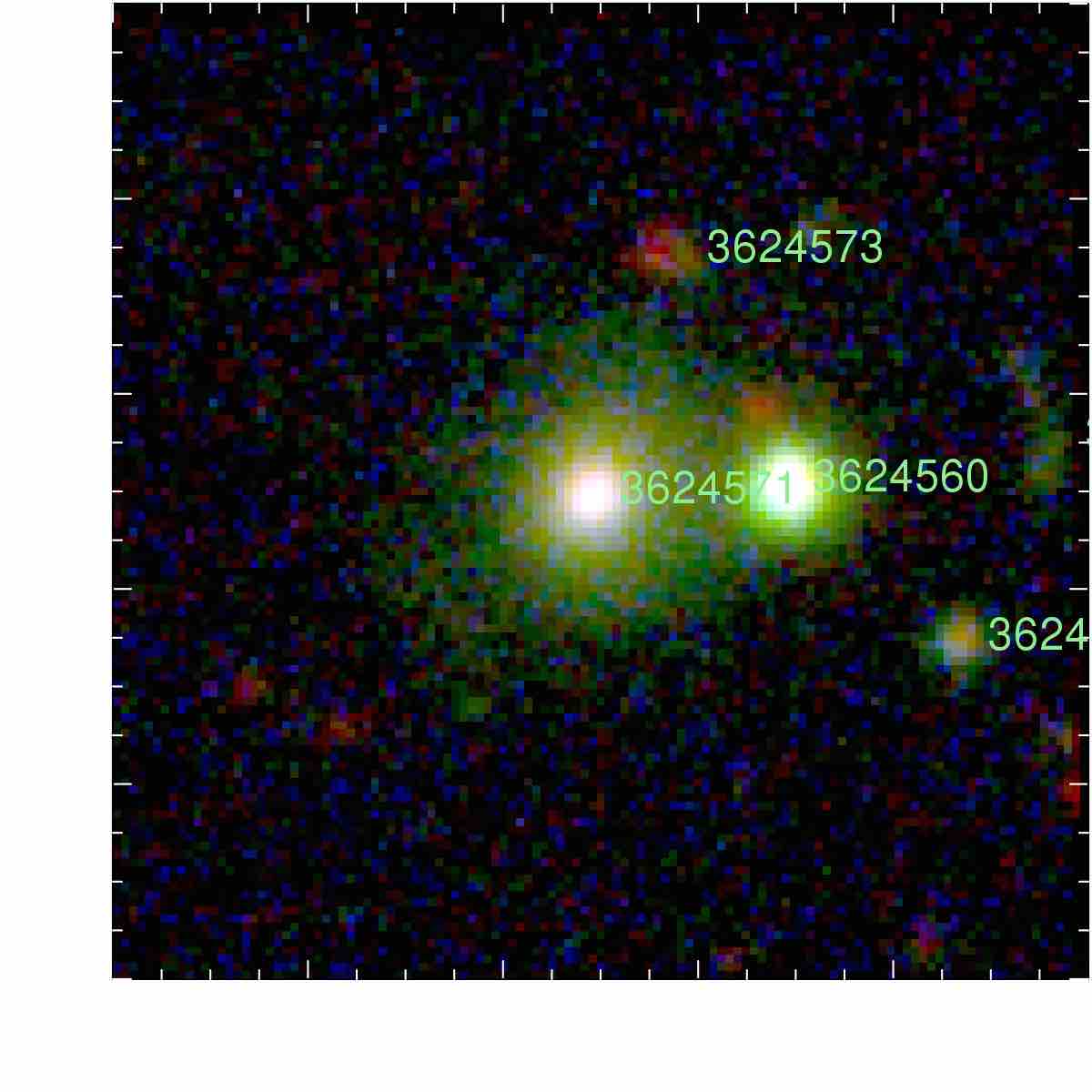} \put (9,85) { \begin{fitbox}{2.25cm}{0.2cm} \color{white}$\bf BC$ \end{fitbox}} \end{overpic} \\ 

\end{array}
$
{\textbf{Figure~\ref{pdrdiaglit}.} continued}

\end{figure*}


\begin{figure*}
$
\begin{array}{ccc}
\includegraphics[width=8.4cm]{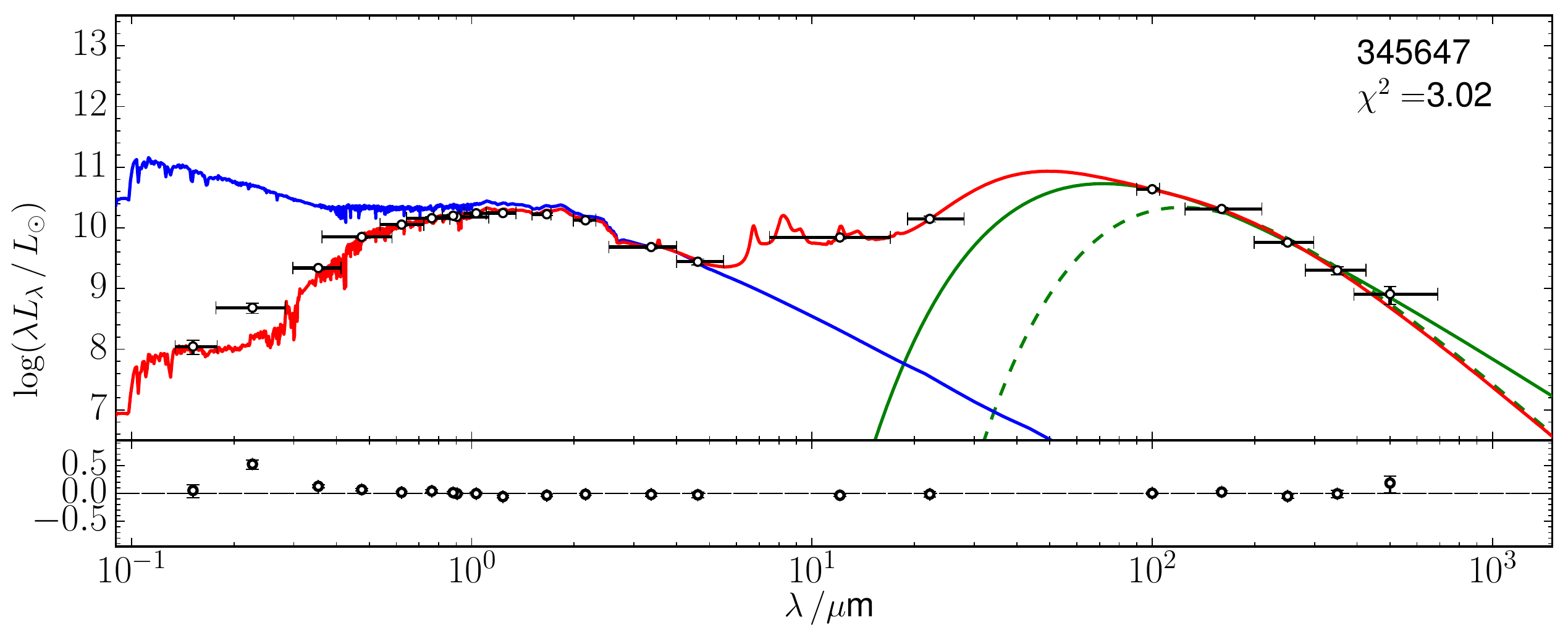} &
\includegraphics[width=5.0cm]{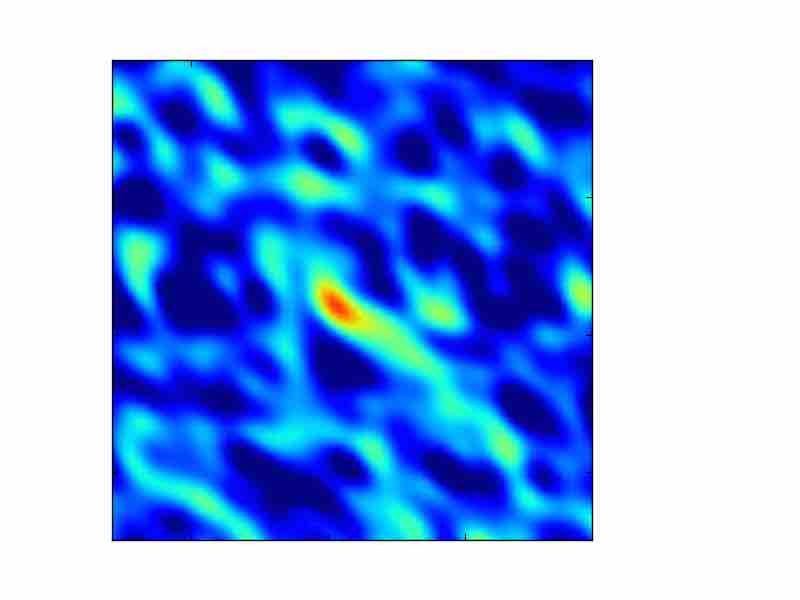} &
\hspace*{-1.2cm}\begin{overpic}[width=3.4cm]{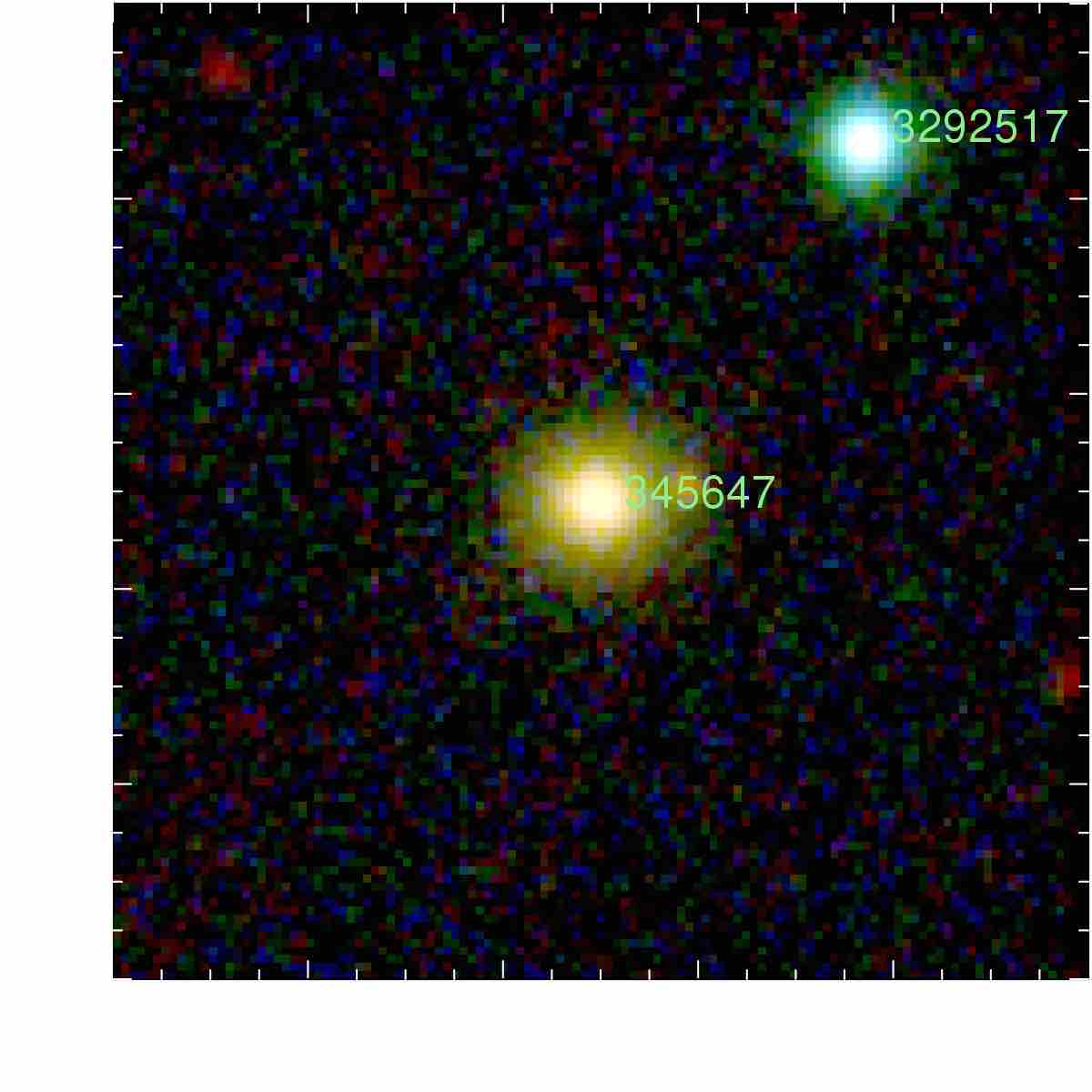} \put (9,85) { \begin{fitbox}{2.25cm}{0.2cm} \color{white}$\bf BDC$ \end{fitbox}} \end{overpic} \\ 	
	
\includegraphics[width=8.4cm]{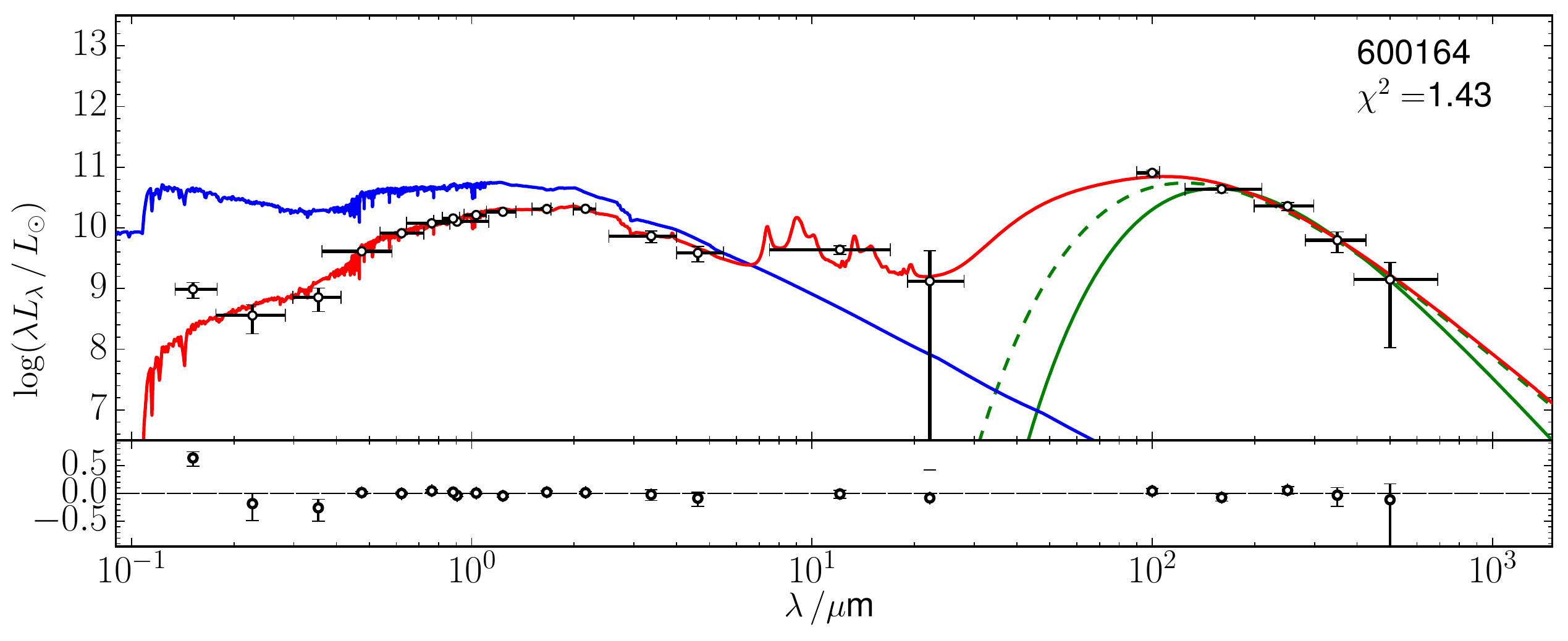} &
\includegraphics[width=5.0cm]{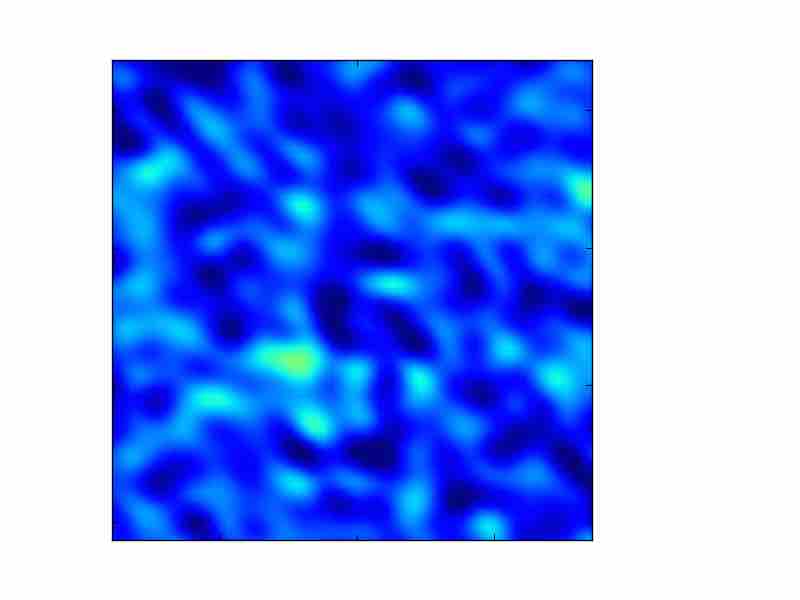} &
\hspace*{-1.2cm}\begin{overpic}[width=3.4cm]{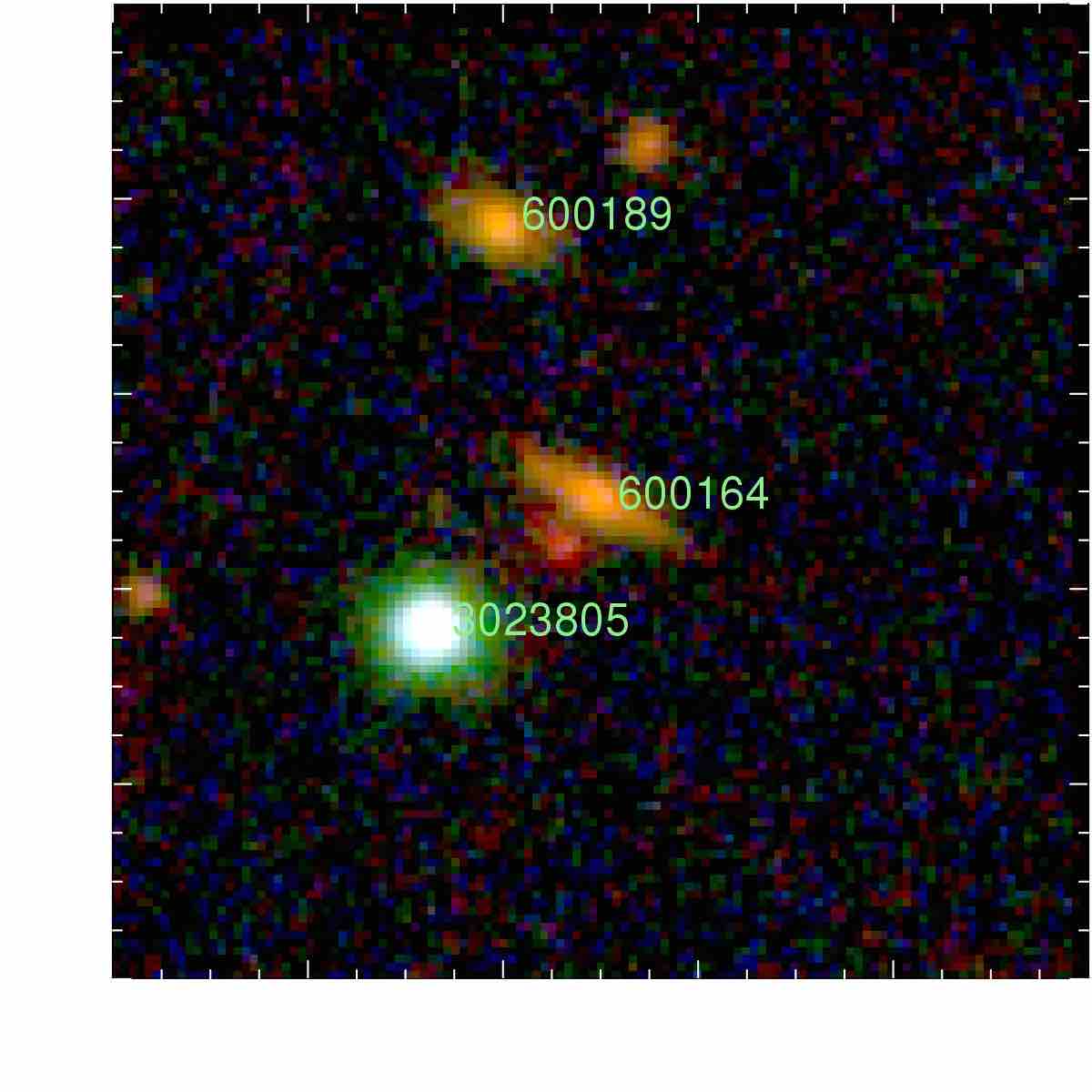} \put (9,85) { \begin{fitbox}{2.25cm}{0.2cm} \color{white}$\bf DBC$ \end{fitbox}} \end{overpic} \\ 	

\includegraphics[width=8.4cm]{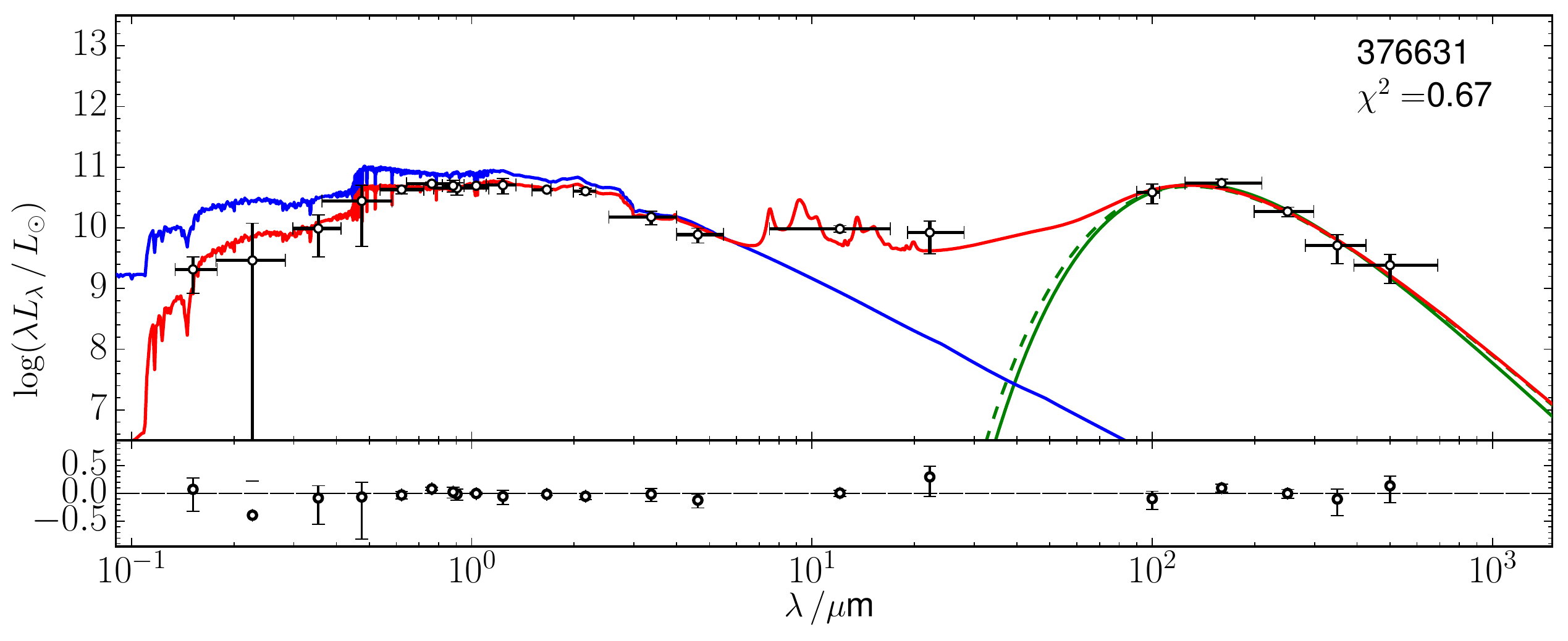} &
\includegraphics[width=5.0cm]{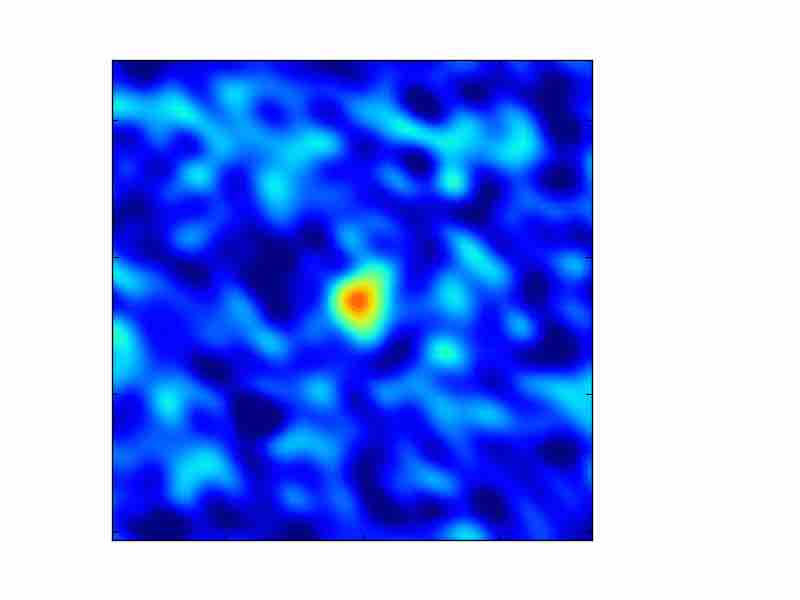} &
\hspace*{-1.2cm}\begin{overpic}[width=3.4cm]{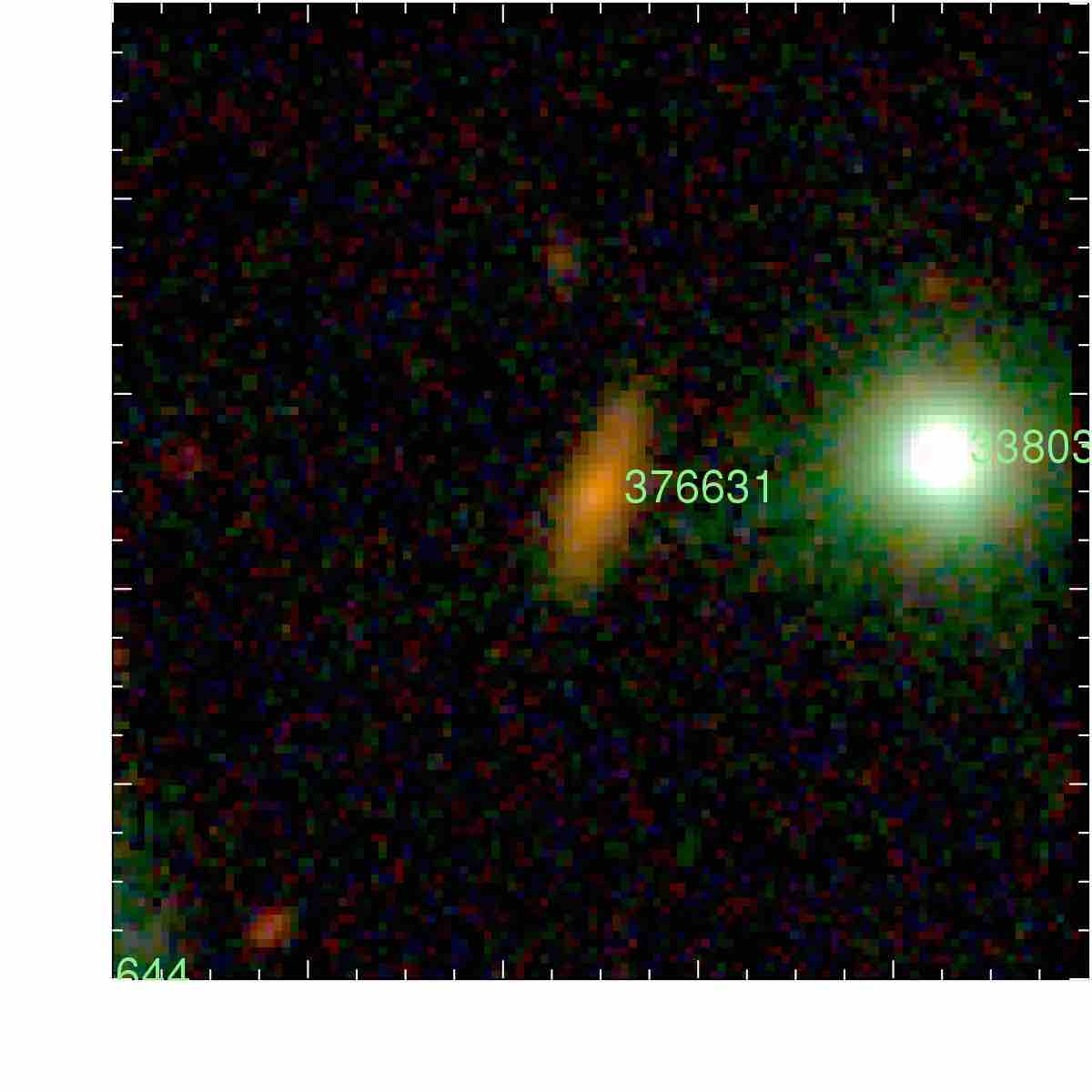} \put (9,85) { \begin{fitbox}{2.25cm}{0.2cm} \color{white}$\bf DB$ \end{fitbox}} \end{overpic} \\ 	

\includegraphics[width=8.4cm]{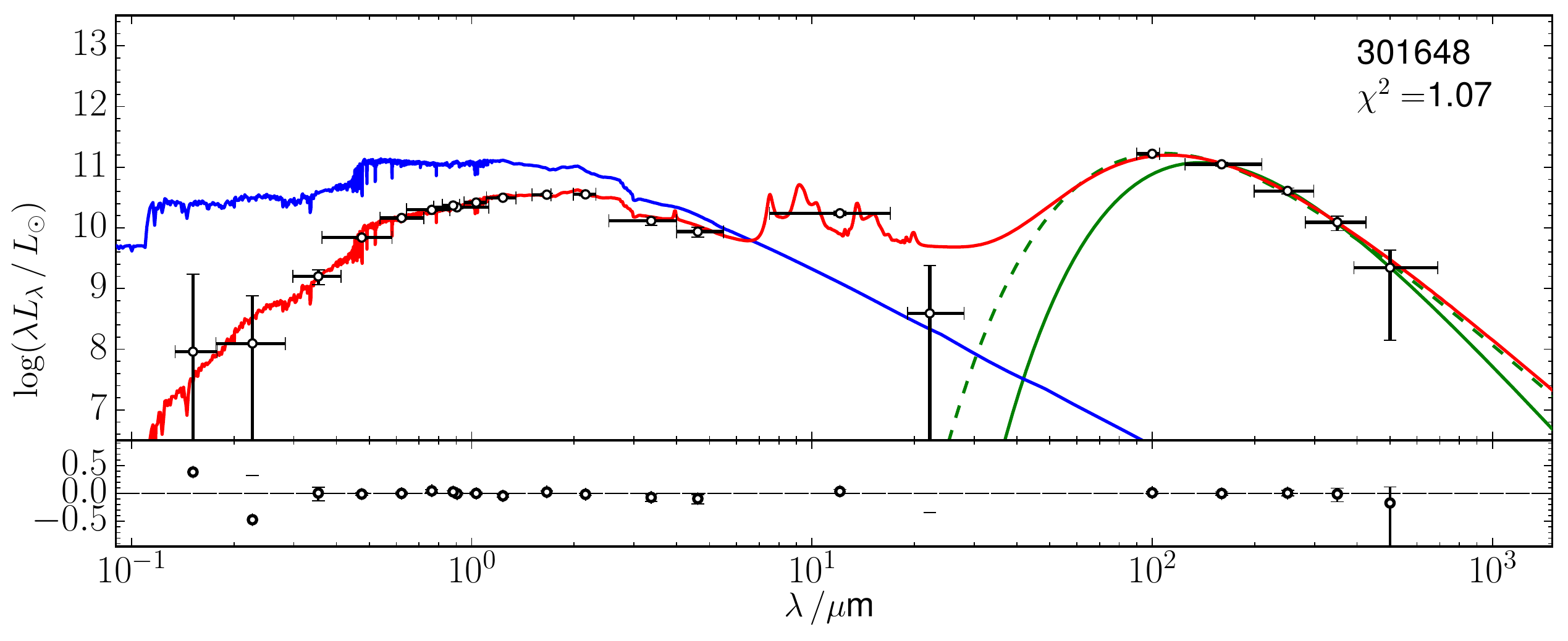} &
\includegraphics[width=5.0cm]{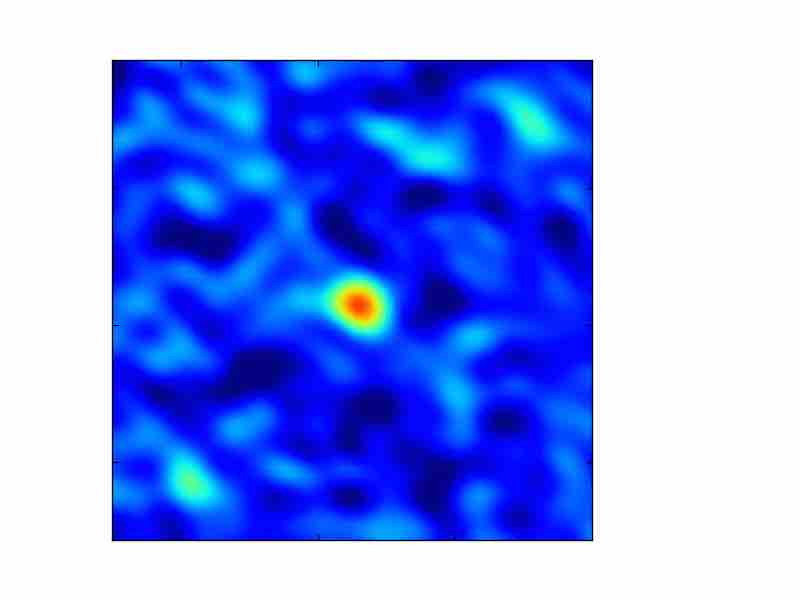} &
\hspace*{-1.2cm}\begin{overpic}[width=3.4cm]{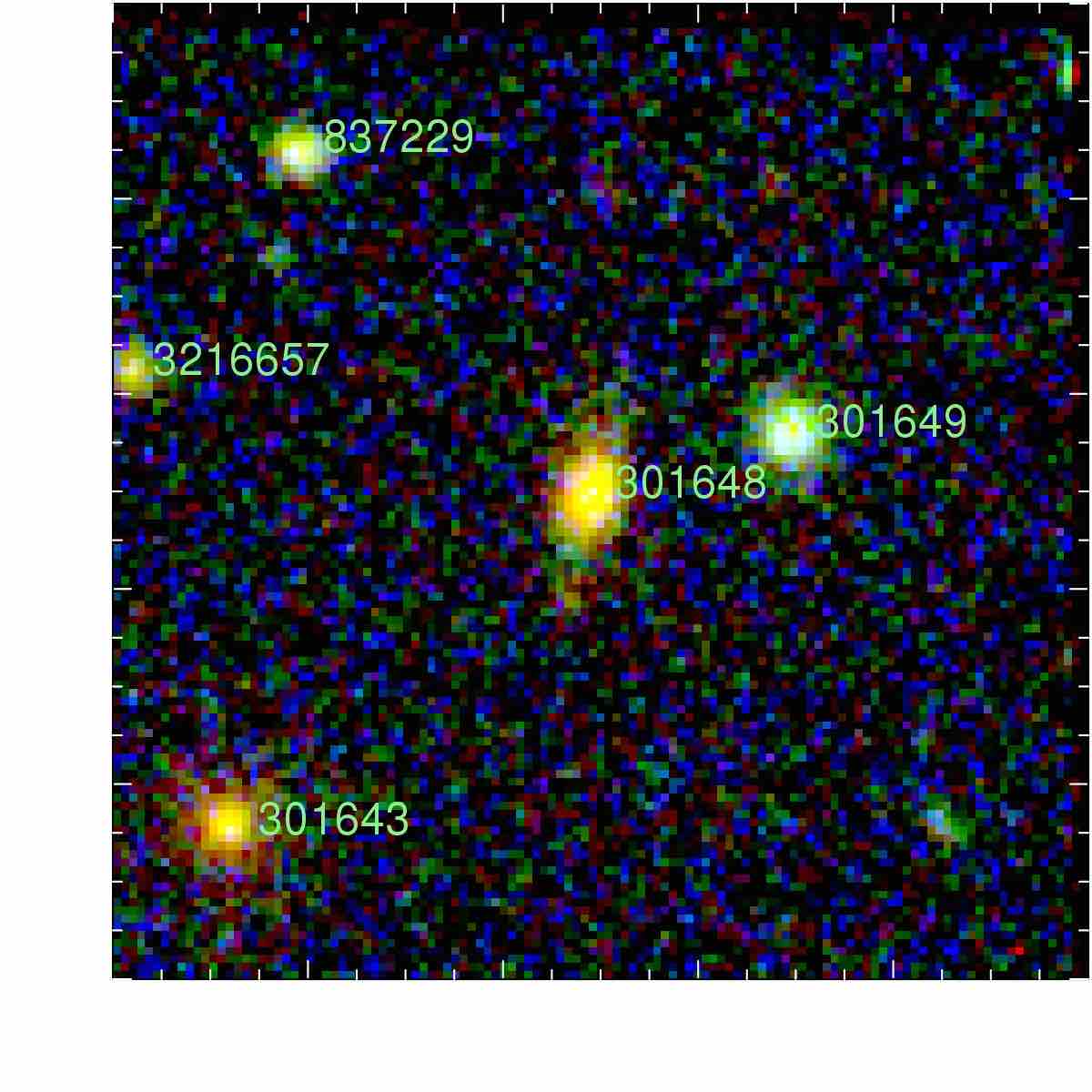} \put (9,85) { \begin{fitbox}{2.25cm}{0.2cm} \color{white}$\bf BD$ \end{fitbox}} \end{overpic} \\ 	

\includegraphics[width=8.4cm]{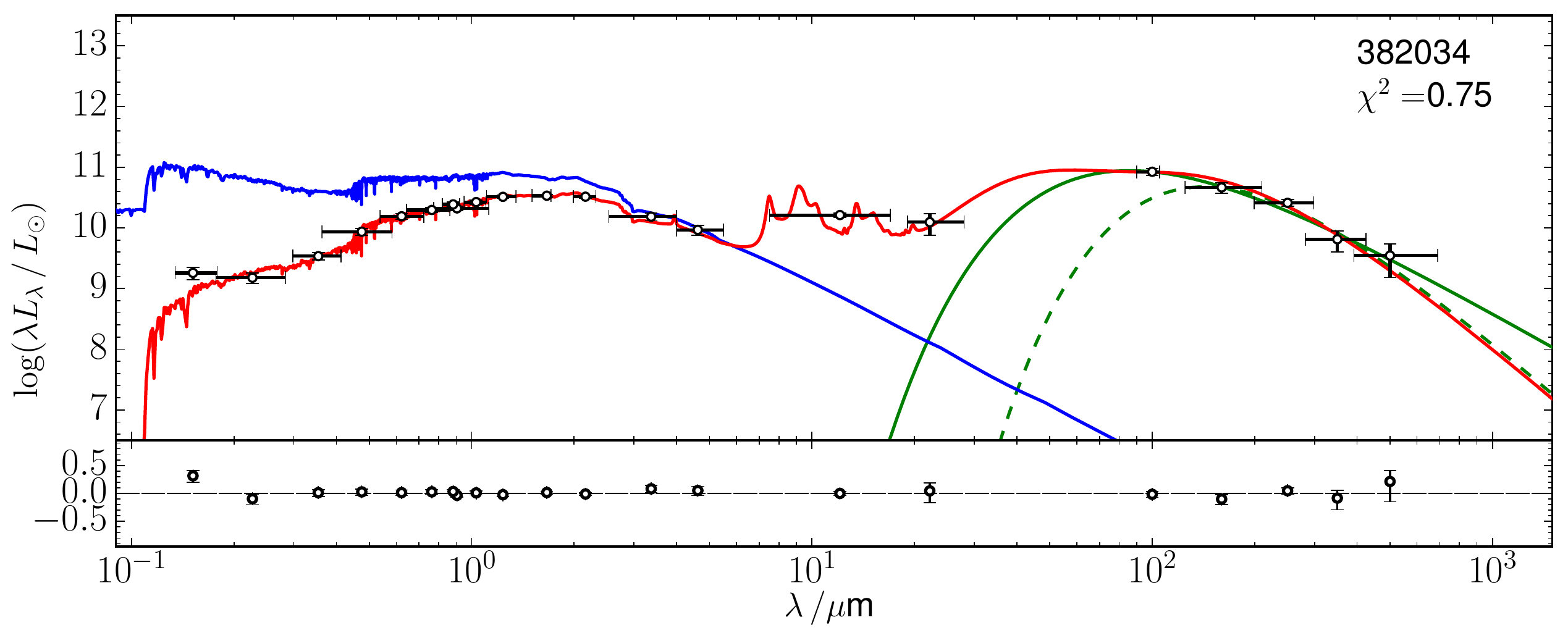} &
\includegraphics[width=5.0cm]{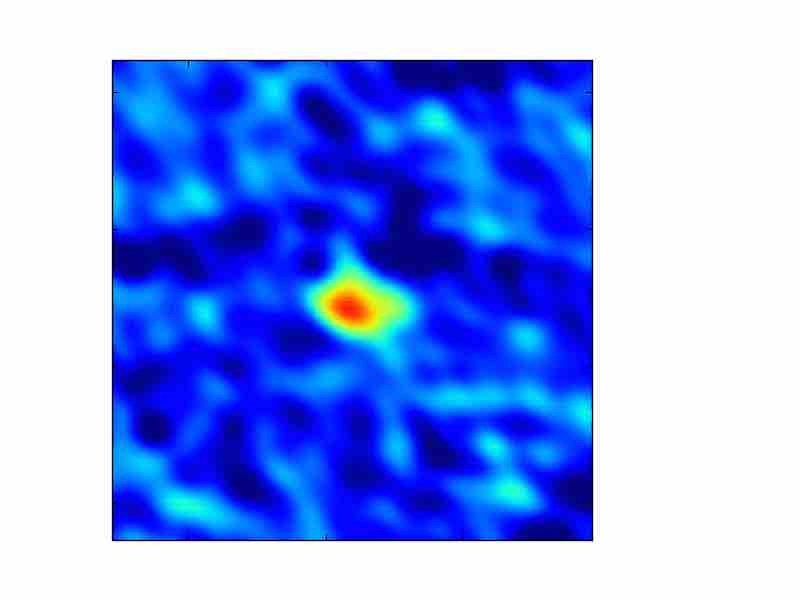} &
\hspace*{-1.2cm}\begin{overpic}[width=3.4cm]{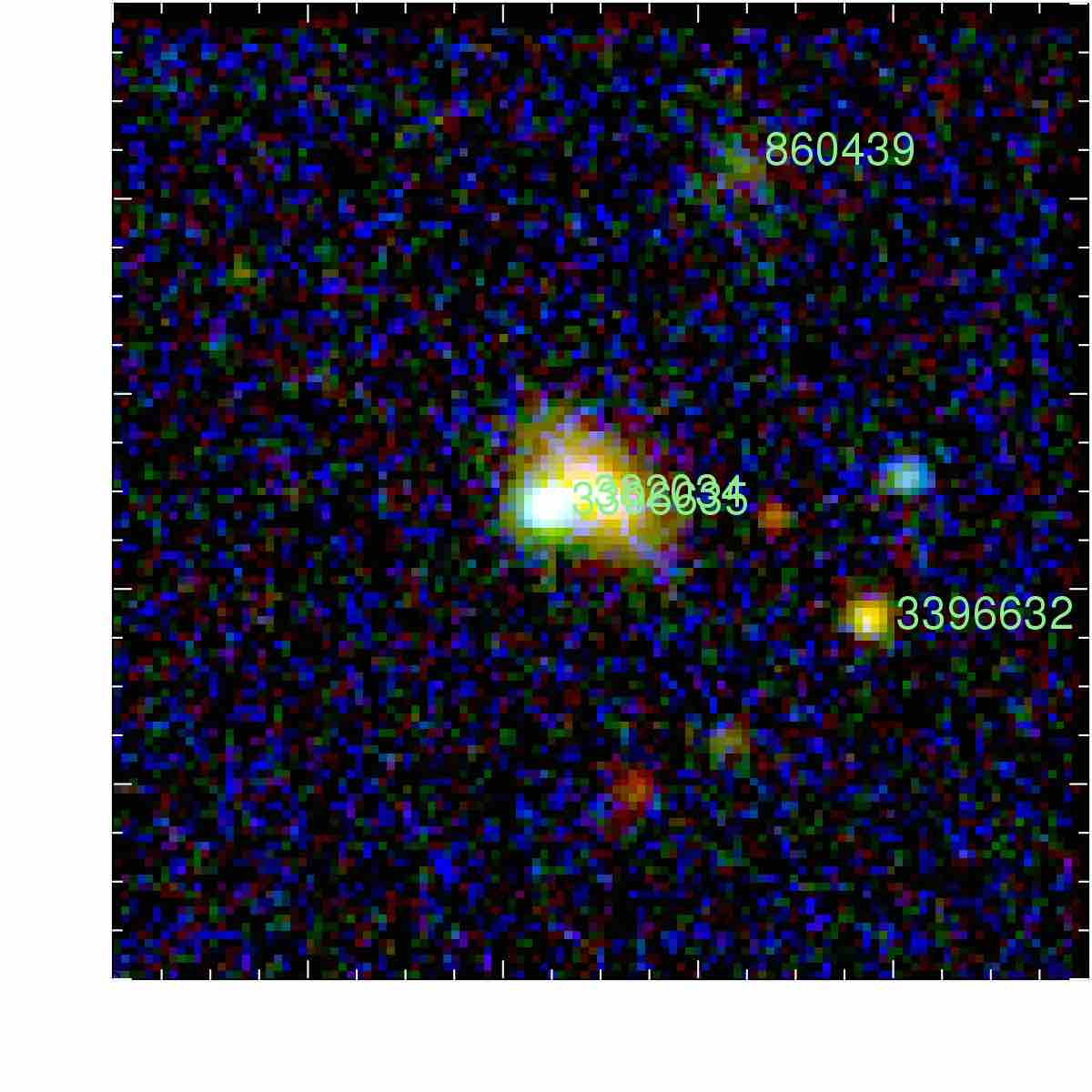} \put (9,85) { \begin{fitbox}{2.25cm}{0.2cm} \color{white}$\bf M$ \end{fitbox}} \end{overpic} \\ 	


\end{array}
$
\caption{Same as Figure~\ref{pdrdiaglit} but for spectrally undetected sources. In these cases the CO cubes are collapsed blindly, between $\pm 250$ km s$^{-1}$ from expected observed frequency.}\label{Undetected}

\end{figure*}


\begin{figure*}
$
\begin{array}{ccc}
\includegraphics[width=8.4cm]{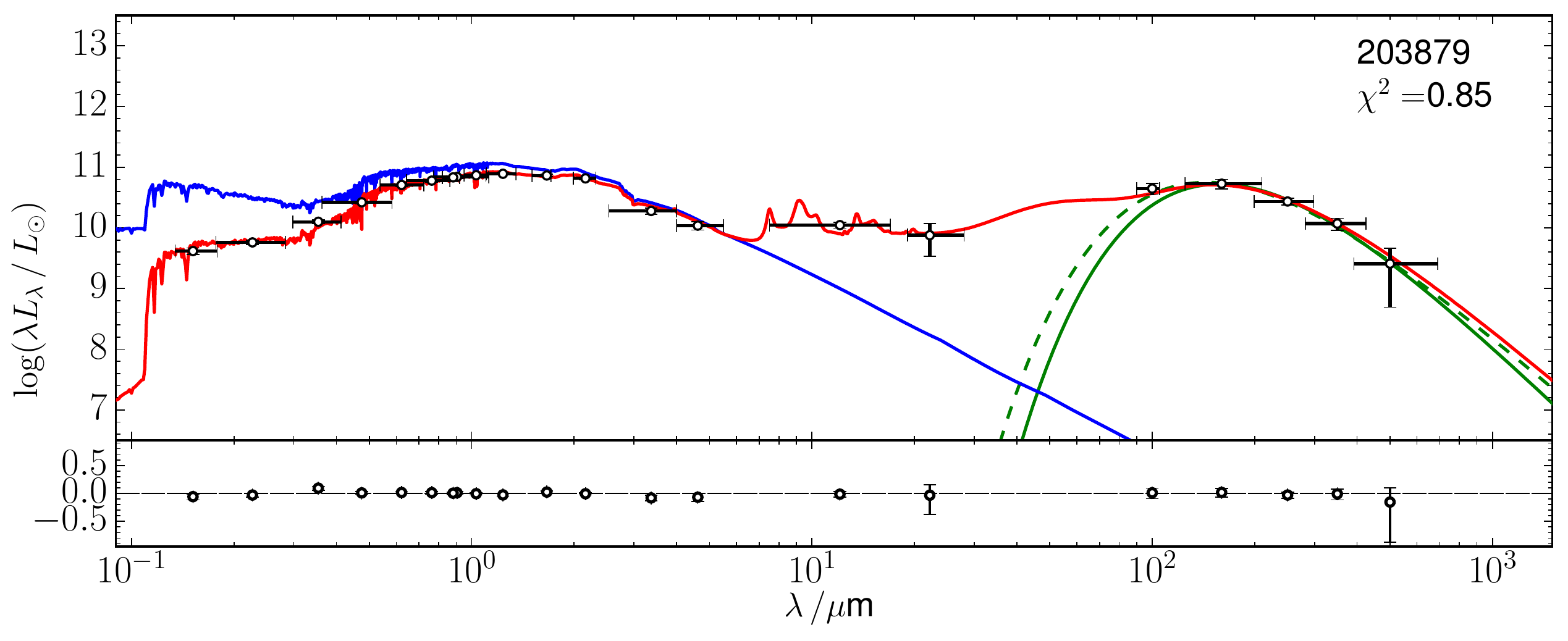} &
\includegraphics[width=5.0cm]{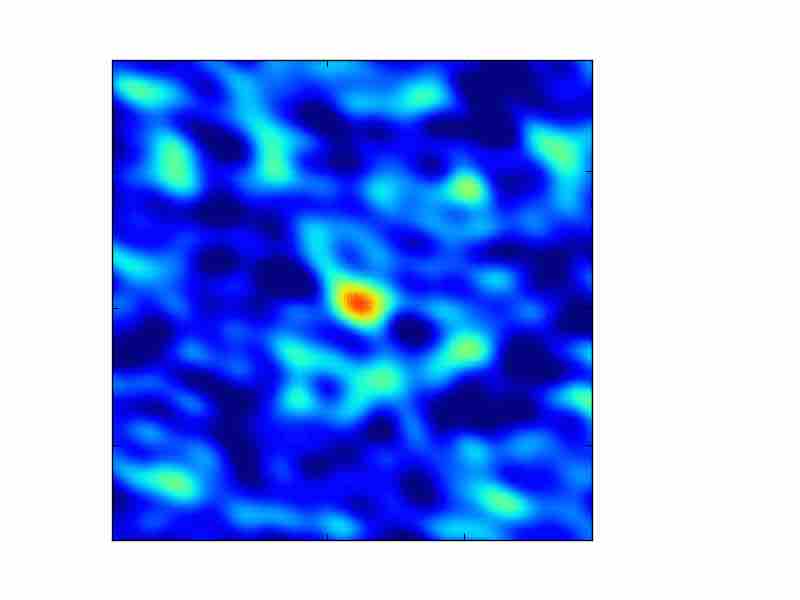} &
\hspace*{-1.2cm}\begin{overpic}[width=3.4cm]{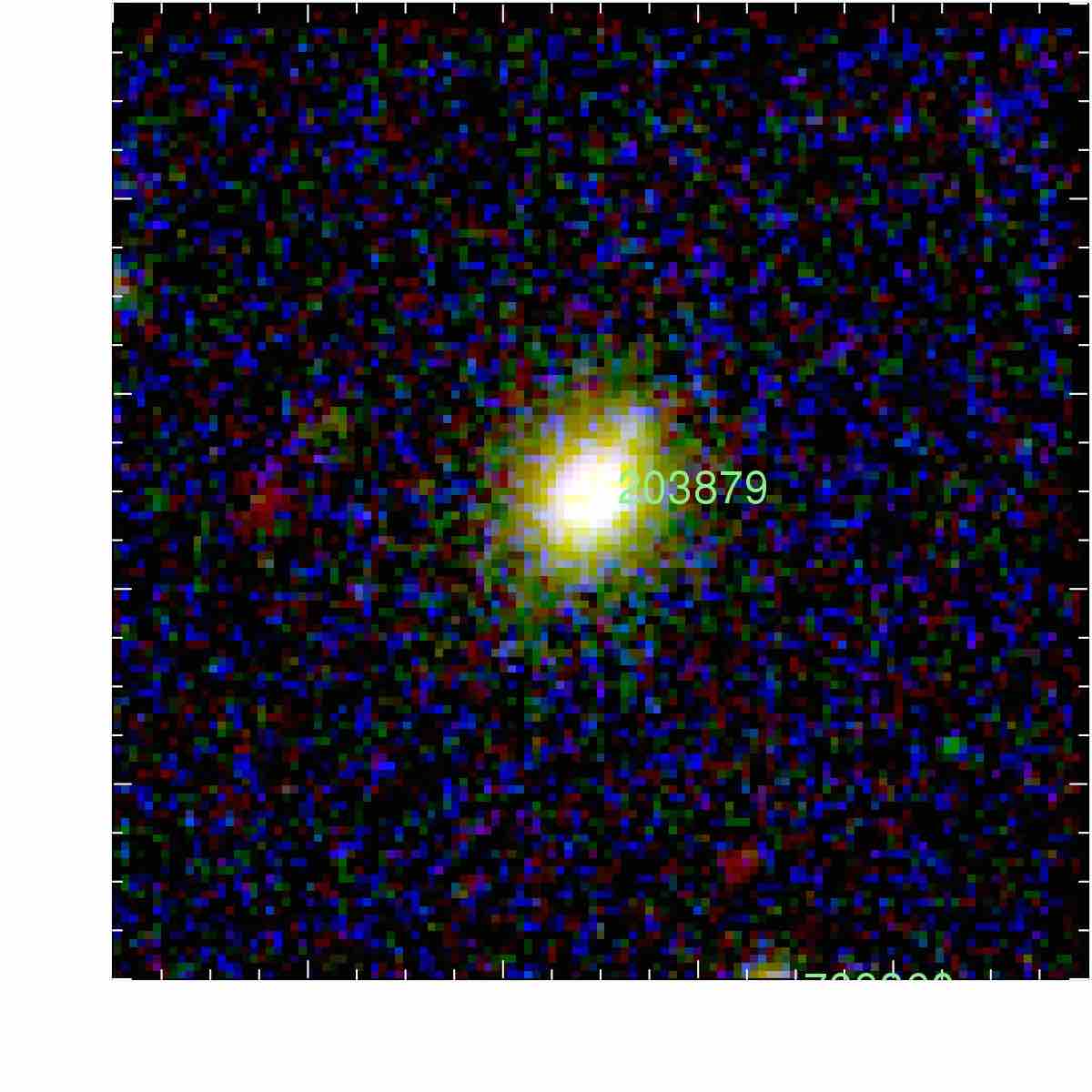} \put (9,85) { \begin{fitbox}{2.25cm}{0.2cm} \color{white}$\bf BD$ \end{fitbox}} \end{overpic} \\ 	
	
\includegraphics[width=8.4cm]{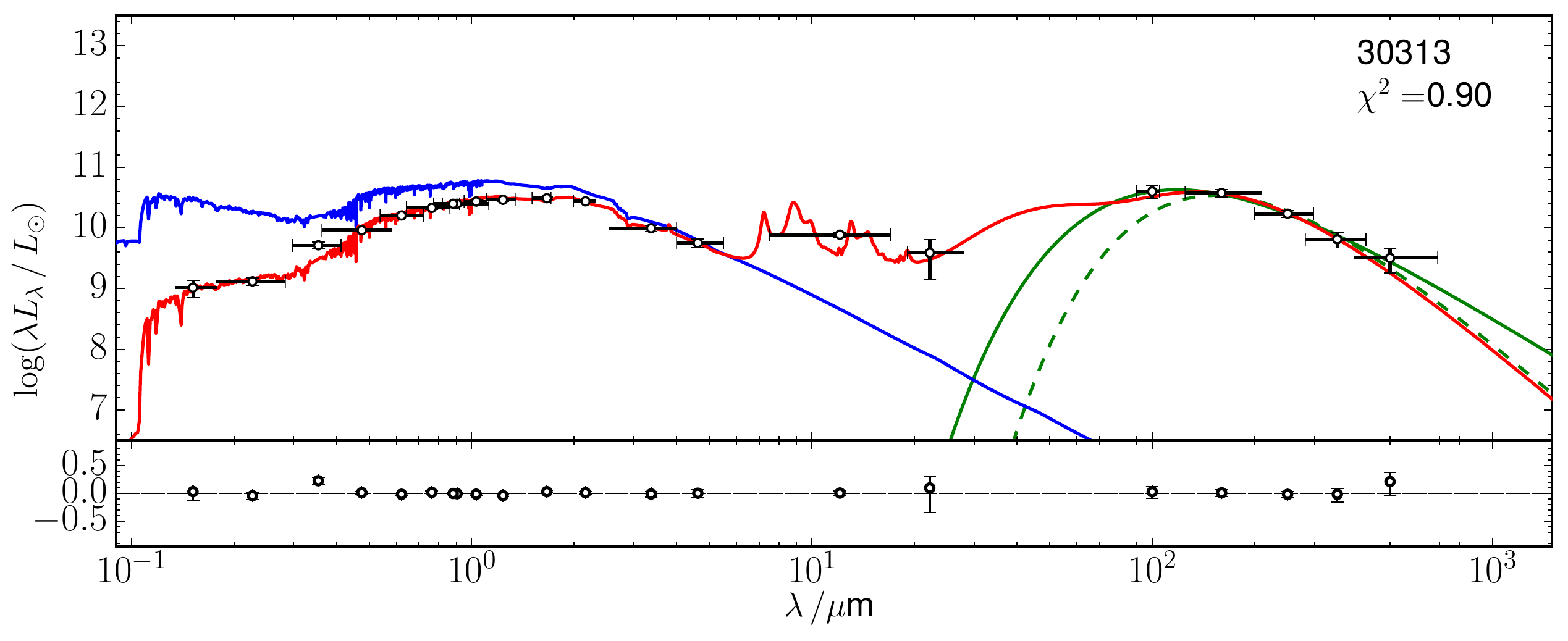} &
\includegraphics[width=5.0cm]{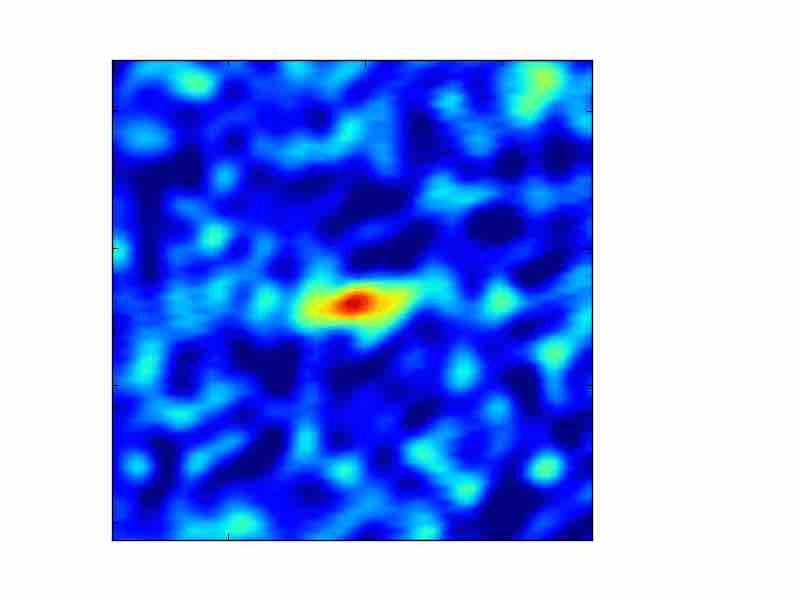} &
\hspace*{-1.2cm}\begin{overpic}[width=3.4cm]{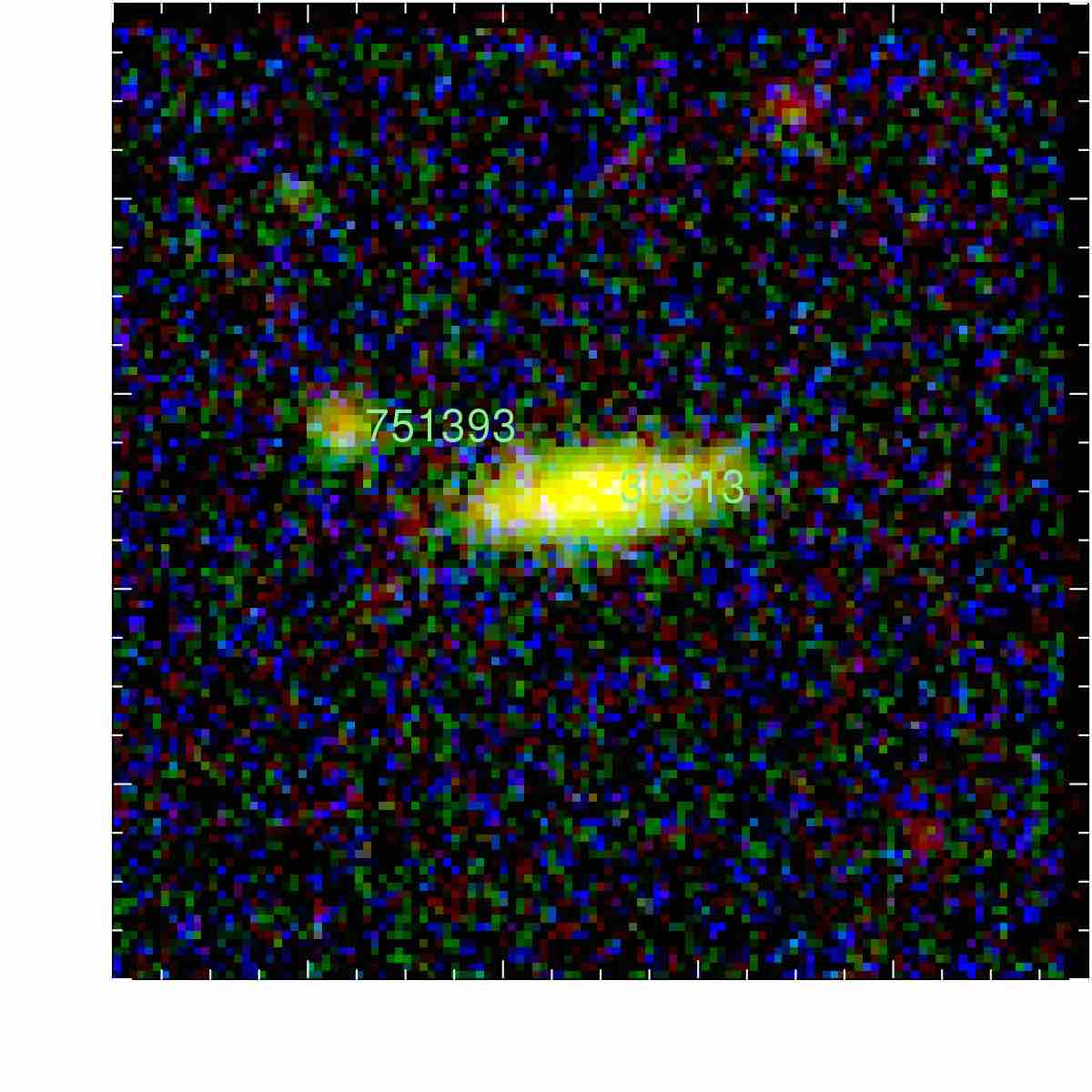} \put (9,85) { \begin{fitbox}{2.25cm}{0.2cm} \color{white}$\bf DBC$ \end{fitbox}} \end{overpic} \\ 	

\includegraphics[width=8.4cm]{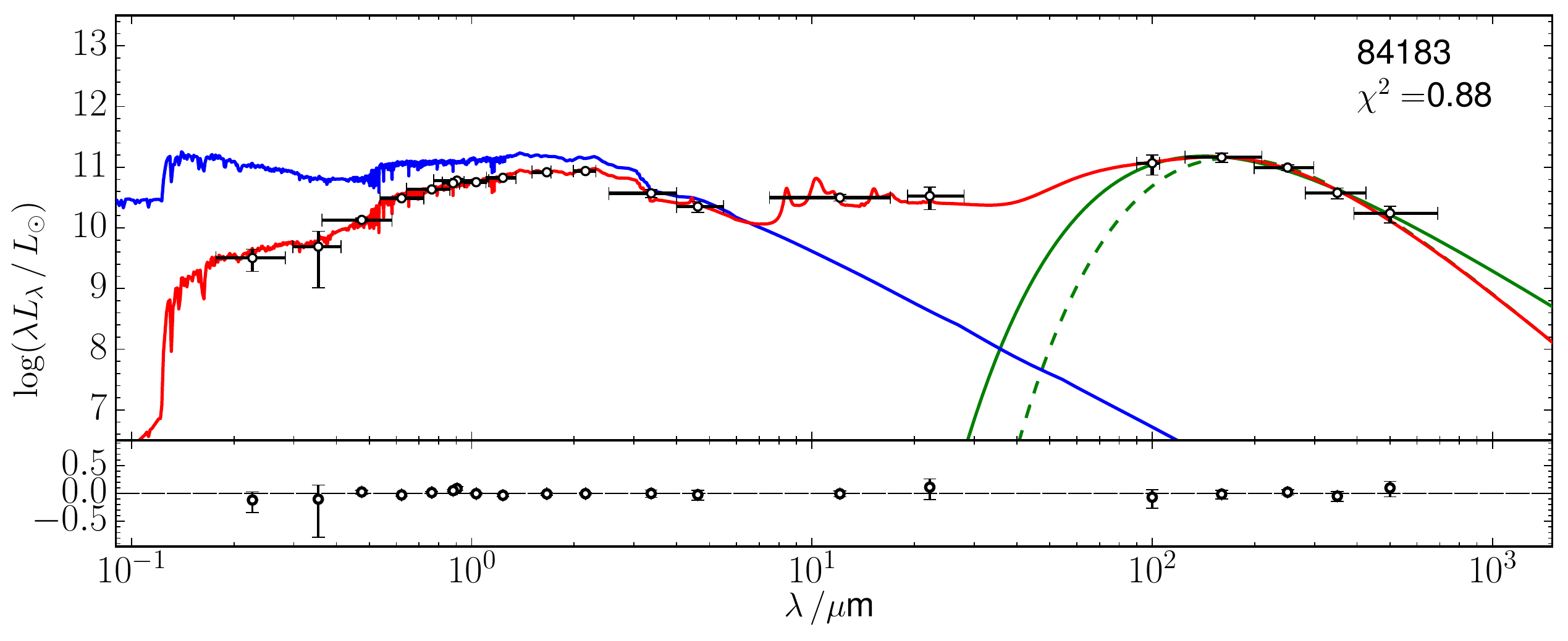} &
\includegraphics[width=5.0cm]{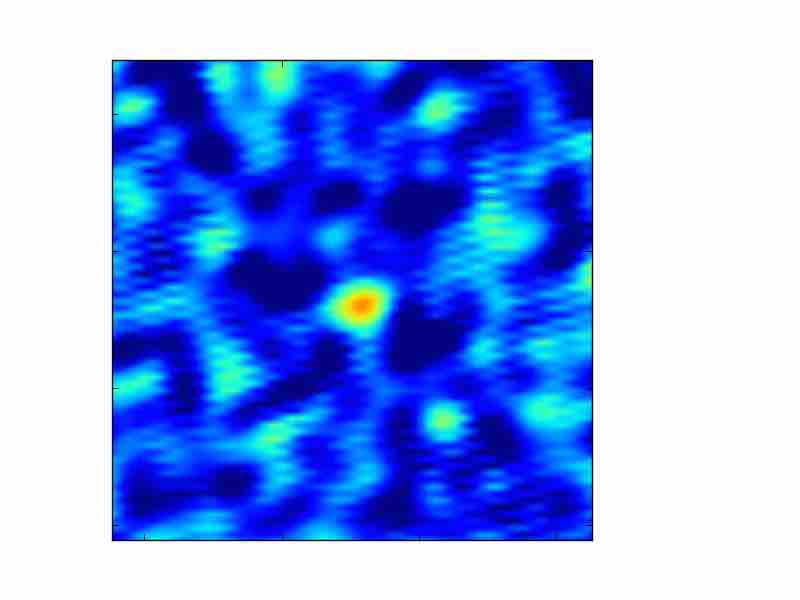} &
\hspace*{-1.2cm}\begin{overpic}[width=3.4cm]{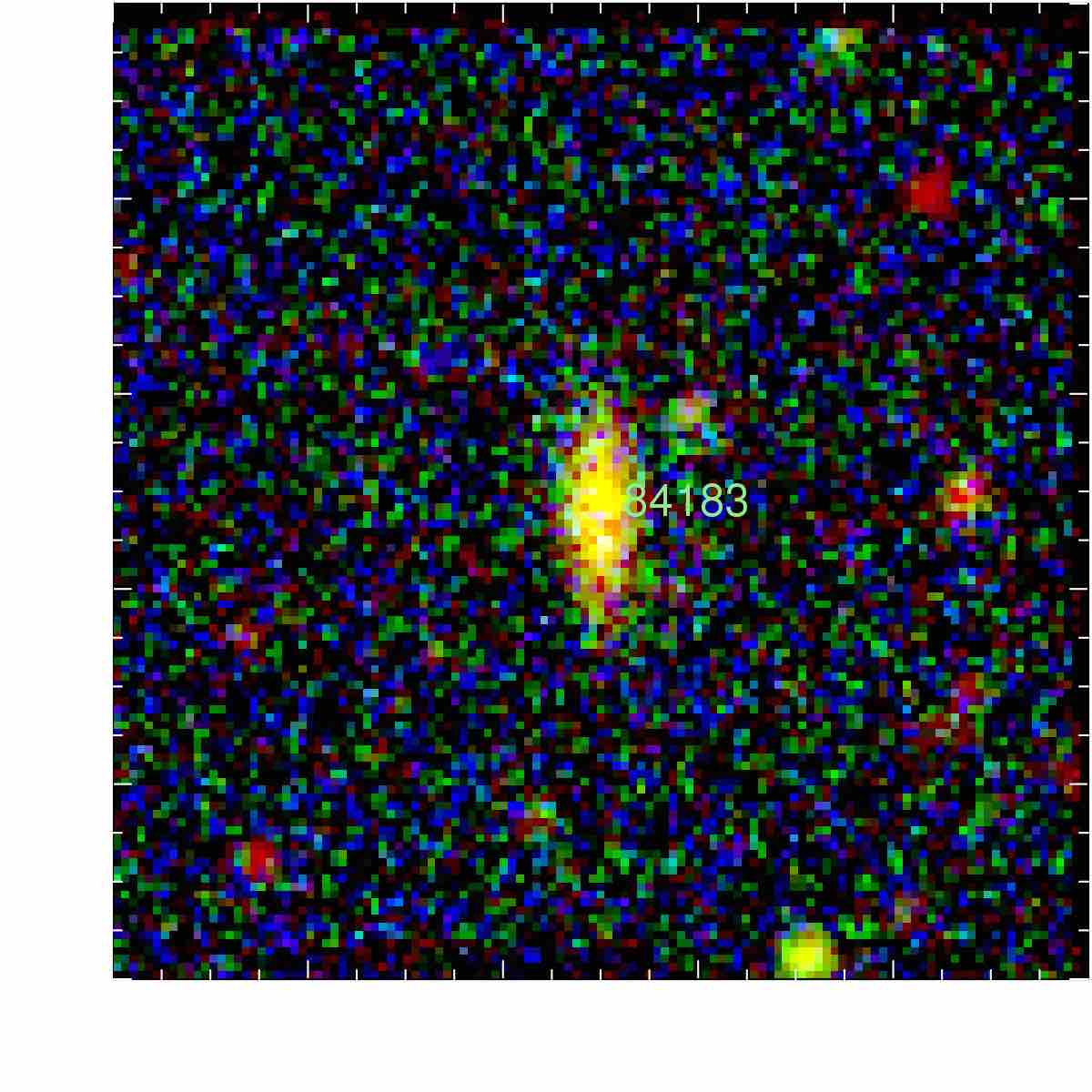} \put (9,85) { \begin{fitbox}{2.25cm}{0.2cm} \color{white}$\bf DB$ \end{fitbox}} \end{overpic} \\ 	

\includegraphics[width=8.4cm]{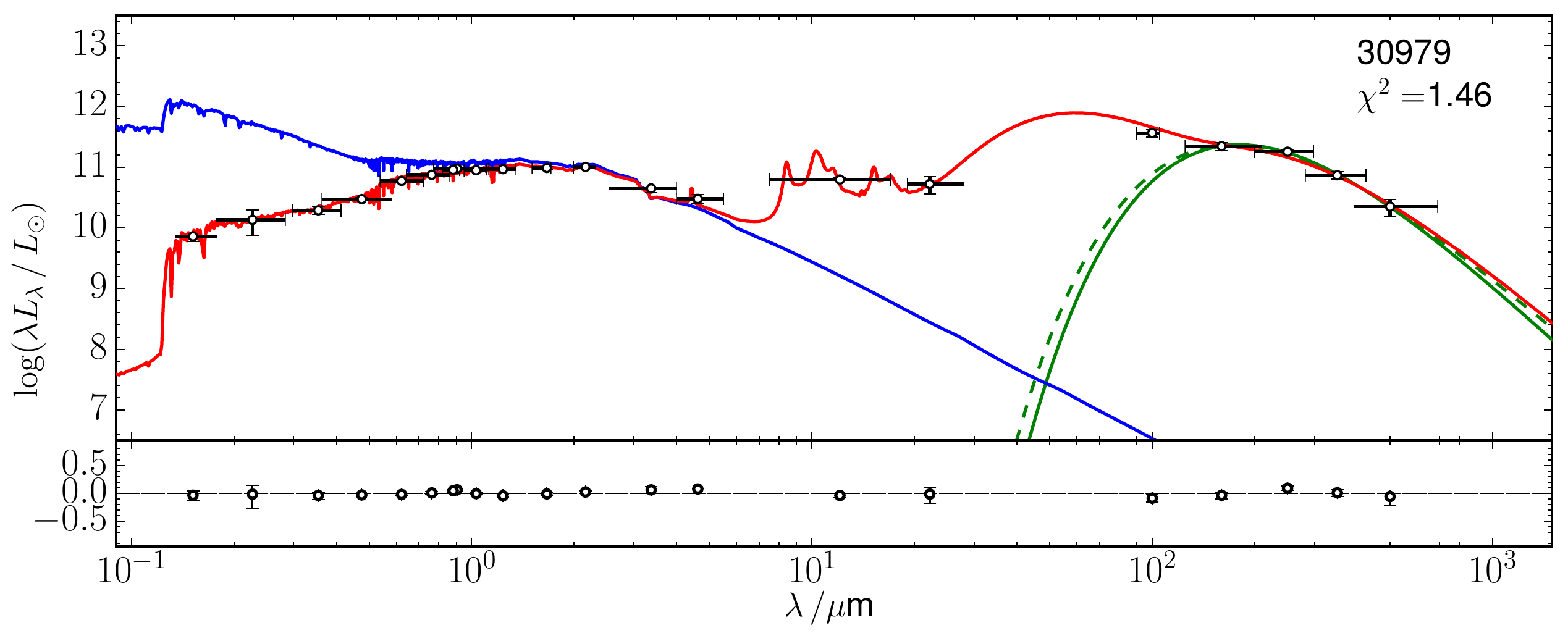} &
\includegraphics[width=5.0cm]{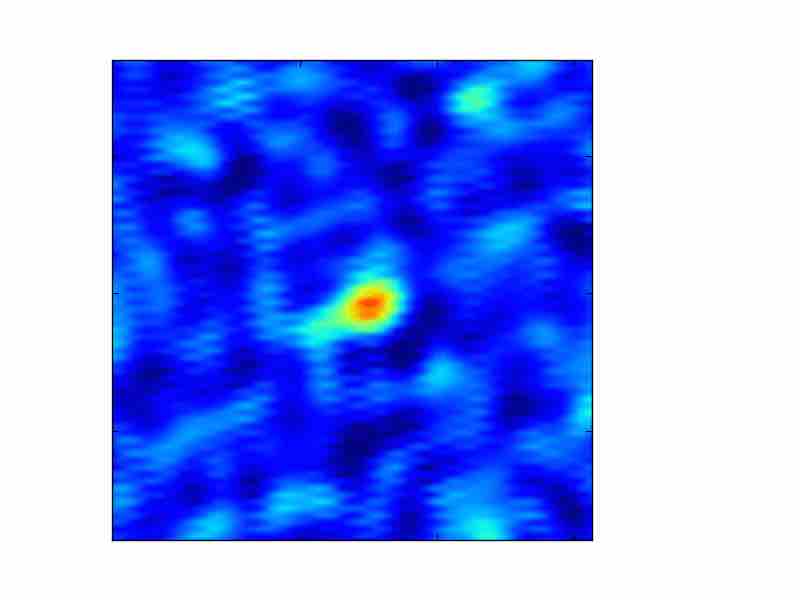} &
\hspace*{-1.2cm}\begin{overpic}[width=3.4cm]{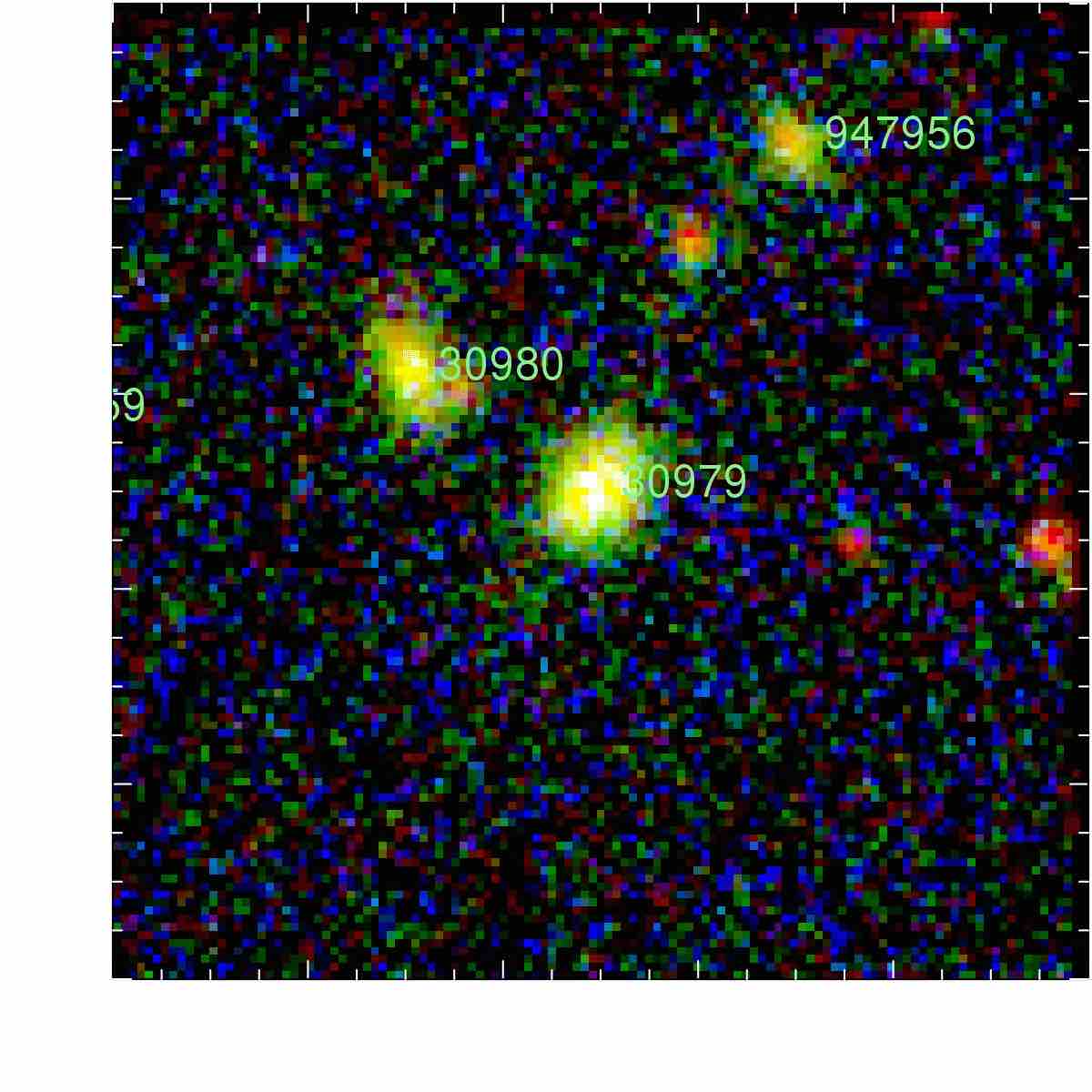} \put (9,85) { \begin{fitbox}{2.25cm}{0.2cm} \color{white}$\bf BC$ \end{fitbox}} \end{overpic} \\ 	

\includegraphics[width=8.4cm]{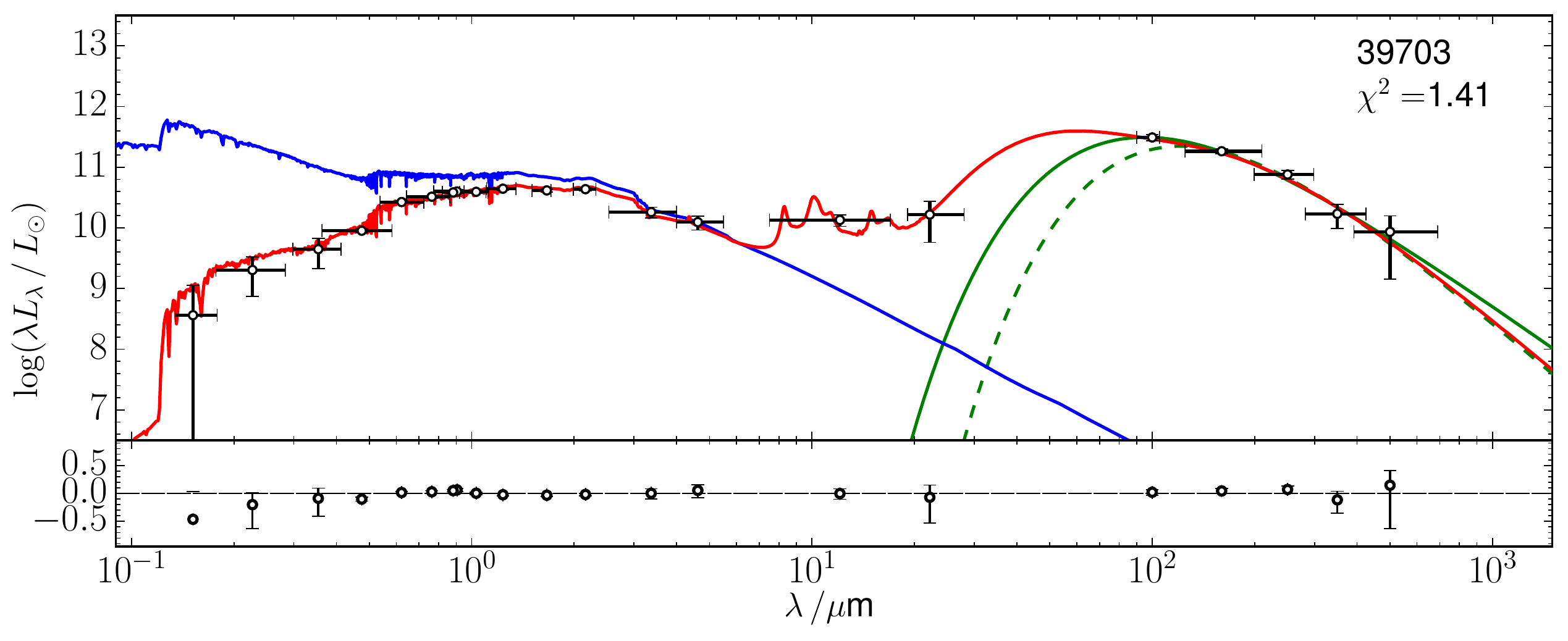} &
\includegraphics[width=5.0cm]{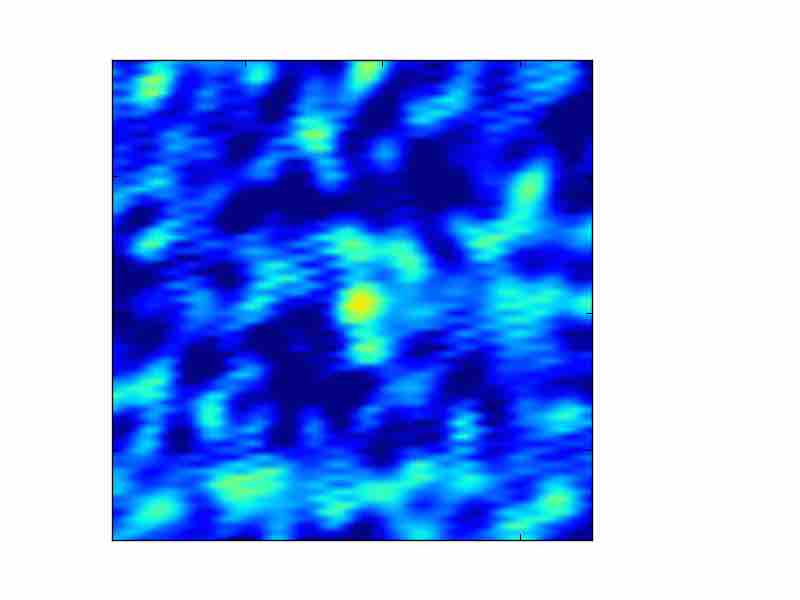} &
\hspace*{-1.2cm}\begin{overpic}[width=3.4cm]{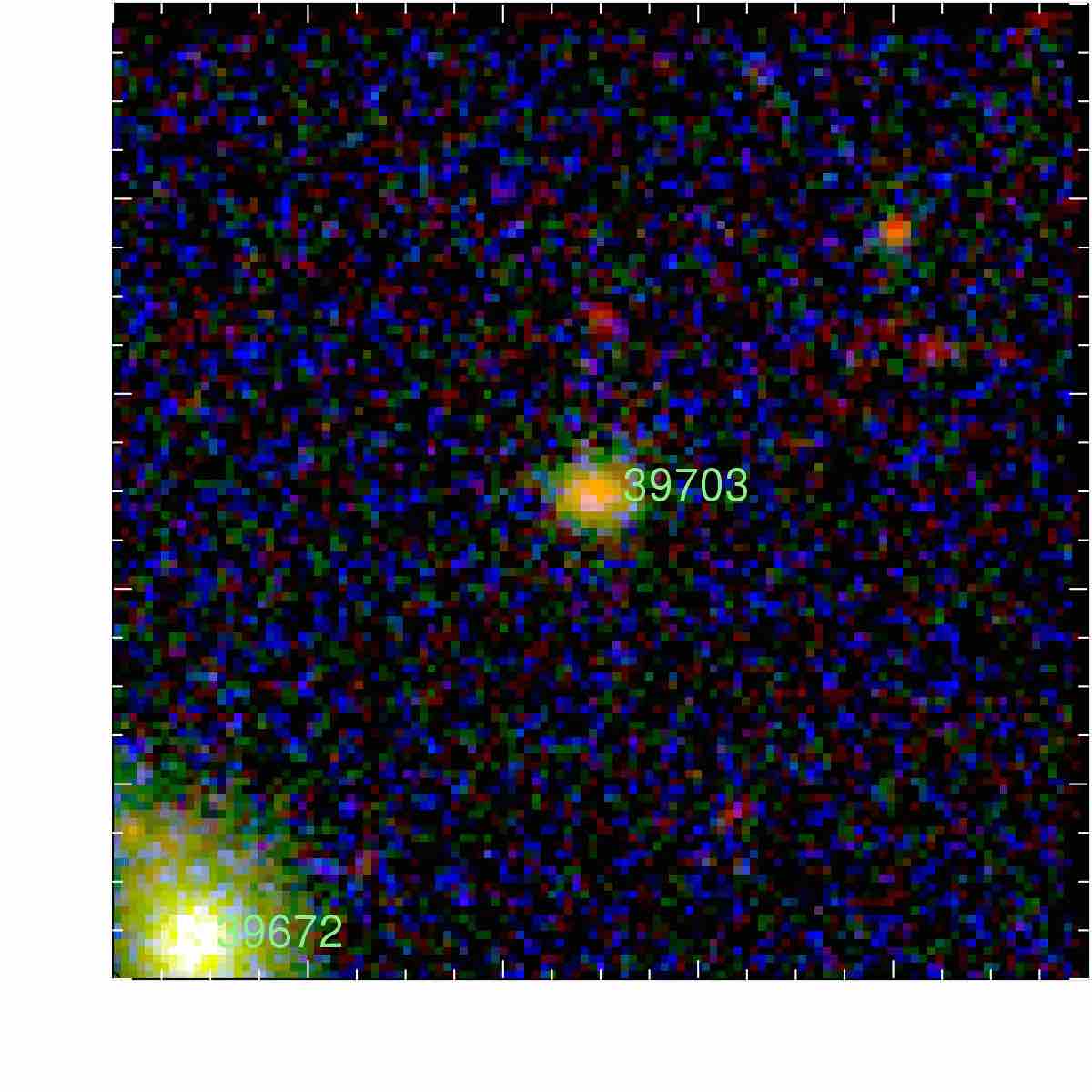} \put (9,85) { \begin{fitbox}{2.25cm}{0.2cm} \color{white}$\bf B$ \end{fitbox}} \end{overpic} \\


\end{array}
$
{\textbf{Figure~\ref{Undetected}.} continued}

\end{figure*}


\begin{figure*}
$
\begin{array}{ccc}
\includegraphics[width=8.4cm]{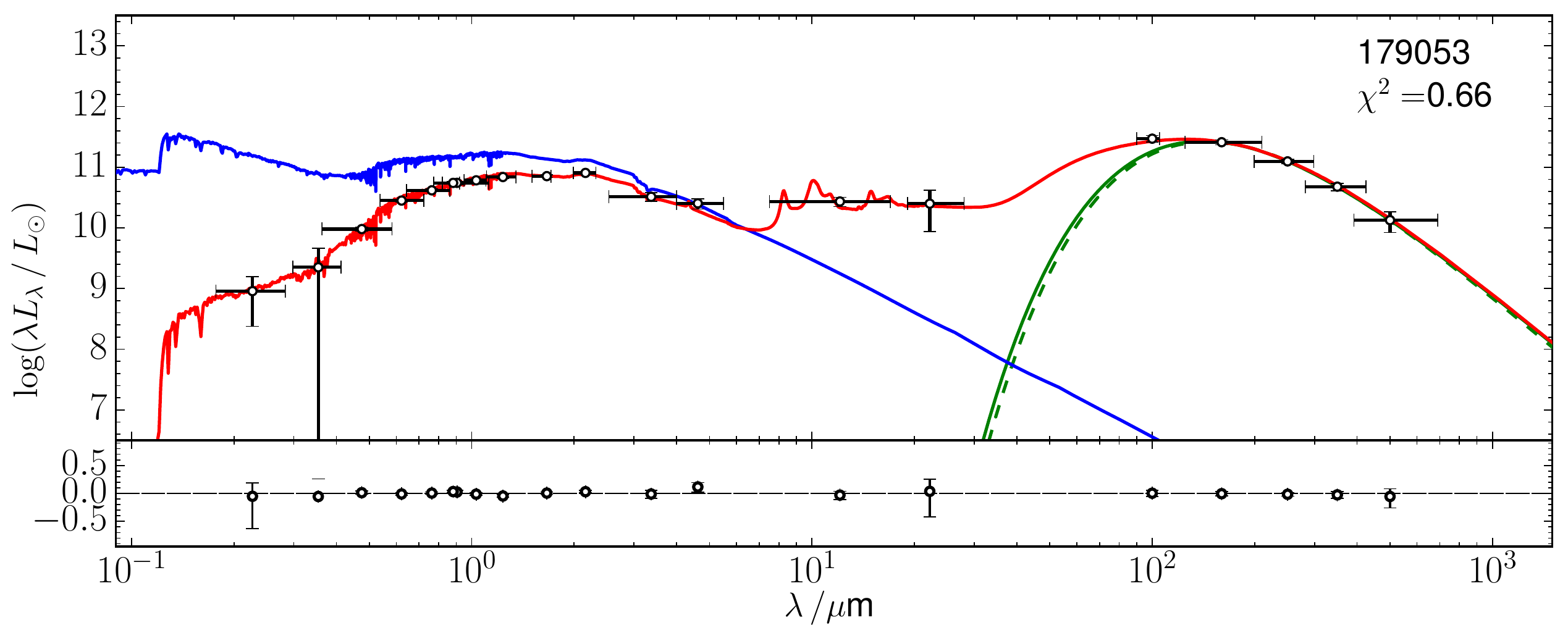} &
\includegraphics[width=5.0cm]{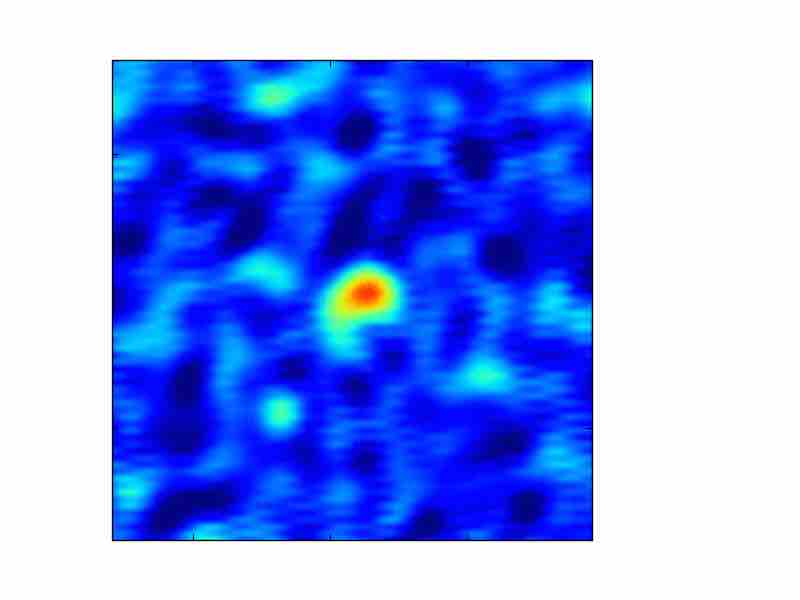} &
\hspace*{-1.2cm}\begin{overpic}[width=3.4cm]{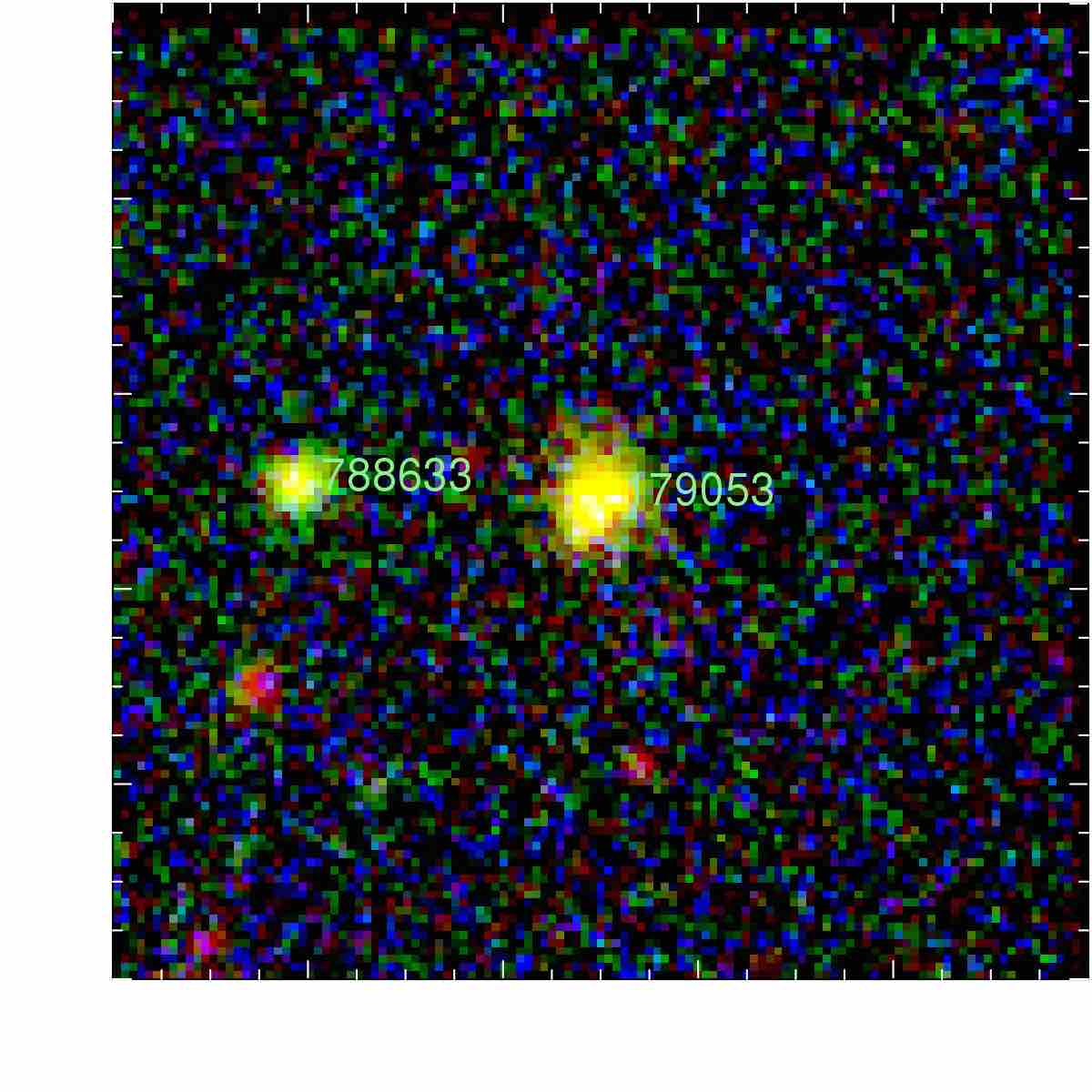} \put (9,85) { \begin{fitbox}{2.25cm}{0.2cm} \color{white}$\bf B$ \end{fitbox}} \end{overpic} \\ 	
	
\includegraphics[width=8.4cm]{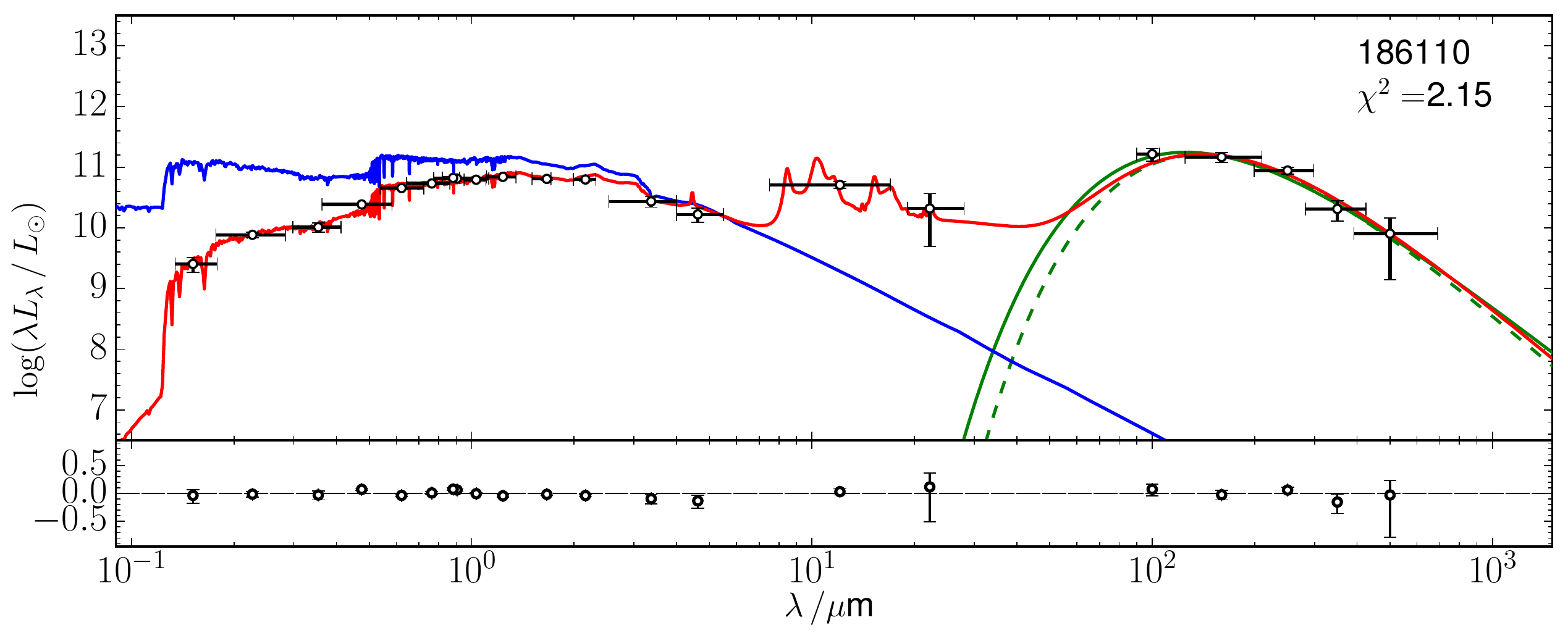} &
\includegraphics[width=5.0cm]{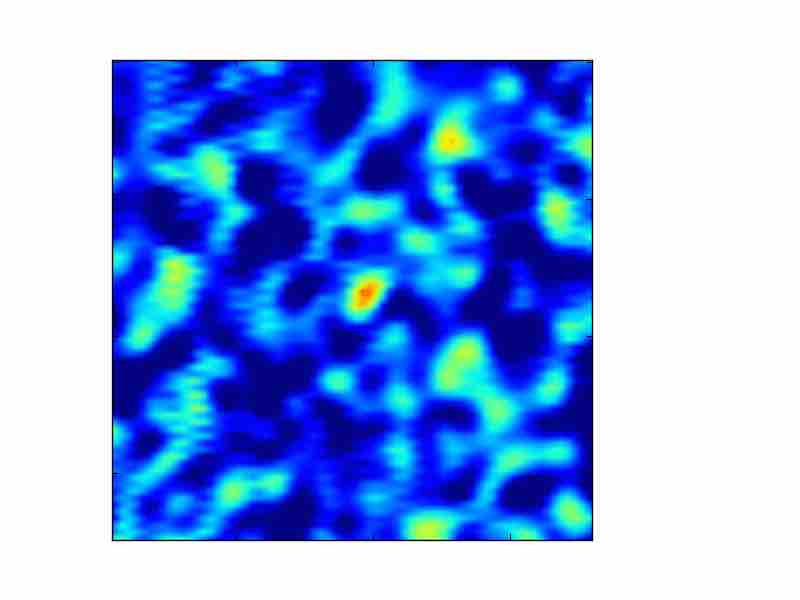} &
\hspace*{-1.2cm}\begin{overpic}[width=3.4cm]{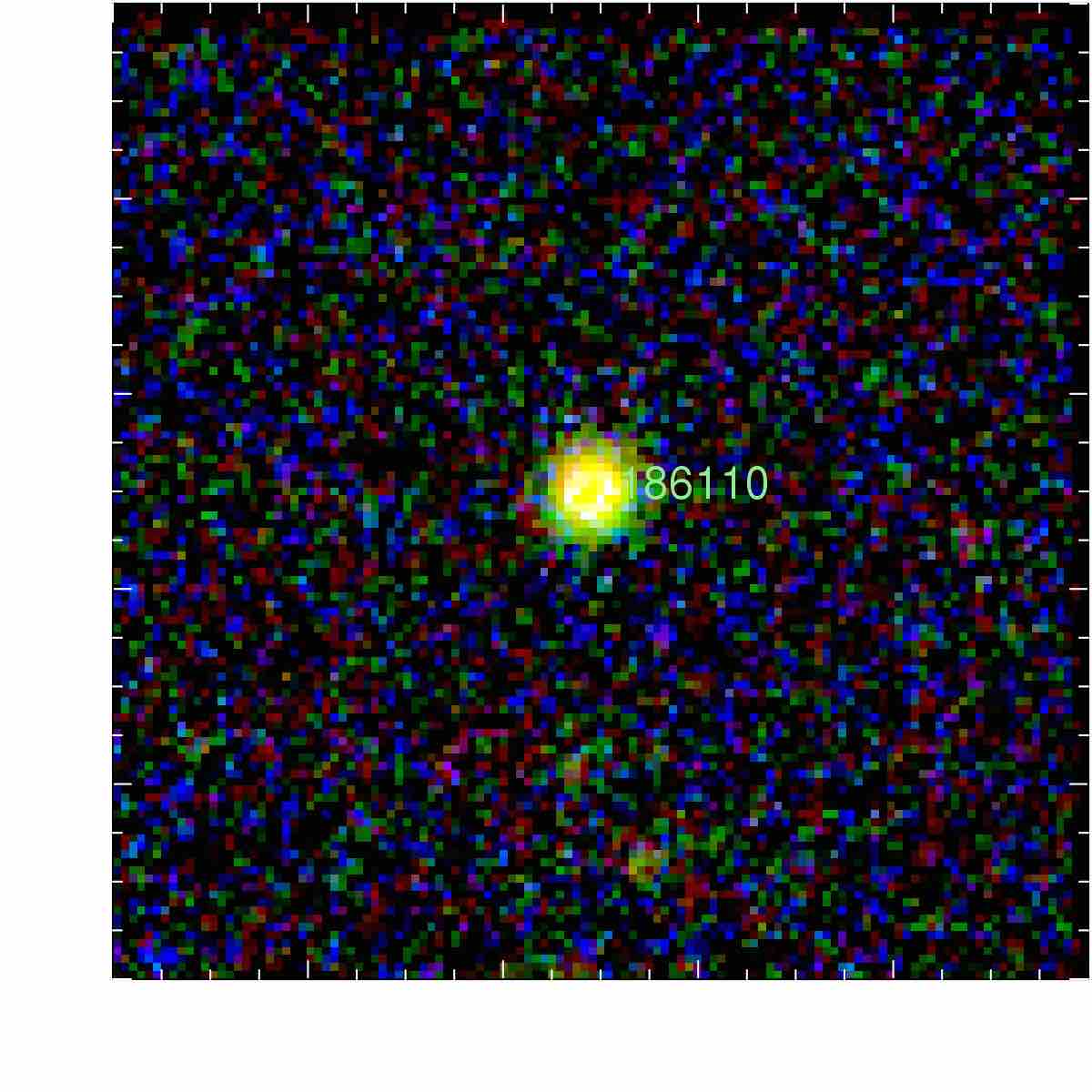} \put (9,85) { \begin{fitbox}{2.25cm}{0.2cm} \color{white}$\bf BD$ \end{fitbox}} \end{overpic} \\ 

\includegraphics[width=8.4cm]{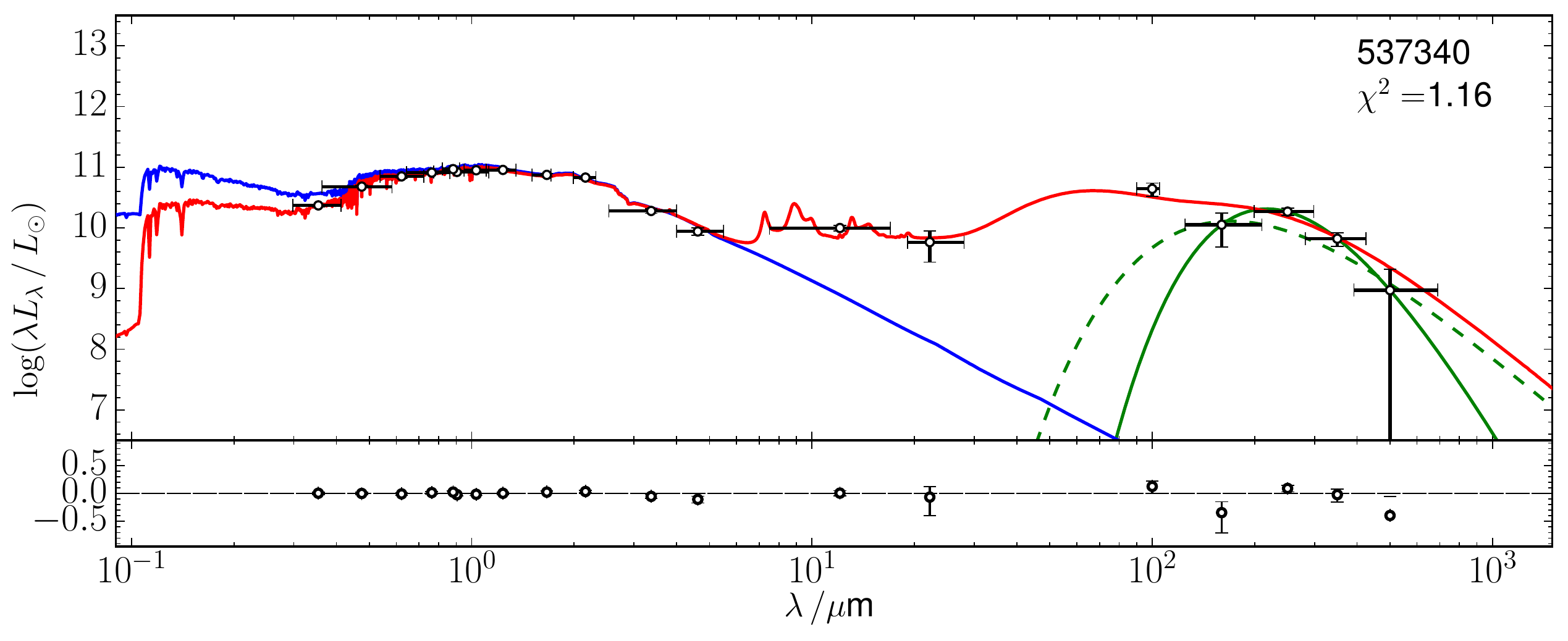} &
\includegraphics[width=5.0cm]{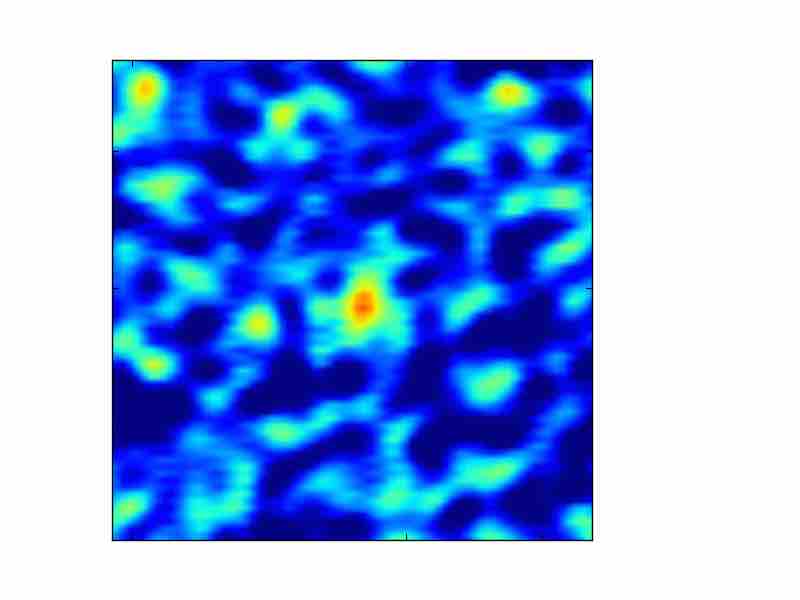} &
\hspace*{-1.2cm}\begin{overpic}[width=3.4cm]{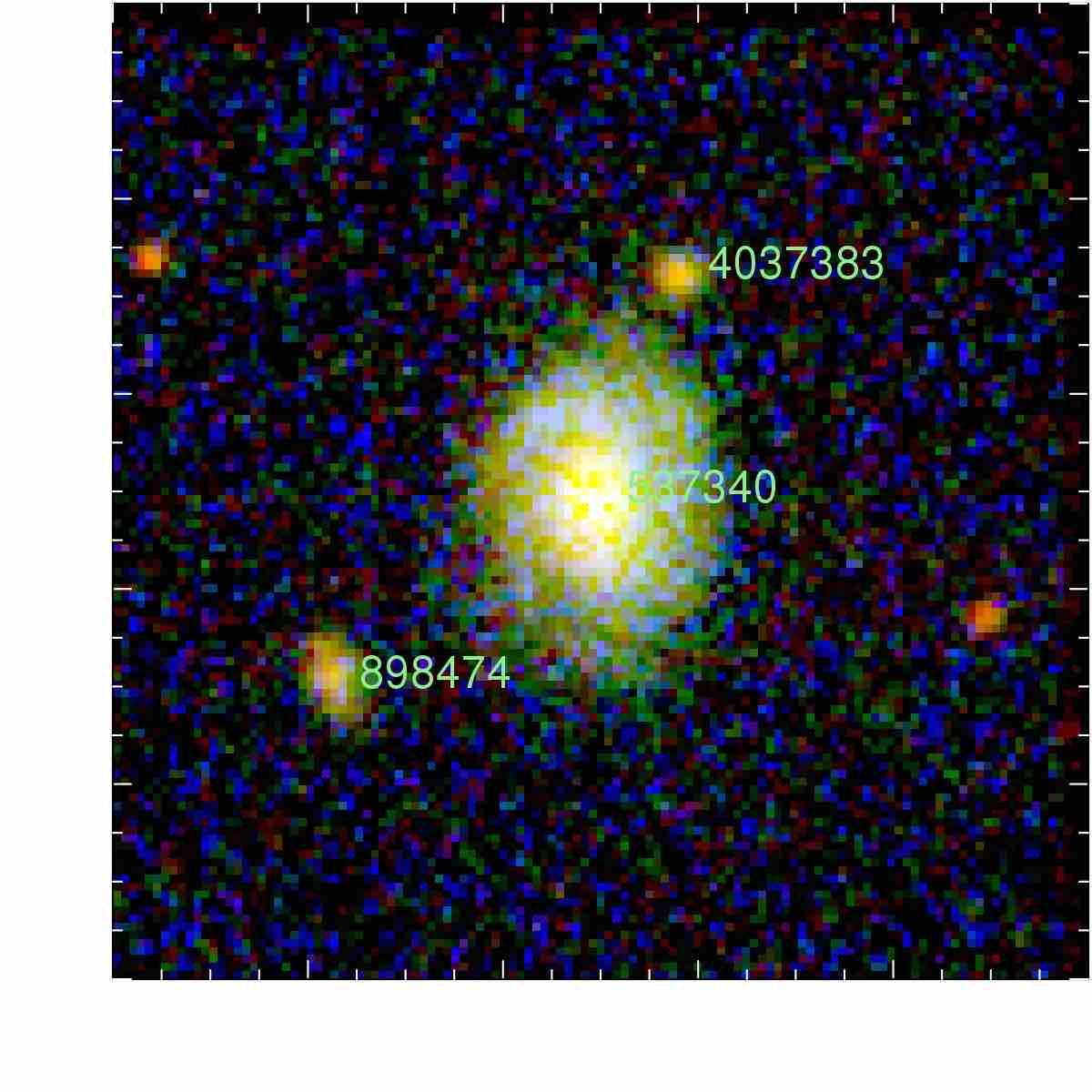} \put (9,85) { \begin{fitbox}{2.25cm}{0.2cm} \color{white}$\bf B$ \end{fitbox}} \end{overpic} \\ 

\includegraphics[width=8.4cm]{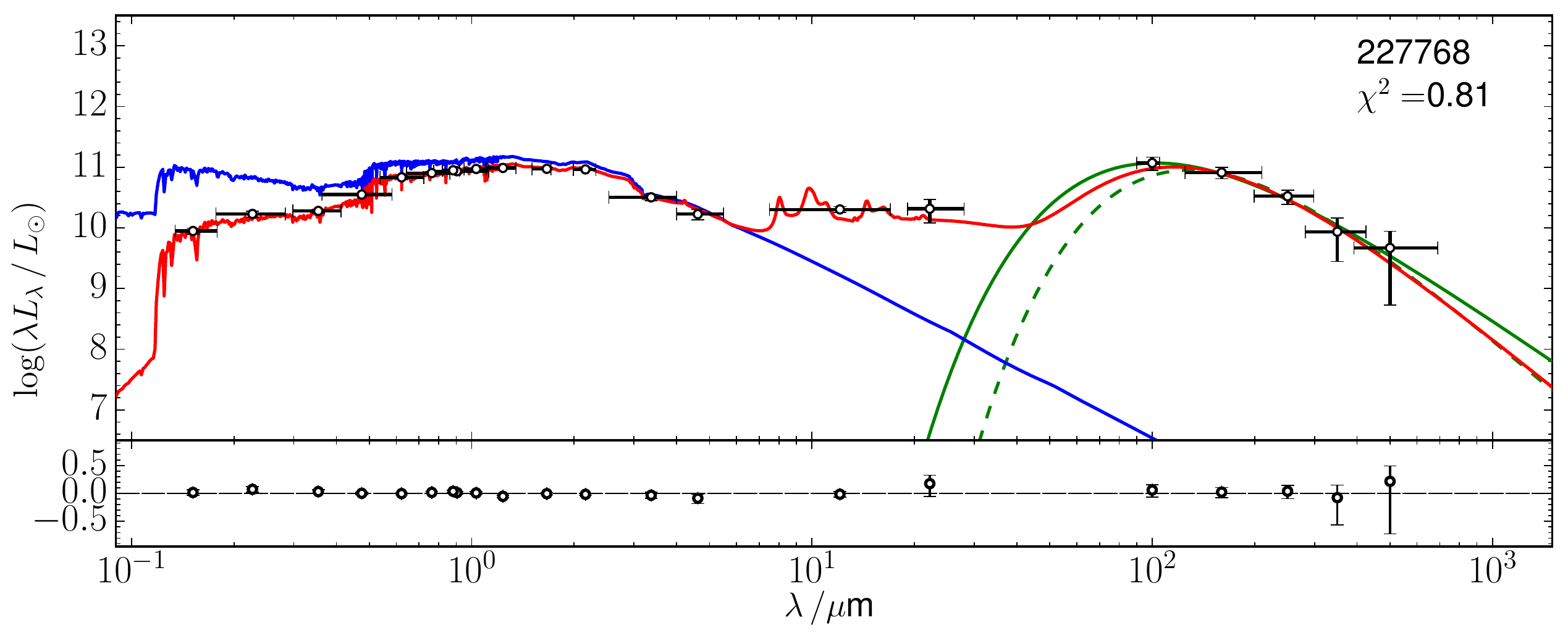} &
\includegraphics[width=5.0cm]{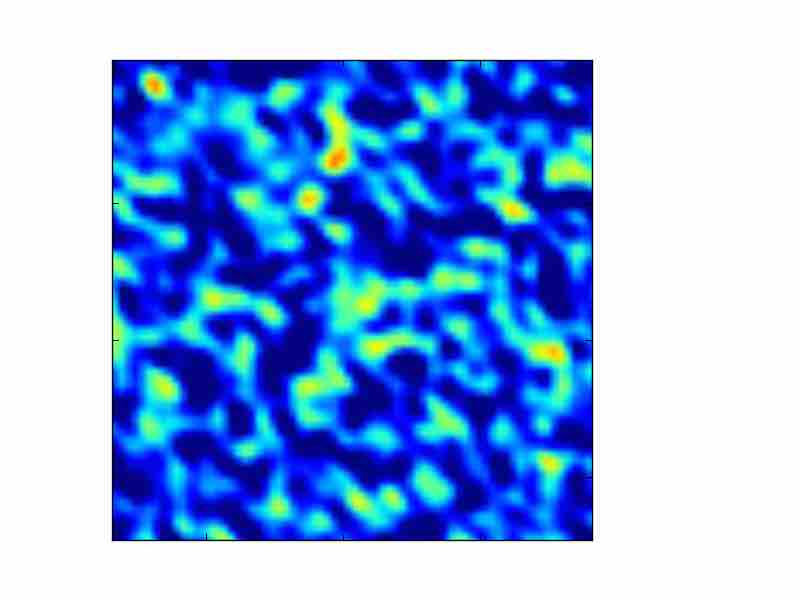} &
\hspace*{-1.2cm}\begin{overpic}[width=3.4cm]{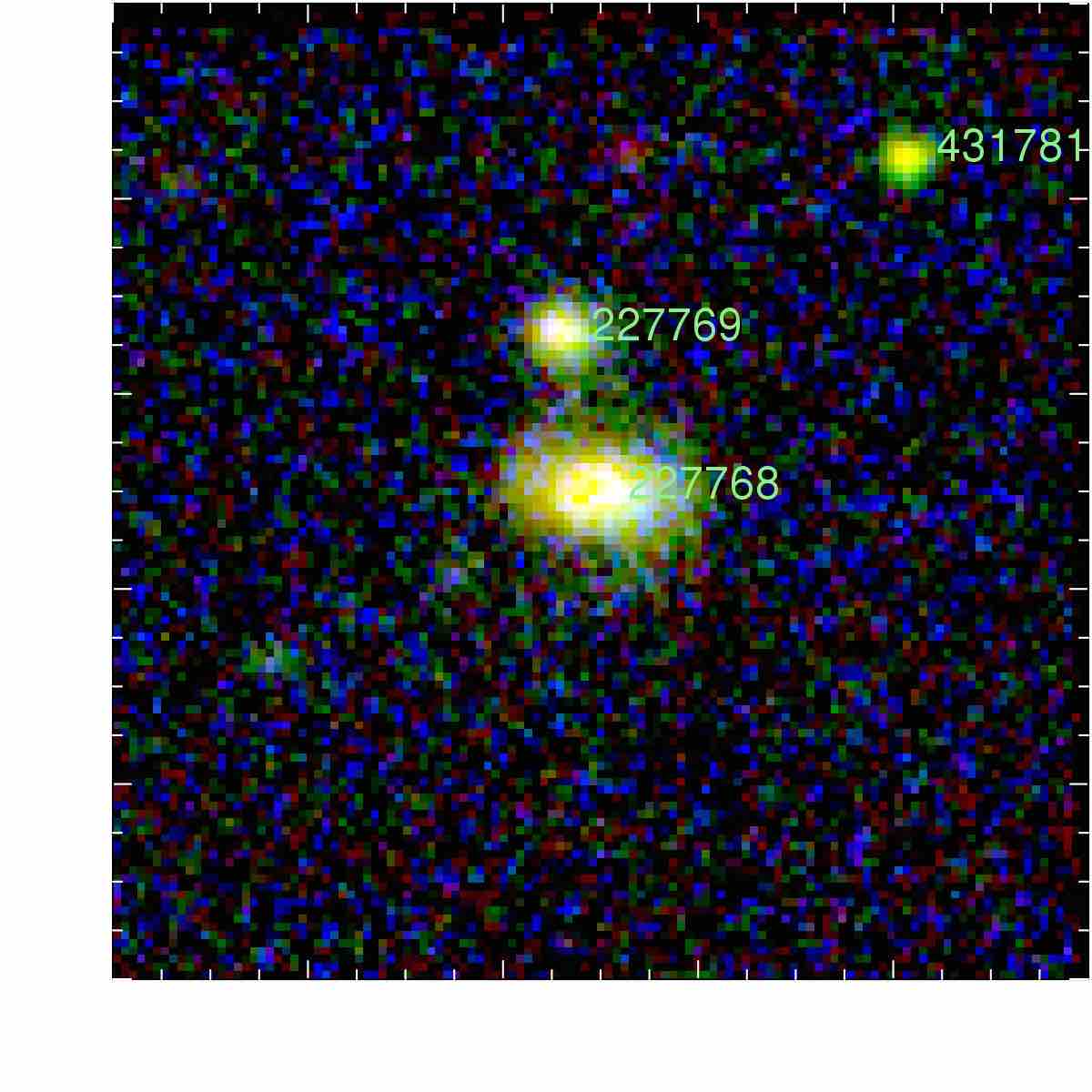} \put (9,85) { \begin{fitbox}{2.25cm}{0.2cm} \color{white}$\bf BDC$ \end{fitbox}} \end{overpic} \\ 

\includegraphics[width=8.4cm]{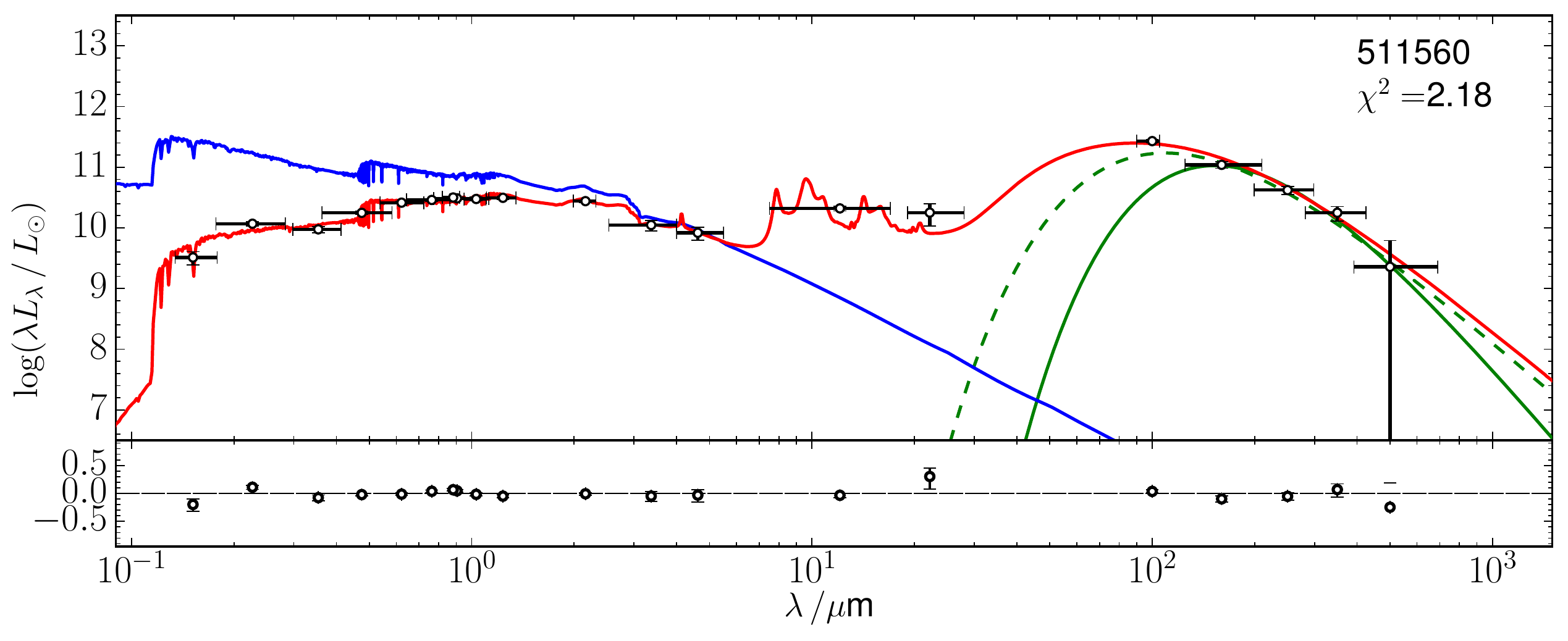} &
\includegraphics[width=5.0cm]{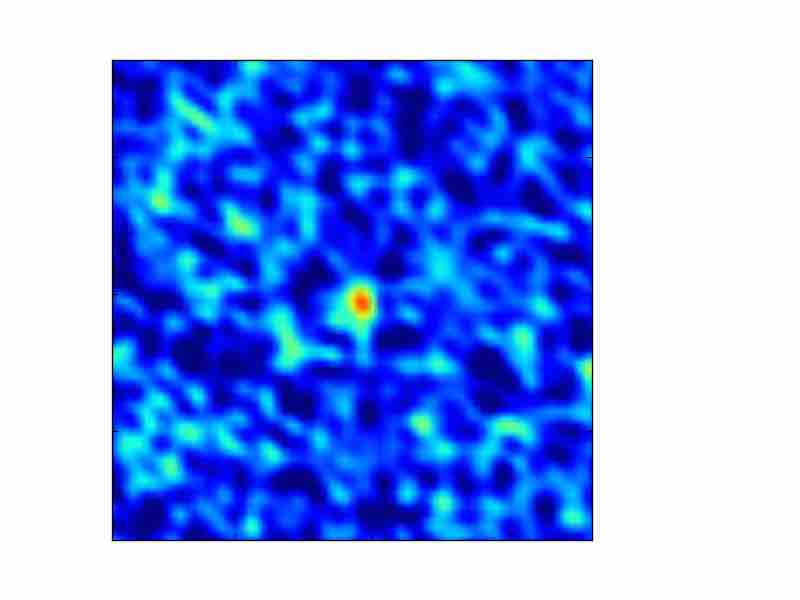} &
\hspace*{-1.2cm}\begin{overpic}[width=3.4cm]{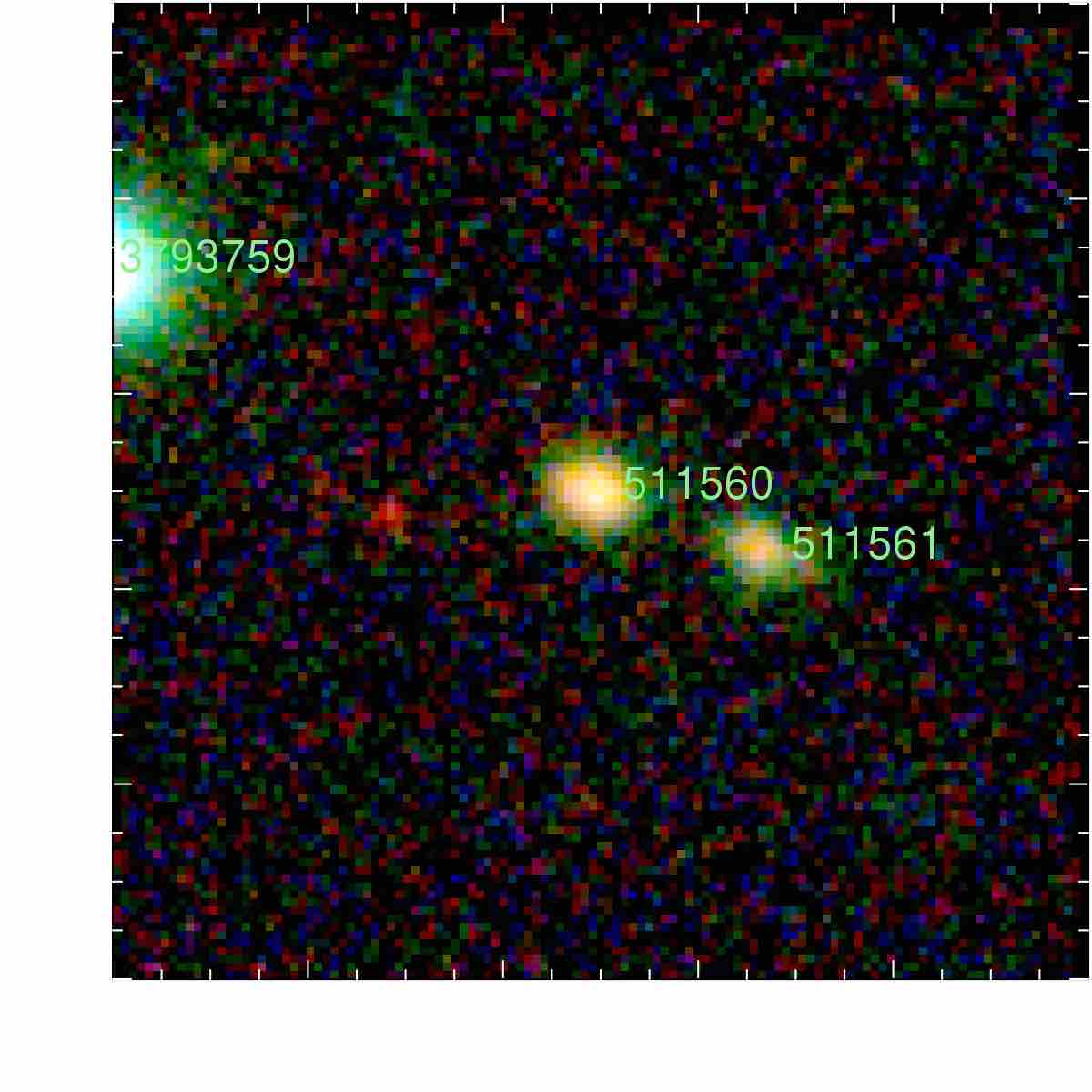} \put (9,85) { \begin{fitbox}{2.25cm}{0.2cm} \color{white}$\bf BDC$ \end{fitbox}} \end{overpic} \\ 


\end{array}
$
{\textbf{Figure~\ref{Undetected}.} continued}

\end{figure*}


\begin{figure*}
$
\begin{array}{ccc}
\includegraphics[width=8.4cm]{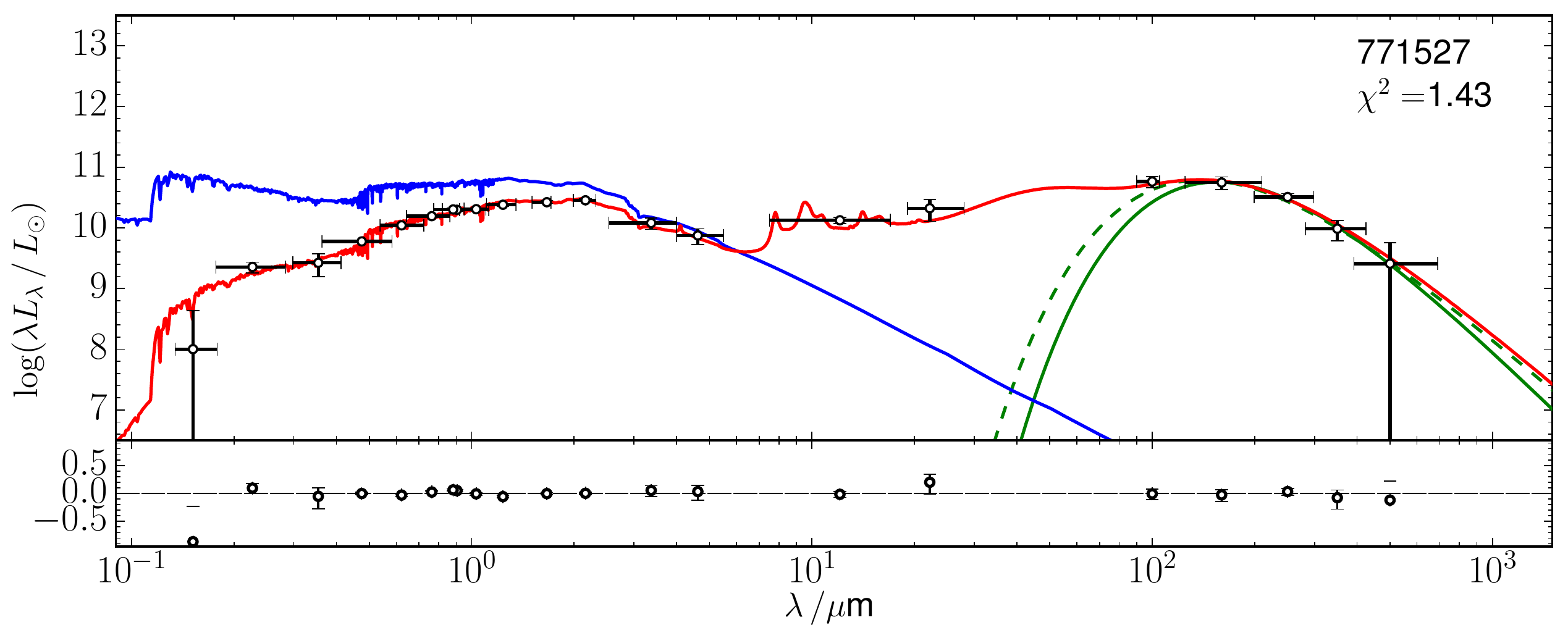} &
\includegraphics[width=5.0cm]{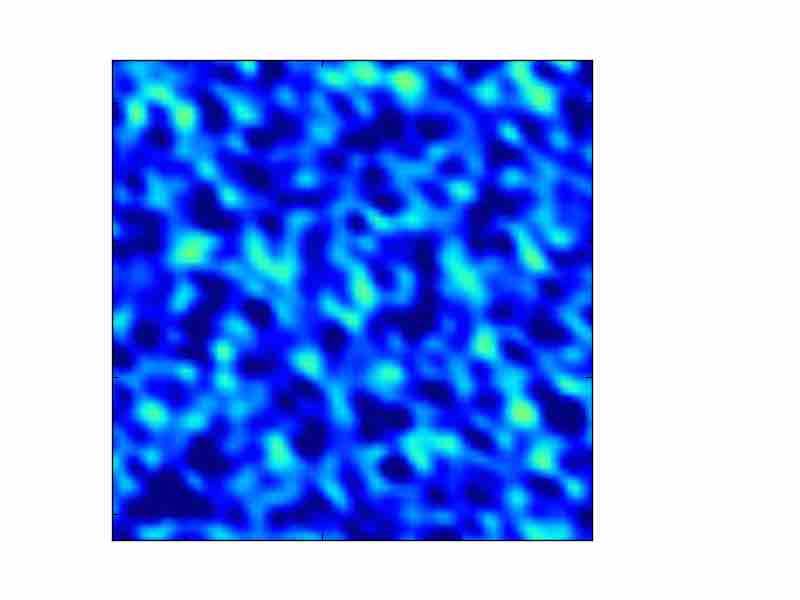} &
\hspace*{-1.2cm}\begin{overpic}[width=3.4cm]{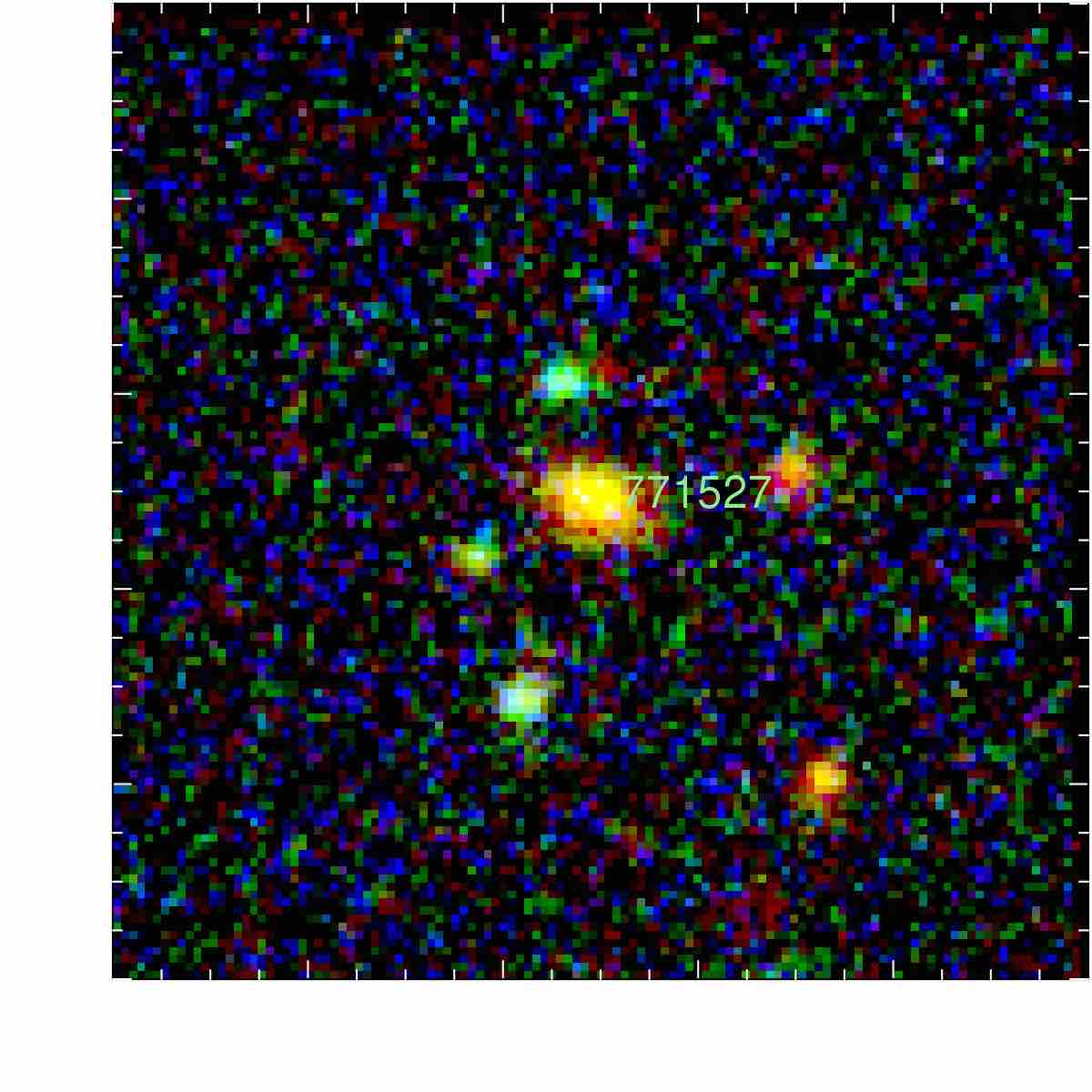} \put (9,85) { \begin{fitbox}{2.25cm}{0.2cm} \color{white}$\bf BDC$ \end{fitbox}} \end{overpic} \\ 	

\includegraphics[width=8.4cm]{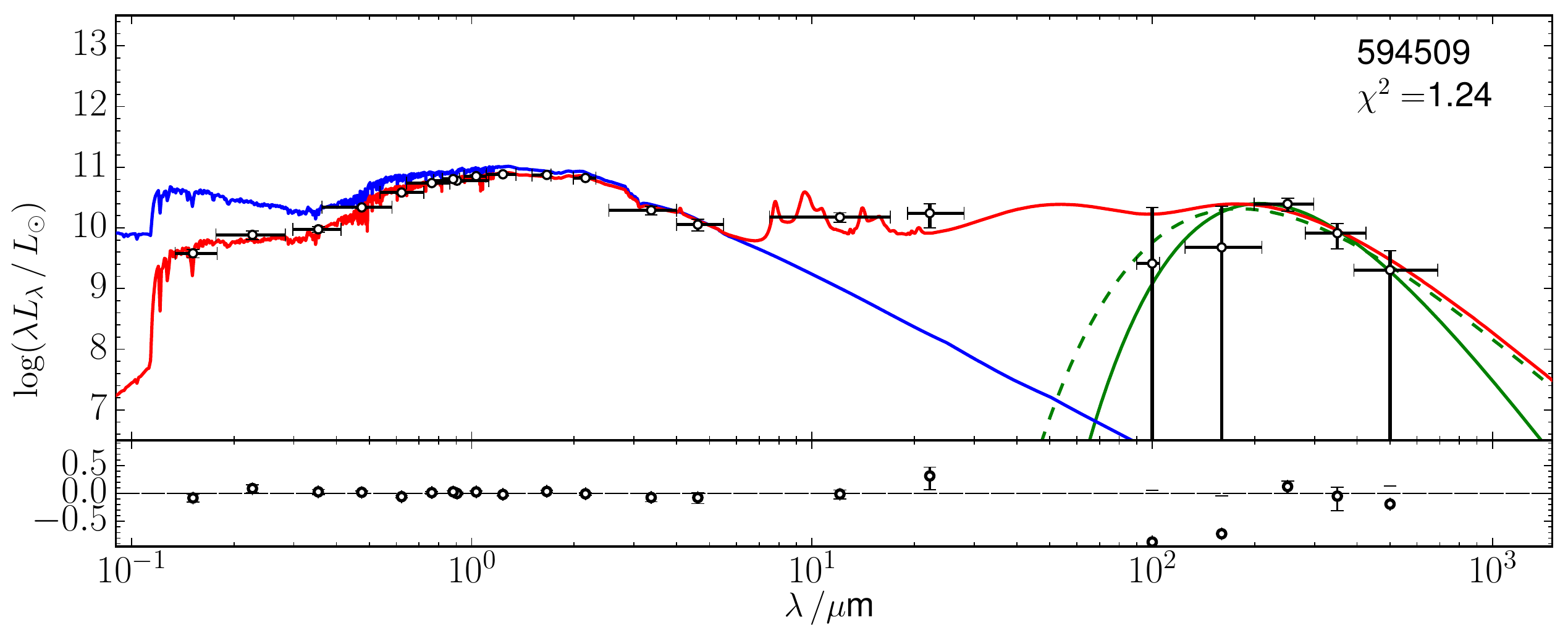} &
\includegraphics[width=5.0cm]{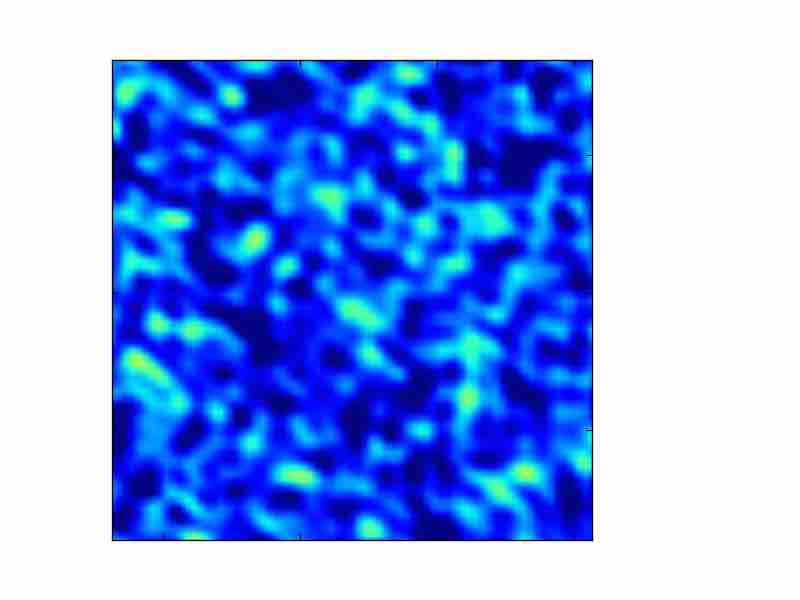} &
\hspace*{-1.2cm}\begin{overpic}[width=3.4cm]{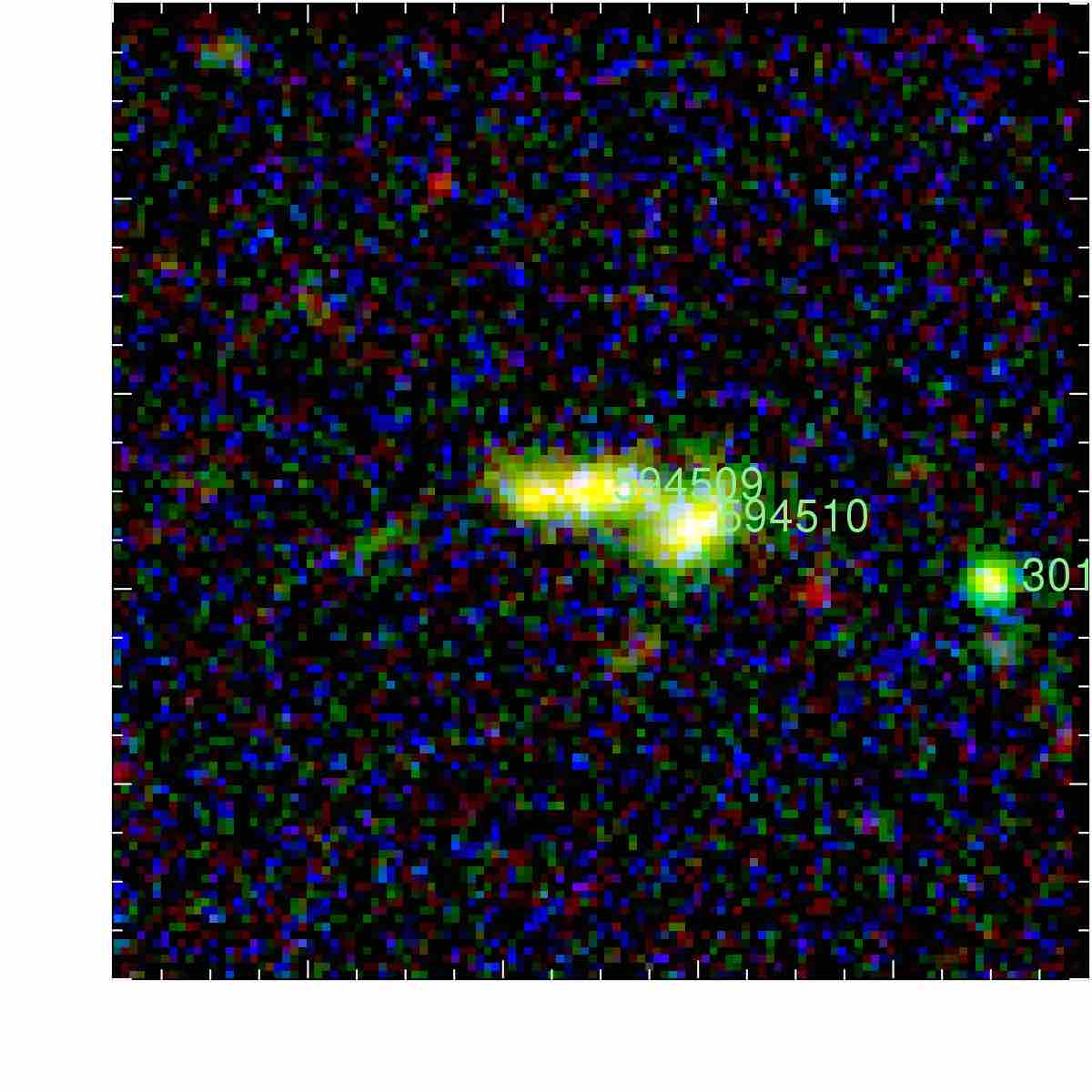} \put (9,85) { \begin{fitbox}{2.25cm}{0.2cm} \color{white}$\bf M$ \end{fitbox}} \end{overpic} \\ 		 

\includegraphics[width=8.4cm]{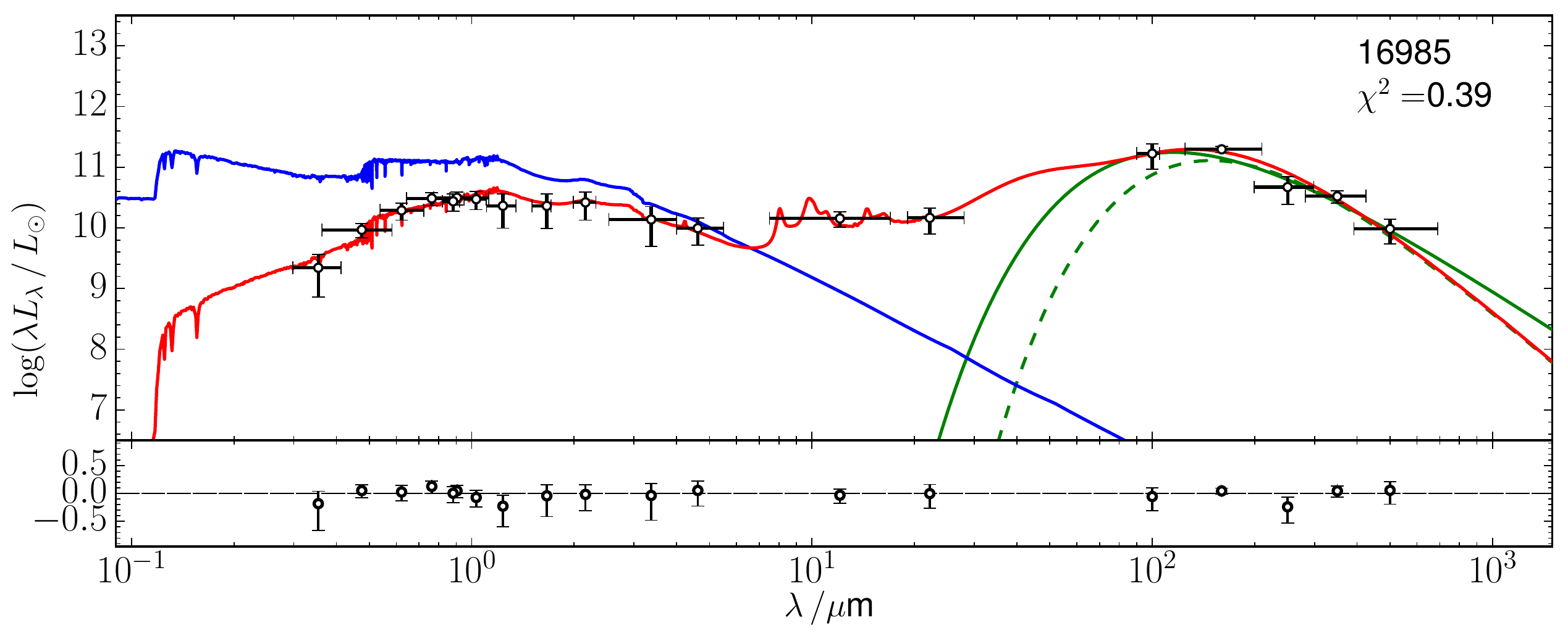} &
\includegraphics[width=5.0cm]{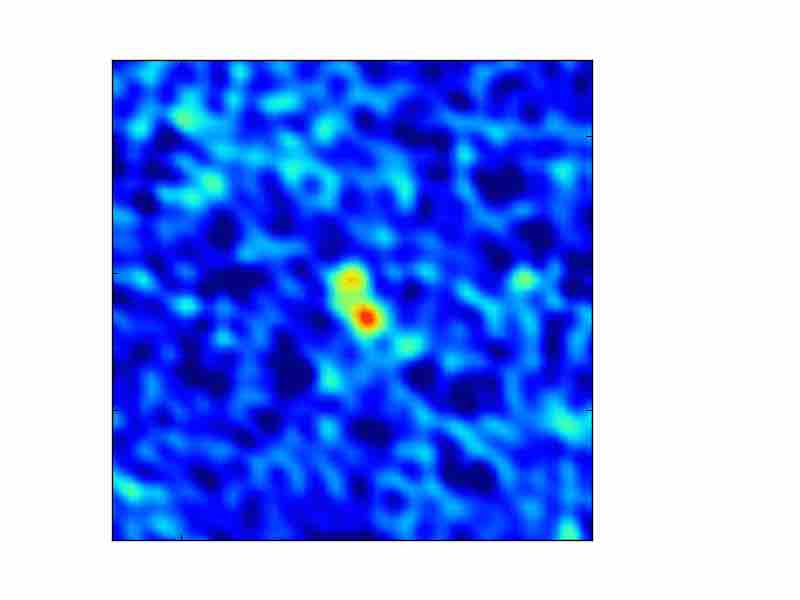} &
\hspace*{-1.2cm}\begin{overpic}[width=3.4cm]{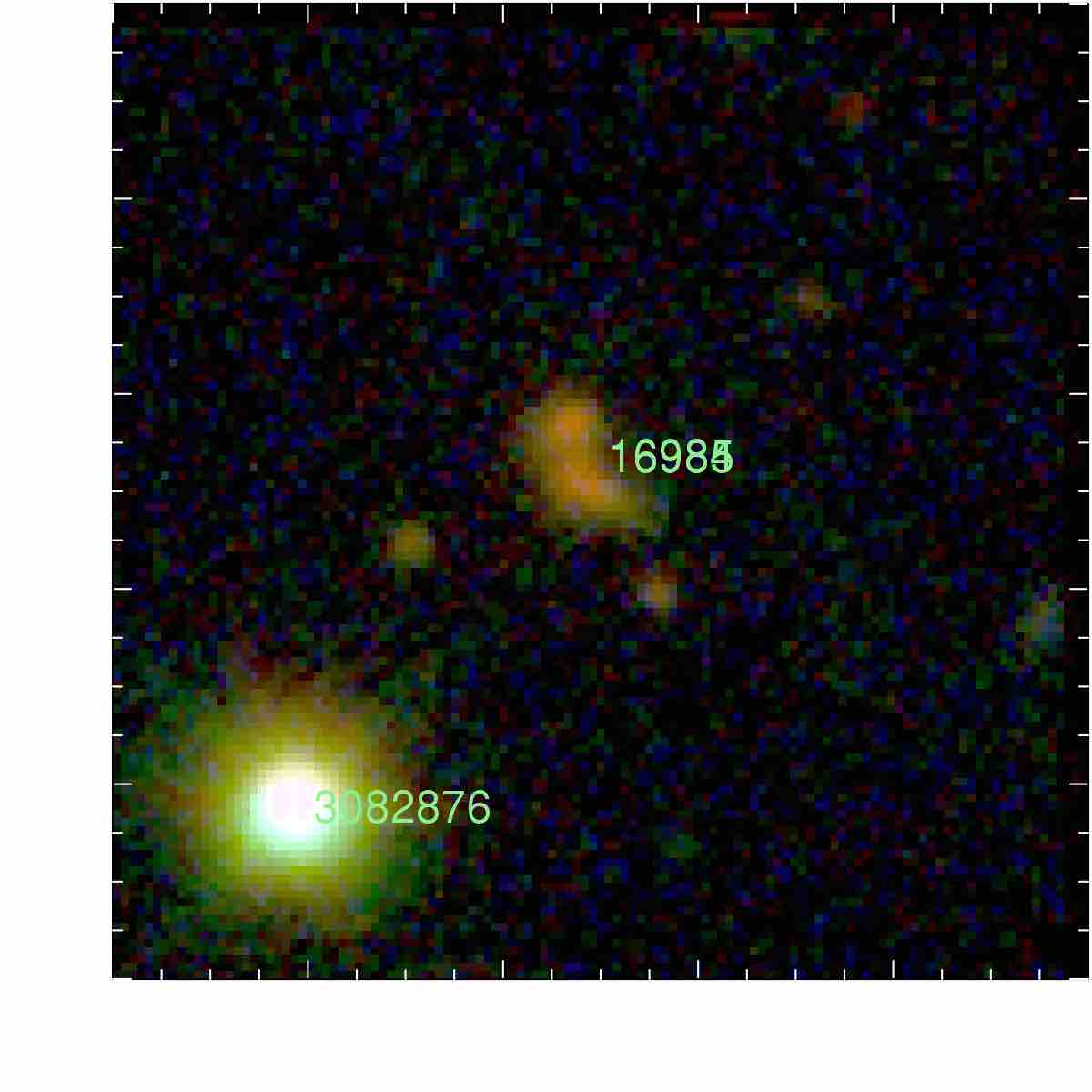} \put (9,85) { \begin{fitbox}{2.25cm}{0.2cm} \color{white}$\bf M$ \end{fitbox}} \end{overpic} \\ 

\end{array}
$
{\textbf{Figure~\ref{Undetected}.} continued}

\end{figure*}



\bsp	
\label{lastpage}
\end{document}